\documentclass[fleqn,usenatbib]{mnras}

\usepackage[T1]{fontenc}
\usepackage{ae,aecompl}
\usepackage{lscape}
\usepackage[T1]{fontenc}

\DeclareRobustCommand{\VAN}[3]{#2}
\let\VANthebibliography\thebibliography
\def\thebibliography{\DeclareRobustCommand{\VAN}[3]{##3}\VANthebibliography}

\usepackage{graphicx}	
\usepackage{epstopdf}
\usepackage{amsmath}	
\usepackage{amssymb}	
\usepackage{multirow}
\usepackage{caption}
\usepackage{longtable}
\usepackage{cprotect}
\usepackage{lscape}
\usepackage{morefloats}
\usepackage{multirow,bigdelim,array,pdflscape,booktabs,psfig}
\usepackage{graphicx} 
\usepackage{subcaption}

\usepackage[slantedGreek]{newtxmath}

\usepackage{newtxtext,newtxmath}

\def\emerlin{$e$-MERLIN}

\title[LeMMINGs IV: The X-ray LeMMINGs sample (nuclei)]{LeMMINGs IV: The X-ray properties of a statistically-complete sample of the nuclei in active and inactive galaxies from the Palomar sample}

\author[D.~R.~A. Williams et al.]{D.~R.~A. Williams$^{1,2,3}$\thanks{E-mail: david.williams-7@manchester.ac.uk},
M. Pahari$^{4,3}$,
R.~D. Baldi$^{5,3}$,
I.~M. McHardy$^{3}$,
S. Mathur$^{6,7,8}$, 
R.~J. Beswick$^{1}$, 
\newauthor A.~Beri$^{9,3}$,
P.~Boorman$^{10}$, 
S.~Aalto$^{11}$,
A.~Alberdi$^{12}$,
M.~K.~Argo$^{13}$,
B.~T.~Dullo$^{14}$,
D.~M.~Fenech$^{15}$,
\newauthor D.~A.~Green$^{15}$,
J.~H.~Knapen$^{16,17}$,
I.~Mart\'i-Vidal$^{18,19}$, 
J.~Moldon$^{12,1}$,
C.~G.~Mundell$^{20}$, 
T.~W.~B.~Muxlow$^{1}$,
\newauthor F.~Panessa$^{21}$,
M.~P\'erez-Torres$^{12}$,
P.~Saikia$^{22}$,
F.~Shankar$^{3}$,
I.~R.~Stevens$^{23}$,
P.~Uttley$^{24}$
\\
$^{1}$ Jodrell Bank Centre for Astrophysics, School of Physics and Astronomy, The University of Manchester, Manchester, M13 9PL, UK\\
$^{2}$Department of Physics, University of Oxford, Denys Wilkinson Building, Keble Road, Oxford, OX1 3RH, UK\\
$^{3}$ School of Physics and Astronomy, University of Southampton, Southampton, SO17 1BJ, UK\\
$^{4}$Department of Physics, Indian Institute of Technology, Hyderabad 502285, India\\
$^{5}$INAF - Istituto di Radioastronomia, Via P. Gobetti 101, I-40129 Bologna, Italy\\
$^{6}$ Astronomy department, The Ohio State University, Columbus, OH, 43210, USA\\
$^{7}$ Center for Astronomy and Astro-particle Physics, The Ohio State University, Columbus, OH 43210, USA.\\
$^{8}$ Eureka Scientific, 2452 DELMER ST STE 100, Oakland, CA, 94602, USA\\
$^{9}$DST-INSPIRE Faculty, Indian Institute of Science Education and Research (IISER), Mohali, Punjab, 140306, India\\
$^{10}$Astronomical Institute, Academy of Sciences, Bo\u{c}n\'{i}' II 1401, CZ-14131 Prague, Czech Republic\\
$^{11}$Department of Space, Earth and Environment, Chalmers University of Technology, Onsala Observatory, SE-439 92 Onsala, Sweden\\
$^{12}$Instituto de Astrof\'isica de Andaluc\'ia (IAA-CSIC), Glorieta de la Astronom\'ia s/n, 18008 Granada, Spain\\
$^{13}$Jeremiah Horrocks Institute, University of Central Lancashire, Preston PR1 2HE, UK\\
$^{14}$Departamento de F\'isica de la Tierra y Astrof\'isica, IPARCOS, Universidad Complutense de Madrid, E-28040 Madrid, Spain\\
$^{15}$ Astrophysics Group, Cavendish Laboratory, 19 J.~J.~Thomson Avenue, Cambridge CB3 0HE, UK\\
$^{16}$Instituto de Astrof\'{i}sica de Canarias, V\'{i}a L\'{a}ctea S/N, E-38205 La Laguna, Spain\\
$^{17}$Departamento de Astrof\'{i}sica, Universidad de La Laguna, E-38206 La Laguna, Spain\\
$^{18}$Observatori Astron\'{o}mic, Universitat de Val\`{e}ncia, Parc Cient\'{i}fic, Paterna, Val\`{e}ncia, Spain\\ 
$^{19}$Departament d'Astronomia i Astrof\'{i}sica, Universitat de Val\`{e}ncia, Burjassot, Val\`{e}ncia, Spain\\
$^{20}$ Department of Physics, University of Bath, Claverton Down, Bath, BA2 7AY, UK\\
$^{21}$INAF - Istituto di Astrofisica e Planetologia Spaziali, via Fosso del Cavaliere 100, I-00133 Roma, Italy\\
$^{22}$Center for Astro, Particle and Planetary Physics (CAP$^3$), New York University Abu Dhabi, PO Box 129188, Abu Dhabi, UAE\\
$^{23}$ School of Physics and Astronomy, University of Birmingham, Edgbaston, Birmingham B15 2TT, UK\\
$^{24}$ Anton Pannekoek Institute for Astronomy (API), University of Amsterdam, Science Park 904, 1098 XH Amsterdam, the Netherlands\\
}

\date{Accepted XXX. Received YYY; in original form ZZZ}

\pubyear{2021}

\begin{document}
\label{firstpage}
\pagerange{\pageref{firstpage}--\pageref{lastpage}}
\maketitle

\begin{abstract}
All 280 of the statistically-complete Palomar sample of nearby ($<$120\,Mpc) galaxies $\delta >$20$^{\circ}$ have been observed at 1.5\,GHz as part of the LeMMINGs \emerlin{} legacy survey. 
Here, we present \textit{Chandra} X-ray observations of the nuclei of 213 of these galaxies, including a statistically-complete sub-set of 113 galaxies in the declination range 40$^{\circ}$ $<\delta<$ 65$^{\circ}$. 
We observed galaxies of all optical spectral types, including `active' galaxies (e.g., LINERs and Seyferts) and `inactive' galaxies like \ion{H}{ii} galaxies and absorption line galaxies (ALG). 
The X-ray flux limit of our survey is 1.65$\times$10$^{-14}$~erg s$^{-1}$ cm$^{-2}$ (0.3$-$10\,keV). 
We detect X-ray emission coincident within 2-arcsec of the nucleus in 150/213 galaxies, including 13/14 Seyferts, 68/77 LINERs, 13/22 ALGs and 56/100 \ion{H}{ii} galaxies, but cannot completely rule out contamination from non-AGN processes in sources with nuclear luminosities $\la$10$^{39}$~erg s$^{-1}$. 
We construct an X-ray Luminosity function (XLF) and find that the local galaxy XLF, when including all AGN types, can be represented as a single power-law of slope $-0.54 \pm 0.06$.  
The Eddington ratio of the Seyferts is usually 2$-$4 decades higher than that of the LINERs, ALGs and \ion{H}{ii} galaxies, which are mostly detected with Eddington ratios $\la$10$^{-3}$. 
Using [\ion{O}{iii}] line measurements and BH masses from the literature, we show that LINERs, \ion{H}{ii} galaxies and ALGs follow similar correlations to low luminosities, suggesting that some `inactive' galaxies may harbour AGN.
\end{abstract}



\clearpage
\section{Introduction}

In the nuclear regions of nearby galaxies, optical emission line ratio diagrams (called BPT diagrams) are commonly used to discriminate star formation (SF) from accretion onto super-massive black holes (SMBHs), known as active galactic nuclei \citep[AGN, e.g.,][]{BaldwinPhillipsTerlevich81,Kewley06,Buttiglione2010}. At the lowest luminosities, optical emission lines can be too weak to provide a reliable interpretation of the nuclear activity. 
This issue is most prevalent in low-luminosity AGN (LLAGN), defined by H$\alpha$ luminosity, \textit{L}$_{\rm H_{\alpha}}$ $<$ 10$^{40}$~erg s$^{-1}$ \citep{ho97a} or by their X-ray luminosities \textit{L}$_{\rm X-ray} <$ 10$^{42}$~erg s$^{-1}$ \citep{Ptak2001}. These definitions commonly encompass nearby AGN such as Seyferts \citep{Seyfert1941} and Low-Ionisation Nuclear Emission Line regions \cite[LINERs, first defined in][]{heckman80}, which are likely powered by a central AGN engine. Other galaxies, such as the nuclei in star forming galaxies known as \ion{H}{ii} galaxies and absorption line galaxies (ALGs) do not have strong enough emission lines to be unequivocally powered by an AGN, but may include a weak or dormant SMBH, in which case they may harbour a  LLAGN.

The study of LLAGN is important for several reasons. They represent the most numerous type of AGN in the Universe \citep{Ptak2001,nagao02,filho06} and their low luminosities are thought to be caused by a combination of low accretion rates \citep{KauffmannHeckman2009} and low radiative efficiency of inefficient accretion processes \citep{ho99,maoz07,panessa07,HoReview}. LLAGN are often associated with SMBHs of lower masses ($<10^{7}~M_{\odot}$) and represent the most common accretion state in SMBHs \citep{HoReview}. Therefore, LLAGN provide the best opportunity to understand the bulk of the population of AGN, which is important for cosmological models of SMBH evolution, the physics and efficiency of accretion at the lowest luminosities, the triggering mechanisms of accretion and the local SMBH luminosity functions. 

In order to detect the presence of a LLAGN in a nearby galaxy when optical emission lines are weak or not present, multi-wavelength data must be used instead. AGN are often detected at cm wavelengths because dust in the interstellar medium is transparent, allowing for unobscured views of the nuclei of nearby galaxies. 
However the radio luminosity from LLAGN is often very low (10$^{-4}$  of the bolometric output; \citealt{Condon92}) and can be contaminated by SF. Thus to distinguish AGN we require both high sensitivity and high angular resolution. We have therefore observed all 280 galaxies above declination, $\delta >$ +20$^{\circ}$ from the Palomar bright spectroscopic sample of nearby galaxies \citep[][hereafter referred to as the `Palomar sample']{Filippenko85,Ho95,ho97a,ho97b,ho97c,ho97d,Ho03,Ho09} with the \emerlin{} radio interferometer array as part of the \textbf{L}egacy \textbf{e}-\textbf{M}ERLIN \textbf{M}ulti-band \textbf{I}maging of \textbf{N}earby \textbf{G}alaxies \textbf{S}urvey (LeMMINGs; \citealt{BeswickLemmings, BaldiLeMMINGs,BaldiLeMMINGs2,BaldiLeMMINGs3}) programme. The Palomar sample is generally accepted to be the most statistically complete sample of nearby galaxies. LeMMINGs provides sub-mJy sensitivity with resolution of 0.15$\arcsec$ at 1.5\,GHz \citep{BaldiLeMMINGs,BaldiLeMMINGs2,BaldiLeMMINGs3} and 0.05$\arcsec$ at 5\,GHz (analysis still is still on-going and will be presented in future papers).

Radio observations on their own cannot always distinguish the AGN and additional observations in other wavebands are required. The X-ray waveband is particularly valuable: compact X-ray emission with a steep powerlaw spectrum (photon index, $\Gamma$=1.3$-$2.1, where $\rm n(E)dE \propto E^{-\Gamma}$) is commonly interpreted as a `smoking gun' for an AGN \citep{NandraPounds1994,Piconcelli2005,Ishibashi}. However, X-ray emission in the centres of galaxies can be contaminated by X-ray binaries (XRBs) and Ultra-luminous X-ray sources (ULXs) below X-ray luminosities of 10$^{39}$~erg s$^{-1}$. Previous \textit{ROSAT} X-ray observations have insufficient angular resolution to remove the contribution of these sources \citep{RobertsWarwick}, necessitating the sub-arcsecond resolution X-ray imaging only possible with the \textit{Chandra} X-ray observatory \citep[hereafter \textit{Chandra},][]{Chandra}. The combination of sub-arcsecond angular resolution and high sensitivity provided by \textit{Chandra} allows for the detection of faint nuclei down to X-ray luminosities of 10$^{39-42}$~erg s$^{-1}$ \citep{Fabbiano2006}.

X-ray studies of nearby LLAGN have focussed on the known `active' galaxies like Seyferts and LINERs \citep{HoLINERs,ho01b,panessa06,Gonzalez2006,akylas09,HernandezGarcia2014,Gonzalez2015}, often returning detection fractions of X-ray nuclei $\gtrsim60$ per cent. However, these studies have often missed the  \ion{H}{ii} and ALG galaxies which may also harbour an AGN. Other studies have prioritised larger samples in which these `inactive' sources may be selected, but the inhomogeneous nature of these samples makes statistical comparisons between different types of source difficult: \citet{Zhang09} used \textit{Chandra}, \textit{XMM-Newton} and \textit{ROSAT} to study 187 objects within a distance of 15\,Mpc but employ their own multi-wavelength criteria; \citet{Liu2010} analysed 383 objects from the entire \textit{Chandra} archive which is biased to the well-known AGN and interesting off-nuclear sources such as ULXs; \citet{She2016} made a volume limited (d$<$50\,Mpc) \textit{Chandra} sample, finding a detection fraction of 44 per cent for all galaxies, including the `inactive' sources, but the sample was limited by the number of observations in the \textit{Chandra} archive. In an effort to overcome these issues and compile a statistically-complete sample of LLAGN, we have constructed a catalogue of nuclear X-ray emission in nearby galaxies selected from the Palomar survey covered by LeMMINGs data, compiling all available data in the \textit{Chandra} archive. We obtained 48 new observations of nearby galaxies to complete the sample in a declination range. Our catalogue has sub-arcsecond imaging in the X-ray band and is unbiased towards `inactive' galaxies due to the parent sample selection.

This paper is structured as follows: in Section 2, we describe the observations and data reduction, in Section 3 we show the Chandra X-ray data results and present the sources detected. In Section 4, we discuss the results and implications of X-ray emission in the nearby Universe. Finally, in Section 5 we summarise our results and present our conclusions.

\begin{table*}
    \centering
    \begin{tabular}{c c c c c c c c c c c c c}
    
\hline
Galaxy &      Right &    Dec. &  Gal. Lat. &  Dist  &  Obs. & Exposure & Sample & Det. &  Mass &  O$_{[\rm III]}$ & AGN & Hubble\\
Name &      Asc. &  $\delta$   &  |$b$| &  (Mpc) &  ID. & (secs) & Status & Sig. &  log(M$_{\odot}$) & log(Lum.) & Class & Type\\
(1) &      (2) &    (3) &  (4) &  (5) &  (6) & (7) & (8) & (9) &  (10) & (11) & (12) & (13)\\
\hline

NGC7817 &    0.995 &  20.752 &  $-$40.76 &        31.5 &      - &        - &      -   &           Not obs. &         $6.21 \pm 0.22$ &  39.29    & HII &       Spi.\\
IC10 &    5.096 &  59.293 &    3.34 &         1.3 &   8458 &    43571 &      -   &             Undet &         $5.11 \pm 0.82$ &  37.13    & HII &    Irr.\\
NGC147 &    8.299 &  48.507 &  $-$14.25 &         0.7 &      - &        - &      -   &           Not obs. &         $4.28 \pm 0.40$ &  -    & ALG &   Ell.\\
NGC185 &    9.739 &  48.337 &  $-$14.48 &         0.7 &      - &        - &      -   &           Not obs. &         $4.10 \pm 0.21$ &  34.63    & LINER &   Ell.\\
NGC205 &   10.092 &  41.683 &  $-$21.14 &         0.7 &   4691 &     9870 &      C   &             Undet &         $3.83 \pm 1.84$ &  -    & ALG &   Ell.\\
NGC221 &   10.674 &  40.866 &  $-$21.98 &         0.7 &   5690 &   113027 &      C   &             Undet &         $6.39 \pm 0.19$ &  -    & ALG &   Ell.\\
NGC224 &   10.685 &  41.269 &  $-$21.57 &         0.7 &  14196 &    42848 &      C   &                 146.78 &         $7.84 \pm 0.05$ &  -    & ALG &       Spi.\\
NGC266 &   12.449 &  32.278 &  $-$30.59 &        62.4 &  16013 &    84950 &      -   &                 204.06 &         $8.37 \pm 0.07$ &  39.43    & LINER &       Spi.\\
NGC278 &   13.018 &  47.550 &  $-$15.32 &        11.8 &   2055 &    38259 &      -   &                  11.87 &         $5.62 \pm 0.33$ &  37.47    & HII &       Spi.\\
NGC315 &   14.454 &  30.352 &  $-$32.50 &        65.8 &   4156 &    55016 &      -   &                  33.96 &         $8.92 \pm 0.31$ &  39.44    & LINER &   Ell.\\
         
\hline
    \end{tabular}
    \caption{First 10 rows of the table of basic properties of the X-ray sample presented in this paper. The full table can be found in the online supplementary material. We show (1) the galaxy name; (2) Right Ascension in decimal degrees; (3) Declination in decimal degrees; (4) the Galactic Latitude ($\delta$); (5) the distance in Mpc, which is obtained from \citep[][and references therein]{ho97a}; (6) the \textit{Chandra} observation ID used for this analysis; (7) the exposure length in seconds; (8) denoted `C' if part of the `Complete' sample described in Section~\ref{sec:sample}, denoted `Y' if this is a new \textit{Chandra} observation obtained in observing cycle 17 (programme ID 19708646 and 18620515, PI:McHardy), or denoted `C+Y' if part of the `Complete' sample and a new \textit{Chandra} observation; (9) the detection significance where a source is observed or detected, `Undet.' if the source is not detected in the observations or `Not obs.' if there is no data in the \textit{Chandra} archive; (10) a black hole mass measurement taken from the literature where possible \citep[e.g.,][]{vanderbosch16}, but where no dynamical measurements exist, we use the $M{-}\sigma$ relationship of \citet{tremaine02}, using the stellar velocity dispersions of \citep{ho97a}; (11) a measurement of the [\ion{O}{iii}] line luminosity from \citep{BaldiLeMMINGs} or \citet{BaldiLeMMINGs2}, taken from the literature or \citep{ho97a} in erg s$^{-1}$; (12) the new AGN classification given to the source based on the AGN re-classifcation scheme described in \citet{BaldiLeMMINGs} and \citet{BaldiLeMMINGs2}; (13) a simplified version of the Hubble types originally shown in \citet{ho97a}.}\label{tab:basicappendix}
\end{table*}

\section{Sample Selection and \textit{Chandra} Observations}
\label{sec:sample}
To build a statistically-complete sample of nearby LLAGN, we started with the Palomar sample \citep{Filippenko85,Ho95,ho97a,ho97b,ho97c,ho97d,Ho03,Ho09}. The Palomar sample is statistically-complete to a brightness limit of $B_T < 12.5 \rm mag$. We selected sources with $\delta >20^\circ$, to ensure that the synthesized beams in the radio observations were not highly elliptical, which is important for detecting small scale (sub-arcsecond, sub-kpc) jets. These 280 galaxies represent the LeMMINGs radio survey of nearby galaxies \citep{BeswickLemmings,BaldiLeMMINGs,BaldiLeMMINGs2,BaldiLeMMINGs3}. The goal of LeMMINGs is to probe accretion and star formation in nearby galaxies in the radio waveband at 1.5 and 5\,GHz with resolutions of up to 150\,mas and 50\,mas respectively, in concert with ancillary multi-wavelength data. Thus far, all 280 of the LeMMINGs objects have been observed by \emerlin{} at 1.5\,GHz \citep[see][]{BaldiLeMMINGs,BaldiLeMMINGs2,BaldiLeMMINGs3}. Observations at 5\,GHz have been completed and the data are being analysed (Williams et al., in prep.). To diagnose the central engine, all of the objects in the LeMMINGs survey were re-classified using their optical spectra from the Palomar sample into the Seyfert, LINER, ALG or \ion{H}{ii} galaxy classifications according to updated diagnostic diagrams \citep{Kewley06,Buttiglione2010}. The re-classification scheme can be found in \citet{BaldiLeMMINGs,BaldiLeMMINGs2}, but for convenience, we list these classifications in Table~\ref{tab:basicappendix}. 

Of the 280 Palomar galaxies above $\delta$ $= 20^{\circ}$, 125/280 had been observed by \textit{Chandra} observing programmes as of June 2015 and the data were publicly available\footnote{\hfill{Data were obtained from the HEASarc website: \url{https://heasarc.gsfc.nasa.gov/db-perl/W3Browse/w3browse.pl}}}. An additional 48 objects were observed as part of \textit{Chandra} observing cycle 17 (programme ID 19708646 and 18620515, PI: McHardy), to provide complete \textit{Chandra} coverage of all Palomar galaxies, selected in the declination range 40$-$65$^{\circ}$ and Galactic latitude $\left | \rm \textit{b} \right |$ $>$ 20$^{\circ}$. By selecting in this range, we ensured that the sample was not biased towards known `active' galaxies (e.g., LINERs and Seyferts) and avoided significant extinction from the Milky Way. We observed any previously unobserved sources for 10\,ks, sufficient for detecting a source at an X-ray luminosity of 10$^{39}$~erg s$^{-1}$ at the median distance of the LeMMINGs sample ($\sim$20\,Mpc), with a Hydrogen absorbing column ($N_{\rm H}$) of up to 10$^{23}$ cm$^{-2}$. The additional 48 objects included some objects with existing short \textit{Chandra} observations, to increase the combined exposure on each of these galaxies to 10\,ks. Unfortunately, one object, NGC~2685, was missed from the original observing list, but other than this object, the LeMMINGs and Palomar surveys are now complete for the declination range 40$-$65$^{\circ}$ and galactic latitude cut. 

\begin{figure}
\includegraphics[width=\columnwidth]{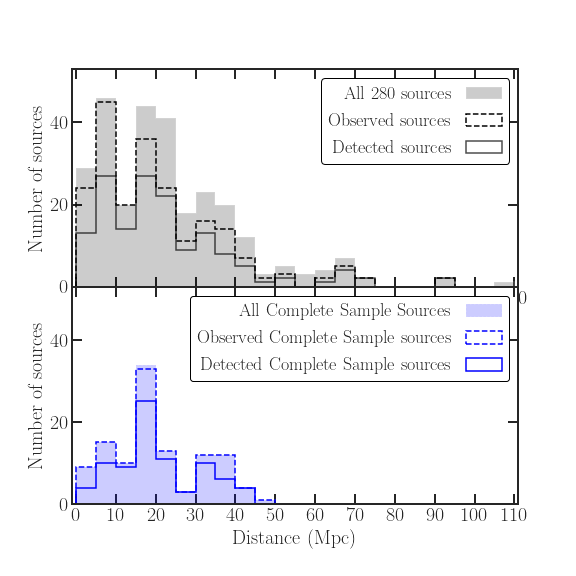}
\caption{Histogram showing the full number of sources, observed sources and detected sources in the entire (\textit{top} panel) and `Complete' (\textit{bottom} panel) Palomar X-ray data from \textit{Chandra} as a function of distance. In the \textit{top} panel, the light grey histogram shows all 280 sources, the black dashed line shows the observed (213) sources and the black dot dash line represents the detected sources (150). In the \textit{bottom} panel, the light blue histogram shows all 113 sources in the `Complete' sample, the blue dashed line shows the observed sources and the blue dot dash line represents the detected sources (82). }.
\label{fig:SourcesbyDistance}

\end{figure}

Additional observations of the Palomar objects in our sample have been observed as part of other \textit{Chandra} programmes, and these have also been included in our final sample. Therefore, we analyse all publicly available \textit{Chandra} X-ray data up to June 2018, and now \textit{Chandra} X-ray data exist for 213/280 objects ($\sim$76 per cent) in the LeMMINGs sample. For sources with multiple observations, we choose those with the longest exposure in order to improve the chance of source detection or for improved signal-to-noise for spectral modelling. In addition, we do not use any grating observations, choosing only ACIS-S and ACIS-I observations. A future manuscript (LeMMINGs VI, Pahari et al. in prep) will analyse all publicly available ACIS datasets for the purposes of finding all X-ray sources in the LeMMINGs fields and perform variability analyses. 
The 213 galaxies observed by \textit{Chandra} constitute the `X-ray LeMMINGs sample', which we hereafter refer to as the `entire' sample. We plot this sample as a function of distance in the \textit{top} panel of Figure~\ref{fig:SourcesbyDistance}. Furthermore, the 113 objects in the 40$-$65$^{\circ}$ declination range are referred to as the `Complete' sample (\textit{bottom} panel of Figure~\ref{fig:SourcesbyDistance}). We performed a Kolmogorov$-$Smirnov (KS) test on the distance distributions between the two samples and could not conclude that the two samples were statistically different. Future \textit{Chandra} proposals will be submitted with the goal of observing all 280 galaxies with at least 10\,ks exposure times, to complete the entire sample.

\begin{table*}
\centering
    \begin{tabular}{ c c c c c c c c}
\hline Galaxy & Detection & log Flux & log Flux & log Flux & log Lum. & log Lum. & log Lum.\\
Name & Signif. & 0.3$-$10\,keV & 0.3$-$2\,keV & 2$-$10\,keV & 0.3$-$10\,keV & 0.3$-$2\,keV & 2$-$10\,keV\\
 &  & erg s$^{-1}$ cm$^{-2}$ & erg s$^{-1}$ cm$^{-2}$ & erg s$^{-1}$ cm$^{-2}$ & erg s$^{-1}$ & erg s$^{-1}$ & erg s$^{-1}$\\
(1) & (2) & (3) & (4) & (5) & (6) & (7) & (8)\\
\hline

IC10 & < & $-$13.92 & - & - & 36.38 & - & -\\
NGC205 & < & $-$14.00 & - & - & 35.77 & - & -\\
NGC221 & < & $-$14.24 & - & - & 35.53 & - & -\\
NGC224 & 146.78 & $-12.59 \pm 0.05$ & $-12.77 \pm 0.10$ & $-13.04 \pm 0.10$ & $37.18 \pm 0.05$ & $37.00 \pm 0.10$ & $36.73 \pm 0.10$\\
NGC266 & 204.06 & $-12.81 \pm 0.06$ & $-13.28 \pm 0.10$ & $-12.99 \pm 0.10$ & $40.86 \pm 0.06$ & $40.39 \pm 0.10$ & $40.68 \pm 0.10$\\
NGC278 & 11.87 & $-14.18 \pm 0.10$ & $-14.29 \pm 0.11$ & $-14.80 \pm 0.20$ & $38.05 \pm 0.10$ & $37.93 \pm 0.11$ & $37.42 \pm 0.20$\\
NGC315 & 33.96 & $-11.86 \pm 0.03$ & $-12.58 \pm 0.04$ & $-11.95 \pm 0.04$ & $41.86 \pm 0.03$ & $41.13 \pm 0.04$ & $41.77 \pm 0.04$\\
NGC404 & 89.63 & $-13.53 \pm 0.07$ & $-13.71 \pm 0.16$ & $-14.00 \pm 0.16$ & $37.31 \pm 0.07$ & $37.13 \pm 0.16$ & $36.84 \pm 0.16$\\
NGC410 & 26.74 & $-12.21 \pm 0.06$ & $-12.35 \pm 0.03$ & $-12.77 \pm 0.28$ & $41.57 \pm 0.06$ & $41.43 \pm 0.03$ & $41.00 \pm 0.28$\\
NGC507 & 58.35 & $-12.95 \pm 0.02$ & $-12.98 \pm 0.02$ & $-14.08 \pm 0.02$ & $40.76 \pm 0.02$ & $40.73 \pm 0.02$ & $39.63 \pm 0.02$\\
\hline
\end{tabular}
\caption{First 10 rows of the table showing flux and luminosity measurements obtained from X-ray spectral fitting (see Section~\ref{sec:Xrayspec} for the sources that have been observed in the \textit{Chandra} archive. The full table can be found in the online supplementary material. All fluxes in this table are calculated using the \textsc{cflux} command in \texttt{XSpec}. In the table we show (1) the galaxy name; (2) the detection significance if detected, else a "<" denotes a non detected source where the flux and luminosity measurements given in the 0.3$-$10\,keV band are 3$\sigma$ upper limits; (3) The logarithm of the flux in the 0.3$-$10.0\,keV band, if detected. If undetected then the model-independent flux upper limit is given from the \textsc{srcflux} tools (see text); (4) The logarithm of the flux in the 0.3$-$2.0\,keV band; (5) The logarithm of the flux in the 2.0$-$10.0\,keV band; (6) The logarithm of the X-ray luminosity in the 0.3$-$10.0\,keV band, if detected. If undetected then the model-independent flux upper limit from the \textsc{srcflux} tools is used to calculate a luminosity upper limit; (7) The logarithm of the X-ray luminosity in the 0.3$-$2.0\,keV band; (8) The logarithm of the X-ray luminosity in the 2.0$-$10.0\,keV band. All uncertainties are shown at the 1 $\sigma$ level. All fluxes are measured in erg s$^{-1}$ cm$^{-2}$ and all luminosities are measured in erg s$^{-1}$.}\label{tab:long}
\end{table*}

A large number of the observed LeMMINGs objects in the \textit{Chandra} archive are known `active' galaxies based on their optical diagnostic emission-line `BPT' classifications \citep{BaldwinPhillipsTerlevich81,Kewley06,Buttiglione2010}. However, with the addition of the new observations in the `Complete' sample, we are able to ameliorate the imbalance of observations by AGN type. Most of the objects in the newly obtained data have never been observed with \textit{Chandra} before and over half are \ion{H}{ii} galaxies (29/48), which greatly helps the statistical completeness of the entire sample. Now, 100/141 HII galaxies (71 per cent), 22/28 ALGs (79 per cent), 14/18 Seyferts (78 per cent) and 77/93 LINERs (83 per cent), have been observed with \textit{Chandra} for our sample, which is now more consistent with the parent sample split.
Out of the 213 objects observed with \textit{Chandra} by June 2018, the numbers are now more equitable: 100 \ion{H}{ii}s, 22 ALGs, 14 Seyferts and 77 LINERs, comparing to the original ratio of 70 \ion{H}{ii}:22 ALG:14 Seyfert:59 LINER. When comparing to the overall fraction of AGN in the LEMMINGs radio sample (48 per cent \ion{H}{ii} galaxies, 33 per cent LINERs, 12 per cent ALGs, 7 per cent Seyferts), the entire X-ray sample corresponds to 47 per cent \ion{H}{ii} galaxies, 36 per cent LINERs, 10 per cent ALGs and 7 per cent Seyferts, which are very similar. 
Therefore, the entire X-ray sample currently represents one of the most complete and unbiased surveys of high-resolution X-ray emission from nearby galaxies.

\section{\textit{Chandra} Data Reduction}
We performed standard reduction procedures using the \textit{Chandra} Interactive Analysis of Observations (\textsc{CIAO 4.11})\footnote{\hfill\url{http://cxc.harvard.edu/ciao/}} software to reduce the ACIS data using the updated calibration database (\textsc{CALDB 4.8.2}) in \textsc{CIAO}.
For the purpose of core detection significance, we used the {\texttt wavdetect} tool in \textsc{CIAO}, which is based on a wavelet transform algorithm and used frequently for point source detection with \textit{Chandra} observations \citep[e.g.][]{Freeman2002, Liu2010}. This tool is able to resolve sources which are closely separated on the scale of the point spread function (PSF) and able to distinguish diffuse emission due to its advanced treatment of the background. The average background level was equivalent to 1.65$\times$10$^{-14}$~erg cm$^{-2}$ s$^{-1}$. We extracted flux images and ran the {\texttt wavdetect} task on each on-axis chip with scales of 1, ~2, ~4, ~8 and ~16 arcsec in the 0.3$-$10 keV band. We also performed manual examination of each source reported by {\texttt wavdetect} for possible false detections. 

For the sources detected by {\texttt wavdetect}, with the observed flux higher than 5$\times$ the background level, circular regions of radius 2 arcsec around the \textit{Hubble} Space Telescope optical centre of the galaxy were used to extract the flux and spectra of the objects. \textit{Hubble} data exist for 173/280 LeMMINGs objects and will be presented in an upcoming work (LeMMINGs V; Dullo et al., in prep). If \textit{Hubble} data were not available, then the galaxy centroid positions were obtained from the literature or from the NASA/IPAC Extragalactic Database\footnote{\hfill\url{https://ned.ipac.caltech.edu/}}. The background region was selected as an annular region with radii between 2 and 5 arcsec for most of the observations. However, for objects where an additional X-ray source was present within 4 arcsec of the core, we selected a 5 arcsec circular region close to the core but free of X-ray sources for the background. We checked and found that such a change in the background selection does not affect the background count rate significantly. Where pile-up is suspected in the spectra, we use an annulus to remove the highly affected inner region and only use the outer region that was unaffected by pile-up. We then use this data to fit a spectrum and report those values here. We checked the fits with a pile-up model in the spectra and do not find significant differences in the reported fluxes between the two methods. Source and background spectra were extracted using the source and background regions using the {\texttt specextract} tool in \textsc{CIAO}. Auxiliary Response Files (ARFs) and Redistribution Matrix Files (RMFs) were computed using observation specific aspect solution, mask and bad pixel files; dead area corrections were applied and ARFs were corrected to account for X-rays falling outside the finite size and the shape of the aperture.

For faint sources with an observed flux between 3$-$5 times the background flux level, a circular region of 10 arcsec was chosen to extract the source spectrum while an annulus region with radii of 10 and 20 arcsec, centered on the source was chosen to extract the background spectrum. Such a choice is made to ensure that both regions contain enough counts to extract observation specific ARFs for the source as well as the background. We checked to see if the difference in annulus size made a significant change in the extracted flux values, but it did not. However, we were careful to choose such regions so that they contain no additional X-ray sources with count rates more than 5 per cent of the typical background level. NGC~4826 and NGC~5907 were two such examples where bright X-ray sources are present within 10$\arcsec$ radius of the core position and they were avoided by choosing a suitable size of the extraction region. 

For sources where there was no X-ray counterpart detected in the central region, the \texttt{srcflux} tool was used to provide a model-independent estimate of the net count rates and fluxes including uncertainties in 0.3$-$10\,keV energy band. Due to very low X-ray count rates, 20 arcsec circular source and background regions free of off-nuclear emission were chosen for this estimation. The PSF fractions in the source and background regions are estimated from the model PSF using \texttt{arfcorr} tool in the 0.3$-$10 keV band.

\label{sec:Xrayspec}

We extracted and fitted the X-ray spectra of all the nuclear sources using version 12.11.1 of the \textsc{XSPEC} software \citep{xspec}. For bright sources, the extracted background-subtracted spectra are binned so that the signal-to-noise in each spectral bin is 3 or higher. For faint sources, the spectral binning is performed so that each bin contains at least 10 counts. For a few sources which have a detection significance of 4 or less, a binning of 5 counts per spectral bin was used. For spectral fitting of faint sources in \textsc{XSPEC}, the Gehrels weighting method \citep{Gehrels1986} was used which is suitable for Poisson data. For spectral modelling of faint sources, we used W-statistics, which is C-statistics with the background spectrum provided in \textsc{XSPEC}, while the Anderson Darling statistic \citep{andersondarling,andersondarling2} was used as the test statistic. While fitting, if some parameters could not be constrained within the acceptable range, they were fixed to the typical values, e.g., the width of the narrow Gaussian Fe emission line was fixed to 0.01\,keV in some cases, and if the power-law photon index was not constrained with the range of 0.5 and 6, it was fixed to 1.8 while fitting.  

Due to the low count rates in many of the observations, we fit simple models to the data. The base model used was \textsc{phabs} $\times$ \textsc{ztbabs}$ \times$ \textsc{zpowerlaw}. The Galactic absorption (\textsc{phabs}) was obtained from \citet{Kalberla2005}\footnote{An online tool exists at:\hfill\url{ https://heasarc.gsfc.nasa.gov/cgi$-$bin/Tools/w3nh/w3nh.pl}} and was fixed. The host galaxy absorption (\textsc{ztbabs}) was allowed to vary, as was the photon index (\textsc{zpowerlaw}). For most sources (102/150 detections, 68 per cent), the low count rates prevented more complex models than the simple absorbed power-law described above, and the parameters for these models are presented in Table.~\ref{tab:basicspeec}. However, in many bright sources, we required additional components to fit the more complicated spectra. These additional models included: \textsc{zgauss} for the Iron K alpha fluorescence line at 6.4\,keV (present in 8 objects); an additional absorber model, either \textsc{zpcfabs} or \textsc{zxipcf} (required in the models of 30 objects); \textsc{apec}, a collisionally ionised plasma model (a necessary additional model in 25 objects, note that NGC\,4151 required a second \textsc{apec} model; \textsc{gabs} a Gaussian absorption model (2 objects)). 
NGC\,5194 is a well-studied bright Compton-thick AGN \citep[see e.g.,][]{Brightman17}. Correspondingly, the high-quality \textit{Chandra} spectrum is flat $\gtrsim$\,3\,keV, and displays a very strong and narrow neutral Fe\,K$\alpha$ line: common features of reflection-dominated Compton-thick AGN. To account for these reflection features in the \textit{Chandra} band, we fit a more complex model. In addition to an intrinsic powerlaw approximated with \textsc{cutoffpl}, we use \textsc{pexrav} to reproduce the reflection spectrum by freezing the \textsc{relrefl} parameter to $-$1 (since spectra $>$\,10\,keV are needed to constrain this parameter well). 
We then included a fraction of the intrinsic powerlaw scattered through a lower column than the absorber \citep[the `warm mirror'; see e.g.,][]{Matt00}. Finally, we included a \textsc{zgauss} component to reproduce the neutral Fe\,K$\alpha$ line and \textsc{apec} to parameterise the ionised gas component visible $\lesssim$\,3\,keV. 
Fluxes for all sources were then extracted from the best fit spectra in the 0.3$-$10.0, 0.3$-$2.0 and 2.0$-$10.0\,keV bands.

\begin{center}
 \begin{figure}
	\includegraphics[width=0.95\columnwidth]{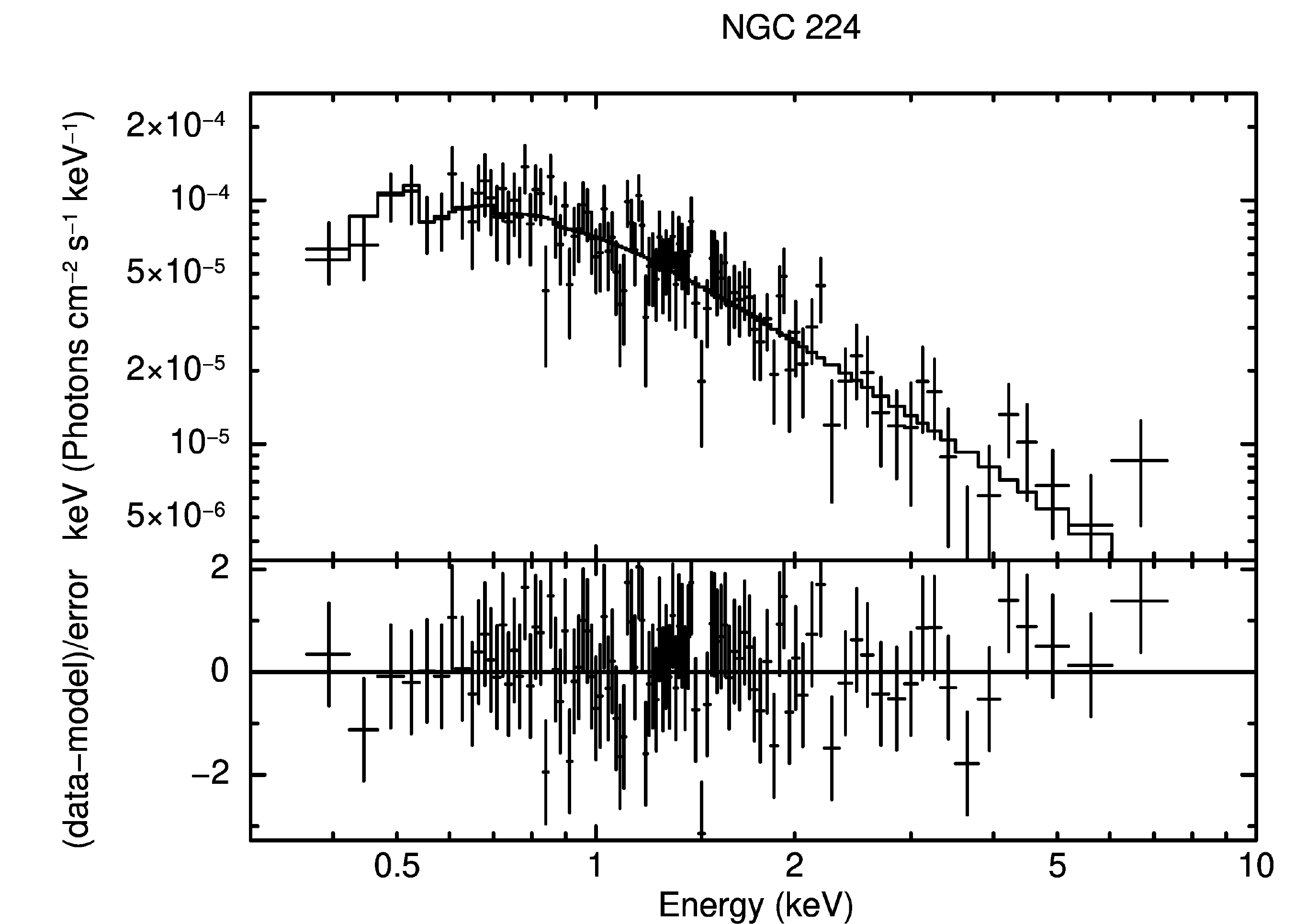}
	\caption{Example X-ray spectrum of one of the X-ray detected sources, NGC224. All other X-ray spectra are shown in the online materials. The \textit{top panel} shows the number of photons cm$^{-2}$ s$^{-1}$ keV$^{-1}$ plotted against the energy in keV across the whole 0.3$-$10.0\,keV band. The \textit{bottom panel} shows the model subtracted from the data, divided by the error. The fit parameters to make these plots are shown in Table~\ref{tab:basicspeec}.}
	 
\end{figure}
\end{center}

 \begin{figure}
	\includegraphics[width=0.48\textwidth]{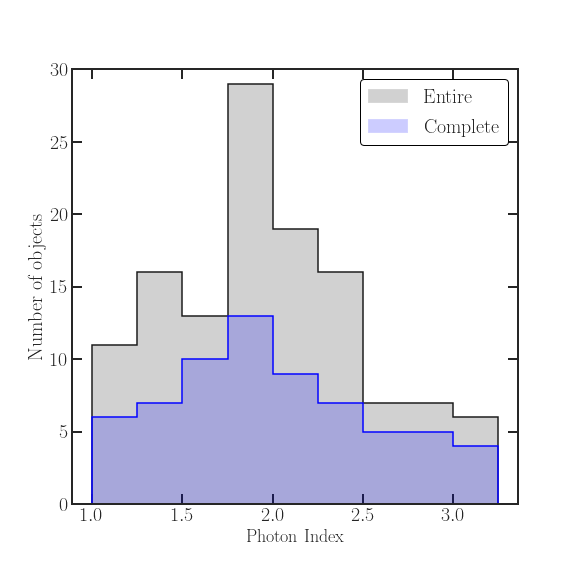}
    \caption{Histogram of the Photon Index from the best fit models of the detected sources in the entire (black) and Complete (blue) X-ray samples, presented here. }
    \label{fig:phot}
\end{figure}

\begin{table*}
\centering
\begin{tabular}{ c c c | c | c | c | c c | c c }

\hline  & Det.	 &  & 
\multicolumn{1}{c}{cstat/} & \multicolumn{1}{c}{\textsc{phabs}} &  \multicolumn{1}{c}{\textsc{ztbabs}} &  \multicolumn{2}{c}{\textsc{zpowerlw}}  &   \multicolumn{2}{c}{\textsc{apec}}\\ 
Name & Sig.	& mod. &  $\chi^{2}$ &   $N_{\rm H}$ &   $N_{\rm H}$ &   Phot.I. &  norm.  &   $kT$ &   norm.\\
(1) & (2) & (3) & (4) & (5) & (6) & (7) & (8) & (9) & (10)\\
\hline

IC10 & - & $\alpha$ & - & 50.6 & - & - & - & - & -\\
NGC205 & - & $\alpha$ & - & 5.83 & - & - & - & - & -\\
NGC221 & - & $\alpha$ & - & 18.3 & - & - & - & - & -\\
NGC224 & 146.78 & $\beta$ & 0.87 & 16.9 &  $<$0.04 & 2.82$_{-0.13}^{+0.14}$ & 9.96$_{-0.84}^{+0.95}\times$10$^{-5}$  & - & -\\
NGC266 & 204.06 & $\beta$  & 0.69 & 5.68 &  0.23$_{-0.06}^{+0.10}$ & 1.91$_{-0.14}^{+0.14}$ & 	3.05$_{-0.17}^{+0.51}\times$10$^{-5}$ & - & -\\
NGC278 & 11.87 & $\beta$ & 10.13/9$^{\rm c}$ & 12.9 &  $<$0.15              & 2.25$_{-0.98}^{+1.22}$ & 	2.30$_{-1.08}^{+1.34}\times$10$^{-6}$ & - & -\\
NGC315 & 33.96 &  $\zeta$  & 1.06 & 5.90 &  0.08$_{-0.06}^{+0.07}$ & 1.49$_{-0.08}^{+0.06}$ & 	1.84$_{-0.32}^{+0.15}\times$10$^{-4}$ &   0.54$_{-0.04}^{+0.02}$ &   2.62$_{-0.61}^{+0.54}\times$10$^{-4}$\\
NGC404 & 89.63 & $\gamma$ & 15.57/20$^{\rm c}$ & 5.71 &  0.44$_{-0.29}^{+0.24}$ & 1.88$_{-0.23}^{+0.25}$ & 	4.89$_{-1.02}^{+1.29}\times$10$^{-6}$ &   0.24$_{-0.05}^{+0.07}$ &   5.17$_{-4.06}^{+16.96}\times$10$^{-5}$\\
NGC410 & 26.74 & $\gamma$ & 0.79 & 5.11 &  0.06$_{-0.06}^{+0.07}$ & $<$2.18               & 	1.95$_{-1.01}^{+2.11}\times$10$^{-5}$  &   0.78$_{-0.05}^{+0.04}$ &   2.07$_{-0.39}^{+0.41}\times$10$^{-4}$\\
NGC507 & 58.35 & $\gamma$ & 1.02 & 5.25 &  0.07$_{-0.06}^{+0.07}$ & $>$2.77 & 	$<$3.71$\times$10$^{-6}$ &   0.91$_{-0.05}^{+0.04}$ &   5.89$_{-0.64}^{+0.77}\times$10$^{-5}$ 	\\
\hline

\end{tabular}
\caption{First ten rows and ten columns of the basic parameters from the \textit{Chandra} X-ray spectral fits, including those where the source was undetected. In this table, we only report the most basic parameters, e.g., the photon index and some of the neutral absorbers, but the table is continued for the additional parameters in the supplementary material for all observed galaxies. In this table, we show the 
(1) Galaxy name; 
(2) detection significance if detected, else a dash is used; 
(3) the X-ray spectral fit model used (see below for list of spectral models); 
(4) the reduced $\chi$ squared value (number), or where the source is faint and Poissonian statistical treatment is required of the data, the C-statistic divided by the number of degrees of freedom e.g., the cstat parameter reported by XSpec, denoted with a superscript letter `c'
(5) the Galactic absorption e.g., the \textsc{phabs} parameter, obtained from \citet{Kalberla2005} in unit of 10$^{20}$~cm$^{-2}$; 
(6) the additional absorption column density in cm$^{-2}$ found in the \text{ztbabs} component, if any, divided by 10$^{22}$; 
(7) and (8) the \textsc{zpowerlw} model photon index and normalisation respectively.
(9) and (10) the \textsc{apec} energy in $kT$ and association normalisation, respectively.
We note that not all spectral components are listed here, but the full list can be found on the online supplementary material.
The spectral models used are defined as follows:
$\alpha$ : undetected; 
$\beta$ : \textsc{phabs} $\times$ \textsc{ztbabs} $\times$ \textsc{zpowerlw}; 
$\gamma$ : \textsc{phabs} $\times$ \textsc{ztbabs}(\textsc{apec} + \textsc{zpowerlw}); 
$\delta$ : \textsc{phabs} $\times$ \textsc{ztbabs} $\times$ \textsc{zxipcf} $\times$ \textsc{zpowerlw}; 
$\zeta$ : \textsc{phabs} $\times$ \textsc{zxipcf} $\times$ \textsc{ztbabs}(\textsc{apec} + \textsc{zpowerlw}); 
$\omega$ : \textsc{phabs} $\times$ \textsc{zxipcf} $\times$ \textsc{zxipcf} $\times$ \textsc{ztbabs}(\textsc{apec} + \textsc{gabs} $\times$ \textsc{zpowerlw});
$\kappa$ : \textsc{phabs} $\times$ \textsc{ztbabs}(\textsc{zgauss} + \textsc{zpowerlw});
$\iota$ : \textsc{phabs} $\times$ \textsc{zxipcf} $\times$ \textsc{ztbabs}(\textsc{zgauss} + \textsc{zpowerlw}); 
$\epsilon$ : \textsc{phabs} $\times$ \textsc{ztbabs} $\times$ \textsc{zpcfabs} $\times$ \textsc{zpowerlw}; 
$\eta$ : \textsc{phabs} $\times$ \textsc{zxipcf} $\times$ \textsc{ztbabs}(\textsc{apec} + \textsc{zgauss} + \textsc{zpowerlw}); 
$\theta$ : \textsc{phabs} $\times$ \textsc{zpcfabs} $\times$ \textsc{ztbabs}(\textsc{zgauss} + \textsc{zpowerlw});
$\mu$ :	\textsc{phabs} $\times$ \textsc{zxipcf} $\times$ \textsc{zxipcf} $\times$ \textsc{ztbabs} $\times$ \textsc{zpowerlw}; 
$\pi$ :	\textsc{phabs} $\times$ \textsc{zpcfabs} $\times$ \textsc{ztbabs}(\textsc{apec} + \textsc{zpowerlw}); 
$\tau$ : \textsc{phabs}(\textsc{ztbabs} $\times$ \textsc{cabs} $\times$ \textsc{cutoffpl} + \textsc{pexrav} + \textsc{const} $\times$ \textsc{cutoffpl} + \textsc{apec} + \textsc{zgauss});
$\rho$ : \textsc{phabs} $\times$ \textsc{ztbabs} $\times$ \textsc{zxipcf} $\times$ \textsc{zxipcf} $\times$ \textsc{gabs}(\textsc{zgauss} + \textsc{apec} + \textsc{apec} + \textsc{zpowerlw}); 
} 
\label{tab:basicspeec}
\end{table*}

The photon indices, \textit{$\Gamma$}, typically range from $\sim$1.0 to 3.5. Figure~\ref{fig:phot} shows a histogram of the distribution of photon indices in both the Complete and entire samples presented in this work. Note that we only plot sources where a photon index was constrained (see below) and not an upper limit. In both samples, the peak of this histogram is in the bin of sources between $\Gamma$=1.75$-$2.00. In some faint sources with fluxes between 3-5 $\sigma$ detection limits in their images the photon index is highly unconstrained. We therefore fixed the photon index to $\Gamma$=1.8 for these faint sources and re-fitted the spectra in order to estimate the host galaxy absorption. These additional fits are provided in the Appendix in table~\ref{tab:photfix}. We describe the results of all our X-ray spectral fitting in Section~\ref{sec:spectralfitsresults}.

\begin{table}
    \centering
    \begin{tabular}{c c c}
    \hline Name &	\textsc{ztbabs} N$_{\rm H}$ & log Lum. 0.3$-$10\,keV\\
(1) & (2) & (3)\\
    \hline
NGC2276& $<$0.10 & 39.49\\
NGC2500& $<$0.27 & 38.43\\
NGC2541& $<$1.41 & 38.24\\
NGC2832& $<$0.07 & 41.79\\
NGC2976& $<$0.20 & 36.49\\
NGC3610& $<$0.07 & 39.96\\
NGC3992& $<$1.42 & 38.45\\
NGC4096& $<$0.14 & 38.28\\
NGC4605 & $<$0.10 & 36.85\\
NGC5308& $<$0.97 & 39.45\\
NGC5371& $<$0.02 & 40.08\\
NGC5473& 1.31$_{-1.02}^{+1.85}$ & 39.63\\
NGC6643& $<$0.21 & 39.13\\
\hline
    \end{tabular}
    \caption{Table for sources with 0.3$-$10\,keV X-ray fluxes between 3$-$5$\times$ the image detection significance level, where the photon index was fixed to 1.8 in order to obtain a value for host galaxy absorption with \textsc{ztbabs}. The errors shown in this table are all at the 1$\sigma$ level. } \label{tab:photfix}
\end{table}

\section{Results}

Here, we describe the results of the X-ray spectral fitting, fluxes and luminosities, and compare them to the ancillary information, e.g., BH mass, as well as complementary optical data and the LeMMINGs radio data. We compute an X-ray Luminosity Function (XLF) and compare the X-ray luminosity to the accretion rates in the X-ray LeMMINGs sample.

 \begin{figure*}
	\includegraphics[width=\textwidth]{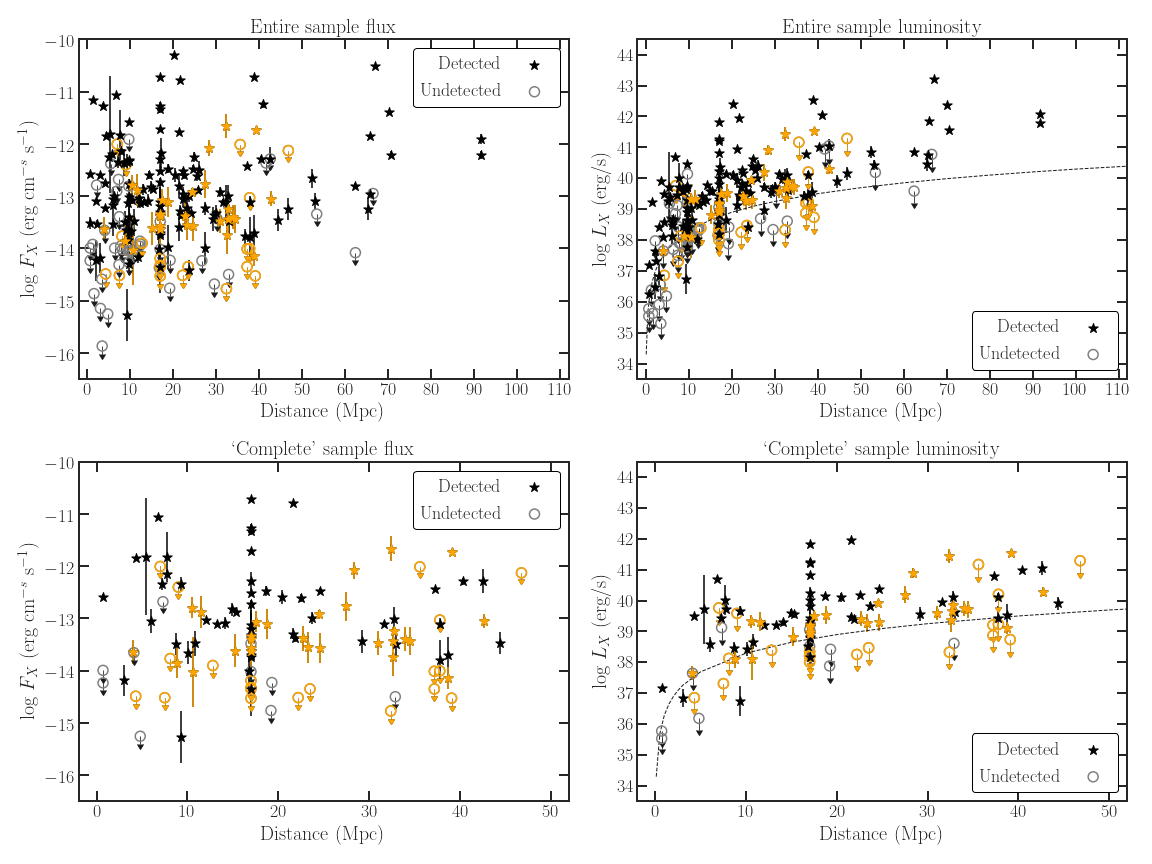}
    \caption{Unabsorbed X-ray flux density (\textit{left} plots) in erg cm$^{-2}$ s$^{-1}$ and luminosity (\textit{right} plots) in erg s$^{-1}$ the 0.3$-$10\,keV band, for the entire sample (\textit{top} row) and `Complete' sample (\textit{bottom} row), as a function of distance (Mpc). The dashed line corresponds to the limiting flux density of the X-ray sample, corresponding to the average background flux flux level of 1.65$\times$10$^{-14}$~erg cm$^{-2}$ s$^{-1}$. Some upper limits above this line may be due to short exposure times for some sources. The different symbols correspond to detected sources (stars) and upper limits for undetected sources (open circles). The orange data points represent the 48 new \textit{Chandra} observations obtained in proposals 19708646 and 18620515.
    }
    \label{fig:xdistance}
\end{figure*}

\subsection{Properties of the X-ray sources}
\label{sec:properties}

Out of the 213 galaxies that have been observed with \textit{Chandra} for our sample, 150 show X-ray emission above a detection threshold of 3$\sigma$ and are co-incident with the optical core of the galaxy.
As the optical positions of these sources are correct to 0.3 arcsec, the FWHM of the Chandra point spread function is 0.5-arcsec and we extracted spectra from a 2-arcsec aperture, they are referred to as ‘detected’ sources. The other 63 sources are considered `undetected' yielding a detection fraction of 150/213 ($\sim$70 per cent). We plot these detection fractions as a function of the distance of the sources in the \textit{top} panel of Figure~\ref{fig:SourcesbyDistance}.
For the `Complete' sample, we arrive at similar detection fractions: 82 detected, 30 non detected, corresponding to a detection fraction $\sim$73 per cent and plot them in the \textit{bottom} panel of Figure~\ref{fig:SourcesbyDistance}.
For the detected sources, the 0.3$-$10.0\,keV X-ray fluxes range from 5.3$\times$10$^{-16}$ to 5.0$\times$10$^{-11}$~erg cm$^{-2}$ s$^{-1}$, while the luminosities span from 1.8$\times$10$^{36}$ to 1.6$\times$10$^{43}$~erg s$^{-1}$.
These flux and luminosity ranges are similar to previous studies of nearby galaxies made with X-ray observatories \citep{RobertsWarwick,Liu2010}.

We note that sources with nuclear luminosities $\la$10$^{39}$~erg s$^{-1}$ could be ULXs/XRBs near the optical nucleus and not necessarily X-ray emission related to the central supermassive black hole \citep{Fabbiano2006,Swartz2011, KaaretULX}. Of the detected sources in the entire sample, 38 fall below this limit, of which 24 are \ion{H}{ii} galaxies, 12 are LINERs and 2 are ALGs. In the Complete sample, the number of objects with $L_{X-ray} <$ 10$^{39}$~erg s$^{-1}$ are: 12 \ion{H}{ii} galaxies, 5 LINERs and 1 ALG. However, one of the benefits of our sample is the overlapping high-resolution radio data from LeMMINGs, which can help disentangle non-AGN sources from real AGN. As accreting SMBHs are more radio-loud than XRBs, the radio data can help distinguish non-AGN activity from true AGN. A future publication (Pahari et al. in prep), will include radio luminosity from LeMMINGs \citep{BaldiLeMMINGs,BaldiLeMMINGs2} as a discriminant for all detected sources in our \textit{Chandra} data, including the nuclear sources. In addition, transient objects in the archive (such as the tidal disruption event, TDE, in NGC~3690) will have large changes in flux over year to decadal time scales. Large changes in X-ray flux over short time periods is unlikely to affect the majority of the sample and the X-ray emission should remain constant over long periods of time. For example, for Seyferts and LINERs, X-ray variability over the course of several years is observed  \citep{HernandezGarcia2014,Hernandez-Garcia2016}. We removed any sources that are known to be transient nuclear events or non-AGN activity at the optical centre of the galaxy. Therefore, three objects - the TDE in NGC~3690 \citep{Arp299TDE}, known nuclear ULXs in IC~342 \citep{Liu2010} and in NGC 3034 (M82) \citep[][]{MuxlowM82} - are removed from the discussions of the detected sources for the rest of this work.

In the \textit{top} row of Figure~\ref{fig:xdistance}, we plot the flux and luminosity values obtained for the entire sample as a function of distance, with the new observations for 48 galaxies plotted in orange for comparison. In the \textit{bottom} row of Figure~\ref{fig:xdistance}, we do the same for the `Complete' sample. We witness no apparent distance dependence of the sources, with a wide range of fluxes and luminosities observed at all distances. 
However, there is an exposure length difference as a function of distance, as objects at distances greater than 40\,Mpc have a mean exposure of 35\,ks, whereas less distant objects have a 30\,ks exposure on average. Exposure times will play a role in the likelihood of detection. It is not possible to quantify the effect this has on the detection likelihood in the sample, however we note that most of the objects in the `Complete' sample reside closer than 40\,Mpc, so this exposure length disparity has less of an effect on the `Complete' sample. 

\begin{table}
	\centering
	\caption{X-ray detected, undetected and unobserved sources by morphological classification breakdown in the entire X-ray Palomar sample. We note that three sources, NGC~3690, NGC~3034 and IC~342, all H{\sc ii} galaxies and noted with the asterisk, are not related to AGN activity, as discussed in Section~\ref{sec:properties}. }
	\label{tab:xfraction}
	\begin{tabular}{l|cccc|c} 
		\hline
                                   & \multicolumn{4}{c|}{optical class}\\
\hline
X-ray       &   LINER & ALG & Seyfert  &  H{\sc ii}  &  Tot\\
		\hline

detected         &   68   & 13  &    13   &  56*   &  150\\
undetected          &   9   & 9  &    1   &  44   &  63\\      
unobserved       &   16   & 6  &    4   &  41   &  67\\
	\hline
Tot                 &   93   & 28 &    18  &  141   &  280\\	
\hline
\end{tabular}

\end{table}

\begin{table}
	\centering
	\caption{X-ray detected, undetected and unobserved sources by morphological classification breakdown in the `Complete' X-ray Palomar sample. We note that one source in the `Complete' sample, NGC~3690, a H{\sc ii} galaxy and noted with the asterisk, is not related to AGN activity, as discussed in Section~\ref{sec:properties}. }
	\label{tab:xfraction2}
	\begin{tabular}{l|cccc|c} 
		\hline
                                   & \multicolumn{4}{c|}{optical class}\\
\hline
X-ray          & LINER & ALG & Seyfert  &  H{\sc ii}  &  Tot\\
		\hline

detected      &   34   & 8  &    5 &  35*   &  82\\
undetected    &  4   & 3  &    0   &  23   &  30\\      
unobserved    &   1   &  0 &  0     &   0  &  1\\
	\hline
Tot          &   39   & 11  &    5  &  58   & 113\\	
\hline
\end{tabular}

\end{table}

For the entire X-ray LeMMINGs sample, the median exposure time of detected sources is 20\,ks, whereas the median exposure for undetected sources is 10\,ks. However, for the `Complete' sample, the detected sources have a median exposure of 13\,ks, and the non-detected sources 10\,ks. This again shows that the `Complete' sample does not suffer much from this potential bias as the entire sample does.

\subsection{X-ray luminosity versus optical properties}

Table~\ref{tab:xfraction} and Table~\ref{tab:xfraction2} show the number of detected and undetected sources divided by optical class, for the entire and `Complete' samples, respectively. Seyferts are the most detected optical AGN type (Entire: 13/14, 93 per cent, Complete:5/5, 100 per cent), followed by LINERs (Entire: 68/77, 88 per cent, Complete: 34/38, 89 per cent), then ALGs (Entire: 13/22, 59 per cent, Complete: 8/11, 73 per cent) and with \ion{H}{ii} galaxies being the least detected (Entire: 56/100, 56 per cent, Complete: 35/58, 60 per cent).

 \begin{figure}
	\includegraphics[width=0.48\textwidth]{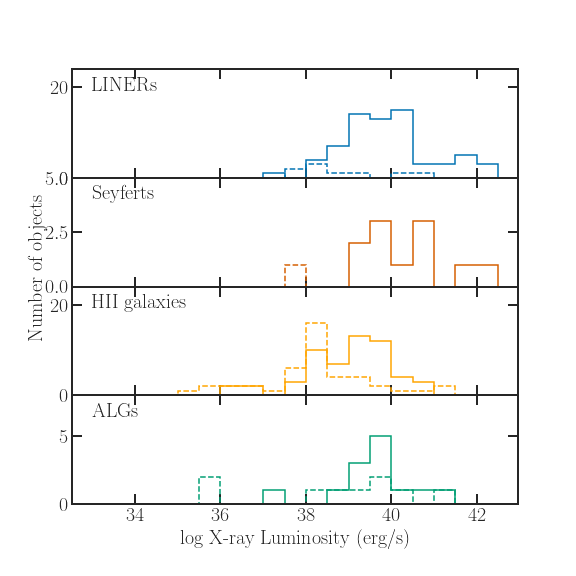}
    \includegraphics[width=0.48\textwidth]{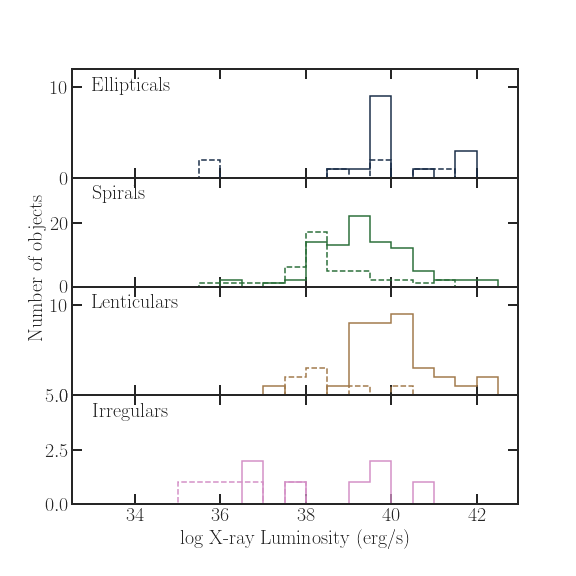}
    \caption{Histograms of the X-ray luminosity (erg s$^{-1}$) per optical class (top plot) and host morphological type (bottom plot). The solid-line histogram represents the X-ray core luminosity distribution of the detected sources and the dashed line corresponds to the upper limits obtained from the non-detected objects.}
    \label{xhist}
\end{figure}

In general, Seyferts have been observed with longer exposures, with a mean exposure of 53\,ks, compared to 29\,ks for LINERs, 32\,ks for \ion{H}{ii} galaxies and 25\,ks for ALGs. A similar exposure length disparity is observed in the `Complete' sample. As a consequence, Seyferts have been observed for longer than other galaxies and are therefore more likely to be detected. However, the distribution of X-ray luminosity is similar between Seyferts and LINERs (see Figure~\ref{xhist}), but the \ion{H}{ii} galaxies and ALGs have different luminosity distibrutions. It is possible that the difference in exposure length could have an affect on the completeness of our sample or the results, but cannot quantify this disparity further.

\begin{figure}
\includegraphics[width=0.95\columnwidth]{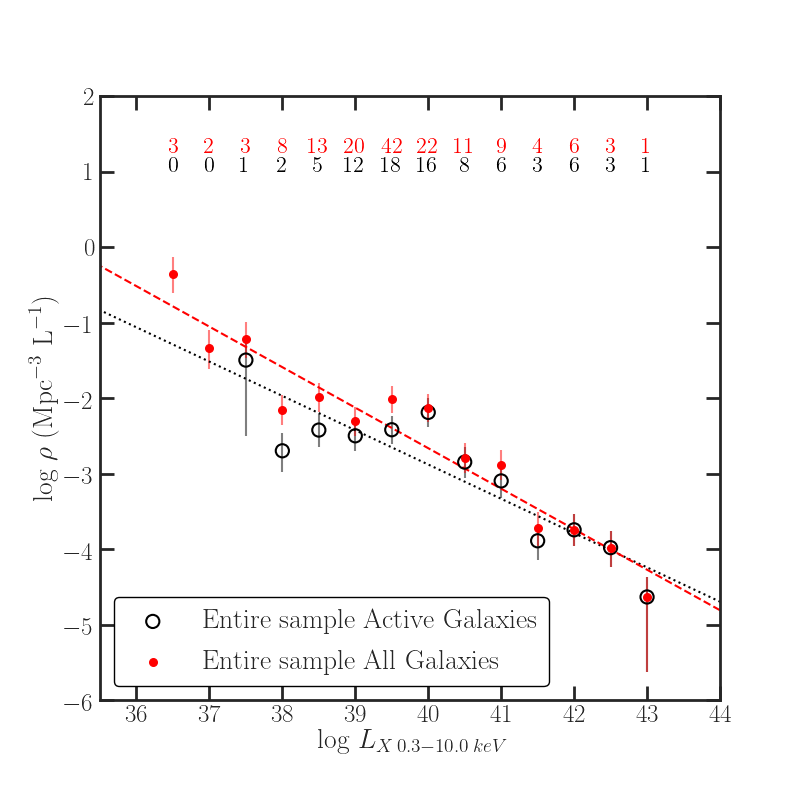}
\caption{The X-ray luminosity function of the entire X-ray Palomar sample. The red circles show the X-ray detections in the entire sample. The black unfilled circles indicate the `active' galaxies based on the BPT diagnostic plots in \citet{BaldiLeMMINGs, BaldiLeMMINGs2}, e.g., the LINERs and Seyferts in the entire sample. The dotted lines are fits to the data, which are in the same colour as the sub-sample used. For the entire sample, the power-law fit yields slope of $-$0.54$\pm$0.06, and for the `active' galaxies the power-law slope is $-$0.45$\pm$0.05. See  Section~\ref{sec:XLF} for further details. In addition the number of objects in each bin is written above each bin in the same colour as the sample used. The uncertainties are drawn from a Poisson distribution. }
\label{fig:XLF}
\end{figure}

\subsubsection{X-ray luminosity distributions vs optical spectroscopic class}

The distribution of X-ray luminosity as a function of AGN optical class is presented in the upper panel of Figure~\ref{xhist}. The detected LINERs are the most commonly observed AGN type with a median X-ray luminosity in the 0.3$-$10.0\,keV band of 5.1$\times$10$^{39}$~erg s$^{-1}$. The Seyferts occupy a similar region in this plot to the LINERs and have a median luminosity of 4.8$\times$10$^{40}$~erg s$^{-1}$. The \ion{H}{ii} galaxies have slightly lower X-ray luminosities, with a broad range between $\sim$10$^{36}$ and 10$^{41}$~erg s$^{-1}$, with a median luminosity of 2.0$\times$10$^{39}$~erg s$^{-1}$. The ALGs have a median luminosity of 3.5$\times$10$^{39}$~erg s$^{-1}$ but show a similar distribution to \ion{H}{ii} galaxies at lower luminosities. As for the upper limits (dashed lines in Figure~\ref{xhist}), \ion{H}{ii} galaxies, which have the lowest detection fraction, follow a broader distribution at a luminosities 1$-$2 decades below the detected sources. 

\subsubsection{X-ray luminosity distributions vs galaxy morphological class}

The lower panel of Figure~\ref{xhist} shows the distribution of X-ray luminosity as a function of galaxy morphological type. We note that this figure shows no overall dependence of X-ray luminosity as a function of the galaxy morphological type, suggesting that nuclear X-ray emission can be found in all Hubble types. The median luminosity values for the sample are as follows: 5.1$\times$10$^{39}$~erg s$^{-1}$ for ellipticals, 9.6$\times$10$^{39}$~erg s$^{-1}$ for lenticulars, 3.3$\times$10$^{39}$~erg s$^{-1}$ for irregulars and 2.5$\times$10$^{39}$~erg s$^{-1}$ for spirals.  Approximately two-thirds of the elliptical galaxies are detected (15/22, 68 per cent). They mostly cluster around $\sim$10$^{40}$~erg s$^{-1}$ with a few more luminous exceptions. The spiral galaxies, which are the most common type in the LeMMINGs sample, have a similar detection rate (93/135, 69 per cent) and follow a broad distribution of X-ray luminosities. The lenticular galaxies show a distribution between $\sim$10$^{39}$~erg s$^{-1}$ and $\sim$10$^{41}$~erg s$^{-1}$ and are the most detected Hubble type (37/44, 84 per cent). The irregulars have the lowest detection fraction (7/12, 58 per cent) and are a mixed bag of X-ray luminosities, likely due to the inhomogeneous nature of these galaxies, but they do include a couple of very low luminosity ($\la$10$^{38}$~erg s$^{-1}$) detected sources.

\begin{figure}
\includegraphics[width=0.95\columnwidth]{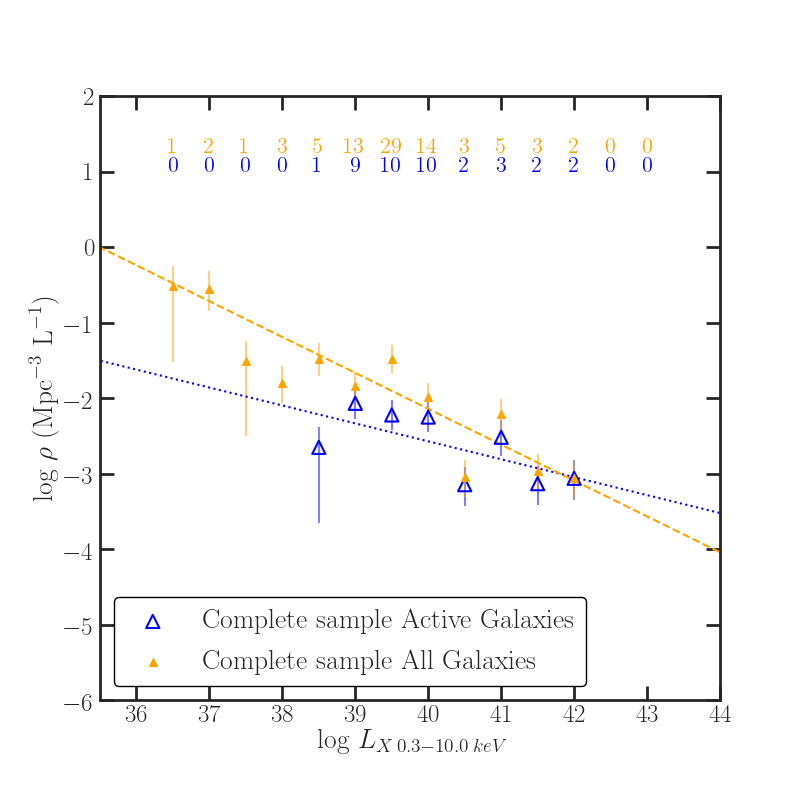}
\caption{As in Figure~\ref{fig:XLF}, showing the X-ray luminosity function of the X-ray `Complete' sample, a declination restricted sub-sample between 40$^{\circ}$ to 65$^{\circ}$. The orange triangles show the X-ray detections in the `Complete' sample. The blue unfilled triangles indicate the `active' galaxies based on the BPT diagnostic plots in \citet{BaldiLeMMINGs,BaldiLeMMINGs2}, e.g., the LINERs and Seyferts in the `Complete' sample. For all of the `Complete' sample, the power-law fit yields slope of $-$0.48$\pm$0.12, and for the `active' galaxies in the `Complete' sample, the power-law slope is $-$0.24$\pm$0.11. See  Section~\ref{sec:XLF} for further details. In addition the number of objects in each bin is written above each bin in the same colour as the sample used. The uncertainties are drawn from a Poisson distribution. }
\label{fig:XLF2}
\end{figure}

\begin{figure}
\includegraphics[width=0.48\textwidth]{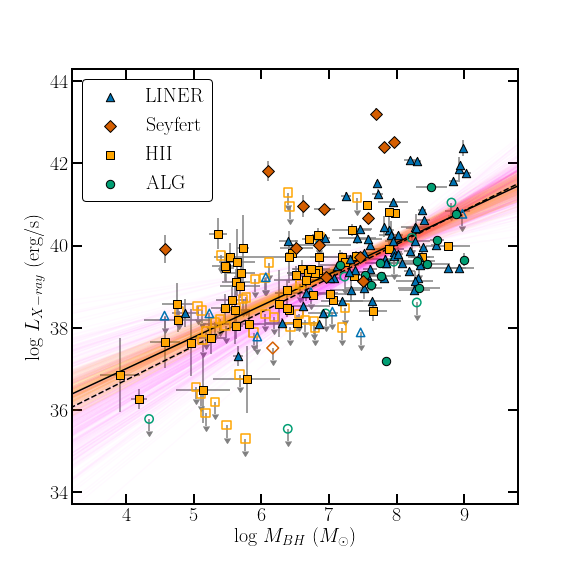}
\caption{Scatter plot showing the unabsorbed X-ray luminosity (0.3$-$10.0\,keV) as a function of the BH masses for the entire sample, divided per optical class (symbol and color coded as in the legend). The filled symbols refer to the detected X-ray sources, while the empty symbols refer to undetected X-ray sources. The solid line represents the linear correlation found for all galaxies, when taking into account the upper limits using the \textsc{linmix} package. The orange lines represent 400 draws from (see text) this fit for all of the galaxies. The dashed line is a fit to all of the sources with a mass $\ga$10$^{7}$M$_{\odot}$ and the purple lines represent 400 draws from these sources. 
}
\label{fig:XrayLumBlackHoleMassScatterPlota}

\end{figure}

\begin{figure}
\includegraphics[width=0.48\textwidth]{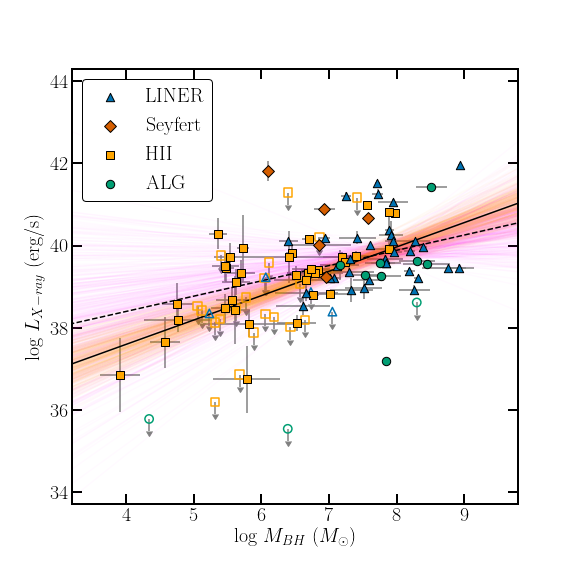}
\caption{Scatter plot showing the unabsorbed X-ray luminosity (0.3$-$10.0\,keV) as a function of the BH masses for the Complete sample, divided per optical class (symbol and color coded as in the legend). The filled symbols refer to the detected X-ray sources, while the empty symbols refer to undetected X-ray sources. The solid line represents the linear correlation found for all galaxies, when taking into account the upper limits using the \textsc{linmix} package. The orange lines represent 400 draws from (see text) this fit for all of the galaxies. The dashed line is a fit to all of the sources with a mass $\ga$10$^{7}$M$_{\odot}$ and the purple lines represent 400 draws from these sources. 
}
\label{fig:XrayLumBlackHoleMassScatterPlotb}

\end{figure}

\subsubsection{Combinations of optical spectroscopic class and galaxy morphological class}

We investigated the connection between the different Hubble morphological types and the optical spectroscopic classes. For the detected LINERs, we find that 35/68 are spirals or irregulars, whereas 33/68 are lenticulars or ellipticals, suggesting LINERs can be found in all morphological types of galaxy. Detected Seyferts, by comparison, are detected only in spiral galaxies (9/15) or lenticulars (4/15). The detected \ion{H}{ii} galaxies are almost exclusively associated with spiral galaxies, 48/56, with 6/56 being irregular galaxies and 2/56 associated with lenticulars. Finally, the ALGs are associated with all galaxy morphological types: 7 lenticulars, 4 ellipticals, 1 spiral and 1 irregular.

Comparing the morphological types to the optical classifications shows that all of the detected ellipticals (15) have LINER (11/15) or ALG (4/11) nuclei. The detected lenticulars are mostly associated with LINERs (22/35) and ALGs (7/35), but also in a 4 Seyferts and 2 \ion{H}{ii} galaxies. Spiral galaxies are found to mostly to have \ion{H}{ii} nuclei (48/91) or LINERs (33/91), with 9 Seyferts and 1 ALG. Of the nine detected irregular galaxies, 6 are associated with \ion{H}{ii} nuclei, 2 with LINERs and 1 with an ALG.

\begin{table}
    \centering
    \renewcommand{\arraystretch}{1.2}
    \begin{tabular}{p{1.2cm} c c c c}
    \hline
    Correlation & $\beta$ & $\alpha$ & $\sigma ^{2}$ & $\hat{\rho}$\\
    (1) & (2) & (3) & (4) & (5)\\
         \hline
        \multicolumn{5}{c}{\textit{log(X-ray):log(M$_{\odot}$})}\\
    \hline
         Entire, all~~~~ & 33.83$^{+0.49}_{-0.50}$ & 0.78$^{+0.07}_{-0.07}$ & 1.16$^{+0.13}_{-0.11}$ & 0.64$^{+0.04}_{-0.05}$\\
         Entire, $>$10$^{7}$M$_{\odot}$ & 33.40$^{+1.51}_{-1.46}$ & 0.83$^{+0.18}_{-0.19}$ & 0.99$^{+0.16}_{-0.13}$ & 0.41$^{+0.08}_{-0.09}$\\
         `Complete', all & 35.02$^{+0.72}_{-0.74}$ & 0.62$^{+0.11}_{-0.10}$ & 1.10$^{+0.18}_{-0.15}$ & 0.53$^{+0.08}_{-0.09}$\\
         `Complete', $>$10$^{7}$M$_{\odot}$ & 36.91$^{+2.17}_{-2.16}$ & 0.37$^{+0.28}_{-0.28}$ & 0.83$^{+0.21}_{-0.15}$ & 0.20$^{+0.15}_{-0.15}$\\
         Entire, Seyferts & 34.98$^{+3.82}_{-3.98}$ & 0.81$^{+0.57}_{-0.55}$  & 2.68$^{+1.69}_{-0.92}$ & 0.42$^{+0.24}_{-0.28}$\\
         Entire, LINERs & 33.33$^{+0.91}_{-0.89}$ & 0.84$^{+0.12}_{-0.12}$  & 0.69$^{+0.13}_{-0.11}$ & 0.67$^{+0.06}_{-0.08}$\\
         Entire, ALGs & 26.00$^{+2.57}_{-2.74}$ & 1.66$^{+0.34}_{-0.32}$  & 0.94$^{+0.48}_{-0.28}$ & 0.83$^{+0.07}_{-0.11}$\\
         Entire, \ion{H}{ii} gal. & 33.43$^{+0.85}_{-0.84}$ & 0.85$^{+0.13}_{-0.14}$ & 1.02$^{+0.20}_{-0.16}$ & 0.61$^{+0.08}_{-0.09}$\\
         \hline
         \multicolumn{5}{c}{\textit{log(X-ray):log([\ion{O}{iii}]})}\\
         \hline
         Entire, all~~~~ & $-$7.07$^{+3.89}_{-4.05}$ & 1.22$^{+0.11}_{-0.10}$ & 0.56$^{+0.12}_{-0.11}$ & 0.83$^{+0.04}_{-0.05}$\\
         `Complete', all & $-$5.58$^{+6.27}_{-6.74}$ & 1.19$^{+0.18}_{-0.17}$ & 0.46$^{+0.14}_{-0.12}$ & 0.79$^{+0.07}_{-0.08}$\\
         Entire, Seyferts & $-$13.72$^{+12.05}_{-15.11}$ & 1.37$_{-0.38}^{+0.30}$ & 0.43$^{+0.63}_{-0.31}$ & 0.93$^{+0.05}_{-0.13}$\\
         Entire, LINERs & $-$2.55$^{+4.90}_{-5.14}$ & 1.11$^{+0.13}_{-0.13}$ & 0.27$^{+0.12}_{-0.09}$ & 0.88$^{+0.04}_{-0.06}$\\
         Entire, \ion{H}{ii} gal. & $-$17.37$^{+12.47}_{-14.92}$ & 1.49$^{-0.33}_{+0.40}$ & 0.84$^{+0.25}_{-0.23}$ & 0.67$^{+0.11}_{-0.13}$\\
         \hline
    \end{tabular}
    \caption{Table of \textsc{LINMIX} fit parameters, as described in the text, for the X-ray:M$_{\odot}$ correlations (\textit{top}) and the X-ray:[\ion{O}{iii}] correlations (\textit{bottom}). The columns are as follows: (1) correlation fit description; (2) y-intercept, $\alpha$; (3) gradient, $\beta$; (4) intrinsic scatter squared, $\sigma^{2}$ - note that this is the direct output from \textsc{LINMIX}, but throughout the text we report the intrinsic scatter i.e. the square root of the value printed above; (5) correlation coefficient, $\hat{\rho}$. The uncertainty values are the 16th and 84th percentile level of the fit for each parameter.}
    \label{tab:tabfits}
\end{table}

\subsection{X-ray luminosity function}
\label{sec:XLF}
We now compute the X-ray luminosity function (XLF), using the $V/V_{max}$ method \citep{Schmidt1968}. The XLF, $\Phi$(log L$_{\rm X}$), i.e. the space density of objects per unit logarithmic interval of luminosity is given by:

\begin{equation}
\Phi \left ( \log L_{X} \right ) = \frac{4 \pi}{\sigma} \sum_{i=1}^{n\left ( \log L_{*} \right )} \frac{1}{V_{max\left ( i \right )}}, 
\end{equation}
\noindent where $\frac{4 \pi}{\sigma}$ is the fraction of the sky surveyed, $n\left ( \log L_{*} \right )$ is the number of objects in a given luminosity bin $L_{*}$, and $V_{max\left ( i \right )}$ is the maximum volume in which the object would be observed to, given the limiting magnitudes/fluxes in both the optical parent Palomar sample and the X-ray sample. The smaller of the two volumes is used, to ensure that the object would be detected in both samples. We place detected sources in bins of equal X-ray power and use Poisson statistics to estimate uncertainties in each luminosity bin. As the LeMMINGs survey is for declinations $>$20$^{\circ}$, the fraction of the sky is limited to 1.316$\pi$ sr, but for the `Complete' sample we are limited further to 0.68$\pi$ sr. The optical Palomar survey is limited to 12.5 mag. We use a flux limit of the X-ray sample, i.e., 1.65$\times$10$^{-14}$~erg cm$^{-2}$ s$^{-1}$. 

Figure~\ref{fig:XLF} shows the XLF of the entire X-ray LeMMINGs sample in filled red circles and we fit a power law of slope $\alpha$ = $-0.54 \pm 0.06$ (red dashed line). 
We also fit the `active' galaxies, (open black circles), finding a power law of gradient $-0.45 \pm 0.05$, shown by the black dotted line. These two fits differ by 1.2$\sigma$. The changes in the overall appearance of the luminosity function is more apparent in the `active' sources, where the X-ray slope appears to flatten slightly below 10$^{39}$~erg s$^{-1}$, which may indicate some contamination from non-AGN sources, but this is based off a smaller number of objects and the two fits agree with one another, within the uncertainties. 

We also plot in Figure~\ref{fig:XLF2} the XLF for the `Complete' sample in the declination range 40 to 65$^{\circ}$, in orange for all galaxies, and blue for the `active' galaxies. We recalculate the XLF for both the entire and `active' galaxy samples and find power law fits of $-0.48 \pm 0.12$ and $-0.24 \pm 0.11$, respectively. These two fit values disagree at the 1.5$\sigma$ level. However, comparing the fit for all galaxies between the entire sample and the Complete sample, yields a difference of 0.5$\sigma$, suggesting that the declination-limited Complete sample is not too dissimilar from the entire sample. Furthermore, the fits of the `active' galaxies differ by 1.7$\sigma$ between the entire sample and the `Complete' sample.

Previous studies of the XLF have focussed on sources with X-ray luminosities in excess of $\sim$10$^{41}$~erg s$^{-1}$ or higher, in order to remove potential contamination from XRBs and ULXs, due to lower resolution X-ray telescopes: \textit{MAXI} \citep{Ueda2011}, \textit{Swift BAT} \citep{Tueller2008,Ajello2012}, \textit{INTEGRAL}: \citep{Sazonov2008} and \textit{XMM-Newton} \citep{Foropoulou}. However, our data resolve the nuclear region to 0.5 arcsec with \textit{Chandra}, so it is less likely that we are contaminated by XRBs and ULXs, allowing us to reach X-ray luminosities of 10$^{36}$~erg s$^{-1}$. Previous XLFs (see \citealt[][for a comparison of some of the XLFs in the literature]{Ajello2012,Ueda2014,Ballantyne}) are described by a broken power-law, which is flatter below $\sim$10$^{43}$~erg s$^{-1}$. The power-law slope below 10$^{43}$~erg s$^{-1}$ is found to be between $-0.8$ and $-1.0$, though our data show a shallower power-law. However, \citet{She2016} show that for local AGN using \textit{Chandra} data, the power-law is more consistent with a slope of $-0.38$ for all types of AGN, with the `active' sources like Seyferts and LINERs showing the flattest power-laws ($-0.15$ and $-0.32$, respectively), while \ion{H}{ii} galaxies and ALGs show steeper power-laws ($-0.68$ and $-0.82$, respectively). The general trend of flatter power-laws for `active' galaxies is also apparent in our sample, with the addition of the ALGs and \ion{H}{ii} galaxies showing a steepening of the overall power-law slope. Hence, our fits are qualitatively consistent with \citet{She2016}.

\subsection{X-ray properties vs black hole mass}

We now investigate the X-ray luminosity as a function of BH mass (see Figure~\ref{fig:XrayLumBlackHoleMassScatterPlota} and Figure~\ref{fig:XrayLumBlackHoleMassScatterPlotb} for the entire and `Complete' samples respectively). The black hole masses used in this study are listed in Table~\ref{tab:basicappendix}. When possible, we use dynamical black hole measurements found in the literature \citep[e.g.,][which accounts for fifty BH mass measurements in the full 280 objects of the LeMMINGs sample]{vanderbosch16}. When not available, the black hole masses are calculated using the relationship between black hole mass and stellar velocity dispersion, known as the $M-\sigma$ relation. The stellar velocity dispersions ($\sigma$) are listed in \citet{ho97a}, while we use the $M-\sigma$ relation obtained in \citet{tremaine02}. However, the estimation of black hole masses in irregular galaxies and star-forming galaxies like \ion{H}{ii} galaxies are more uncertain than those in bulge-dominated galaxies, i.e. ellipticals, and those where AGN are known, such as Seyferts and LINERs. The stellar velocity dispersions used to calculate the black hole masses give a standard deviation in $M_{\rm BH}$ fractional error of 0.25 for irregular and star-forming galaxies, whereas it is 0.17 for ellipticals and AGN. These errors are representative of the scatter found in the $M-\sigma$ relation \citep[][]{tremaine02}. Furthermore, the M$-\sigma$ relation used can affect the accuracy of the black hole mass estimate at lower masses and it is uncertain whether a single relation holds across the entire range of galaxy masses (Dullo et al. 2020, submitted). However, we compared the M$-\sigma$ relations of \citet{tremaine02}, \citet{kormendy13} and \citet{graham13} and found that for black hole masses above 10$^{7}$~M$_{\odot}$ the scatter was 0.5 dex, but below 10$^{6}$~M$_{\odot}$ the scatter increased to 1 dex \citep[see][]{BaldiLeMMINGs}. 
\citet{Shankar16,Shankar19} also discuss potential biases in the normalisations of both the $M_{\rm BH}-\sigma$ and $M_{\rm BH}-M_{\rm gal}$ relations. However, these amount to at most a factor of $\sim$2$-$3 in the mean normalization of the $M_{\rm BH}-\sigma$ relation, and they would not alter the already very broad dispersion in the $L_{\rm X}-M_{\rm BH}$ correlation substantially.

In general, the detection rate of X-ray sources increases with BH mass. The detection fraction for the entire sample objects with BH masses $\ga$10$^{8}$M$_{\odot}$ is 88 per cent (36/41), but below 10$^{6}$M$_{\odot}$, the detection fraction falls to 50 per cent (26/52). Of the five undetected objects with $\ga$10$^{8}$M$_{\odot}$, four of them are ALGs. A similar distribution exists for the `Complete' sample: 12/13 (92 per cent) detection fraction for $\ga$10$^{8}$M$_{\odot}$ and 16/29 (55 per cent) for $\la$10$^{6}$M$_{\odot}$.

\begin{figure}
\includegraphics[width=\columnwidth]{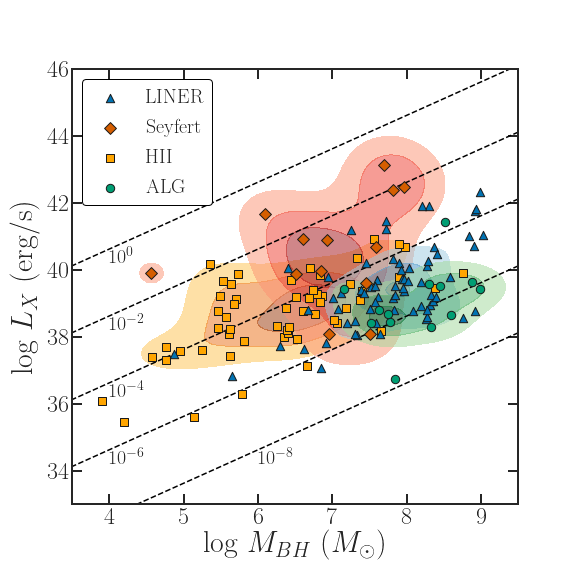}
\caption{The unabsorbed X-ray 0.3$-$10.0\,keV luminosities (L$_{\rm X-ray}$ in erg s$^{-1}$) as a function of the BH masses for the detected sources in the sample, divided per optical class. The dashed lines represent the Eddington ratios $\lambda$ to compare to the X-ray luminosity and BH masses. Each Eddington ratio is labelled next to a dashed line. The uncertainties are not shown but can be seen in Figure~\ref{fig:XrayLumBlackHoleMassScatterPlota}. In addition we plot the 1, 2 and 3 $\sigma$ levels of the distributions for each AGN type in their respective colours.}
\label{fig:XrayLumBlackHoleMassScatterPlot2}
\end{figure}

While the black hole masses range over $\sim$5 orders of magnitude (see Figure~\ref{fig:XrayLumBlackHoleMassScatterPlota}), there are clear distinctions in the different types of AGN nucleus. The detected Seyfert galaxies tend to have the highest X-ray luminosities for each mass bin, when compared to other AGN types. They generally cluster 1$-$2 decades above the other AGN types in X-ray luminosity (see also Figure~\ref{fig:XrayLumBlackHoleMassScatterPlotb} and Figure~\ref{fig:XrayLumBlackHoleMassScatterPlot2}). There is a large distribution in the LINER X-ray luminosity of order 2$-$3 decades. The detected ALGs tend to lie in the same mass bins as the LINERs, but at slightly lower X-ray luminosities ($\sim$10$^{39}$~erg s$^{-1}$). The low mass \ion{H}{ii} galaxies are not often detected, but they appear to follow the same overall trend as the rest of the detected X-ray population, but with a larger distribution towards the lowest BH masses. 

We test the correlation between X-ray luminosity and BH mass, using \textsc{LINMIX}\footnote{A \textsc{Python} module can be obtained from \hfill\url{https://linmix.readthedocs.io/en/latest/index.html}}: a Bayesian framework that folds in uncertainties in both axes as well as upper limits in the y axis \citep{LINMIX}. \textsc{LINMIX} can provide the gradient, y-intercept, the scatter and the correlation coefficient ($\hat{\rho}$) for a given fit. We note that two of the sources (NGC~1003 and NGC~4242) have upper limits on both black hole mass and X-ray luminosity. The \textsc{LINMIX} package is unable to handle upper limits in both axes simultaneously so we remove these two sources from the fits. In Figure~\ref{fig:XrayLumBlackHoleMassScatterPlota} and Figure~\ref{fig:XrayLumBlackHoleMassScatterPlotb}, we display 400 draws from the fitting process to give a visual guide to the scatter in the correlations. As an additional analysis, we computed the correlation for sources with black hole masses greater than 10$^{7}$~M$_{\odot}$, in order to find show any global difference in the slope at lower masses, in analogy to a break found in the radio LeMMINGs sample \citep[][]{BaldiLeMMINGs,BaldiLeMMINGs3}.

We find a correlation for the `Complete' sample sources (solid black line in Figure~\ref{fig:XrayLumBlackHoleMassScatterPlotb}) to be of the form $L_{\rm X-ray}$ $\sim$M$_{\odot}$ $^{0.62 ^{+0.11}_{-0.10}}$, with an intrinsic scatter of 1.05 dex (correlation coefficient, $\hat{\rho}$ = 0.53$^{+0.08}_{-0.09}$). The uncertainty values are the 16th and 84th percentile level of the fit parameters and we show these in full in Table~\ref{tab:tabfits}. The black dashed line shows the correlation for `Complete' sample sources above 10$^{7}$M$_{\odot}$, which is $L_{\rm X-ray}$ $\sim$M$_{\odot}$ $^{0.37 \pm 0.28}$ and an intrinsic scatter of 0.91 dex ($\hat{\rho}$ = 0.20$^{+0.15}_{-0.15}$). The correlation coefficients indicate a positive correlation, but the $>$10$^{7}$M$_{\odot}$ correlation is very weak. For the entire sample, we find slightly different fits, of $L_{\rm X-ray}$ $\sim$M$_{\odot}$ $^{0.78 \pm 0.07}$ (scatter = 1.08 dex, $\hat{\rho}$ = 0.64$^{+0.04}_{-0.05}$) for the entire sample and for those above a black hole mass of 10$^{7}$M$_{\odot}$, we find a correlation of the form $L_{\rm X-ray}$ $\sim$M$_{\odot}$ $^{0.83 ^{+0.18}_{-0.19}}$ (scatter = 1.00 dex, $\hat{\rho}$ = 0.41$^{+0.08}_{-0.09}$). We note that the entire sample includes more galaxies at higher X-ray luminosities which may provide a reason for the steeper gradients and stronger correlation coefficients, but, given the uncertainties in the fits, the entire and Complete sample values agree with one another within $\sim$1$\sigma$. Therefore, we do not find a break X-ray-M$_{\odot}$ relation as in the LeMMINGs radio-M$_{BH}$ relation \citep[][]{BaldiLeMMINGs,BaldiLeMMINGs3}.

We also investigated the correlations in each type of AGN, using the entire sample due to the low numbers of objects in the `Complete' sample for all AGN types. We further note that the entire sample does not appear to differ significantly from the statistically-complete sub-sample. For Seyferts, we find a correlation of $L_{\rm X-ray}$ $\sim$M$_{\odot}$ $^{0.81 ^{+0.57}_{-0.55}}$ (scatter = 1.64 dex and $\hat{\rho}$ = 0.42$^{+0.24}_{-0.28}$). In the literature, correlations between X-ray luminosity and BH mass have led to mixed results \citep[e.g.,][]{Koratkar1991,Kaspi2000,Pellegrini2005,panessa06}. Given the low number of Seyferts in our sample, the large scatter in correlation and the poor correlation coefficient presented here, it is not possible to find any strong correlation between the X-ray luminosity and BH mass for Seyferts.

We performed the same analysis for the LINERs, ALGs and \ion{H}{ii} galaxies and found them to all be correlated between the X-ray luminosity and BH mass. The LINER correlation is $L_{\rm X-ray}\sim$M$_{\rm BH}^{0.84 \pm 0.12}$ with a scatter of 0.83 dex. The \ion{H}{ii} galaxies follow a correlation of $L_{\rm X-ray} \sim$M$_{\rm BH} ^{0.85_{-0.14}^{+0.13}}$ with an intrinsic scatter of 1.01 dex. For the ALGs, the correlation is $L_{\rm X-ray} \sim$M$_{\rm BH} ^{1.66_{-0.32}^{+0.34}}$ with a scatter of 0.91 dex. Given the low detection fraction and the low number of objects, we do not claim a significant correlation for ALGs. For the LINERs and \ion{H}{ii} galaxies, they have remarkably similar fit parameters and intermediate correlation coefficients, suggesting that \ion{H}{ii} galaxies may be similar to LINERs, although we note there is up to 1 dex of scatter in both relations. This finding may represent a continuation of the LINERs down to lower X-ray luminosities, but the uncertainties on sources at lower black hole masses ($<$10$^{6}$M$_{\rm BH})$ is larger, so these correlations are driven by the higher mass objects. Indeed, removing sources below 10$^{6}$M$_{\rm BH}$ results in the same fits, but fitting only sources $<$10$^{6}$M$_{\rm BH}$ leads to unconstrained or very poorly constrained fits in all cases. 

 \begin{figure}
	\includegraphics[width=0.48\textwidth]{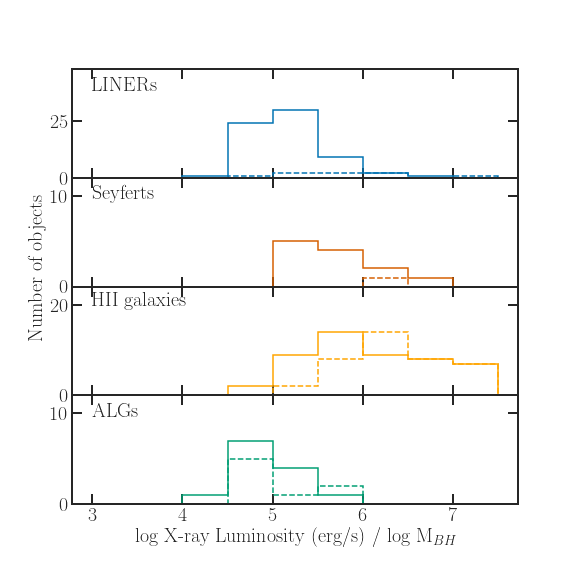}
    \includegraphics[width=0.48\textwidth]{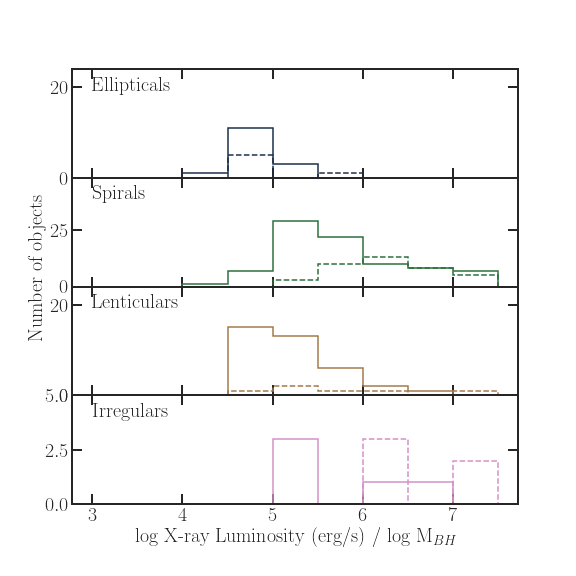}
    \caption{Histograms of the X-ray luminosity divided by the black hole mass - a tracer of the accretion rate - per optical class (top plot) and host morphological type (bottom plot). The solid-line histogram represents the X-ray core luminosity distribution of the detected sources and the dashed line corresponds to the upper limits obtained from the non-detected objects.}
    \label{xhistb}
\end{figure}

\subsection{Eddington Ratio}

In addition to the fits presented above, we compared the 2$-$10\,keV X-ray luminosity and BH mass plot to the Eddington ratio, 
\begin{equation}
    \lambda = \frac{\rm L_{\rm bolometric}}{\rm L_{\rm Eddington}}
\end{equation}
\noindent for the detected sources in the sample in Figure~ \ref{fig:XrayLumBlackHoleMassScatterPlot2}. In order to compare these quantities, we assume that the bolometric luminosity = 30$\times$X-ray luminosity in the 2.0$-$10.0\,keV band following \citep{panessa06}. This assumption is very simplistic, as the bolometric luminosity relies on the shape of the spectral energy distribution for the AGN, which could differ amongst LLAGN types. Furthermore, the value of 30 is valid for more powerful AGN, but observationally, this value ranges from 3 to 16 \citep{ho99}, approximately consistent with the theoretical calculations of optically-thick and geometrically thin accretion disks \citep{Netzer2019}. Hence, as discussed in \citet{panessa06}, for a lower scaling value of $L_{\rm  bol}$ to $L_{\rm  X-ray}$, for example 10, the lines on Figure~\ref{fig:XrayLumBlackHoleMassScatterPlot2} would drop by a factor of 3. However, even with these approximations, the 1, 2 and 3$\sigma$ shaded distributions\footnote{The contour plots are made with \texttt{corner} \citep{corner}.} in Figure~\ref{fig:XrayLumBlackHoleMassScatterPlot2} show that the Seyferts are associated with the higher a mixture of Eddington ratios, from values $\gtrsim$10$^{-3}$, but with a number below this dividing line region with Eddington ratios of $\lesssim$10$^{-3}$. 
Furthermore, the contour plots show the similarity of the LINERs and ALGs, suggesting that the ALG population are similar to low X-ray luminosity LINERs. We performed a two-dimensional, two-sample KS test on these regions for ALGs and LINERs, and found a p-value of 0.06. This p-value suggests evidence for the null hypothesis, indicating that these two samples are not statistically different from one another. For all other combinations, the p-values returned were $\ll 0.05$ and so we can reject the null hypothesis, i.e., the distributions are not drawn from the same sample. For the \ion{H}{ii} galaxies, they are found in a similar X-ray luminosity region to the LINERs (and ALGs), but at lower masses. 

As a final analysis, Figure~\ref{xhistb} shows histograms of the ratio between the X-ray luminosity and black hole mass, which can be used as a tracer of the accretion rate, split once more by AGN and Hubble types. For the different AGN-types, including the `inactive' galaxies, \ion{H}{ii} galaxies have the largest ratios of X-ray luminosity to black hole mass, assuming the emission is from AGN-activity. The next largest ratio are found in Seyferts, suggesting that they are higher accretion rate objects. LINERs and ALGs have lower ratios of X-ray luminosity to black hole mass, consistent with the lower accretion rates in Figure~\ref{fig:XrayLumBlackHoleMassScatterPlot2}. However, it should be noted that \ion{H}{ii} galaxies generally fall in the lower mass bins in Figure~\ref{xhistb} and have larger error bars. To test whether the black hole masses may be driving this correlation, we removed all sources with mass errors in log space $>$0.1, so as to only include those with robust mass measurements. This cut leaves seventy sources and also has the effect of removing sources with black hole masses $<$10$^{6}$M$_{\rm BH}$. In-so-doing, the \ion{H}{ii} galaxies appear to have more similar accretion rates to those of Seyferts and LINERs, with all \ion{H}{ii} galaxies with ratios of X-ray luminosity to black hole mass larger than six being removed. In terms of the Hubble types, spirals and irregular galaxies show higher X-ray to black hole mass ratios, which may be correlated with the prevalence of \ion{H}{ii} galaxies having larger ratios. Ellipticals and lenticulars are more prevalent in the lower X-ray luminosity to black hole mass bins, suggesting lower accretion rates.

\subsection{Spectral Properties of the X-ray sources}
\label{sec:spectralfitsresults}

We now look to the best fit \textit{Chandra} spectra described in Section~\ref{sec:Xrayspec}. The spectra and best fit flux values are reported in the Appendix for each galaxy in Table~\ref{tab:basicspeec}. Here, we analyse the spectra in more detail, comparing them to the AGN type and galaxy morphological types. 

\begin{figure}
    \centering
    \includegraphics[width=0.98\columnwidth]{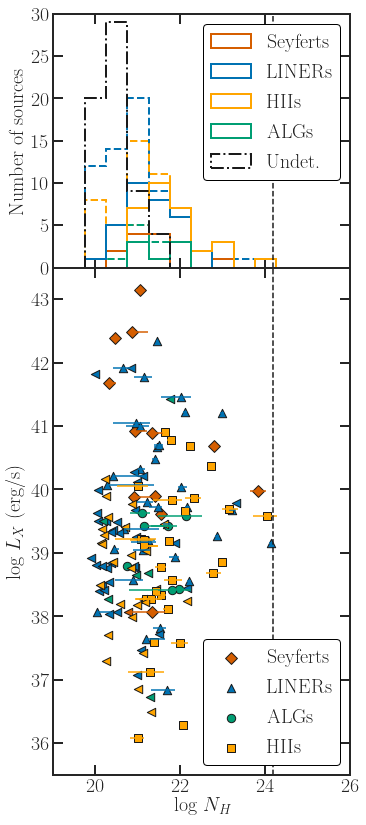}
    \caption{The X-ray luminosity of the entire X-ray sample as a function of absorbing column (N$_{\rm H}$). For sources with detection significances between 3$-$5$\sigma$ in the image, we fixed the photon index to 1.8 and obtained a host galactic absorption (see Section~\ref{sec:Xrayspec}). For the objects with luminosities $>$5$\times$ the detection significance in their image, we used the best fit values, summing together all neutral absorbers and report in this plot the total absorption from the fits. We also show in the \textit{top panel} the Galactic absorption along the line-of-sight to the galaxy for the undetected sources. We show histograms of the N$_{\rm H}$ distribution for detected sources where we could contrain the absorbing column by AGN type (solid lines) and all sources including upper limits by AGN type (dashed lines). In the \textit{bottom panel}, we split the sources by AGN Type. In all cases, left pointing triangles denote upper limits. In both panels we draw a black dashed line which denotes our "Compton-thick" definition, of which no sources are to the right of this line, although some other sources do have values which are close to or their uncertainties pass over this line. }
    \label{fig:phabs}
\end{figure}

\subsubsection{Absorption}

In Section~\ref{sec:Xrayspec} we fitted all spectra with an absorbed power-law and additional simple models where an absorbed power-law was insufficient. For objects detected with a significance of $>$5 $\sigma$, we summed all the neutral absorbers and report a total absorption from these additional models. For sources detected between 3$-$5 $\times$ the detection significance in their image, we fixed the photon index to 1.8 to find the total absorbing column. From this analysis, we are able to report the total absorbing column, which we show in Figure~\ref{fig:phabs}, binned by AGN type. In this figure, we also show the Galactic line-of-sight $N_{\rm H}$ contribution for the undetected sources in the upper histogram. For a large number of the detected sources (80/150, 53 per cent), only upper limits to the host galaxy absorbing columns were found. Excluding sources with upper limits on their host galaxy absorption, we have 70 remaining galaxies, from which we could ascertain a value of $N_{\rm H}$. Of these sources, 49 (70 per cent) have total absorbing columns of $<$10$^{22}$ cm$^{-2}$. These sources therefore are not heavily obscured as they have an Hydrogen column densities less than the average Galactic value, in contrast with studies of other AGN \citep{Burlon2011,Ricci2015,Boorman}. A source with an absorbing column $>$1.5$\times$10$^{24}$ cm$^{-2}$ is called "Compton-thick", which denotes that it is significantly obscured \citep{Comastri,Boorman}. In our sample, we find no sources which can be considered Compton-thick, although four sources (NGC~1161, NGC~3729, NGC~4111 and NGC~7640) have an absorbing column of $>$1.0$\times$10$^{23}$ cm$^{-2}$. We therefore confirm that the nuclei of our sample are typically unobscured, with a large fraction (129/150, 86 per cent) have host galaxy absorbing columns of less than that of the Galaxy, e.g., $<$10$^{22}$ cm$^{-2}$.

Obtaining a reliable estimate of the obscuring column density from X-ray spectra requires a reliable estimate of photoelectric absorption as well as reprocessing (collectively Compton scattering and fluorescence) of the intrinsic AGN emission. For heavily obscured and Compton-thick AGN, this reprocessing can dominate the observed spectral emission, manifesting as a flat spectrum at $E$\,$\lesssim$\,10\,keV, a strong neutral Fe\,K$\alpha$ fluorescence line at rest energy 6.4\,keV and a broad Compton hump peaking at $E$\,$\sim$\,30\,keV (e.g., \citealt{Lightman1988, Reynolds99, Matt00, Murphy09}). However, the Fe\,K$\alpha$ fluorescence line is not always found to be strong in Compton-thick AGN (e.g., \citealt{Gandhi17, Boorman}) and our spectral coverage provided by \textit{Chandra} does not cover the Compton hump $>$\,8\,keV. It is thus difficult to constrain high columns with our phenomenological modelling presented here. In fact, a number of our sources have been classified as Compton-thick by previous works using spectra above 10\,keV: NGC\,2273 \citep{Brightman17}; NGC\,3079 \citep{Brightman17,Marchesi2018}; NGC\,3982 \citep{Kammoun}; NGC\,4102 \citep{Ricci2015,Marchesi2018} and NGC\,5194 \citep{Brightman17}.

Interestingly, NGC\,3079 and NGC\,4102 are found to display non-Seyfert BPT line ratios in our sample (LINER and ALG, respectively), yet are intrinsically-luminous heavily obscured Seyferts. In addition, some of the LINERs in our sample (NGC\,2639, NGC\,4589, NGC\,5005, NGC\,5866 and NGC\,7331) may also be Compton Thick as shown from a sample of X-ray spectra of LINERs \citep{Gonzalez2015}. This hints to a population of obscured Seyferts amongst the optically-classified ALGs, HIIs and LINERs, that hard X-ray spectra could elucidate. However, due to a significant lack of \textit{NuSTAR} coverage in our sample (41/280\,$\sim$\,15\,per cent), a statistically-complete spectral analysis combining \textit{Chandra} and \textit{NuSTAR} is currently not possible. Future observations with \textit{NuSTAR} will enable broadband spectral fits with \textit{Chandra} using physically-motivated obscurer models which are capable of constraining high columns, even into the Compton-thick regime (e.g., \citealt{Masini19, Kammoun19, LaMassa19}). Such observations will shed light on the true proportion of Seyferts in our sample, as well as the Compton-thick fraction in the local Universe.

\subsubsection{Photon index}

\begin{figure}
	\includegraphics[width=0.48\textwidth]{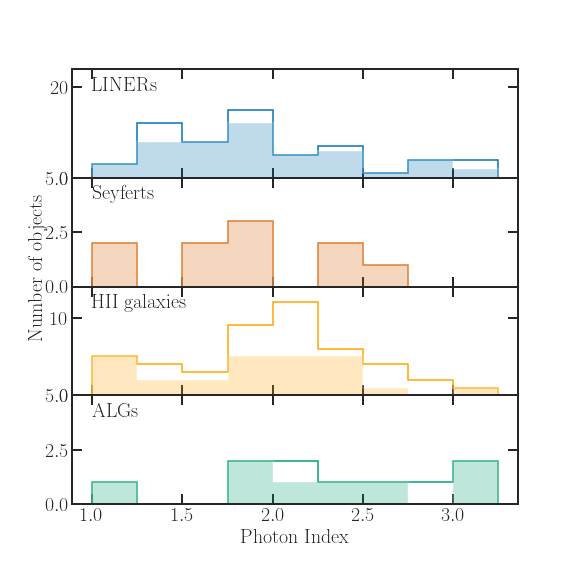}\\
    \includegraphics[width=0.48\textwidth]{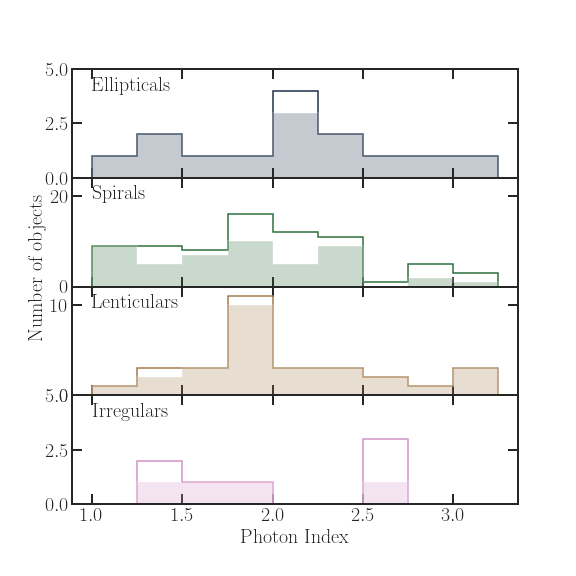}
    \caption{Histograms of the Photon Index from the best fit models of the detected sources in the entire sample, separated by AGN Type (top panel) and Hubble type (bottom panel). The solid lines in all cases represent the best spectral fits, whereas the filled histograms represent the sources with luminosities greater than 10$^{39}$~erg s$^{-1}$.}
    \label{fig:photb}
\end{figure}

Figure~\ref{fig:photb} shows the photon index from the spectral fits (see Section~\ref{sec:Xrayspec}) of the entire sample, broken down into AGN types and Hubble types. In both cases we separate all the data (solid line histograms) from those that had luminosities $>$10$^{39}$~erg s$^{-1}$ (filled histograms). While most of these classifications have a peak number of sources within the photon index range 1.75$-$2.00 or in adjacent bins, there is no clear distinction in the distributions between any AGN type or Hubble type. LINER and Seyfert galaxies, spirals and lenticulars show a peak in the 1.75$-$2.0 bin. The distributions for the photon indices in all classification types span the range of fit values found in Section~\ref{sec:Xrayspec} (see Figure~\ref{fig:phot}). Furthermore, removing sources with luminosities $<$10$^{39}$~erg s$^{-1}$ does not reduce the breadth of these histograms. We therefore find no evidence that spectral index is significantly affected by optical spectral or morphological type. This finding hints at the variety of different X-ray emitting nuclei in the local Universe and that if interpreted as AGN, that these nuclei can be found across a wide range of galaxy types and AGN classifications.

\subsubsection{Eddington Ratio}

\begin{figure}
	\includegraphics[width=\columnwidth]{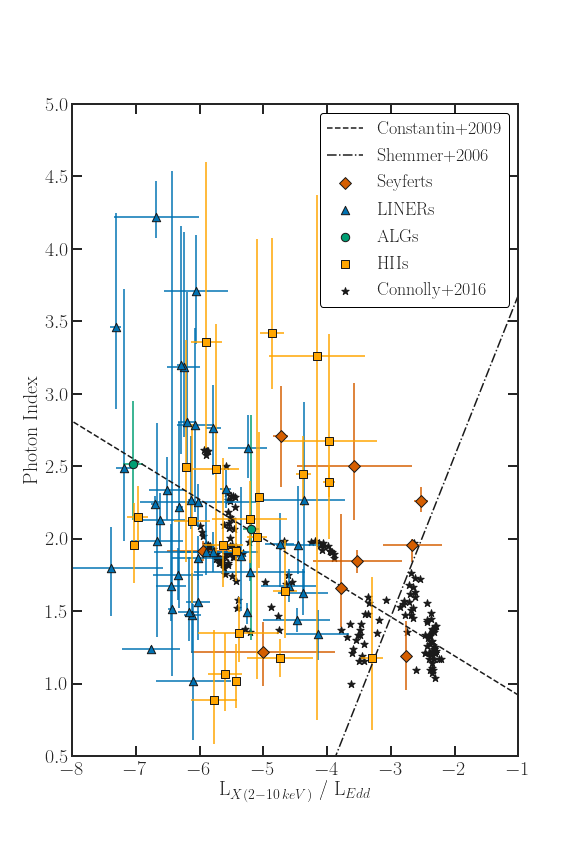}
    \caption{Photon index of the entire sample where reliable photon indices were extracted from the spectra and the source luminosity is $>$10$^{39}$~erg s$^{-1}$ plotted against the tracer of the accretion rate, defined by the ratio between $L_{\rm X-ray (2--10\,keV)}$ and $L_{\rm Edd}$, split by different AGN types. We also plot the results from the study by \citet{Connolly2016}, which includes a sample of 24 Palomar galaxies, but we include averaged values of the data points included in that study. The lines correspond to the fits from \citet{Shemmer2006} for higher Eddington rate radio-quiet AGN and \citet{Constantin2009} for LLAGN.}
\label{fig:connolly}
\end{figure}

The accretion rate can be approximated using the ratio between the X-ray luminosity and the Eddington luminosity, as described in the previous Section. We now plot the photon index as a function of this ratio in Figure~\ref{fig:connolly}, similar to that performed by \citet{Connolly2016} for a sample of 24 Palomar galaxies (their Figure~8). We only plot sources with luminosities $>$10$^{39}$~erg s$^{-1}$ and have a reliable photon index fit value i.e., no upper limits. This cut limits our sample to 63 sources. To compare directly to \citet{Connolly2016}, we use the harder 2$-$10\,keV band fluxes for this figure. We also include the fits for higher Eddington rate radio-quiet AGN \citep{Shemmer2006} and for LLAGN \citep{Constantin2009}, as shown in \citet{Connolly2016}. Our data generally probe the lower ($<$10$^{-3}$ $L_{X}$/$L_{Edd}$) regime, although a handful of objects (notably NGC 4051, NGC 4395 and NGC 5548), are above this threshold. The overall trend follows that of \citet{Constantin2009} for the lower accretion rate objects ($<$10$^{-3}$ L$_{X}$/L$_{Edd}$). We do not have enough objects which have accretion rates $>$10$^{-3}$ $L_{X}$/$L_{Edd}$ to compare with the results of \citet{Shemmer2006}, but we do note that most of the objects that fall near this line are Seyferts and broadly follow the correlation. However, given the large errors and scatter in the range of photon indices for this sample, it is difficult to draw significant conclusions on these data. Longer exposure observations are required to further constrain the photon indices in these objects to further probe these relations.

\subsection{X-ray compared to [\ion{O}{iii}] line luminosity}
\label{sec:OIII}
 \begin{figure}
\includegraphics[width=\columnwidth]{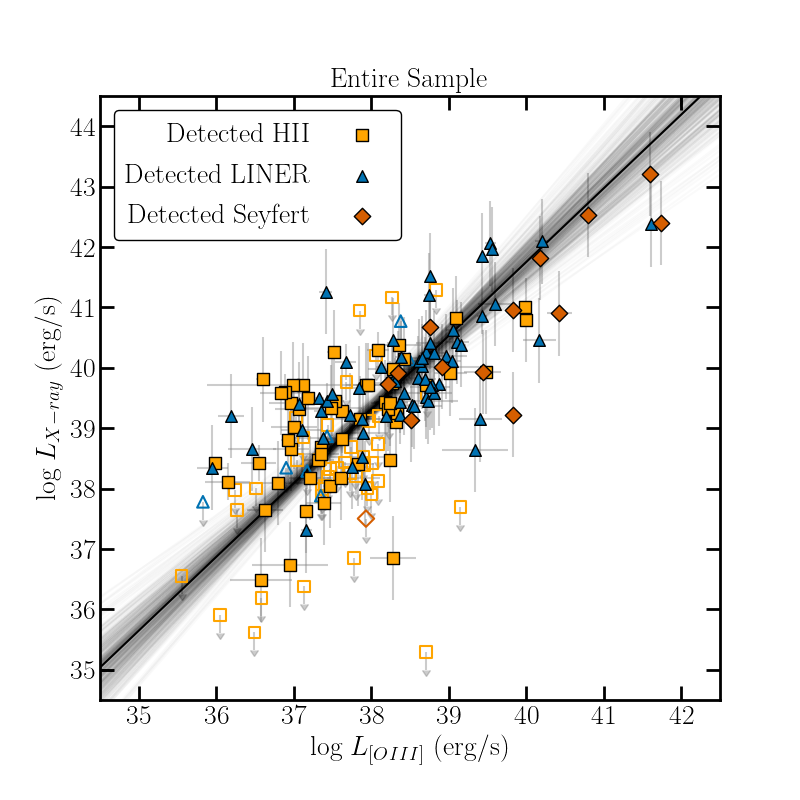}
\includegraphics[width=\columnwidth]{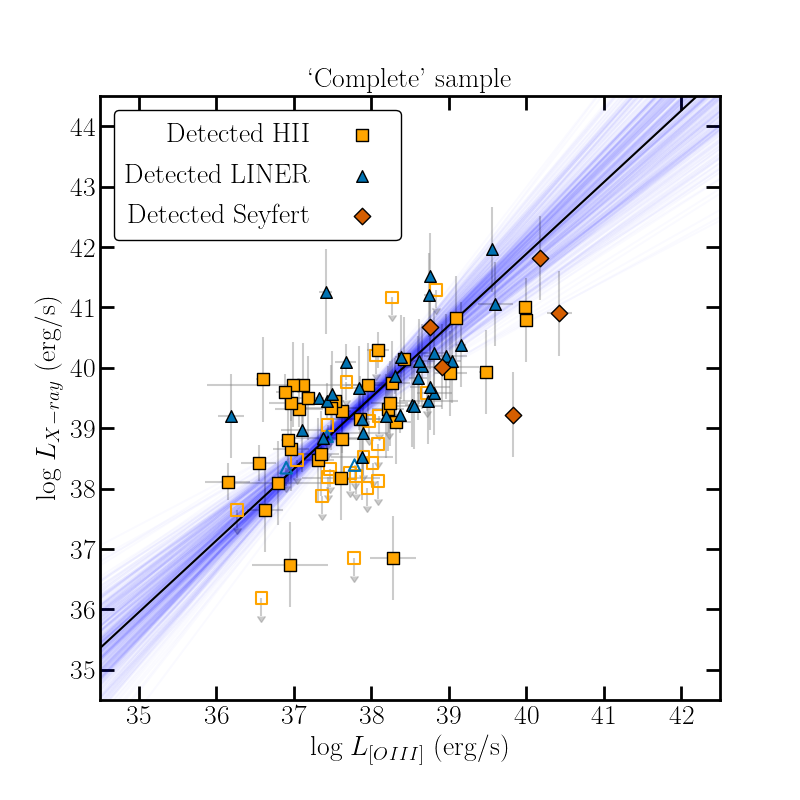}
        \caption{The [\ion{O}{iii}] luminosity vs unabsorbed 0.3$-$10.0\,keV luminosities for the entire (\textit{top} panel) and `Complete' (\textit{bottom} panel) X-ray sample. The different optical classes are coded (symbol and colour) in the plot according to the legend. The filled symbols refer to the detected X-ray sources, while the empty symbols refer to undetected X-ray sources. 
The black line represents the correlation discussed in Section~\ref{sec:OIII} and the black/blue (entire/Complete) region shows 400 draws from the fit, to give an idea of the fit uncertainty. The correlation information is given in the text.}
\label{fig:X-rayOIII}
\end{figure}

When X-ray observations are not available, the forbidden [\ion{O}{iii}] line luminosity, which is easier to measure from ground based instruments, is used as a proxy for the X-ray luminosity \citep[e.g.,][]{panessa06,hardcastle09,gonzalez09,Saikia,Saikia2018}. The above relationships have generally been obtained at relatively high luminosities. Given the usefulness of this relationship it is important to take advantage of our present \textit{Chandra} data to derive the relationship with to lower X-ray luminosities. Future observations can provide [\ion{O}{iii}] luminosities of similarly improved spatial resolution as the Palomar line data used here from \citet{ho97a} is not of particularly high spatial resolution and may be contaminated by some level of SF. 

The \textit{top} panel of Figure~\ref{fig:X-rayOIII} shows the [\ion{O}{iii}] line luminosity obtained from the Palomar survey of all the detected and undetected X-ray sources plotted against the X-ray luminosity in the 0.3$-$10\,keV band. This correlation for the `Complete' sample is shown in \textit{bottom} panel of Figure~\ref{fig:X-rayOIII}. Once more, we use the \textsc{LINMIX} package to include upper limits in the X-ray luminosity and show all of our correlation results in Table~\ref{tab:tabfits}. The correlation is of the form L$_{\rm X-ray} \sim$L$_{\rm [\ion{O}{iii}]} ^{1.22 ^{+0.11}_{-0.10}}$ with a scatter of 0.75 dex for the entire sample ($\hat{\rho}$ = 0.83$^{+0.04}_{-0.05}$) and L$_{\rm X-ray} \sim$L$_{\rm [\ion{O}{iii}]}^{1.19^{+0.18}_{-0.17}}$ with a scatter of 0.83 dex for the `Complete' sample ($\hat{\rho}$ = 0.79$^{+0.07}_{-0.08}$).

The scatter in the correlation is larger at lower X-ray and [\ion{O}{iii}] line luminosities. The LINERs and Seyferts have the highest [\ion{O}{iii}] and X-ray luminosities of the sample, whereas the \ion{H}{ii} galaxies are at the lowest X-ray and [\ion{O}{iii}] luminosities. However, there is a lot of mixing between all the classes, so no clear region for each optical AGN class emerges. Fitting each of the three optical AGN types separately for the entire sample, we arrive at fits of L$_{\rm X-ray} \sim$L$_{\rm [\ion{O}{iii}]} ^{1.37_{-0.30}^{+0.38}}$ with a scatter of 0.66 dex for Seyferts ($\hat{\rho}$ = 0.93$^{+0.05}_{-0.13}$), L$_{\rm X-ray} \sim$L$_{\rm [\ion{O}{iii}]} ^{1.11_{-0.13}^{+0.13}}$ with a scatter of 0.52 dex for LINERs ($\hat{\rho}$ = 0.88$^{+0.04}_{-0.06}$) and L$_{\rm X-ray} \sim$L$_{\rm [\ion{O}{iii}]} ^{1.49^{+0.40}_{-0.33}}$ with a scatter of 0.92 dex for \ion{H}{ii} galaxies ($\hat{\rho}$ = 0.67$^{+0.11}_{-0.13}$). These fits are all consistent with one another, within the uncertainties and show a strong positive correlation. Furthermore, the low scatter in the LINERs and Seyferts shows a tight correlation between the [\ion{O}{iii}] line luminosity and X-ray luminosity in these objects, whereas the larger scatter in the \ion{H}{ii} galaxies add scatter to the overall relation, especially at the lowest luminosities.

Previous work has shown that the X-ray-[\ion{O}{iii}] correlations tend to show a $\sim$1:1 ratio, similar to that found here for the Seyferts \citep{panessa06,hardcastle09}. The X-ray LeMMINGs data corroborates these findings, indicating that the X-ray and [\ion{O}{iii}] ionising radiation are coupled down to very low luminosities, albeit with some additional scatter. A similar correlation is found for LINERs \citep[e.g.,][]{gonzalez09} and it is interesting that the \ion{H}{ii} galaxies follow a similar correlation to the Seyferts and LINERs. However, we note that the X-ray emission from the LINERs may be coming from the jet rather than accretion flow \citep[e.g., see ][]{BalmaverdeCapetti}. 

\section{Discussion}

We have presented X-ray observations of 213/280 objects in the Palomar survey above $\delta$ = 20$^{\circ}$, that as of June 2018 have been observed with \textit{Chandra}. Altogether, 150/213 (70 per cent) objects were detected, which is higher than previous studies of a subset of the Palomar sample using poorer resolution/quality data from \textit{ROSAT} \citep[54 per cent,][]{RobertsWarwick} and \textit{Chandra} \citep[62 per cent,][]{HoUlvestad}. For matching resolution X-ray observations with \textit{Chandra} presented by \citet{She2016}, we find similar detection fractions for all types of AGN. We find that almost all Seyferts, ALGs and LINERs have a nuclear X-ray core, co-incident with the optical nucleus of the galaxy. In addition, around half of the \ion{H}{ii} galaxies in the sample are detected in the X-rays at the optical nucleus. As to the nature of these X-ray sources, the LINERs and Seyferts appear to be associated with luminosities $\ga$10$^{39}$~erg s$^{-1}$ in most cases, but the \ion{H}{ii} galaxies and ALGs represent a less-luminous population of nuclear X-ray emission that may be due to LLAGN, or potentially other X-ray sources in the galaxy such as ULXs or XRBs. To ascertain the nature of the nuclear emission, we used the other multi-wavelength data available, which we discuss below in the case of each of the AGN types.

\subsection{Seyferts}
There are 14 Seyferts in the entire X-ray LeMMINGs sample, of which 13 are detected (93 per cent), but this detection fraction is 100 per cent in the Complete sample (5/5 objects). The only Seyfert that is not detected is NGC~3486, which has only a $\sim$2\,ks observation and is a type II Seyfert which may be obscured by large absorption along the line of sight. The Seyferts have the highest X-ray luminosities ($>$10$^{39}$~erg s$^{-1}$), occupying regions of higher X-ray luminosity per black hole mass bin and [\ion{O}{iii}] emission line luminosity. We do not report a correlation between the X-ray luminosity and the black hole mass for Seyferts. The lack of correlation is likely caused by the large scatter in any relation due to the wide range of Eddington values in this class (see Figure~\ref{fig:XrayLumBlackHoleMassScatterPlot2}), with some Seyferts showing Eddington ratios as low as 10$^{-6}$. The low number of Seyferts in this sample prevents us from making firmer conclusions on this correlation.
The higher Eddington ratio ($\lambda\sim$10$^{-2}$) Seyferts  obtained from the X-ray data indicate that in general, Seyferts accrete efficiently, likely in the form of an optically-thick, geometrically thin accretion disk \citep{ShakuraSunyaev}. They are likely scaled down versions of the more powerful AGN in quasars, accreting at lower Eddington ratios \citep{panessa06}. However, for the Seyferts that have lower Eddington ratios ($\lesssim$10$^{-3}$ \citealt{Merloni03}), they could possibly be powered by some form of radiatively inefficient accretion flow (see Section~\ref{sec:52}).

\subsection{LINERs}
\label{sec:52}
Of the 77 LINERs in the entire X-ray sample, 69 of them are detected (90 per cent), which is similar to the detection rate in the `Complete' sample (35/38, 92 per cent). They are mostly detected with luminosities $\gtrsim$10$^{39}$~erg s$^{-1}$ and are associated with the highest BH masses. They follow a similar X-ray/[\ion{O}{iii}] gradient to the Seyferts.
LINERs are often described by some form of radiatively inefficient accretion flow \citep[RIAF e.g.,][]{Narayan,NarayanADAF}. Our X-ray observations support this interpretation as the inferred Eddington ratios in Figure~\ref{fig:XrayLumBlackHoleMassScatterPlot2} indicate that most LINERs have $\lambda$ $\lesssim$10$^{-4}$ and much weaker X-ray luminosities than those of Seyferts. 
X-ray emission from shocks or post-AGB stars would not be able to explain such high nuclear X-ray luminosities \citep{Allen08,sarzi10,CapettiBaldi2011,Singh2013}. We also note that some of the X-ray emission in the LINERs may come from the jet rather than the accretion flow \citep[see][]{BalmaverdeCapetti,balmaverde08}. We therefore suggest that most of the LINERs in our sample may be powered by a form of RIAF, but further follow-up observations are required of this sample to unequivocally determine the accretion mechanism in LINERs. 

\subsection{Absorption Line Galaxies}
Absorption Line galaxies (ALGs) have often been missed in previous X-ray surveys, as they are not considered `active', like the Seyferts and LINERs. 
However, they have a reasonable detection fraction 13/22 (59 per cent) in the entire sample and 8/11 (73 per cent) in the `Complete' sample. They are associated with X-ray luminosities similar to the LINERs, around 10$^{39}$~erg s$^{-1}$. ALGs tend to be better detected in the higher mass bins but appear in similar regions to LINERs in terms of Eddington ratio (Figure~\ref{fig:XrayLumBlackHoleMassScatterPlot2}), it is possible that they may have a common central engine \citep[][]{baldi10b,BaldiLeMMINGs,BaldiLeMMINGs3}. 
Nonetheless, the ALGs are mostly associated with elliptical galaxies and the lack of a detection of the [\ion{O}{iii}] line limits the interpretation of the central engine. Furthermore, the low implied X-ray luminosities in some of the ALGs (see upper limits in Figure~\ref{xhist}), indicates that they may not be identical to LINERs.
An alternative explanation for the central engines of ALGs is the nuclear recurrence scenario due to an intermittent accretion phenomenon \citep{reynolds97,czerny09}. Further dedicated studies at higher sensitivities of ALGs should be undertaken to ascertain the true cause of their X-ray and multi-wavelength emission.

\subsection{\ion{H}{ii} galaxies}
Of the 100 \ion{H}{ii} galaxies observed by \textit{Chandra} in our sample, 57 were detected in the X-ray (57 per cent). In the `Complete' sample, the \ion{H}{ii} galaxies have a similar detection fraction: 35/58 (60 per cent). Of the detected objects, 32/57 (56 per cent) are of X-ray luminosities $\ga$10$^{39}$~erg s$^{-1}$, similar to that of LINERs and Seyferts. The \ion{H}{ii} galaxies span a range of accretion rates and BH masses, but follow similar correlations to the Seyferts and LINERs, specifically the X-ray/$\rm M_{BH}$ and X-ray/[\ion{O}{iii}] relations, and especially at the higher black hole masses. Therefore the \ion{H}{ii} galaxies with a detected X-ray core, and X-ray luminosity $\ga$10$^{39}$~erg s$^{-1}$ and a BH mass $\gtrsim$10$^{7}$M$_{\odot}$, are likely powered by an AGN. But, there could still be some contribution from SF processes to both the [\ion{O}{iii}] and X-ray luminosities. It is not clear whether the majority of these objects are powered by ineffcient flows, similar to LINERs, or a more efficient accretion mode, similar to Seyferts. However, \citep[][]{BaldiLeMMINGs,BaldiLeMMINGs2} found some `jetted' \ion{H}{ii} galaxies which could be more consistent with LINER-like activity. For the 25 \ion{H}{ii} galaxies that do not fulfil these requirements, e.g., detected \ion{H}{ii} galaxies at lower masses and X-ray luminosities, their central engines are more uncertain, and further investigation is required to classify them as genuine LLAGN or imposters in the form of XRBs and ULXs (see next Section). 
Furthermore, it should be noted that the central object of 43 per cent of the \ion{H}{ii} galaxies are undetected, so deeper X-ray observations are needed in order to fully understand the central engines in these objects.

\subsection{Is there any contamination from XRBs and ULXs?}

Given the 0.5 arcsec on-axis resolution of \textit{Chandra}, we considered X-ray sources which lie within 2 arcsec of the optically defined nucleus as likely being powered by a central SMBH. However, some of these sources may not be an AGN, but may be a ULX/XRB, as is observed in NGC~3034 \citep{MuxlowM82}. In addition, nuclear star formation from O/B star associations can be as luminous as 10$^{35}$~erg s$^{-1}$ \citep{Oskinova} and could cause further contamination. We computed XLFs for the sample, and showed that it is unlikely we are heavily biased to non-AGN sources contaminating the nucleus, as our XLFs are shallower than those expected of ULXs and XRBs \citep[see][and references there-in]{She2016}. However, the XLFs are steeper when the `inactive' galaxies are included in both the entire and `Complete' samples, so some small contamination may be possible in those galaxies. If a discriminating X-ray luminosity of 10$^{39}$~erg s$^{-1}$ is used as a criterion to remove non-AGN sources, then 13/13 detected Seyferts, 56/68 LINERs, 11/13 ALGs and 32/56 \ion{H}{ii} galaxies would be considered AGNs for the entire sample. By Hubble types, 14/15 ellipticals, 59/91 spirals, 33/35 lenticulars and 6/9 irregulars would be considered AGNs. This definition shows that of all types, \ion{H}{ii} galaxies, spirals and irregulars are the most likely to be contaminated with ULX/XRBs. Unsurprisingly, this link has been observed previously between star formation and prevalence of ULXs in a galaxy \citep{King2004, Gilfanov2004, Swartz2009}. But, it is also possible for a $<$10$^{39}$~erg s$^{-1}$ LLAGN to co-exist with ULX/XRBs in a galactic nucleus. Therefore, additional information is needed to categorically remove spurious non-AGN sources from the sample. However, SMBHs are more radio loud than stellar mass sized black holes and in a future publication (Pahari et al. in prep), we will include radio luminosity in our decision as to whether nuclear sources are SMBHs or a contaminating source.

\section{Conclusions}

We have presented archival and new \textit{Chandra} X-ray observatory data for 213/280 (76 per cent) objects in the declination limited ($\delta$ $>20^{\circ}$) Palomar sample, 
and 112/113 of the Palomar galaxies in the sub-sample between 40$^{\circ}$ $<$ $\delta$ $<$ $65^{\circ}$, which we refer to as the 'Complete' sample. Although most galaxies have a considerably longer exposure time, we achieve a minimum observation time of 10\,ks on all observed targets. Using these data, we achieve a background flux level of $\sim$1.65$\times$10$^{-14}$~erg cm$^{-2}$ s$^{-1}$. The entire X-ray LeMMINGs sample has detected X-ray emission co-incident with the optical centre of 150/213 (70 per cent) of the observed galaxies across all optical AGN types and galaxy morphologies. 
The 150 X-ray detected galactic nuclei were fit with simple spectral models in \textsc{XSPEC} and their fluxes computed in the 0.3$-$2.0, 2.0$-$10.0 and 0.3$-$10.0\,keV bands. The detection rate of our \textit{Chandra} data of nuclear X-ray emission compares favourably to previous X-ray studies of the Palomar sample: 70 per cent detections compared to 54 per cent in \cite{RobertsWarwick} and 62 per cent in \cite{HoUlvestad}. Comparing to a previous work based on \textit{Chandra} data by \citet{She2016}, we find broadly similar results: X-ray emission associated with the AGN is observed $\sim$80 per cent of the time in Seyferts, LINERs and ALGs, but also in \ion{H}{ii} galaxies in 50 per cent of objects. 

We determined an X-ray luminosity function (XLF) from the data in our sample and fit a simple power-law of -0.54$\pm$0.06 for the entire sample and -0.45$\pm$0.05 for the `Complete' sample, for all galaxies. Our data probes lower X-ray luminosities than most previous studies \citep[e.g.][]{Ajello2012}, extending the X-ray luminosity function down to 10$^{36}$~erg s$^{-1}$, two orders of magnitude lower than the previous \textit{Chandra} observations \citep{She2016}. We further split the entire and `Complete' samples into `active' sources such as LINERs and Seyferts and all sources, in order to show the differences in XLFs when `inactive' galaxies such as HII and absorption line galaxies are included. We found an XLF power-law of $-0.48\pm0.12$ and $-0.24\pm0.11$ using all galaxies, for the entire and `Complete' samples samples, respectively. In both the entire and `Complete' samples samples, the inclusion of the `inactive' galaxies increased the gradient of the XLF, which may suggest contamination from non-AGN objects, though we note that the power-law fit values are consistent with those found for all galaxies. Furthermore, our single power-law fits are consistent with previous \textit{Chandra} studies of local galaxies \citealt[][]{She2016}. 

In terms of the empirical correlations between the X-ray luminosity and other diagnostics of SMBH activity, e.g., BH mass and [\ion{O}{iii}] line luminosity, correlations were obtained for different optical AGN classes. We fitted the data including upper limits to the X-ray luminosity and black hole masses obtained from the $M-\sigma$ relation or dynamical mass measurements, finding an overall relationship for the entire sample to be $L_{\rm X-ray} \propto {\rm M}_{BH}^{0.78 \pm 0.07}$ with a scatter of 1.08 dex, and for the `Complete' sample of $L_{\rm X-ray} \propto {\rm M}_{BH}^{0.62 ^{+0.11}_{-0.10}}$, with an scatter of 1.05 dex. No strong correlation was observed for Seyferts between the X-ray luminosity and BH mass, which may be due to the low number of Seyferts in this sample. The \ion{H}{ii} galaxies, ALGs and LINERs follow similar correlations in the X-ray$-$BH mass plane. We also note that the detection fraction is much higher amongst higher mass objects (88 per cent for $M_{\rm BH}>$10$^{8}$M$_{\odot}$) than in lower mass objects  (50 per cent for $M_{\rm BH}<$10$^{6}$M$_{\odot}$).
By comparing the black hole masses to the Eddington luminosity, and assuming a bolometric correction factor of 30$\times$L$_{X}$, we showed that the LINERs and ALGs all have very low Eddington ratios ($\lambda$ $<$10$^{-3}$). The Seyfert galaxies tend to have higher Eddington ratios but there are some notable exceptions at lower accretion rates. \ion{H}{ii} galaxies can have a mixture of accretion rates, which may be due to their prevalence with lower mass nuclei, skewing their numbers to higher accretion rates artificially. 

We also fitted the spectra of all the detected sources, using simple absorbed power-law models in most cases. We found that the best fit spectra indicated a preference for a photon index in the range 1.75$-$2.0, consistent with previous studies of brighter AGN. We therefore fixed the photon index to 1.8 for the faintest sources and re-fitted the spectra to ascertain the host galaxy absorption across the sample. We found that for 53 per cent of sources only upper limits to the host galaxy absorption were possible, and for those where reliable fits to the photon index were found, 70 per cent have absorbing column densities of less than the value through our Galaxy. 
As a final analysis with the spectra, we plotted the photon index as a function of the accretion rate. We found that the vast majority of sources in our sample with X-ray luminosities above 10$^{39}$~erg s$^{-1}$ followed the relationship for LLAGN proposed by \citet{Constantin2009}, but we note that further observations are required for refinement of this analysis and to better probe the radio-quiet AGN relationship defined by \citet{Shemmer2006}. This behaviour shows harder when brighter X-ray spectra for the low-luminosity AGNs we present here, with some notable exceptions of powerful AGNs, such as NGC 4051, NGC 4385 and NGC 5548.

A correlation is observed for all sources between the X-ray and [\ion{O}{iii}] line luminosities, down to 10$^{36}$~erg s$^{-1}$, lower than that probed in previous studies \citep{panessa06,hardcastle09}. When including upper limits, we find $L_{\rm X-ray} \propto$L$_{\rm [\ion{O}{iii}]} ^{1.22 ^{+0.11}_{-0.10}}$ for the entire sample and L$_{\rm X-ray} \propto$L$_{\rm [\ion{O}{iii}]} ^{1.19^{+0.18}_{-0.17}}$ for the `Complete' sample, but with significant scatter about the best fit lines. This correlation compares favourably with the correlations ~$\sim$1:1 ratio between the X-rays and [\ion{O}{iii}] line found in previous studies. However, we note that different AGN types follow slightly different tracks in the X-ray:[\ion{O}{iii}] plane, with the LINERs showing the smallest scatter of all classes, suggesting a strong coupling between the two variables. 

In conclusion, our sample provides the most statistically-complete and unbiased surveys of accretion in the nearby Universe performed to date, for both `active' and `inactive' galaxies. Further work is ongoing to characterise the off-nuclear X-ray sources, including their timing, spectral and multi-wavelength properties (Pahari et al. in prep) and future work will include establishing the fundamental plane of black hole activity with this data \citep[][]{Saikia2018B} with the wider LeMMINGs sample \citep{BaldiLeMMINGs} at sub-arcsecond resolution and the forthcoming 5\,GHz equivalent LeMMINGs \emerlin{} study (Williams et al. in prep.).

\section*{Data Availability}

All of the \textit{Chandra} X-ray data presented here can be downloaded from the public heasarc archives, noted in the manuscript in Section~\ref{sec:sample}. The values from the fitting procedures, fluxes, luminosities and spectra can be obtained from the online supplementary material. 

\section*{Acknowledgements}

We would like to acknowledge the support of the LeMMINGs project, upon which this study is based. We acknowledge funding from the Mayflower Scholarship from the University of Southampton afforded to David Williams to complete this work. This work was also supported by the Oxford Centre for Astrophysical Surveys, which is funded through generous support from the Hintze Family Charitable Foundation. 
MP acknowledges Royal Society-SERB Newton International Fellowship support funded jointly by the Royal Society, UK and the Science and Engineering Board of India (SERB) through Newton-Bhabha Fund. IMcH thanks the Royal Society for the award of a Royal Society Leverhulme Trust Senior Research Fellowship. RDB and IMcH also acknowledge the support of STFC under grant [ST/M001326/1]. 
SM gratefully acknowledges the support from the Chandra grant GO7-18080X  issued by the Chandra X-ray Center, which is operated by the Smithsonian Astrophysical Observatory for and on behalf of the National Aeronautics Space Administration under contract NAS8-03060.
A.B is grateful to the Royal Society and SERB (Science and Engineering Research Board, India). A.B. is supported by an INSPIRE Faculty grant (DST/INSPIRE/04/2018/001265) by the Department of
Science and Technology, Govt. of India.
P. G. B. acknowledges financial support from the STFC and the Czech Science Foundation project No. 19- 05599Y.
S. A. gratefully acknowledges support from an ERC Adva. nced Grant 789410
J.H.K. acknowledges financial support from the European Union's Horizon 2020 research and innovation programme under Marie Sk\l odowska-Curie grant agreement No 721463 to the SUNDIAL ITN network, from the State Research Agency (AEI-MCINN) of the Spanish Ministry of Science and Innovation under the grant "The structure and evolution of galaxies and their central regions" with reference PID2019-105602GB-I00/10.13039/501100011033, and from IAC project P/300724, financed by the Ministry of Science and Innovation, through the State Budget and by the Canary Islands Department of Economy, Knowledge and Employment, through the Regional Budget of the Autonomous Community.
B.T.D acknowledges support from a Spanish postdoctoral fellowship `Ayudas 1265 para la atracci\'on del talento investigador. Modalidad 2: j\'ovenes investigadores.' funded by Comunidad de Madrid under grant number 2016-T2/TIC-2039 and support from  grant `Ayudas para la realizaci\'on de proyectos de I+D para j\'ovenes doctores 2019.' funded by Comunidad de Madrid and Universidad Complutense de Madrid under grant number PR65/19-22417.
AA and MAPT acknowledge support from the Spanish MCIU through grant PGC2018-098915-B-C21. JM acknowledges support from the grant RTI2018-096228-B-C31 (MICIU/FEDER, EU). AA, MAPT and JM acknowledge financial support from the State Agency for Research of the Spanish MCIU through the ``Center of Excellence Severo Ochoa'' award to the Instituto de Astrof\'isica de Andaluc\'ia (SEV-2017-0709).  
CGM acknowledges support from Mrs Hiroko and Mr Jim Sherwin. 
FS acknowledges partial support from a Leverhulme Trust Research Fellowship.
We also acknowledge Jodrell Bank Centre for Astrophysics, which is funded by the STFC. \emerlin{} and formerly, MERLIN, is a National Facility operated by the University of Manchester at Jodrell Bank Observatory on behalf of STFC. The scientific results reported in this article are based to a significant degree on observations made by the Chandra X-ray Observatory and on data obtained from the Chandra Data Archive. This research has made use of software provided by the Chandra X-ray Center (CXC) in the application package CIAO.
DW would also like to thank Sam Connolly, Dimitrios Emmanolopoulos and Sara Motta for useful discussions.




\bibliographystyle{mnras}

\bibliography{LeMMINGs_IV_Final_Tex_File.bib}

\begin{thebibliography}{}
\makeatletter
\relax
\def\mn@urlcharsother{\let\do\@makeother \do\$\do\&\do\#\do\^\do\_\do\%\do\~}
\def\mn@doi{\begingroup\mn@urlcharsother \@ifnextchar [ {\mn@doi@}
  {\mn@doi@[]}}
\def\mn@doi@[#1]#2{\def\@tempa{#1}\ifx\@tempa\@empty \href
  {http://dx.doi.org/#2} {doi:#2}\else \href {http://dx.doi.org/#2} {#1}\fi
  \endgroup}
\def\mn@eprint#1#2{\mn@eprint@#1:#2::\@nil}
\def\mn@eprint@arXiv#1{\href {http://arxiv.org/abs/#1} {{\tt arXiv:#1}}}
\def\mn@eprint@dblp#1{\href {http://dblp.uni-trier.de/rec/bibtex/#1.xml}
  {dblp:#1}}
\def\mn@eprint@#1:#2:#3:#4\@nil{\def\@tempa {#1}\def\@tempb {#2}\def\@tempc
  {#3}\ifx \@tempc \@empty \let \@tempc \@tempb \let \@tempb \@tempa \fi \ifx
  \@tempb \@empty \def\@tempb {arXiv}\fi \@ifundefined
  {mn@eprint@\@tempb}{\@tempb:\@tempc}{\expandafter \expandafter \csname
  mn@eprint@\@tempb\endcsname \expandafter{\@tempc}}}

\bibitem[\protect\citeauthoryear{{Ajello}, {Alexander}, {Greiner}, {Madejski},
  {Gehrels}  \& {Burlon}}{{Ajello} et~al.}{2012}]{Ajello2012}
{Ajello} M.,  {Alexander} D.~M.,  {Greiner} J.,  {Madejski} G.~M.,  {Gehrels}
  N.,   {Burlon} D.,  2012, \mn@doi [\apj] {10.1088/0004-637X/749/1/21}, \href
  {https://ui.adsabs.harvard.edu/abs/2012ApJ...749...21A} {749, 21}

\bibitem[\protect\citeauthoryear{{Akylas} \& {Georgantopoulos}}{{Akylas} \&
  {Georgantopoulos}}{2009}]{akylas09}
{Akylas} A.,  {Georgantopoulos} I.,  2009, \mn@doi [\aap]
  {10.1051/0004-6361/200811371}, \href
  {http://adsabs.harvard.edu/abs/2009A%26A...500..999A} {500, 999}

\bibitem[\protect\citeauthoryear{{Allen}, {Groves}, {Dopita}, {Sutherland}  \&
  {Kewley}}{{Allen} et~al.}{2008}]{Allen08}
{Allen} M.~G.,  {Groves} B.~A.,  {Dopita} M.~A.,  {Sutherland} R.~S.,
  {Kewley} L.~J.,  2008, \mn@doi [\apjs] {10.1086/589652}, \href
  {http://adsabs.harvard.edu/abs/2008ApJS..178...20A} {178, 20}

\bibitem[\protect\citeauthoryear{Anderson \& Darling}{Anderson \&
  Darling}{1954}]{andersondarling}
Anderson T.~W.,  Darling D.~A.,  1954, \mn@doi [Journal of the American
  Statistical Association] {10.1080/01621459.1954.10501232}, 49, 765

\bibitem[\protect\citeauthoryear{{Arnaud}}{{Arnaud}}{1996}]{xspec}
{Arnaud} K.~A.,  1996, in {Jacoby} G.~H.,  {Barnes} J.,  eds,  Astronomical
  Society of the Pacific Conference Series Vol. 101, Astronomical Data Analysis
  Software and Systems V. p.~17

\bibitem[\protect\citeauthoryear{{Baldi} \& {Capetti}}{{Baldi} \&
  {Capetti}}{2010}]{baldi10b}
{Baldi} R.~D.,  {Capetti} A.,  2010, \mn@doi [\aap]
  {10.1051/0004-6361/201014446}, \href
  {http://adsabs.harvard.edu/abs/2010A%26A...519A..48B} {519, A48}

\bibitem[\protect\citeauthoryear{{Baldi} et~al.,}{{Baldi}
  et~al.}{2018}]{BaldiLeMMINGs}
{Baldi} R.~D.,  et~al., 2018, \mn@doi [\mnras] {10.1093/mnras/sty342}, \href
  {http://adsabs.harvard.edu/abs/2018MNRAS.476.3478B} {476, 3478}

\bibitem[\protect\citeauthoryear{{Baldi} et~al.,}{{Baldi}
  et~al.}{2021a}]{BaldiLeMMINGs2}
{Baldi} R.~D.,  et~al., 2021a, \mn@doi [\mnras] {10.1093/mnras/staa3519}, \href
  {https://ui.adsabs.harvard.edu/abs/2021MNRAS.500.4749B} {500, 4749}

\bibitem[\protect\citeauthoryear{{Baldi} et~al.,}{{Baldi}
  et~al.}{2021b}]{BaldiLeMMINGs3}
{Baldi} R.~D.,  et~al., 2021b, \mn@doi [\mnras] {10.1093/mnras/stab2613}, \href
  {https://ui.adsabs.harvard.edu/abs/2021MNRAS.508.2019B} {508, 2019}

\bibitem[\protect\citeauthoryear{{Baldwin}, {Phillips}  \&
  {Terlevich}}{{Baldwin} et~al.}{1981}]{BaldwinPhillipsTerlevich81}
{Baldwin} J.~A.,  {Phillips} M.~M.,   {Terlevich} R.,  1981, \mn@doi [PASP]
  {10.1086/130766}, \href {http://adsabs.harvard.edu/abs/1981PASP...93....5B}
  {93, 5}

\bibitem[\protect\citeauthoryear{{Ballantyne}}{{Ballantyne}}{2014}]{Ballantyne}
{Ballantyne} D.~R.,  2014, \mn@doi [\mnras] {10.1093/mnras/stt2095}, \href
  {https://ui.adsabs.harvard.edu/abs/2014MNRAS.437.2845B} {437, 2845}

\bibitem[\protect\citeauthoryear{{Balmaverde} \& {Capetti}}{{Balmaverde} \&
  {Capetti}}{2006}]{BalmaverdeCapetti}
{Balmaverde} B.,  {Capetti} A.,  2006, \mn@doi [A\&A]
  {10.1051/0004-6361:20054031}, \href
  {http://adsabs.harvard.edu/abs/2006A%26A...447...97B} {447, 97}

\bibitem[\protect\citeauthoryear{{Balmaverde}, {Baldi}  \&
  {Capetti}}{{Balmaverde} et~al.}{2008}]{balmaverde08}
{Balmaverde} B.,  {Baldi} R.~D.,   {Capetti} A.,  2008, \mn@doi [A\&A]
  {10.1051/0004-6361:200809810}, \href
  {http://adsabs.harvard.edu/abs/2008A%26A...486..119B} {486, 119}

\bibitem[\protect\citeauthoryear{{Beswick}, {Argo}, {Evans}, {McHardy},
  {Williams}  \& {Westcott}}{{Beswick} et~al.}{2014}]{BeswickLemmings}
{Beswick} R.,  {Argo} M.~K.,  {Evans} R.,  {McHardy} I.,  {Williams} D.~R.~A.,
   {Westcott} J.,  2014, in Proceedings of the 12th European VLBI Network
  Symposium and Users Meeting (EVN 2014). 7-10 October 2014. Cagliari, Italy..
  p.~10

\bibitem[\protect\citeauthoryear{{Boorman}, {Gandhi}, {Balokovi{\'c}},
  {Brightman}, {Harrison}, {Ricci}  \& {Stern}}{{Boorman}
  et~al.}{2018}]{Boorman}
{Boorman} P.~G.,  {Gandhi} P.,  {Balokovi{\'c}} M.,  {Brightman} M.,
  {Harrison} F.,  {Ricci} C.,   {Stern} D.,  2018, \mn@doi [\mnras]
  {10.1093/mnras/sty861}, \href
  {https://ui.adsabs.harvard.edu/abs/2018MNRAS.477.3775B} {477, 3775}

\bibitem[\protect\citeauthoryear{{Brightman} et~al.,}{{Brightman}
  et~al.}{2017}]{Brightman17}
{Brightman} M.,  et~al., 2017, \mn@doi [] {10.3847/1538-4357/aa75c9}, \href
  {https://ui.adsabs.harvard.edu/abs/2017ApJ...844...10B} {844, 10}

\bibitem[\protect\citeauthoryear{{Burlon}, {Ajello}, {Greiner}, {Comastri},
  {Merloni}  \& {Gehrels}}{{Burlon} et~al.}{2011}]{Burlon2011}
{Burlon} D.,  {Ajello} M.,  {Greiner} J.,  {Comastri} A.,  {Merloni} A.,
  {Gehrels} N.,  2011, \mn@doi [\apj] {10.1088/0004-637X/728/1/58}, \href
  {https://ui.adsabs.harvard.edu/abs/2011ApJ...728...58B} {728, 58}

\bibitem[\protect\citeauthoryear{{Buttiglione}, {Capetti}, {Celotti}, {Axon},
  {Chiaberge}, {Macchetto}  \& {Sparks}}{{Buttiglione}
  et~al.}{2010}]{Buttiglione2010}
{Buttiglione} S.,  {Capetti} A.,  {Celotti} A.,  {Axon} D.~J.,  {Chiaberge} M.,
   {Macchetto} F.~D.,   {Sparks} W.~B.,  2010, \mn@doi [\aap]
  {10.1051/0004-6361/200913290}, \href
  {http://adsabs.harvard.edu/abs/2010A%26A...509A...6B} {509, A6}

\bibitem[\protect\citeauthoryear{{Capetti} \& {Baldi}}{{Capetti} \&
  {Baldi}}{2011}]{CapettiBaldi2011}
{Capetti} A.,  {Baldi} R.~D.,  2011, \mn@doi [\aap]
  {10.1051/0004-6361/201016388}, \href
  {http://adsabs.harvard.edu/abs/2011A%26A...529A.126C} {529, A126}

\bibitem[\protect\citeauthoryear{{Comastri}}{{Comastri}}{2004}]{Comastri}
{Comastri} A.,  2004, {Compton-Thick AGN: The Dark Side of the X-Ray
  Background}.
p.~245

\bibitem[\protect\citeauthoryear{{Condon}}{{Condon}}{1992}]{Condon92}
{Condon} J.~J.,  1992, \mn@doi [ARA\&A] {10.1146/annurev.aa.30.090192.003043},
  \href {http://adsabs.harvard.edu/abs/1992ARA%26A..30..575C} {30, 575}

\bibitem[\protect\citeauthoryear{{Connolly}, {McHardy}, {Skipper}  \&
  {Emmanoulopoulos}}{{Connolly} et~al.}{2016}]{Connolly2016}
{Connolly} S.~D.,  {McHardy} I.~M.,  {Skipper} C.~J.,   {Emmanoulopoulos} D.,
  2016, \mn@doi [\mnras] {10.1093/mnras/stw878}, \href
  {http://adsabs.harvard.edu/abs/2016MNRAS.459.3963C} {459, 3963}

\bibitem[\protect\citeauthoryear{{Constantin}, {Green}, {Aldcroft}, {Kim},
  {Haggard}, {Barkhouse}  \& {Anderson}}{{Constantin}
  et~al.}{2009}]{Constantin2009}
{Constantin} A.,  {Green} P.,  {Aldcroft} T.,  {Kim} D.-W.,  {Haggard} D.,
  {Barkhouse} W.,   {Anderson} S.~F.,  2009, \mn@doi [\apj]
  {10.1088/0004-637X/705/2/1336}, \href
  {https://ui.adsabs.harvard.edu/abs/2009ApJ...705.1336C} {705, 1336}

\bibitem[\protect\citeauthoryear{{Czerny}, {Siemiginowska}, {Janiuk},
  {Nikiel-Wroczy{\'n}ski}  \& {Stawarz}}{{Czerny} et~al.}{2009}]{czerny09}
{Czerny} B.,  {Siemiginowska} A.,  {Janiuk} A.,  {Nikiel-Wroczy{\'n}ski} B.,
  {Stawarz} {\L}.,  2009, \mn@doi [\apj] {10.1088/0004-637X/698/1/840}, \href
  {http://adsabs.harvard.edu/abs/2009ApJ...698..840C} {698, 840}

\bibitem[\protect\citeauthoryear{{Fabbiano}}{{Fabbiano}}{2006}]{Fabbiano2006}
{Fabbiano} G.,  2006, \mn@doi [\araa] {10.1146/annurev.astro.44.051905.092519},
  \href {https://ui.adsabs.harvard.edu/abs/2006ARA%26A..44..323F} {44, 323}

\bibitem[\protect\citeauthoryear{{Filho}, {Barthel}  \& {Ho}}{{Filho}
  et~al.}{2006}]{filho06}
{Filho} M.~E.,  {Barthel} P.~D.,   {Ho} L.~C.,  2006, \mn@doi [\aap]
  {10.1051/0004-6361:20054510}, \href
  {http://adsabs.harvard.edu/abs/2006A%26A...451...71F} {451, 71}

\bibitem[\protect\citeauthoryear{{Filippenko} \& {Sargent}}{{Filippenko} \&
  {Sargent}}{1985}]{Filippenko85}
{Filippenko} A.~V.,  {Sargent} W.~L.~W.,  1985, \mn@doi [ApJS]
  {10.1086/191012}, \href {http://adsabs.harvard.edu/abs/1985ApJS...57..503F}
  {57, 503}

\bibitem[\protect\citeauthoryear{Foreman-Mackey}{Foreman-Mackey}{2016}]{corner}
Foreman-Mackey D.,  2016, \mn@doi [The Journal of Open Source Software]
  {10.21105/joss.00024}, 24

\bibitem[\protect\citeauthoryear{{Fotopoulou} et~al.,}{{Fotopoulou}
  et~al.}{2016}]{Foropoulou}
{Fotopoulou} S.,  et~al., 2016, \mn@doi [\aap] {10.1051/0004-6361/201424763},
  \href {https://ui.adsabs.harvard.edu/abs/2016A&A...587A.142F} {587, A142}

\bibitem[\protect\citeauthoryear{{Freeman}, {Kashyap}, {Rosner}  \&
  {Lamb}}{{Freeman} et~al.}{2002}]{Freeman2002}
{Freeman} P.~E.,  {Kashyap} V.,  {Rosner} R.,   {Lamb} D.~Q.,  2002, \mn@doi
  [\apjs] {10.1086/324017}, \href
  {https://ui.adsabs.harvard.edu/abs/2002ApJS..138..185F} {138, 185}

\bibitem[\protect\citeauthoryear{{Gandhi} et~al.,}{{Gandhi}
  et~al.}{2017}]{Gandhi17}
{Gandhi} P.,  et~al., 2017, \mn@doi [\mnras] {10.1093/mnras/stx357}, \href
  {https://ui.adsabs.harvard.edu/abs/2017MNRAS.467.4606G} {467, 4606}

\bibitem[\protect\citeauthoryear{{Gehrels}}{{Gehrels}}{1986}]{Gehrels1986}
{Gehrels} N.,  1986, \mn@doi [\apj] {10.1086/164079}, \href
  {https://ui.adsabs.harvard.edu/abs/1986ApJ...303..336G} {303, 336}

\bibitem[\protect\citeauthoryear{{Gilfanov}, {Grimm}  \& {Sunyaev}}{{Gilfanov}
  et~al.}{2004}]{Gilfanov2004}
{Gilfanov} M.,  {Grimm} H.-J.,   {Sunyaev} R.,  2004, \mn@doi [Nuclear Physics
  B Proceedings Supplements] {10.1016/j.nuclphysbps.2004.04.065}, \href
  {http://adsabs.harvard.edu/abs/2004NuPhS.132..369G} {132, 369}

\bibitem[\protect\citeauthoryear{{Gonz{\'a}lez-Mart{\'\i}n}, {Masegosa},
  {M{\'a}rquez}, {Guerrero}  \& {Dultzin-Hacyan}}{{Gonz{\'a}lez-Mart{\'\i}n}
  et~al.}{2006}]{Gonzalez2006}
{Gonz{\'a}lez-Mart{\'\i}n} O.,  {Masegosa} J.,  {M{\'a}rquez} I.,  {Guerrero}
  M.~A.,   {Dultzin-Hacyan} D.,  2006, \mn@doi [\aap]
  {10.1051/0004-6361:20054756}, \href
  {https://ui.adsabs.harvard.edu/abs/2006A&A...460...45G} {460, 45}

\bibitem[\protect\citeauthoryear{{Gonz{\'a}lez-Mart{\'{\i}}n}, {Masegosa},
  {M{\'a}rquez}, {Guainazzi}  \&
  {Jim{\'e}nez-Bail{\'o}n}}{{Gonz{\'a}lez-Mart{\'{\i}}n}
  et~al.}{2009}]{gonzalez09}
{Gonz{\'a}lez-Mart{\'{\i}}n} O.,  {Masegosa} J.,  {M{\'a}rquez} I.,
  {Guainazzi} M.,   {Jim{\'e}nez-Bail{\'o}n} E.,  2009, \mn@doi [\aap]
  {10.1051/0004-6361/200912288}, \href
  {http://adsabs.harvard.edu/abs/2009A%26A...506.1107G} {506, 1107}

\bibitem[\protect\citeauthoryear{{Gonz{\'a}lez-Mart{\'\i}n}
  et~al.,}{{Gonz{\'a}lez-Mart{\'\i}n} et~al.}{2015}]{Gonzalez2015}
{Gonz{\'a}lez-Mart{\'\i}n} O.,  et~al., 2015, \mn@doi [\aap]
  {10.1051/0004-6361/201425254}, \href
  {https://ui.adsabs.harvard.edu/abs/2015A&A...578A..74G} {578, A74}

\bibitem[\protect\citeauthoryear{{Graham} \& {Scott}}{{Graham} \&
  {Scott}}{2013}]{graham13}
{Graham} A.~W.,  {Scott} N.,  2013, \mn@doi [\apj]
  {10.1088/0004-637X/764/2/151}, \href
  {http://adsabs.harvard.edu/abs/2013ApJ...764..151G} {764, 151}

\bibitem[\protect\citeauthoryear{{Hardcastle}, {Evans}  \&
  {Croston}}{{Hardcastle} et~al.}{2009}]{hardcastle09}
{Hardcastle} M.~J.,  {Evans} D.~A.,   {Croston} J.~H.,  2009, \mn@doi [\mnras]
  {10.1111/j.1365-2966.2009.14887.x}, \href
  {http://adsabs.harvard.edu/abs/2009MNRAS.396.1929H} {396, 1929}

\bibitem[\protect\citeauthoryear{{Heckman}}{{Heckman}}{1980}]{heckman80}
{Heckman} T.~M.,  1980, A\&A, \href
  {http://adsabs.harvard.edu/abs/1980A%26A....87..152H} {87, 152}

\bibitem[\protect\citeauthoryear{{Hern{\'a}ndez-Garc{\'{\i}}a},
  {Gonz{\'a}lez-Mart{\'{\i}}n}, {Masegosa}  \&
  {M{\'a}rquez}}{{Hern{\'a}ndez-Garc{\'{\i}}a}
  et~al.}{2014}]{HernandezGarcia2014}
{Hern{\'a}ndez-Garc{\'{\i}}a} L.,  {Gonz{\'a}lez-Mart{\'{\i}}n} O.,  {Masegosa}
  J.,   {M{\'a}rquez} I.,  2014, \mn@doi [\aap] {10.1051/0004-6361/201424140},
  \href {http://adsabs.harvard.edu/abs/2014A%26A...569A..26H} {569, A26}

\bibitem[\protect\citeauthoryear{{Hern{\'a}ndez-Garc{\'\i}a}, {Masegosa},
  {Gonz{\'a}lez-Mart{\'\i}n}, {M{\'a}rquez}  \&
  {Perea}}{{Hern{\'a}ndez-Garc{\'\i}a} et~al.}{2016}]{Hernandez-Garcia2016}
{Hern{\'a}ndez-Garc{\'\i}a} L.,  {Masegosa} J.,  {Gonz{\'a}lez-Mart{\'\i}n} O.,
   {M{\'a}rquez} I.,   {Perea} J.,  2016, \mn@doi [\apj]
  {10.3847/0004-637X/824/1/7}, \href
  {https://ui.adsabs.harvard.edu/abs/2016ApJ...824....7H} {824, 7}

\bibitem[\protect\citeauthoryear{{Ho}}{{Ho}}{1999a}]{HoLINERs}
{Ho} L.~C.,  1999a, \mn@doi [Advances in Space Research]
  {10.1016/S0273-1177(99)00211-2}, \href
  {http://adsabs.harvard.edu/abs/1999AdSpR..23..813H} {23, 813}

\bibitem[\protect\citeauthoryear{{Ho}}{{Ho}}{1999b}]{ho99}
{Ho} L.~C.,  1999b, \mn@doi [\apj] {10.1086/307137}, \href
  {http://adsabs.harvard.edu/abs/1999ApJ...516..672H} {516, 672}

\bibitem[\protect\citeauthoryear{{Ho}}{{Ho}}{2008}]{HoReview}
{Ho} L.~C.,  2008, \mn@doi [ARA\&A] {10.1146/annurev.astro.45.051806.110546},
  \href {http://adsabs.harvard.edu/abs/2008ARA%26A..46..475H} {46, 475}

\bibitem[\protect\citeauthoryear{{Ho} \& {Ulvestad}}{{Ho} \&
  {Ulvestad}}{2001}]{HoUlvestad}
{Ho} L.~C.,  {Ulvestad} J.~S.,  2001, \mn@doi [ApJS] {10.1086/319185}, \href
  {http://adsabs.harvard.edu/abs/2001ApJS..133...77H} {133, 77}

\bibitem[\protect\citeauthoryear{{Ho}, {Filippenko}  \& {Sargent}}{{Ho}
  et~al.}{1995}]{Ho95}
{Ho} L.~C.,  {Filippenko} A.~V.,   {Sargent} W.~L.,  1995, \mn@doi [ApJS]
  {10.1086/192170}, \href {http://adsabs.harvard.edu/abs/1995ApJS...98..477H}
  {98, 477}

\bibitem[\protect\citeauthoryear{{Ho}, {Filippenko}  \& {Sargent}}{{Ho}
  et~al.}{1997a}]{ho97a}
{Ho} L.~C.,  {Filippenko} A.~V.,   {Sargent} W.~L.~W.,  1997a, \mn@doi [ApJS]
  {10.1086/313041}, \href {http://adsabs.harvard.edu/abs/1997ApJS..112..315H}
  {112, 315}

\bibitem[\protect\citeauthoryear{{Ho}, {Filippenko}, {Sargent}  \& {Peng}}{{Ho}
  et~al.}{1997b}]{ho97b}
{Ho} L.~C.,  {Filippenko} A.~V.,  {Sargent} W.~L.~W.,   {Peng} C.~Y.,  1997b,
  \mn@doi [ApJS] {10.1086/313042}, \href
  {http://adsabs.harvard.edu/abs/1997ApJS..112..391H} {112, 391}

\bibitem[\protect\citeauthoryear{{Ho}, {Filippenko}  \& {Sargent}}{{Ho}
  et~al.}{1997c}]{ho97c}
{Ho} L.~C.,  {Filippenko} A.~V.,   {Sargent} W.~L.~W.,  1997c, ApJ, \href
  {http://adsabs.harvard.edu/abs/1997ApJ...487..568H} {487, 568}

\bibitem[\protect\citeauthoryear{{Ho}, {Filippenko}  \& {Sargent}}{{Ho}
  et~al.}{1997d}]{ho97d}
{Ho} L.~C.,  {Filippenko} A.~V.,   {Sargent} W.~L.~W.,  1997d, \mn@doi [ApJ]
  {10.1086/304643}, \href {http://adsabs.harvard.edu/abs/1997ApJ...487..591H}
  {487, 591}

\bibitem[\protect\citeauthoryear{{Ho} et~al.,}{{Ho} et~al.}{2001}]{ho01b}
{Ho} L.~C.,  et~al., 2001, \apjl, \href
  {http://adsabs.harvard.edu/cgi-bin/nph-bib_query?bibcode=2001\apj...549L..51H&db_key=AST}
  {549, L51}

\bibitem[\protect\citeauthoryear{{Ho}, {Filippenko}  \& {Sargent}}{{Ho}
  et~al.}{2003}]{Ho03}
{Ho} L.~C.,  {Filippenko} A.~V.,   {Sargent} W.~L.~W.,  2003, \mn@doi [ApJ]
  {10.1086/345354}, \href {http://adsabs.harvard.edu/abs/2003ApJ...583..159H}
  {583, 159}

\bibitem[\protect\citeauthoryear{{Ho}, {Greene}, {Filippenko}  \&
  {Sargent}}{{Ho} et~al.}{2009}]{Ho09}
{Ho} L.~C.,  {Greene} J.~E.,  {Filippenko} A.~V.,   {Sargent} W.~L.~W.,  2009,
  \mn@doi [ApJS] {10.1088/0067-0049/183/1/1}, \href
  {http://adsabs.harvard.edu/abs/2009ApJS..183....1H} {183, 1}

\bibitem[\protect\citeauthoryear{{Ishibashi} \& {Courvoisier}}{{Ishibashi} \&
  {Courvoisier}}{2010}]{Ishibashi}
{Ishibashi} W.,  {Courvoisier} T.~J.-L.,  2010, \mn@doi [\aap]
  {10.1051/0004-6361/200913587}, \href
  {http://adsabs.harvard.edu/abs/2010A%26A...512A..58I} {512, A58}

\bibitem[\protect\citeauthoryear{{Kaaret}, {Feng}  \& {Roberts}}{{Kaaret}
  et~al.}{2017}]{KaaretULX}
{Kaaret} P.,  {Feng} H.,   {Roberts} T.~P.,  2017, \mn@doi [\araa]
  {10.1146/annurev-astro-091916-055259}, \href
  {https://ui.adsabs.harvard.edu/abs/2017ARA%26A..55..303K} {55, 303}

\bibitem[\protect\citeauthoryear{{Kalberla}, {Burton}, {Hartmann}, {Arnal},
  {Bajaja}, {Morras}  \& {P{\"o}ppel}}{{Kalberla} et~al.}{2005}]{Kalberla2005}
{Kalberla} P.~M.~W.,  {Burton} W.~B.,  {Hartmann} D.,  {Arnal} E.~M.,  {Bajaja}
  E.,  {Morras} R.,   {P{\"o}ppel} W.~G.~L.,  2005, \mn@doi [\aap]
  {10.1051/0004-6361:20041864}, \href
  {http://adsabs.harvard.edu/abs/2005A%26A...440..775K} {440, 775}

\bibitem[\protect\citeauthoryear{{Kammoun} et~al.,}{{Kammoun}
  et~al.}{2019}]{Kammoun19}
{Kammoun} E.~S.,  et~al., 2019, \mn@doi [\apj] {10.3847/1538-4357/ab1c5f},
  \href {https://ui.adsabs.harvard.edu/abs/2019ApJ...877..102K} {877, 102}

\bibitem[\protect\citeauthoryear{{Kammoun} et~al.,}{{Kammoun}
  et~al.}{2020}]{Kammoun}
{Kammoun} E.~S.,  et~al., 2020, \mn@doi [\apj] {10.3847/1538-4357/abb29f},
  \href {https://ui.adsabs.harvard.edu/abs/2020ApJ...901..161K} {901, 161}

\bibitem[\protect\citeauthoryear{{Kaspi}, {Smith}, {Netzer}, {Maoz}, {Jannuzi}
  \& {Giveon}}{{Kaspi} et~al.}{2000}]{Kaspi2000}
{Kaspi} S.,  {Smith} P.~S.,  {Netzer} H.,  {Maoz} D.,  {Jannuzi} B.~T.,
  {Giveon} U.,  2000, \mn@doi [\apj] {10.1086/308704}, \href
  {http://adsabs.harvard.edu/abs/2000ApJ...533..631K} {533, 631}

\bibitem[\protect\citeauthoryear{{Kauffmann} \& {Heckman}}{{Kauffmann} \&
  {Heckman}}{2009}]{KauffmannHeckman2009}
{Kauffmann} G.,  {Heckman} T.~M.,  2009, \mn@doi [\mnras]
  {10.1111/j.1365-2966.2009.14960.x}, \href
  {https://ui.adsabs.harvard.edu/abs/2009MNRAS.397..135K} {397, 135}

\bibitem[\protect\citeauthoryear{{Kelly}}{{Kelly}}{2007}]{LINMIX}
{Kelly} B.~C.,  2007, \mn@doi [\apj] {10.1086/519947}, \href
  {https://ui.adsabs.harvard.edu/abs/2007ApJ...665.1489K} {665, 1489}

\bibitem[\protect\citeauthoryear{{Kewley}, {Groves}, {Kauffmann}  \&
  {Heckman}}{{Kewley} et~al.}{2006}]{Kewley06}
{Kewley} L.~J.,  {Groves} B.,  {Kauffmann} G.,   {Heckman} T.,  2006, \mn@doi
  [MNRAS] {10.1111/j.1365-2966.2006.10859.x}, \href
  {http://adsabs.harvard.edu/abs/2006MNRAS.372..961K} {372, 961}

\bibitem[\protect\citeauthoryear{{King}}{{King}}{2004}]{King2004}
{King} A.~R.,  2004, \mn@doi [\mnras] {10.1111/j.1365-2966.2004.07403.x}, \href
  {http://adsabs.harvard.edu/abs/2004MNRAS.347L..18K} {347, L18}

\bibitem[\protect\citeauthoryear{{Koratkar} \& {Gaskell}}{{Koratkar} \&
  {Gaskell}}{1991}]{Koratkar1991}
{Koratkar} A.~P.,  {Gaskell} C.~M.,  1991, \mn@doi [\apjl] {10.1086/185977},
  \href {http://adsabs.harvard.edu/abs/1991ApJ...370L..61K} {370, L61}

\bibitem[\protect\citeauthoryear{{Kormendy} \& {Ho}}{{Kormendy} \&
  {Ho}}{2013}]{kormendy13}
{Kormendy} J.,  {Ho} L.~C.,  2013, \mn@doi [\araa]
  {10.1146/annurev-astro-082708-101811}, \href
  {http://adsabs.harvard.edu/abs/2013ARA%26A..51..511K} {51, 511}

\bibitem[\protect\citeauthoryear{{LaMassa}, {Yaqoob}, {Boorman}, {Tzanavaris},
  {Levenson}, {Gandhi}, {Ptak}  \& {Heckman}}{{LaMassa}
  et~al.}{2019}]{LaMassa19}
{LaMassa} S.~M.,  {Yaqoob} T.,  {Boorman} P.~G.,  {Tzanavaris} P.,  {Levenson}
  N.~A.,  {Gandhi} P.,  {Ptak} A.~F.,   {Heckman} T.~M.,  2019, \mn@doi [\apj]
  {10.3847/1538-4357/ab552c}, \href
  {https://ui.adsabs.harvard.edu/abs/2019ApJ...887..173L} {887, 173}

\bibitem[\protect\citeauthoryear{{Lightman} \& {White}}{{Lightman} \&
  {White}}{1988}]{Lightman1988}
{Lightman} A.~P.,  {White} T.~R.,  1988, \mn@doi [\apj] {10.1086/166905}, \href
  {https://ui.adsabs.harvard.edu/abs/1988ApJ...335...57L} {335, 57}

\bibitem[\protect\citeauthoryear{{Liu}}{{Liu}}{2011}]{Liu2010}
{Liu} J.,  2011, \mn@doi [\apjs] {10.1088/0067-0049/192/1/10}, \href
  {http://adsabs.harvard.edu/abs/2011ApJS..192...10L} {192, 10}

\bibitem[\protect\citeauthoryear{{Maoz}}{{Maoz}}{2007}]{maoz07}
{Maoz} D.,  2007, \mn@doi [\mnras] {10.1111/j.1365-2966.2007.11735.x}, \href
  {https://ui.adsabs.harvard.edu/abs/2007MNRAS.377.1696M} {377, 1696}

\bibitem[\protect\citeauthoryear{{Marchesi}, {Ajello}, {Marcotulli},
  {Comastri}, {Lanzuisi}  \& {Vignali}}{{Marchesi} et~al.}{2018}]{Marchesi2018}
{Marchesi} S.,  {Ajello} M.,  {Marcotulli} L.,  {Comastri} A.,  {Lanzuisi} G.,
   {Vignali} C.,  2018, \mn@doi [\apj] {10.3847/1538-4357/aaa410}, \href
  {https://ui.adsabs.harvard.edu/abs/2018ApJ...854...49M} {854, 49}

\bibitem[\protect\citeauthoryear{{Masini}, {Comastri}, {Hickox}, {Koss},
  {Civano}, {Brigthman}, {Brusa}  \& {Lanzuisi}}{{Masini}
  et~al.}{2019}]{Masini19}
{Masini} A.,  {Comastri} A.,  {Hickox} R.~C.,  {Koss} M.,  {Civano} F.,
  {Brigthman} M.,  {Brusa} M.,   {Lanzuisi} G.,  2019, \mn@doi [\apj]
  {10.3847/1538-4357/ab3214}, \href
  {https://ui.adsabs.harvard.edu/abs/2019ApJ...882...83M} {882, 83}

\bibitem[\protect\citeauthoryear{{Matt}, {Fabian}, {Guainazzi}, {Iwasawa},
  {Bassani}  \& {Malaguti}}{{Matt} et~al.}{2000}]{Matt00}
{Matt} G.,  {Fabian} A.~C.,  {Guainazzi} M.,  {Iwasawa} K.,  {Bassani} L.,
  {Malaguti} G.,  2000, \mn@doi [\mnras] {10.1046/j.1365-8711.2000.03721.x},
  \href {https://ui.adsabs.harvard.edu/abs/2000MNRAS.318..173M} {318, 173}

\bibitem[\protect\citeauthoryear{{Mattila} et~al.,}{{Mattila}
  et~al.}{2018}]{Arp299TDE}
{Mattila} S.,  et~al., 2018, \mn@doi [Science] {10.1126/science.aao4669}, \href
  {https://ui.adsabs.harvard.edu/abs/2018Sci...361..482M} {361, 482}

\bibitem[\protect\citeauthoryear{{Merloni}, {Heinz}  \& {di Matteo}}{{Merloni}
  et~al.}{2003}]{Merloni03}
{Merloni} A.,  {Heinz} S.,   {di Matteo} T.,  2003, \mn@doi [MNRAS]
  {10.1046/j.1365-2966.2003.07017.x}, \href
  {http://adsabs.harvard.edu/abs/2003MNRAS.345.1057M} {345, 1057}

\bibitem[\protect\citeauthoryear{{Murphy} \& {Yaqoob}}{{Murphy} \&
  {Yaqoob}}{2009}]{Murphy09}
{Murphy} K.~D.,  {Yaqoob} T.,  2009, \mn@doi [\mnras]
  {10.1111/j.1365-2966.2009.15025.x}, \href
  {https://ui.adsabs.harvard.edu/abs/2009MNRAS.397.1549M} {397, 1549}

\bibitem[\protect\citeauthoryear{{Muxlow} et~al.,}{{Muxlow}
  et~al.}{2010}]{MuxlowM82}
{Muxlow} T.~W.~B.,  et~al., 2010, \mn@doi [\mnras]
  {10.1111/j.1745-3933.2010.00845.x}, \href
  {http://adsabs.harvard.edu/abs/2010MNRAS.404L.109M} {404, L109}

\bibitem[\protect\citeauthoryear{{Nagao}, {Murayama}, {Shioya}  \&
  {Taniguchi}}{{Nagao} et~al.}{2002}]{nagao02}
{Nagao} T.,  {Murayama} T.,  {Shioya} Y.,   {Taniguchi} Y.,  2002, \apj, \href
  {http://adsabs.harvard.edu/cgi-bin/nph-bib_query?bibcode=2002\apj...567...73N&amp;db_key=AST}
  {567, 73}

\bibitem[\protect\citeauthoryear{{Nandra} \& {Pounds}}{{Nandra} \&
  {Pounds}}{1994}]{NandraPounds1994}
{Nandra} K.,  {Pounds} K.~A.,  1994, \mn@doi [\mnras]
  {10.1093/mnras/268.2.405}, \href
  {http://adsabs.harvard.edu/abs/1994MNRAS.268..405N} {268, 405}

\bibitem[\protect\citeauthoryear{{Narayan}, {Kato}  \& {Honma}}{{Narayan}
  et~al.}{1997}]{Narayan}
{Narayan} R.,  {Kato} S.,   {Honma} F.,  1997, ApJ, \href
  {http://adsabs.harvard.edu/abs/1997ApJ...476...49N} {476, 49}

\bibitem[\protect\citeauthoryear{{Narayan}, {Mahadevan}  \&
  {Quataert}}{{Narayan} et~al.}{1998}]{NarayanADAF}
{Narayan} R.,  {Mahadevan} R.,   {Quataert} E.,  1998, in {Abramowicz} M.~A.,
  {Bj{\"o}rnsson} G.,   {Pringle} J.~E.,  eds, Theory of Black Hole Accretion
  Disks. p.~148 (\mn@eprint {} {astro-ph/9803141})

\bibitem[\protect\citeauthoryear{{Netzer}}{{Netzer}}{2019}]{Netzer2019}
{Netzer} H.,  2019, \mn@doi [\mnras] {10.1093/mnras/stz2016}, \href
  {https://ui.adsabs.harvard.edu/abs/2019MNRAS.488.5185N} {488, 5185}

\bibitem[\protect\citeauthoryear{{Oskinova}}{{Oskinova}}{2005}]{Oskinova}
{Oskinova} L.~M.,  2005, \mn@doi [\mnras] {10.1111/j.1365-2966.2005.09229.x},
  \href {https://ui.adsabs.harvard.edu/abs/2005MNRAS.361..679O} {361, 679}

\bibitem[\protect\citeauthoryear{{Panessa}, {Bassani}, {Cappi}, {Dadina},
  {Barcons}, {Carrera}, {Ho}  \& {Iwasawa}}{{Panessa} et~al.}{2006}]{panessa06}
{Panessa} F.,  {Bassani} L.,  {Cappi} M.,  {Dadina} M.,  {Barcons} X.,
  {Carrera} F.~J.,  {Ho} L.~C.,   {Iwasawa} K.,  2006, \mn@doi [\aap]
  {10.1051/0004-6361:20064894}, \href
  {http://adsabs.harvard.edu/abs/2006A%26A...455..173P} {455, 173}

\bibitem[\protect\citeauthoryear{{Panessa}, {Barcons}, {Bassani}, {Cappi},
  {Carrera}, {Ho}  \& {Pellegrini}}{{Panessa} et~al.}{2007}]{panessa07}
{Panessa} F.,  {Barcons} X.,  {Bassani} L.,  {Cappi} M.,  {Carrera} F.~J.,
  {Ho} L.~C.,   {Pellegrini} S.,  2007, \mn@doi [\aap]
  {10.1051/0004-6361:20066943}, \href
  {https://ui.adsabs.harvard.edu/abs/2007A&A...467..519P} {467, 519}

\bibitem[\protect\citeauthoryear{{Pellegrini}}{{Pellegrini}}{2005}]{Pellegrini2005}
{Pellegrini} S.,  2005, \mn@doi [\apj] {10.1086/429267}, \href
  {http://adsabs.harvard.edu/abs/2005ApJ...624..155P} {624, 155}

\bibitem[\protect\citeauthoryear{{Piconcelli}, {Jimenez-Bail{\'o}n},
  {Guainazzi}, {Schartel}, {Rodr{\'{\i}}guez-Pascual}  \&
  {Santos-Lle{\'o}}}{{Piconcelli} et~al.}{2005}]{Piconcelli2005}
{Piconcelli} E.,  {Jimenez-Bail{\'o}n} E.,  {Guainazzi} M.,  {Schartel} N.,
  {Rodr{\'{\i}}guez-Pascual} P.~M.,   {Santos-Lle{\'o}} M.,  2005, \mn@doi
  [\aap] {10.1051/0004-6361:20041621}, \href
  {http://adsabs.harvard.edu/abs/2005A%26A...432...15P} {432, 15}

\bibitem[\protect\citeauthoryear{{Ptak}}{{Ptak}}{2001}]{Ptak2001}
{Ptak} A.,  2001, \mn@doi [X-ray Astronomy: Stellar Endpoints, AGN, and the
  Diffuse X-ray Background] {10.1063/1.1434645}, \href
  {http://adsabs.harvard.edu/abs/2001AIPC..599..326P} {599, 326}

\bibitem[\protect\citeauthoryear{{Reynolds}}{{Reynolds}}{1997}]{reynolds97}
{Reynolds} C.~S.,  1997, MNRAS, 286, 513

\bibitem[\protect\citeauthoryear{{Reynolds}, {Young}, {Begelman}  \&
  {Fabian}}{{Reynolds} et~al.}{1999}]{Reynolds99}
{Reynolds} C.~S.,  {Young} A.~J.,  {Begelman} M.~C.,   {Fabian} A.~C.,  1999,
  \mn@doi [\apj] {10.1086/306913}, \href
  {https://ui.adsabs.harvard.edu/abs/1999ApJ...514..164R} {514, 164}

\bibitem[\protect\citeauthoryear{{Ricci}, {Ueda}, {Koss}, {Trakhtenbrot},
  {Bauer}  \& {Gandhi}}{{Ricci} et~al.}{2015}]{Ricci2015}
{Ricci} C.,  {Ueda} Y.,  {Koss} M.~J.,  {Trakhtenbrot} B.,  {Bauer} F.~E.,
  {Gandhi} P.,  2015, \mn@doi [\apjl] {10.1088/2041-8205/815/1/L13}, \href
  {https://ui.adsabs.harvard.edu/abs/2015ApJ...815L..13R} {815, L13}

\bibitem[\protect\citeauthoryear{{Roberts} \& {Warwick}}{{Roberts} \&
  {Warwick}}{2000}]{RobertsWarwick}
{Roberts} T.~P.,  {Warwick} R.~S.,  2000, \mn@doi [\mnras]
  {10.1046/j.1365-8711.2000.03384.x}, \href
  {http://adsabs.harvard.edu/abs/2000MNRAS.315...98R} {315, 98}

\bibitem[\protect\citeauthoryear{{Saikia}, {K{\"o}rding}  \& {Falcke}}{{Saikia}
  et~al.}{2015}]{Saikia}
{Saikia} P.,  {K{\"o}rding} E.,   {Falcke} H.,  2015, \mn@doi [MNRAS]
  {10.1093/mnras/stv731}, \href
  {http://adsabs.harvard.edu/abs/2015MNRAS.450.2317S} {450, 2317}

\bibitem[\protect\citeauthoryear{{Saikia}, {K{\"o}rding}  \& {Dibi}}{{Saikia}
  et~al.}{2018a}]{Saikia2018}
{Saikia} P.,  {K{\"o}rding} E.,   {Dibi} S.,  2018a, \mn@doi [\mnras]
  {10.1093/mnras/sty754}, \href
  {https://ui.adsabs.harvard.edu/abs/2018MNRAS.477.2119S} {477, 2119}

\bibitem[\protect\citeauthoryear{{Saikia}, {K{\"o}rding}, {Coppejans},
  {Falcke}, {Williams}, {Baldi}, {Mchardy}  \& {Beswick}}{{Saikia}
  et~al.}{2018b}]{Saikia2018B}
{Saikia} P.,  {K{\"o}rding} E.,  {Coppejans} D.~L.,  {Falcke} H.,  {Williams}
  D.,  {Baldi} R.~D.,  {Mchardy} I.,   {Beswick} R.,  2018b, \mn@doi [\aap]
  {10.1051/0004-6361/201833233}, \href
  {https://ui.adsabs.harvard.edu/abs/2018A&A...616A.152S} {616, A152}

\bibitem[\protect\citeauthoryear{{Sarzi} et~al.,}{{Sarzi}
  et~al.}{2010}]{sarzi10}
{Sarzi} M.,  et~al., 2010, \mn@doi [\mnras] {10.1111/j.1365-2966.2009.16039.x},
  \href {http://adsabs.harvard.edu/abs/2010MNRAS.402.2187S} {402, 2187}

\bibitem[\protect\citeauthoryear{{Sazonov}, {Krivonos}, {Revnivtsev},
  {Churazov}  \& {Sunyaev}}{{Sazonov} et~al.}{2008}]{Sazonov2008}
{Sazonov} S.,  {Krivonos} R.,  {Revnivtsev} M.,  {Churazov} E.,   {Sunyaev} R.,
   2008, \mn@doi [\aap] {10.1051/0004-6361:20078537}, \href
  {https://ui.adsabs.harvard.edu/abs/2008A&A...482..517S} {482, 517}

\bibitem[\protect\citeauthoryear{{Schmidt}}{{Schmidt}}{1968}]{Schmidt1968}
{Schmidt} M.,  1968, \mn@doi [\apj] {10.1086/149446}, \href
  {http://adsabs.harvard.edu/abs/1968ApJ...151..393S} {151, 393}

\bibitem[\protect\citeauthoryear{{Seyfert}}{{Seyfert}}{1941}]{Seyfert1941}
{Seyfert} C.~K.,  1941, \mn@doi [PASP] {10.1086/125320}, \href
  {http://adsabs.harvard.edu/abs/1941PASP...53..231S} {53, 231}

\bibitem[\protect\citeauthoryear{{Shakura} \& {Sunyaev}}{{Shakura} \&
  {Sunyaev}}{1973}]{ShakuraSunyaev}
{Shakura} N.~I.,  {Sunyaev} R.~A.,  1973, A\&A, \href
  {http://adsabs.harvard.edu/abs/1973A%26A....24..337S} {24, 337}

\bibitem[\protect\citeauthoryear{{Shankar} et~al.,}{{Shankar}
  et~al.}{2016}]{Shankar16}
{Shankar} F.,  et~al., 2016, \mn@doi [\mnras] {10.1093/mnras/stw678}, \href
  {http://adsabs.harvard.edu/abs/2016MNRAS.460.3119S} {460, 3119}

\bibitem[\protect\citeauthoryear{{Shankar} et~al.,}{{Shankar}
  et~al.}{2019}]{Shankar19}
{Shankar} F.,  et~al., 2019, \mn@doi [\mnras] {10.1093/mnras/stz376}, \href
  {https://ui.adsabs.harvard.edu/abs/2019MNRAS.485.1278S} {485, 1278}

\bibitem[\protect\citeauthoryear{{She}, {Ho}  \& {Feng}}{{She}
  et~al.}{2017}]{She2016}
{She} R.,  {Ho} L.~C.,   {Feng} H.,  2017, \mn@doi [\apj]
  {10.3847/1538-4357/835/2/223}, \href
  {https://ui.adsabs.harvard.edu/abs/2017ApJ...835..223S} {835, 223}

\bibitem[\protect\citeauthoryear{{Shemmer}, {Brandt}, {Netzer}, {Maiolino}  \&
  {Kaspi}}{{Shemmer} et~al.}{2006}]{Shemmer2006}
{Shemmer} O.,  {Brandt} W.~N.,  {Netzer} H.,  {Maiolino} R.,   {Kaspi} S.,
  2006, \mn@doi [\apjl] {10.1086/506911}, \href
  {https://ui.adsabs.harvard.edu/abs/2006ApJ...646L..29S} {646, L29}

\bibitem[\protect\citeauthoryear{{Singh} et~al.,}{{Singh}
  et~al.}{2013}]{Singh2013}
{Singh} R.,  et~al., 2013, \mn@doi [\aap] {10.1051/0004-6361/201322062}, \href
  {http://adsabs.harvard.edu/abs/2013A%26A...558A..43S} {558, A43}

\bibitem[\protect\citeauthoryear{Stephens}{Stephens}{1974}]{andersondarling2}
Stephens M.~A.,  1974, \mn@doi [Journal of the American Statistical
  Association] {10.1080/01621459.1974.10480196}, 69, 730

\bibitem[\protect\citeauthoryear{{Swartz}, {Tennant}  \& {Soria}}{{Swartz}
  et~al.}{2009}]{Swartz2009}
{Swartz} D.~A.,  {Tennant} A.~F.,   {Soria} R.,  2009, \mn@doi [\apj]
  {10.1088/0004-637X/703/1/159}, \href
  {http://adsabs.harvard.edu/abs/2009ApJ...703..159S} {703, 159}

\bibitem[\protect\citeauthoryear{{Swartz}, {Soria}, {Tennant}  \&
  {Yukita}}{{Swartz} et~al.}{2011}]{Swartz2011}
{Swartz} D.~A.,  {Soria} R.,  {Tennant} A.~F.,   {Yukita} M.,  2011, \mn@doi
  [\apj] {10.1088/0004-637X/741/1/49}, \href
  {https://ui.adsabs.harvard.edu/abs/2011ApJ...741...49S} {741, 49}

\bibitem[\protect\citeauthoryear{{Tremaine} et~al.,}{{Tremaine}
  et~al.}{2002}]{tremaine02}
{Tremaine} S.,  et~al., 2002, \apj, \href
  {http://adsabs.harvard.edu/cgi-bin/nph-bib_query?bibcode=2002\apj...574..740T&amp;db_key=AST}
  {574, 740}

\bibitem[\protect\citeauthoryear{{Tueller}, {Mushotzky}, {Barthelmy},
  {Cannizzo}, {Gehrels}, {Markwardt}, {Skinner}  \& {Winter}}{{Tueller}
  et~al.}{2008}]{Tueller2008}
{Tueller} J.,  {Mushotzky} R.~F.,  {Barthelmy} S.,  {Cannizzo} J.~K.,
  {Gehrels} N.,  {Markwardt} C.~B.,  {Skinner} G.~K.,   {Winter} L.~M.,  2008,
  \mn@doi [\apj] {10.1086/588458}, \href
  {https://ui.adsabs.harvard.edu/abs/2008ApJ...681..113T} {681, 113}

\bibitem[\protect\citeauthoryear{{Ueda} et~al.,}{{Ueda}
  et~al.}{2011}]{Ueda2011}
{Ueda} Y.,  et~al., 2011, \mn@doi [\pasj] {10.1093/pasj/63.sp3.S937}, \href
  {https://ui.adsabs.harvard.edu/abs/2011PASJ...63S.937U} {63, S937}

\bibitem[\protect\citeauthoryear{{Ueda}, {Akiyama}, {Hasinger}, {Miyaji}  \&
  {Watson}}{{Ueda} et~al.}{2014}]{Ueda2014}
{Ueda} Y.,  {Akiyama} M.,  {Hasinger} G.,  {Miyaji} T.,   {Watson} M.~G.,
  2014, \mn@doi [\apj] {10.1088/0004-637X/786/2/104}, \href
  {https://ui.adsabs.harvard.edu/abs/2014ApJ...786..104U} {786, 104}

\bibitem[\protect\citeauthoryear{{Weisskopf}, {Tananbaum}, {Van Speybroeck}  \&
  {O'Dell}}{{Weisskopf} et~al.}{2000}]{Chandra}
{Weisskopf} M.~C.,  {Tananbaum} H.~D.,  {Van Speybroeck} L.~P.,   {O'Dell}
  S.~L.,  2000, in {Truemper} J.~E.,  {Aschenbach} B.,  eds,  Society of
  Photo-Optical Instrumentation Engineers (SPIE) Conference Series Vol. 4012,
  X-Ray Optics, Instruments, and Missions III. pp 2--16 (\mn@eprint {arXiv}
  {astro-ph/0004127}), \mn@doi{10.1117/12.391545}

\bibitem[\protect\citeauthoryear{{Zhang}, {Soria}, {Zhang}, {Swartz}  \&
  {Liu}}{{Zhang} et~al.}{2009}]{Zhang09}
{Zhang} W.~M.,  {Soria} R.,  {Zhang} S.~N.,  {Swartz} D.~A.,   {Liu} J.~F.,
  2009, \mn@doi [\apj] {10.1088/0004-637X/699/1/281}, \href
  {http://adsabs.harvard.edu/abs/2009ApJ...699..281Z} {699, 281}

\bibitem[\protect\citeauthoryear{{van den Bosch}}{{van den
  Bosch}}{2016}]{vanderbosch16}
{van den Bosch} R.~C.~E.,  2016, \mn@doi [\apj] {10.3847/0004-637X/831/2/134},
  \href {http://adsabs.harvard.edu/abs/2016ApJ...831..134V} {831, 134}

\makeatother
\end{thebibliography}






\section{Online Supplementary Material}

\onecolumn

{\renewcommand{\arraystretch}{1.2}%
\begin{center}
\begin{longtable}{ c c c c c c c c c c c c c }
\caption{Basic properties of the X-ray sample presented in this paper. We show (1) the galaxy name; (2) Right Ascension in decimal degrees; (3) Declination in decimal degrees; (4) the Galactic Latitude ($\delta$); (5) the distance in Mpc, which is obtained from \citep[][and references therein]{ho97a}; (6) the \textit{Chandra} observation ID used for this analysis; (7) the exposure length in seconds; (8) denoted `C' if part of the `Complete' sample described in Section~\ref{sec:sample}, denoted `Y' if this is a new \textit{Chandra} observation obtained in observing cycle 17 (programme ID 19708646 and 18620515, PI:McHardy), or denoted `C+Y' if part of the `Complete' sample and a new \textit{Chandra} observation; (9) the detection significance where a source is observed or detected, `Undet.' if the source is not detected in the observations or `Not obs.' if there is no data in the \textit{Chandra} archive; (10) a black hole mass measurement taken from the literature where possible \citep[e.g.,][]{vanderbosch16}, but where no dynamical measurements exist, we use the $M{-}\sigma$ relationship of \citet{tremaine02}, using the stellar velocity dispersions of \citep{ho97a}; (11) a measurement of the [\ion{O}{iii}] line luminosity from \citep{BaldiLeMMINGs} or \citet{BaldiLeMMINGs2}, taken from the literature or \citep{ho97a} in erg s$^{-1}$; (12) the new AGN classification given to the source based on the AGN re-classifcation scheme described in \citet{BaldiLeMMINGs} and \citet{BaldiLeMMINGs2}; (13) a simplified version of the Hubble types originally shown in \citet{ho97a}.  } \label{tab:basicappendix}\\

\hline
Galaxy &      Right &    Dec. &  Gal. Lat. &  Dist  &  Obs. & Exposure & Sample & Det. &  Mass &  O$_{[\rm III]}$ & AGN & Hubble\\
Name &      Asc. &  $\delta$   &  |$b$| &  (Mpc) &  ID. & (secs) & Status & Sig. &  log(M$_{\odot}$) & log(Lum.) & Class & Type\\
(1) &      (2) &    (3) &  (4) &  (5) &  (6) & (7) & (8) & (9) &  (10) & (11) & (12) & (13)\\
\hline
\endfirsthead

\multicolumn{13}{c}%
{\tablename\ \thetable\ -- \textit{Continued from previous page}}\\
\hline
Galaxy &      Right &    Dec. &  Gal. Lat. &  Dist  &  Obs. & Exposure & Sample & Det. &  Mass &  O$_{[\rm III]}$ & AGN & Hubble\\
Name &      Asc. &  $\delta$   &  |$b$| &  (Mpc) &  ID. & (secs) & Status & Sig. &  log(M$_{\odot}$) & log(Lum.) & Class & Type\\
(1) &      (2) &    (3) &  (4) &  (5) &  (6) & (7) & (8) & (9) &  (10) & (11) & (12) & (13)\\
\hline
\endhead
\hline \multicolumn{13}{r}{\textit{Continued on next page}}\\
\endfoot
\hline
\endlastfoot

NGC7817 &    0.995 &  20.75 &  $-$40.76 &        31.5 &      - &        - &      -   &           Not obs. &         $6.21\pm0.22$ &  39.29    & HII &       Spi.\\
   IC10 &    5.096 &  59.29 &    3.34 &         1.3 &   8458 &    43571 &      -   &             Undet &         $5.11\pm0.82$ &  37.13    & HII &    Irr.\\
 NGC147 &    8.299 &  48.51 &  $-$14.25 &         0.7 &      - &        - &      -   &           Not obs. &         $4.28\pm0.40$ &  -    & ALG &   Ell.\\
 NGC185 &    9.739 &  48.34 &  $-$14.48 &         0.7 &      - &        - &      -   &           Not obs. &         $4.10\pm0.21$ &  34.63    & LINER &   Ell.\\
 NGC205 &   10.092 &  41.69 &  $-$21.14 &         0.7 &   4691 &     9870 &      C   &             Undet &         $3.83\pm1.84$ &  -    & ALG &   Ell.\\
 NGC221 &   10.674 &  40.87 &  $-$21.98 &         0.7 &   5690 &   113027 &      C   &             Undet &         $6.39\pm0.19$ &  -    & ALG &   Ell.\\
 NGC224 &   10.685 &  41.27 &  $-$21.57 &         0.7 &  14196 &    42848 &      C   &                 146.78 &         $7.84\pm0.05$ &  -    & ALG &       Spi.\\
 NGC266 &   12.449 &  32.28 &  $-$30.59 &        62.4 &  16013 &    84950 &      -   &                 204.06 &         $8.37\pm0.07$ &  39.43    & LINER &       Spi.\\
\end{longtable}
\end{center}
}

\clearpage

{\renewcommand{\arraystretch}{1.2}%
\begin{center}
\begin{longtable}{ c c c c c c c c}
\caption{Table showing flux and luminosity measurements obtained from X-ray spectral fitting (see Section~\ref{sec:Xrayspec} for the sources that have been observed in the \textit{Chandra} archive. The full table can be found in the online supplementary material. All fluxes in this table are calculated using the \textsc{cflux} command in \texttt{XSpec}. In the table we show (1) the galaxy name; (2) the detection significance if detected, else a "<" denotes a non detected source where the flux and luminosity measurements given in the 0.3--10\,keV band are 3$\sigma$ upper limits; (3) The logarithm of the flux in the 0.3--10.0\,keV band; (4) The logarithm of the flux in the 0.3--2.0\,keV band; (5) The logarithm of the flux in the 2.0--10.0\,keV band; (6) The logarithm of the X-ray luminosity in the 0.3--10.0\,keV band; (7) The logarithm of the X-ray luminosity in the 0.3--2.0\,keV band; (8) The logarithm of the X-ray luminosity in the 2.0--10.0\,keV band. All uncertainties are shown at the 1 $\sigma$ level. All fluxes are measured in erg s$^{-1}$ cm$^{-2}$ and all luminosities are measured in erg s$^{-1}$.}\\

\hline 
Galaxy & Detection & log Flux & log Flux & log Flux & log Lum. & log Lum. & log Lum.\\
Name & Signif. & 0.3--10\,keV & 0.3--2\,keV & 2--10\,keV & 0.3--10\,keV & 0.3--2\,keV & 2--10\,keV\\
 &  & erg s$^{-1}$ cm$^{-2}$ & erg s$^{-1}$ cm$^{-2}$ & erg s$^{-1}$ cm$^{-2}$ & erg s$^{-1}$ & erg s$^{-1}$ & erg s$^{-1}$\\
(1) & (2) & (3) & (4) & (5) & (6) & (7) & (8)\\
\endfirsthead

\multicolumn{8}{c}%
{{ \tablename\ \thetable{} -- continued from previous page}}\\
\hline 
Galaxy & Detection & log Flux & log Flux & log Flux & log Lum. & log Lum. & log Lum.\\
Name & Signif. & 0.3--10\,keV & 0.3--2\,keV & 2--10\,keV & 0.3--10\,keV & 0.3--2\,keV & 2--10\,keV\\
 &  & erg s$^{-1}$ cm$^{-2}$ & erg s$^{-1}$ cm$^{-2}$ & erg s$^{-1}$ cm$^{-2}$ & erg s$^{-1}$ & erg s$^{-1}$ & erg s$^{-1}$\\
(1) & (2) & (3) & (4) & (5) & (6) & (7) & (8)\\
\endhead

\multicolumn{8}{r}{{Continued on next page}}\\ 
\endfoot

\endlastfoot

IC 10 & < & -13.92 & - & -& 36.38 & - &  -\\
NGC 205 & < & -14.00 & - & -& 35.77 & - &  -\\
NGC 221 & < & -14.24 & - & -& 35.53 & - &  -\\
NGC 224 & 146.78 & -12.59$\pm$0.05 & -12.77$\pm$0.10 & -13.04$\pm$0.10 & 37.18$\pm$0.05 & 37.00$\pm$0.10 & 36.73$\pm$0.10\\
NGC 266 & 204.06 & -12.81$\pm$0.06 & -13.28$\pm$0.10 & -12.99$\pm$0.10 & 40.86$\pm$0.06 & 40.39$\pm$0.10 & 40.68$\pm$0.10\\
NGC 278 & 11.87 & -14.18$\pm$0.10 & -14.29$\pm$0.11 & -14.80$\pm$0.20 & 38.05$\pm$0.10 & 37.93$\pm$0.11 & 37.42$\pm$0.20\\
NGC 315 & 33.96 & -11.86$\pm$0.03 & -12.58$\pm$0.04 & -11.95$\pm$0.04 & 41.86$\pm$0.03 & 41.13$\pm$0.04 & 41.77$\pm$0.04\\
NGC 404 & 89.63 & -13.53$\pm$0.07 & -13.71$\pm$0.16 & -14.00$\pm$0.16 & 37.31$\pm$0.07 & 37.13$\pm$0.16 & 36.84$\pm$0.16\\
\end{longtable}
\end{center}
}

\clearpage

{\renewcommand{\arraystretch}{1.2}%
\begin{center}
\begin{longtable}{ c c c | c | c | c c | c }
\caption{The basic parameters from the \textit{Chandra} X-ray spectral fits, including those where the source was undetected. In this table, we only report the most basic parameters, e.g., the photon index and some of the neutral absorbers, but the table is continued in the additional tables in this appendix (see Table~\ref{tab:otherspec},~\ref{tab:absspec}).  We show the (1) Galaxy name; (2) detection significance if detected, else a dash is used; (3) the X-ray spectral fit model used (see below for list of spectral models); (4) the Galactic absorption e.g., the \textsc{phabs} parameter, obtained from \citet{Kalberla2005} in unit of 10$^{20}$~cm$^{-2}$; (5) the additional absorption column density in cm$^{-2}$ found in the \text{ztbabs} component, if any, divided by 10$^{22}$; (6) and (7) the \textsc{zpowerlw} model photon index and normalisation respectively; (8) the reduced $\chi$-squared value (number), or where the source is faint and Poissonian statistical treatment is required of the data, the C-statistic divided by the number of degrees of freedom e.g., the cstat parameter reported by XSpec, denoted with a superscript letter `c'
The spectral models used are defined as follows:
$\alpha$ : undetected; 
$\beta$ : \textsc{phabs} $\times$ \textsc{ztbabs} $\times$ \textsc{zpowerlw}; 
$\gamma$ : \textsc{phabs} $\times$ \textsc{ztbabs}(\textsc{apec} + \textsc{zpowerlw}); 
$\delta$ : \textsc{phabs} $\times$ \textsc{ztbabs} $\times$ \textsc{zxipcf} $\times$ \textsc{zpowerlw}; 
$\zeta$ : \textsc{phabs} $\times$ \textsc{zxipcf} $\times$ \textsc{ztbabs}(\textsc{apec} + \textsc{zpowerlw}); 
$\omega$ : \textsc{phabs} $\times$ \textsc{zxipcf} $\times$ \textsc{zxipcf} $\times$ \textsc{ztbabs}(\textsc{apec} + \textsc{gabs} $\times$ \textsc{zpowerlw});
$\kappa$ : \textsc{phabs} $\times$ \textsc{ztbabs}(\textsc{zgauss} + \textsc{zpowerlw});
$\iota$ : \textsc{phabs} $\times$ \textsc{zxipcf} $\times$ \textsc{ztbabs}(\textsc{zgauss} + \textsc{zpowerlw}); 
$\epsilon$ : \textsc{phabs} $\times$ \textsc{ztbabs} $\times$ \textsc{zpcfabs} $\times$ \textsc{zpowerlw}; 
$\eta$ : \textsc{phabs} $\times$ \textsc{zxipcf} $\times$ \textsc{ztbabs}(\textsc{apec} + \textsc{zgauss} + \textsc{zpowerlw}); 
$\theta$ : \textsc{phabs} $\times$ \textsc{zpcfabs} $\times$ \textsc{ztbabs}(\textsc{zgauss} + \textsc{zpowerlw});
$\mu$ :	\textsc{phabs} $\times$ \textsc{zxipcf} $\times$ \textsc{zxipcf} $\times$ \textsc{ztbabs} $\times$ \textsc{zpowerlw}; 
$\pi$ :	\textsc{phabs} $\times$ \textsc{zpcfabs} $\times$ \textsc{ztbabs}(\textsc{apec} + \textsc{zpowerlw}); 
$\tau$ : \textsc{phabs}(\textsc{ztbabs} $\times$ \textsc{cabs} $\times$ \textsc{cutoffpl} + \textsc{pexrav} + \textsc{const} $\times$ \textsc{cutoffpl} + \textsc{apec} + \textsc{zgauss});
$\rho$ : \textsc{phabs} $\times$ \textsc{ztbabs}(\textsc{apec} + \textsc{laor} + \textsc{zgauss} + \textsc{zpowerlw}); 
} \label{tab:basicspeec}\\

\hline
  & Det.	 &   & \multicolumn{1}{c}{\textsc{phabs}} &  \multicolumn{1}{c}{\textsc{ztbabs}} &  \multicolumn{2}{c}{\textsc{zpowerlw}}  & 
\multicolumn{1}{c}{cstat/}\\ 
Name & Sig.	& mod. &   $N_{\rm H}$ &   $N_{\rm H}$ &   Phot.I. &  norm. &  $\chi^2$\\
(1) & (2) & (3) & (4) & (5) & (6) & (7) & (8)\\
\hline 
\endfirsthead

\multicolumn{8}{c}%
{{ \tablename\ \thetable{} -- continued from previous page}}\\
\hline
  & Det.	 &   & \multicolumn{1}{c}{\textsc{phabs}} &  \multicolumn{1}{c}{\textsc{ztbabs}} &  \multicolumn{2}{c}{\textsc{zpowerlw}}  & 
\multicolumn{1}{c}{cstat/}\\ 
Name & Sig.	& mod. &   $N_{\rm H}$ &   $N_{\rm H}$ &   Phot.I. &  norm. &  $\chi^2$\\
(1) & (2) & (3) & (4) & (5) & (6) & (7) & (8)\\
\hline  
\endhead

\hline \multicolumn{8}{r}{{Continued on next page}}\\ 
\endfoot

\hline \hline
\endlastfoot

IC10 & - & $\alpha$ & 50.6 & - & - & - & -\\
NGC205 & - & $\alpha$ & 5.83 & - & - & - & -\\
NGC221 & - & $\alpha$ & 18.3 & - & - & - & -\\
NGC224 & 146.78 & $\beta$ & 16.9 &  $<$0.04 & 2.82$_{-0.13}^{+0.14}$ & 9.96$_{-0.84}^{+0.95}\times$10$^{-5}$ & 0.87\\
NGC266 & 204.06 & $\beta$  & 5.68 &  0.23$_{-0.06}^{+0.10}$ & 1.91$_{-0.14}^{+0.14}$ & 	3.05$_{-0.17}^{+0.51}\times$10$^{-5}$ & 0.69\\
NGC278 & 11.87 & $\beta$ & 12.9 &  $<$0.15              & 2.25$_{-0.98}^{+1.22}$ & 	2.30$_{-1.08}^{+1.34}\times$10$^{-6}$ & 10.13/9$^{\rm c}$\\
NGC315 & 33.96 &  $\zeta$  & 5.90 &  0.08$_{-0.06}^{+0.07}$ & 1.49$_{-0.08}^{+0.06}$ & 	1.84$_{-0.32}^{+0.15}\times$10$^{-4}$ & 1.06\\
NGC404 & 89.63 & $\gamma$ & 5.71 &  0.44$_{-0.29}^{+0.24}$ & 1.88$_{-0.23}^{+0.25}$ & 	4.89$_{-1.02}^{+1.29}\times$10$^{-6}$ & 15.57/20$^{\rm c}$\\
\end{longtable}
\end{center}
}

\clearpage

\begin{landscape}

{\renewcommand{\arraystretch}{1.2}%
\begin{center}
\begin{longtable}{ c c | c c | c c c | c c c }
\caption{Additional spectral parameters for more complex spectral fits, including additional \textsc{apec}, \textsc{zgauss} and \textsc{gabs} models. This table is only relevant for spectra fit with spectral models $\zeta$, $\gamma$, $\omega$, $\delta$, $\iota$, $\eta$, $\kappa$, $\theta$ and $\pi$=. Note that where the \textdagger symbol is used, this denotes the parameter was frozen before fitting. The errors shown in this table are all at the 1$\sigma$ level. } \label{tab:otherspec}\\

\hline 
 &	 &   \multicolumn{2}{c}{\textsc{apec}} &  \multicolumn{3}{c}{\textsc{zgauss}} &  \multicolumn{3}{c}{\textsc{gabs}}\\ 
Name &	mod. &   $kT$ &   norm. &   LineE &  sigma &  norm. &  LineE & sigma & Strength\\
(1) & (2) & (3) & (4) & (5) & (6) & (7) & (8) & (9) & (10)\\
\hline 
\endfirsthead

\multicolumn{10}{c}%
{{ \tablename\ \thetable{} -- continued from previous page}}\\
\hline  &	 &   \multicolumn{2}{c}{\textsc{apec}} &  \multicolumn{3}{c}{\textsc{zgauss}} &  \multicolumn{3}{c}{\textsc{gabs}}\\ 
Name &	mod. &   $kT$ &   norm. &   LineE &  sigma &  norm. &  LineE & sigma & Strength\\
(1) & (2) & (3) & (4) & (5) & (6) & (7) & (8) & (9) 7 (10)\\ \hline 
\endhead

\hline \multicolumn{10}{r}{{Continued on next page}}\\ 
\endfoot

\hline \hline
\endlastfoot

NGC315 &	$\zeta$ &   0.54$_{-0.04}^{+0.02}$ &   2.62$_{-0.61}^{+0.54}\times$10$^{-4}$ &  -&  - &  - &  - & - &  -\\
NGC404 &	$\gamma$ &   0.24$_{-0.05}^{+0.07}$ &   5.17$_{-4.06}^{+16.96}\times$10$^{-5}$ &  - &  - & - &  - &  - &  -\\
NGC410 &	$\gamma$ &   0.78$_{-0.05}^{+0.04}$ &   2.07$_{-0.39}^{+0.41}\times$10$^{-4}$ &  -&  - &  - &  - &  - &  -\\
NGC507 &	$\gamma$ &   0.91$_{-0.05}^{+0.04}$ &   5.89$_{-0.64}^{+0.77}\times$10$^{-5}$ &  -&  - &  - &  - &  - &  -\\
NGC1275 &	$\omega$ &   0.79$_{-0.04}^{+0.04}$ &   1.42$_{-0.20}^{+0.46}\times$10$^{-2}$ &  - &  - &  - &  6.64$_{-0.07}^{+0.06}$ & 0.69$_{-0.04}^{+0.04}$ & 1 \textdagger\\
IC342* &	$\gamma$ &   0.82$_{-0.07}^{+0.17}$ &   6.75$_{-1.81}^{+2.77}\times$10$^{-5}$ &  - &  - &  - &  - & - &  -\\
NGC2273 &	$\iota$ &   - &   - &  6.41$_{-0.02}^{+0.02}$ &  9.54$_{-2.96}^{+2.78}\times$10$^{-2}$ &  5.56$_{-1.46}^{+1.22}\times$10$^{-5}$ &  - & - &  -\\ 
NGC2549 &	$\delta$ &   1.80$_{-0.42}^{+0.73}$ &   $<$7.72$\times$10$^{-2}$ &  - &  - &  - &  - & - &  -\\
\end{longtable}
\end{center}
}

\end{landscape}

\begin{landscape}
\clearpage
{\renewcommand{\arraystretch}{1.2}%
\begin{center}
\begin{longtable}{ c c | c c | c c c | c c c | c c}
\caption{Additional spectral parameters for more complex spectral fits, including additional \textsc{zpcfabs} and \textsc{zxipcf} models, as well as the second \textsc{apec} model for NGC\,4151. This table is only relevant for spectra fit with spectral models $\zeta$, $\epsilon$, $\omega$, $\delta$, $\iota$, $\eta$, $\theta$, $\mu$, $\tau$, $\pi$ and $\lambda$. Note that where the \textdagger symbol is used, this denotes the parameter was frozen before fitting. The errors shown in this table are all at the 1$\sigma$ level. }\label{tab:absspec}\\

\hline 
 &	 &   \multicolumn{2}{c}{\textsc{zpcfabs}} &  \multicolumn{3}{c}{\textsc{zxipcf}} &  \multicolumn{3}{c}{\textsc{zxipcf (2)}}  &   \multicolumn{2}{c}{\textsc{apec (2)}}\\ 
Name &	mod. &   $N_{\rm H}$ &   CvrFract. &   $N_{\rm H}$ &  logxi &  CvrFract. &  $N_{\rm H}$ & logxi & CvrFract. $kT$ &   norm.\\
(1) & (2) & (3) & (4) & (5) & (6) & (7) & (8) & (9) & (10) & (11) & (12)\\ \hline 
\endfirsthead

\multicolumn{12}{c}%
{{ \tablename\ \thetable{} -- continued from previous page}}\\
\hline  &	 &   \multicolumn{2}{c}{\textsc{zpcfabs}} &  \multicolumn{3}{c}{\textsc{zxipcf}} &  \multicolumn{3}{c}{\textsc{zxipcf (2)}}  &   \multicolumn{2}{c}{\textsc{apec (2)}}\\ 
Name &	mod. &   $N_{\rm H}$ &   CvrFract. &   $N_{\rm H}$ &  logxi &  CvrFract. &  $N_{\rm H}$ & logxi & CvrFract. $kT$ &   norm.\\
(1) & (2) & (3) & (4) & (5) & (6) & (7) & (8) & (9) & (10) & (11) & (12)\\ \hline 
\endhead

\hline \multicolumn{10}{r}{{Continued on next page}}\\ 
\endfoot

\hline \hline
\endlastfoot

NGC315 &	$\zeta$ &  - &  - & 0.39$_{-0.04}^{+0.03}$ & $<$-1.59 & 0.84$_{-0.07}^{+0.05}$ &  - & - &  -  & - &  -\\
NGC598 &	$\delta$ &  - &  - & 177.28$_{-0.89}^{+0.81}$ & 0.26$_{-0.59}^{+0.03}$ & 0.99$_{-0.01}^{+0.01}$ &  - & - &  -  & - &  -\\
NGC891 &	$\epsilon$ &  9.34$_{-0.86}^{+1.01}$ &  0.95$_{-0.04}^{+0.01}$ & - &  - & - &  - & - &  -  & - &  -\\
NGC1161 &   $\epsilon$ &  20.63$_{-4.40}^{+6.48}$ &  0.97$_{-0.04}^{+0.02}$ & - &  - & - &  - & - &  -  & - &  -\\
NGC1275 &	$\omega$ &  - &  - & 1.79$_{-0.13}^{+0.13}$ & -3 \textdagger & 0.79$_{-0.03}^{+0.04}$ &  61.86$_{-0.55}^{+0.95}$ & -0.55$_{-0.11}^{+0.06}$ & 0.99$_{-0.01}^{+0.01}$  & - &  -\\
NGC2273 &	$\iota$ &  - &  - & 15.69$_{-8.36}^{+5.74}$ & 0.01 \textdagger & 0.90$_{-0.10}^{+0.02}$ &  - & - &  -  & - &  -\\
NGC2549 &	$\delta$ &  - &  - & 65.95$_{-32.77}^{+333.194}$ & $<$1.70 & $>$0.84 &  - & - &  -  & - &  -\\
NGC2683 &	$\delta$ &  - &  - & 46.16$_{-6.64}^{+6.30}$ & -2.00$_{-0.02}^{+0.05}$ & 1 \textdagger &  - & - &  -  & - &  -\\
\end{longtable}
\end{center}
}
\end{landscape}
\clearpage

\begin{landscape}

{\renewcommand{\arraystretch}{1.2}%
\begin{center}
\begin{longtable}{ c c | c | c c c | c c c c c c c c}
\caption{Additional spectral parameters for NGC\,5194. Note that where the \textdagger symbol is used, this denotes the parameter was frozen before fitting. The errors shown in this table are all at the 1$\sigma$ level. The \textsc{cabs} parameter was set equal to the \textsc{ztbabs} parameter in Table~\ref{tab:basicspeec}. In addition, we used a \textsc{cutoffpl} model instead of a powerlaw (see text). The \textsc{pexrav} model parameters were set to those of other parameters e.g. the photon index and normalisations, or frozen. } \label{tab:5194}\\

\hline 
 &	 &   \multicolumn{1}{c}{\textsc{cabs}} &  \multicolumn{3}{c}{\textsc{cutoffpl}} &  \multicolumn{7}{c}{\textsc{pexrav}}\\ 
Name &	mod. &   $N_{\rm H}$ & Phot.I. & HighECut (keV) & norm. &   Phot.I. &  foldE(keV) &  relrefl &  abund & Fe abund & cosIncl &   norm.\\
(1) & (2) & (3) & (4) & (5) & (6) & (7) & (8) & (9) & (10) & (11) & (12)\\ \hline 
\endfirsthead

\hline\\ 
\endfoot

NGC5194 &	$\tau$ &  6601.11$_{-2252.91}^{+3479.39}$ & 1.8 \textdagger & 300 \textdagger & 6.57$^{2.91}_{4.32}\times$10$^{-5}$ &  1.8 \textdagger & 300 \textdagger & -1 \textdagger & 1.0 \textdagger & 1.0 \textdagger &  0.45 \textdagger  & 6.57$^{2.91}_{4.32}\times$10$^{-5}$\\

\end{longtable}
\end{center}
}
\end{landscape}

We plot all of the 150 detected sources in the following images. In all plots, the source is labelled at the top of the image. The \textit{top panel} in each plot shows the number of photons cm$^{-2}$ s$^{-1}$ keV$^{-1}$ plotted against the energy in keV across the whole 0.3--10.0\,keV band. The \textit{bottom panel} in each plot shows the model subtracted from the data, divided by the error. The fit parameters to make these plots are shown in this Online Supplementary Material Tables.

\twocolumn

\setcounter{figure}{0}

\subsection{X-ray Spectra}
We plot all of the 150 detected sources in the following images. In all plots, the source is labelled at the top of the image. The \textit{top panel} in each plot shows the number of photons cm$^{-2}$ s$^{-1}$ keV$^{-1}$ plotted against the energy in keV across the whole 0.3--10.0\,keV band. The \textit{bottom panel} in each plot shows the model subtracted from the data, divided by the error. The fit parameters to make these plots are shown in this Appendix.

\begin{center}
 \begin{figure}
	\includegraphics[width=0.90\columnwidth]{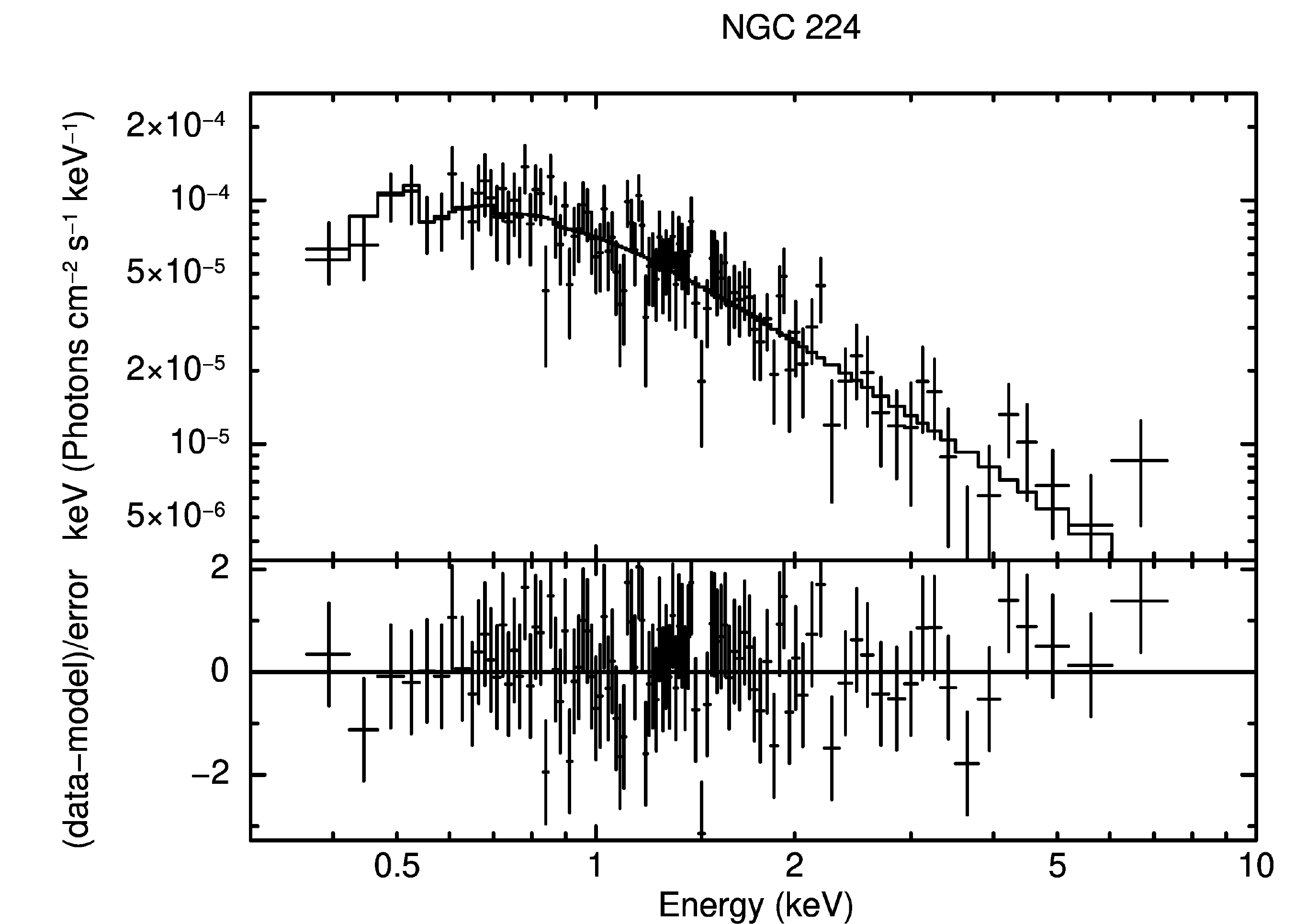}

\end{figure}
\end{center}

\begin{center}
 \begin{figure}
	\includegraphics[width=0.90\columnwidth]{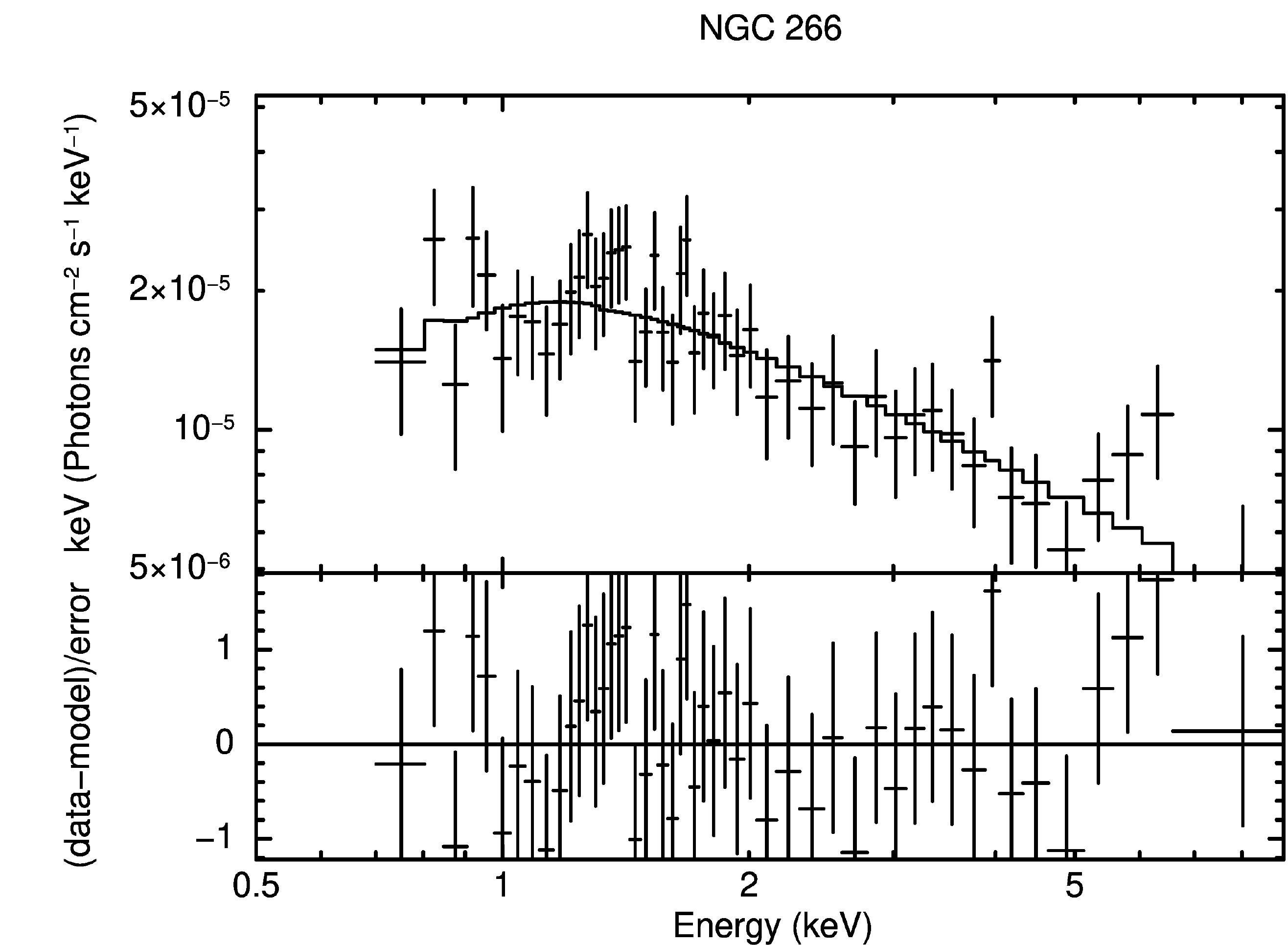}

\end{figure}
\end{center}

\begin{center}
 \begin{figure}
	\includegraphics[width=0.90\columnwidth]{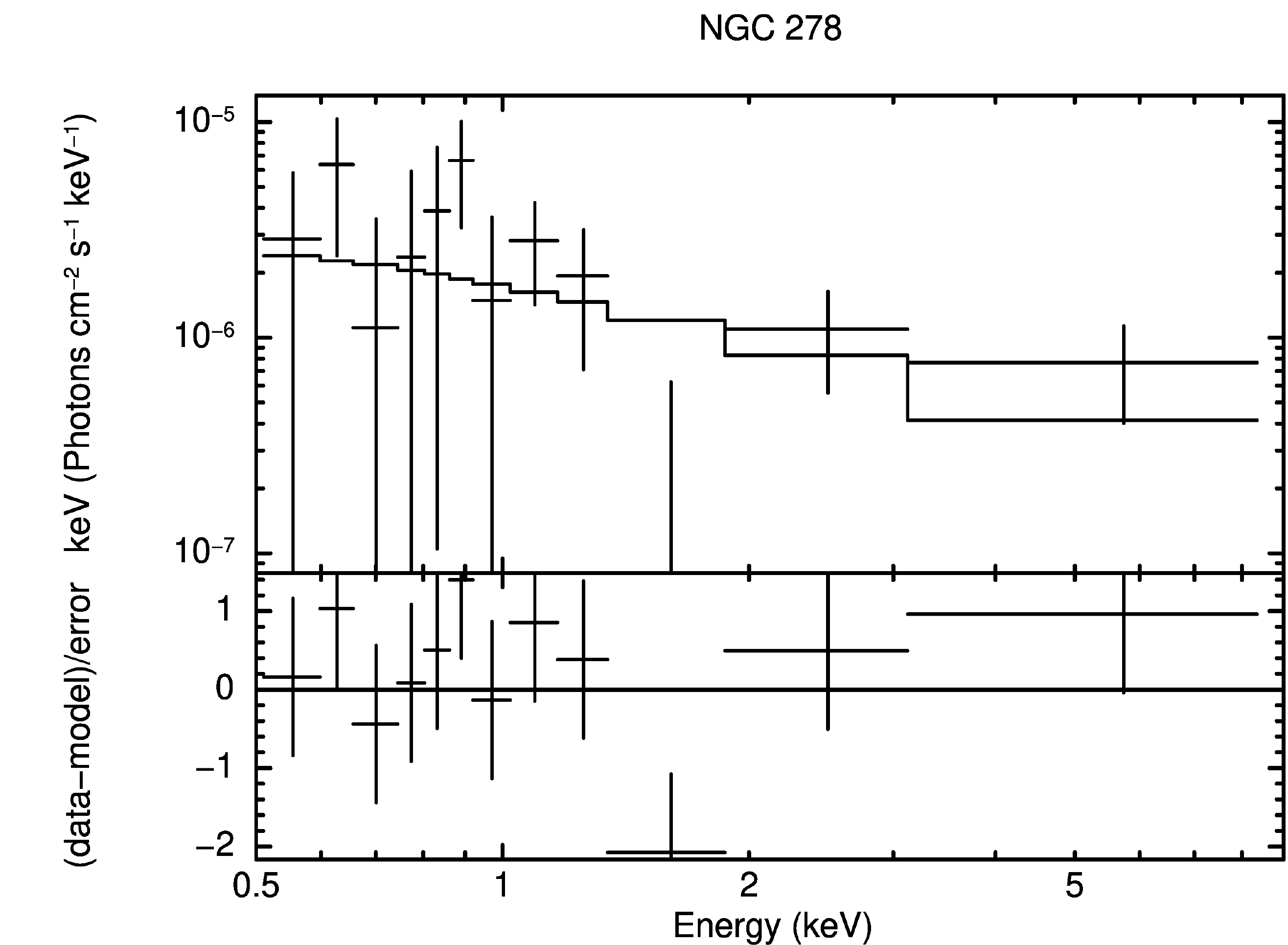}

\end{figure}
\end{center}

\begin{center}
 \begin{figure}
	\includegraphics[width=0.90\columnwidth]{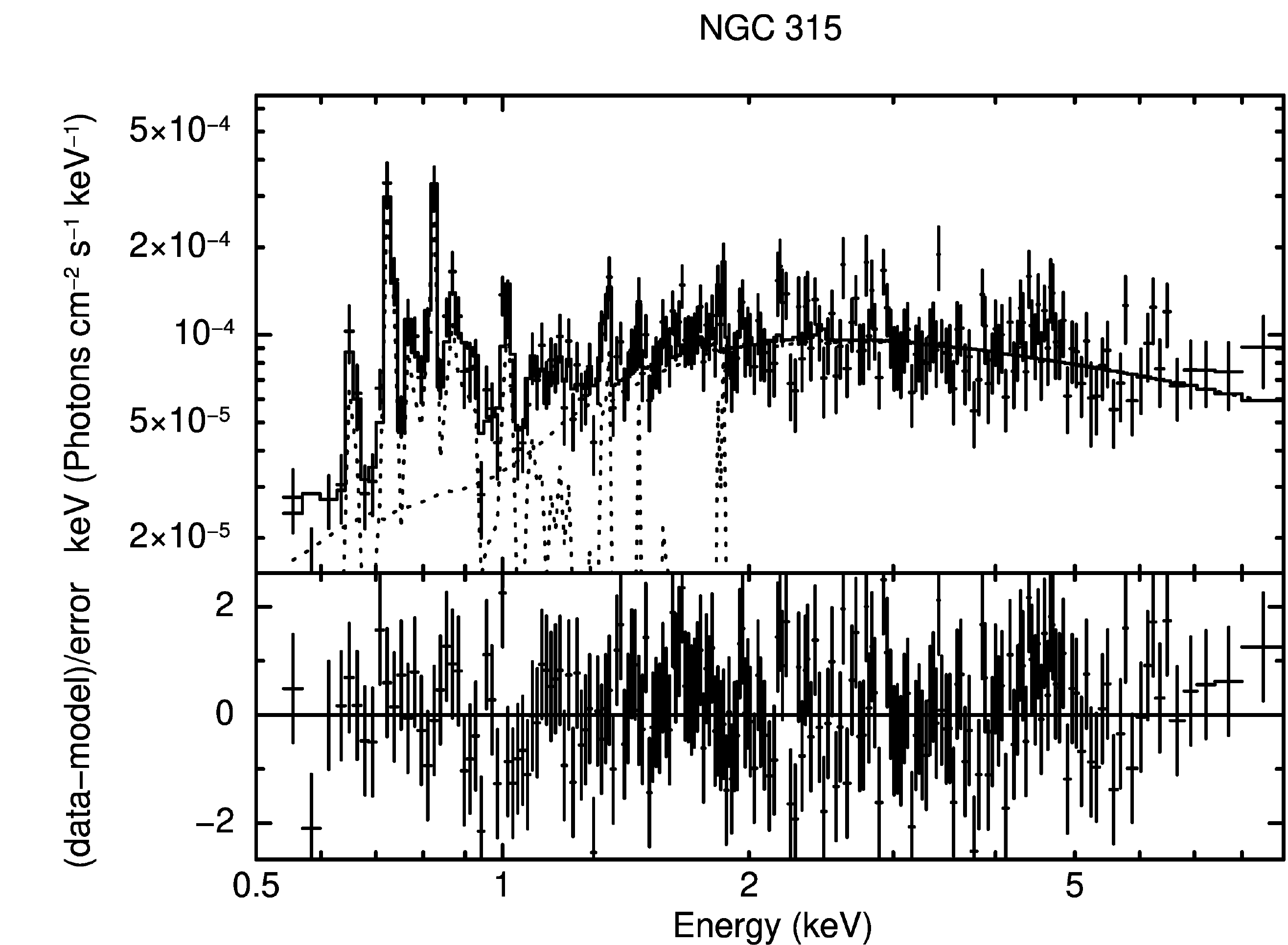}

\end{figure}
\end{center}

\begin{center}
 \begin{figure}
	\includegraphics[width=0.90\columnwidth]{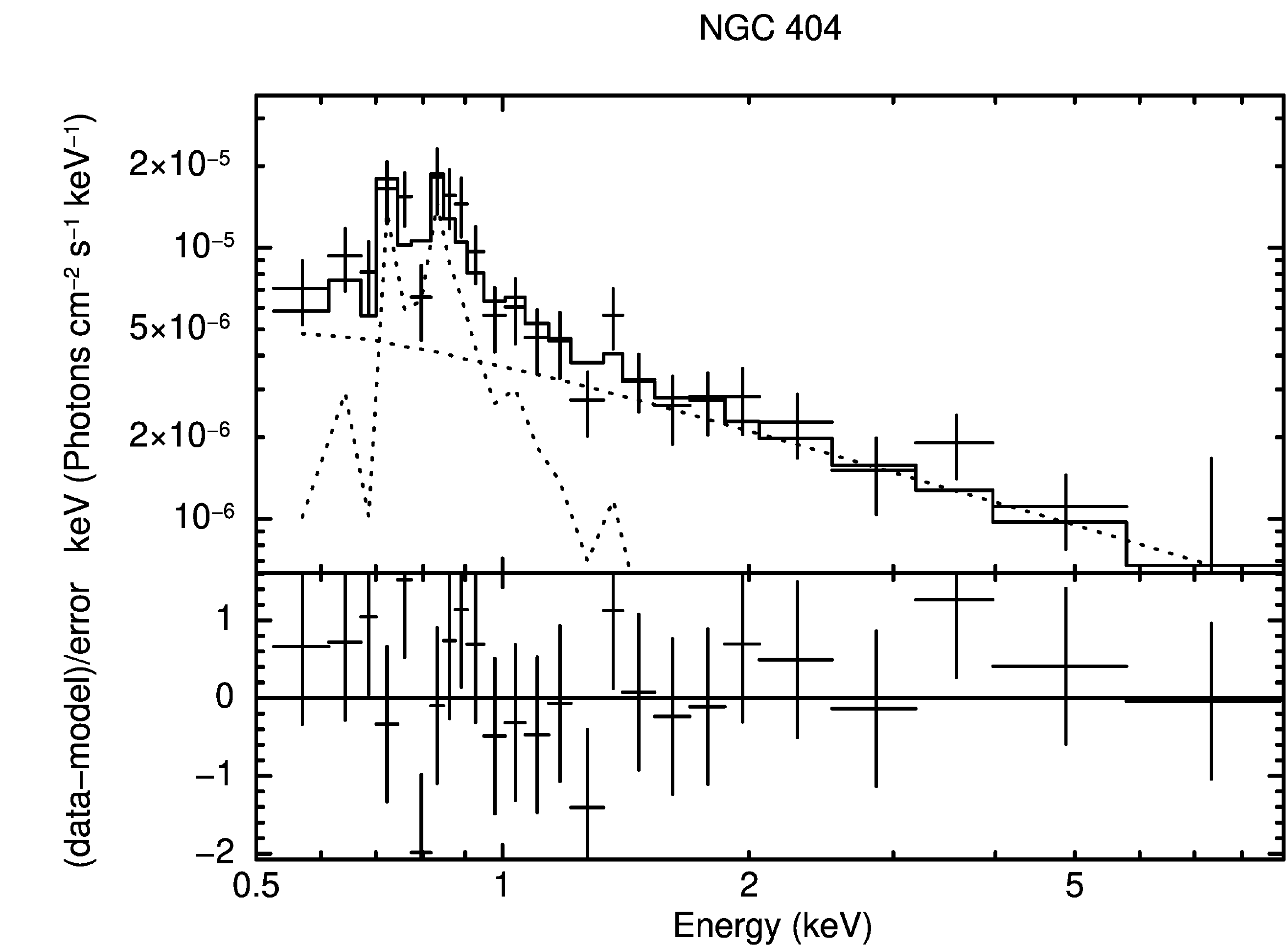}

\end{figure}
\end{center}

\begin{center}
 \begin{figure}
	\includegraphics[width=0.90\columnwidth]{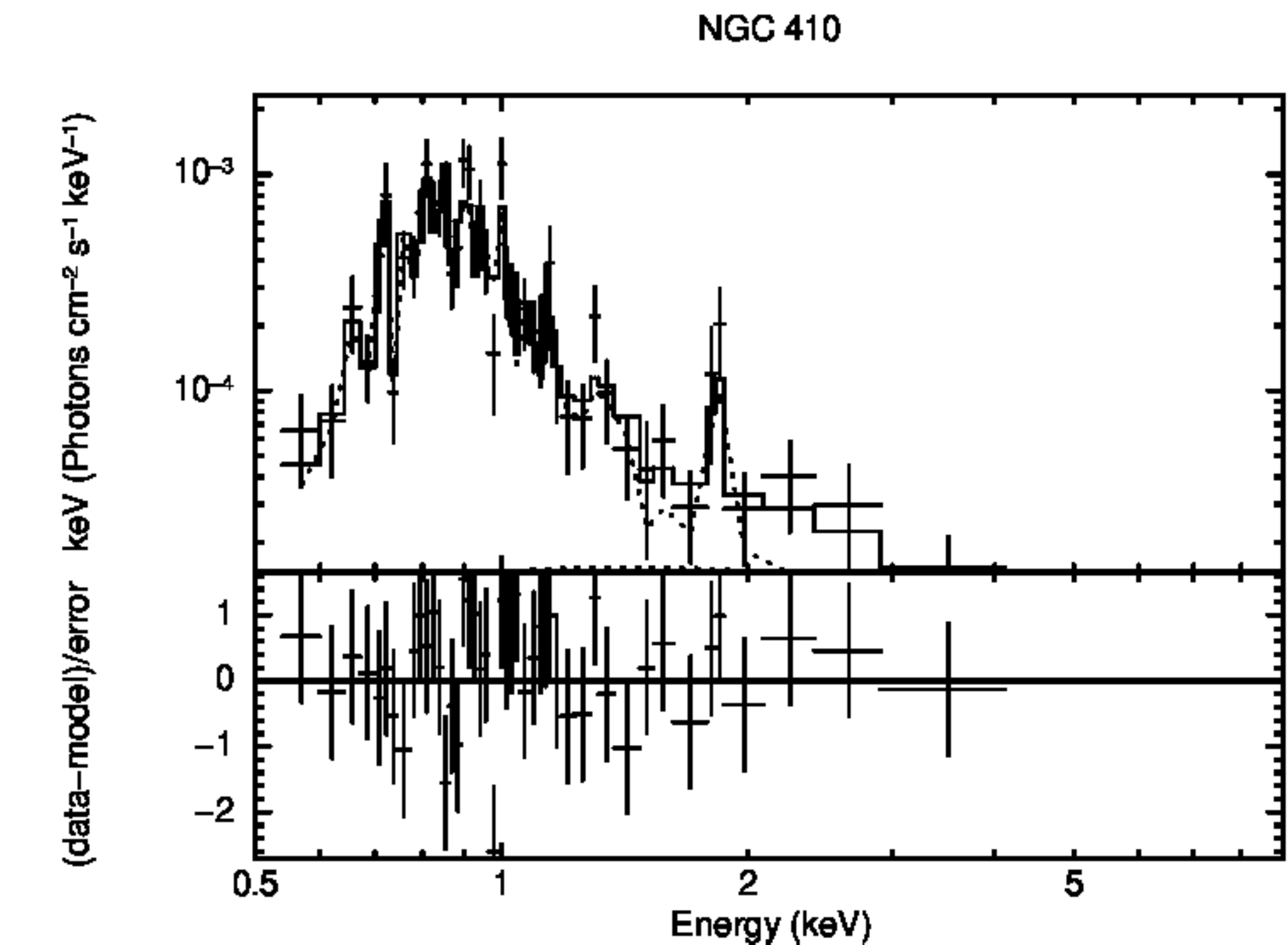}

\end{figure}
\end{center}

%
	 

\begin{center}
 \begin{figure}
	\includegraphics[width=0.89\columnwidth]{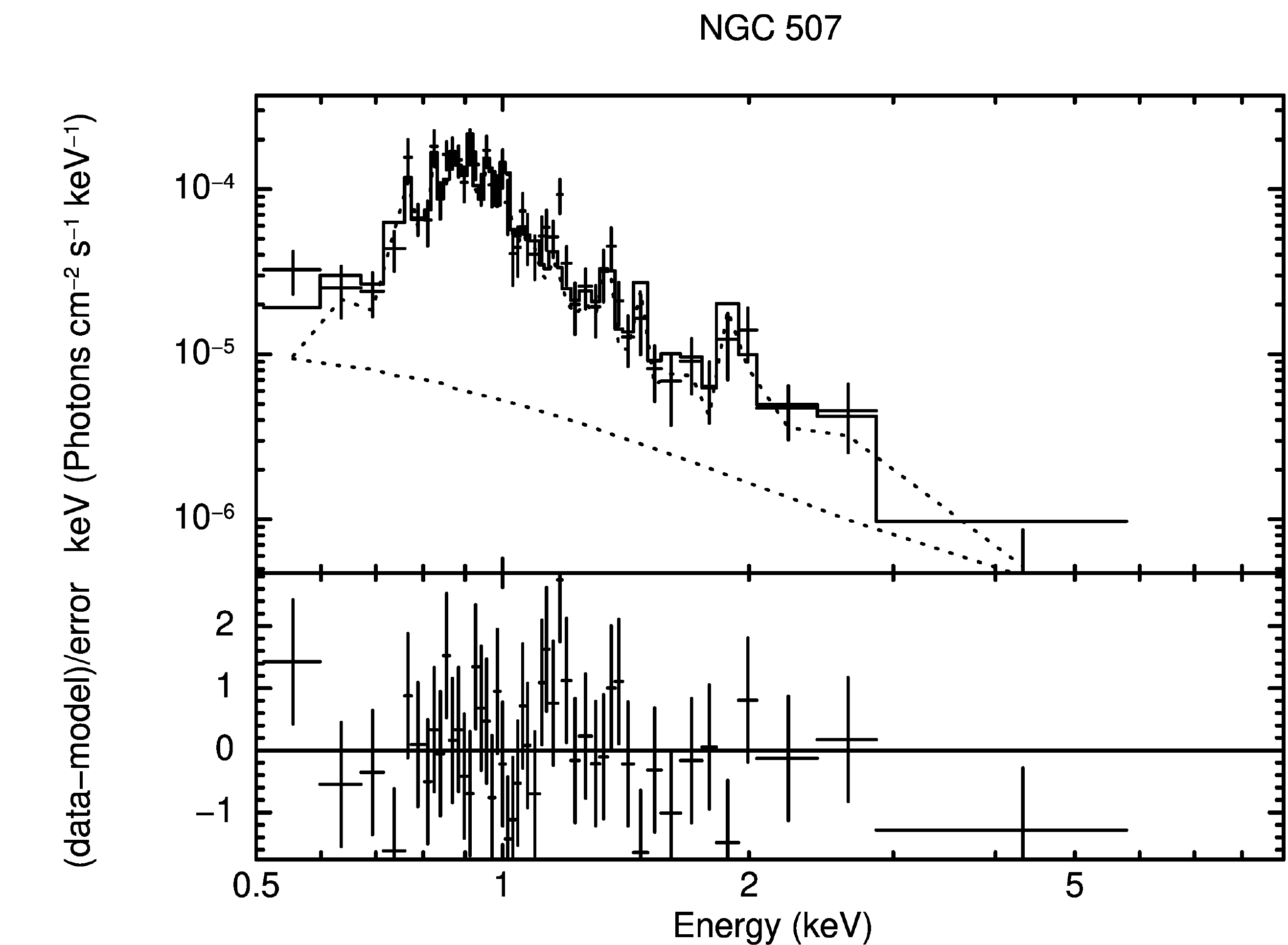}

\end{figure}
\end{center}

\begin{center}
 \begin{figure}
	\includegraphics[width=0.89\columnwidth]{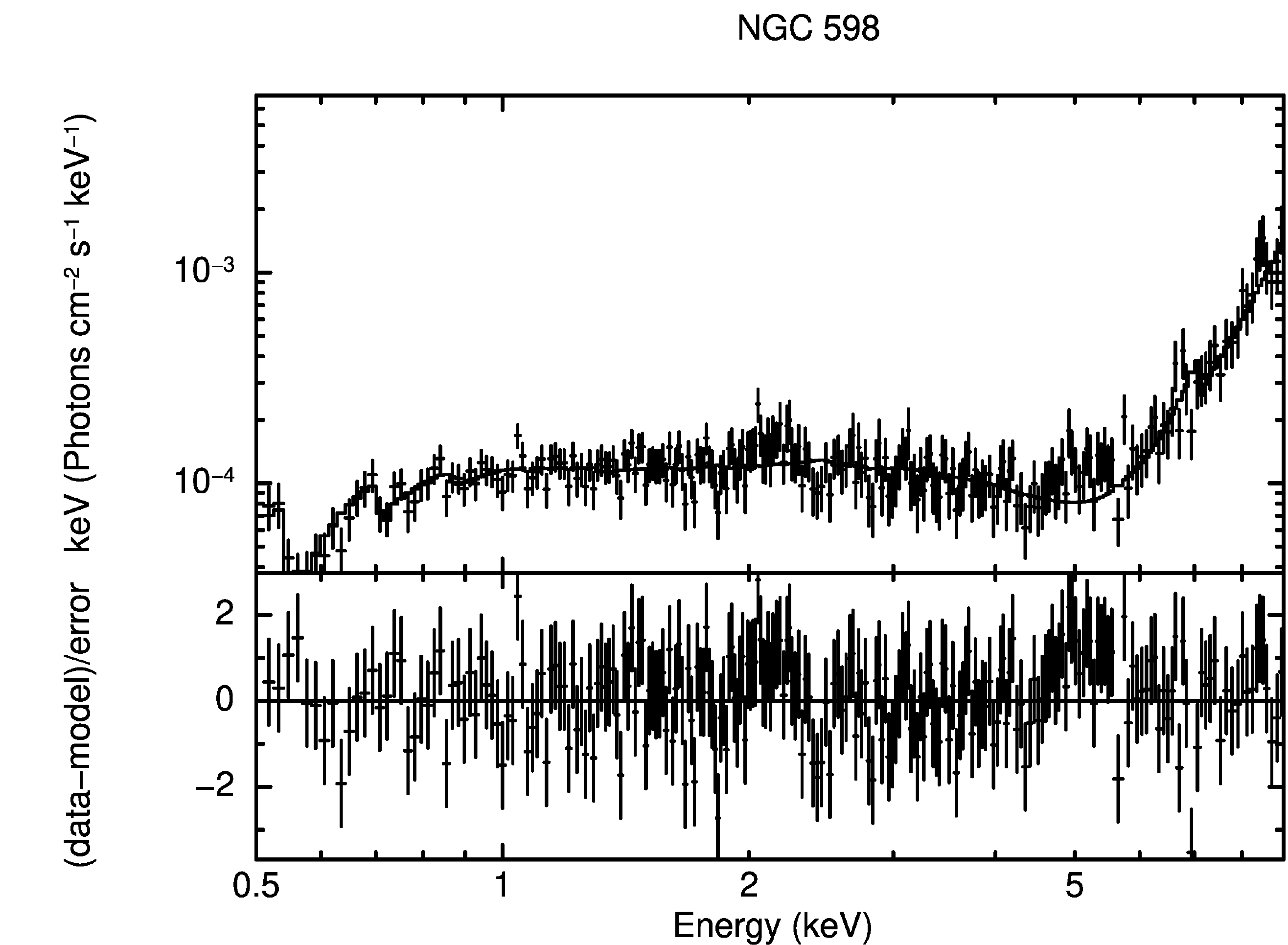}

\end{figure}
\end{center}

\begin{center}
 \begin{figure}
	\includegraphics[width=0.89\columnwidth]{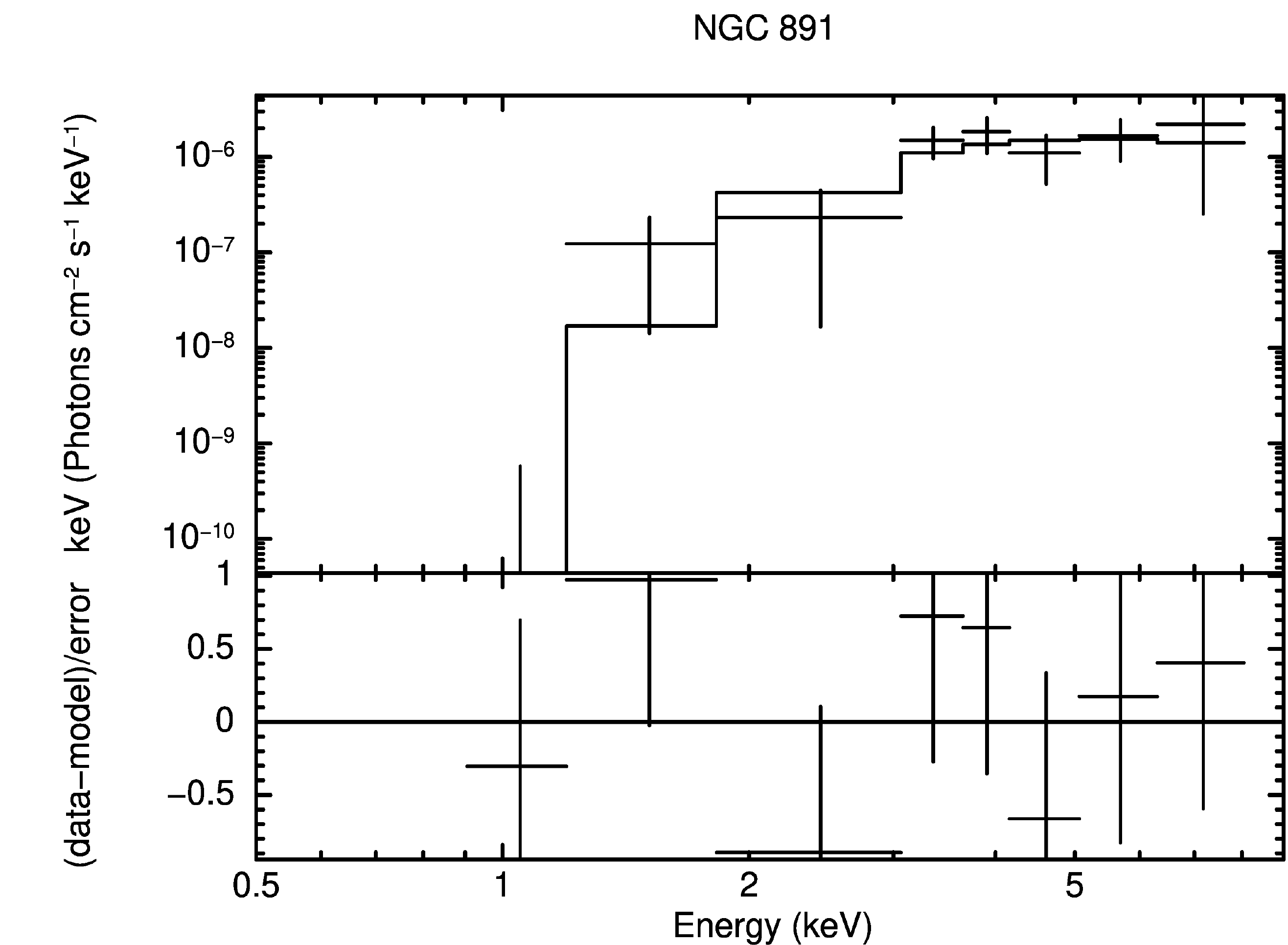}

\end{figure}
\end{center}

\begin{center}
 \begin{figure}
	\includegraphics[width=0.89\columnwidth]{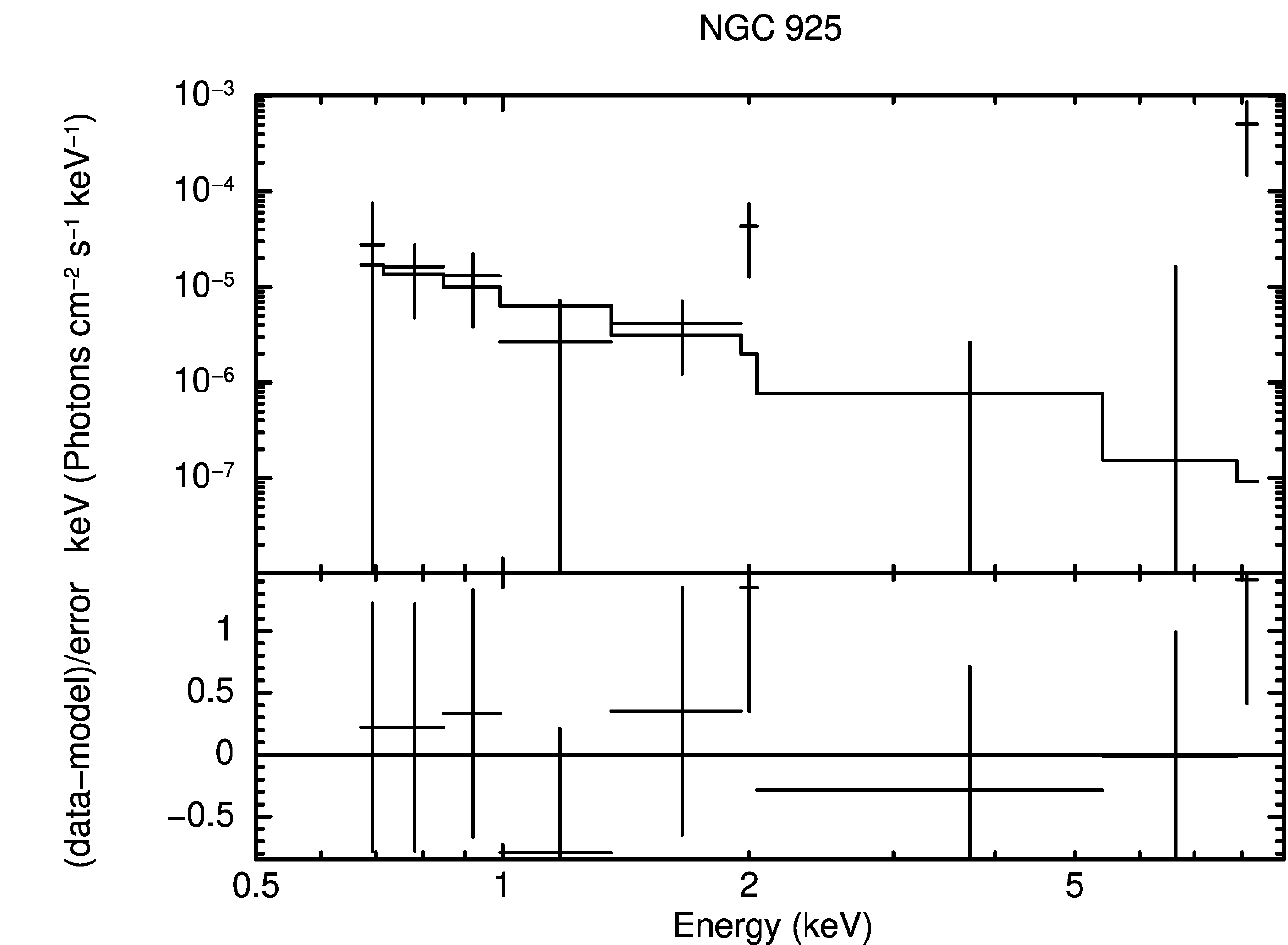}

\end{figure}
\end{center}

\begin{center}
 \begin{figure}
	\includegraphics[width=0.89\columnwidth]{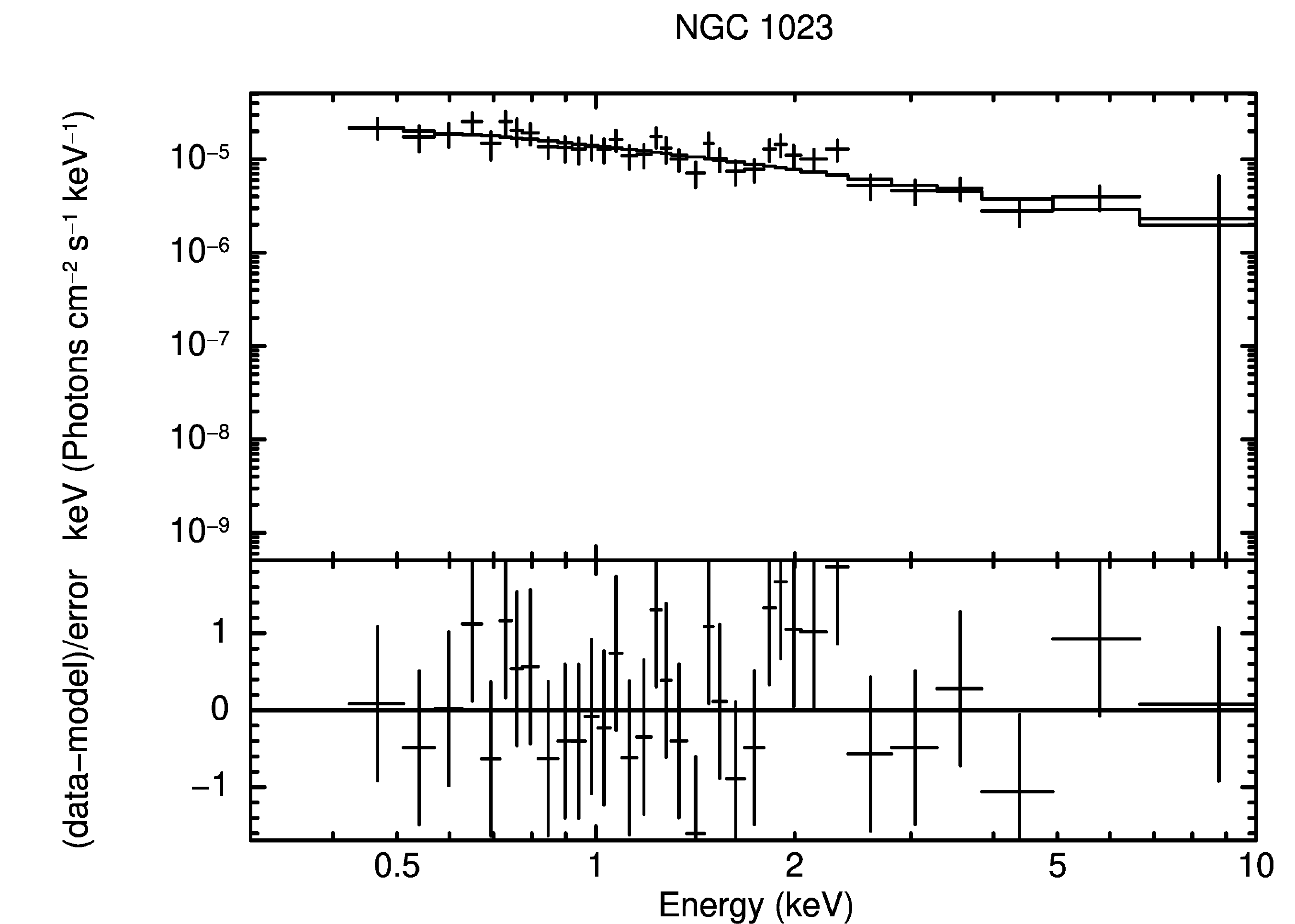}

\end{figure}
\end{center}

\begin{center}
 \begin{figure}
	\includegraphics[width=0.89\columnwidth]{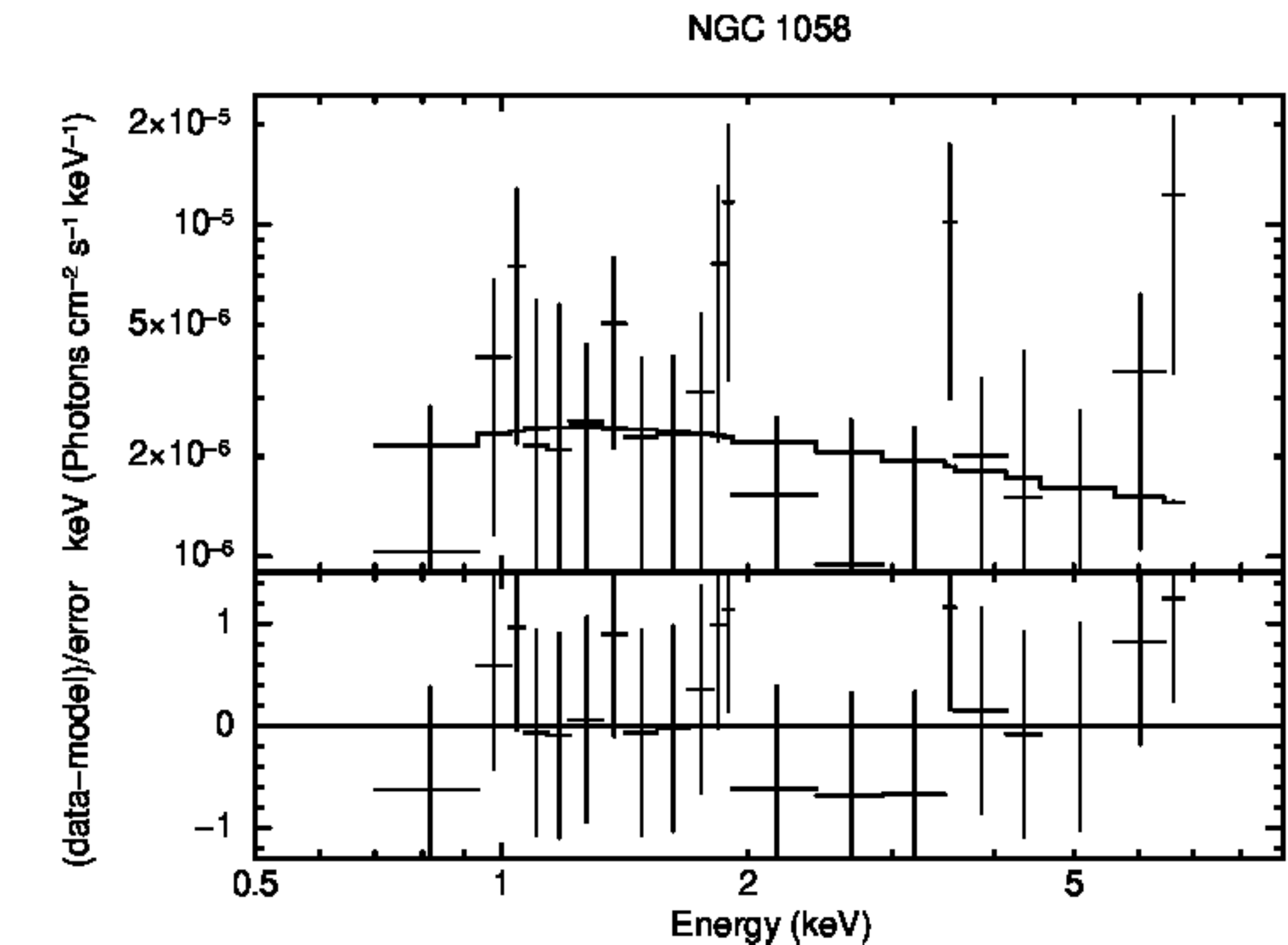}

\end{figure}
\end{center}

%
	 

\begin{center}
 \begin{figure}
	\includegraphics[width=0.89\columnwidth]{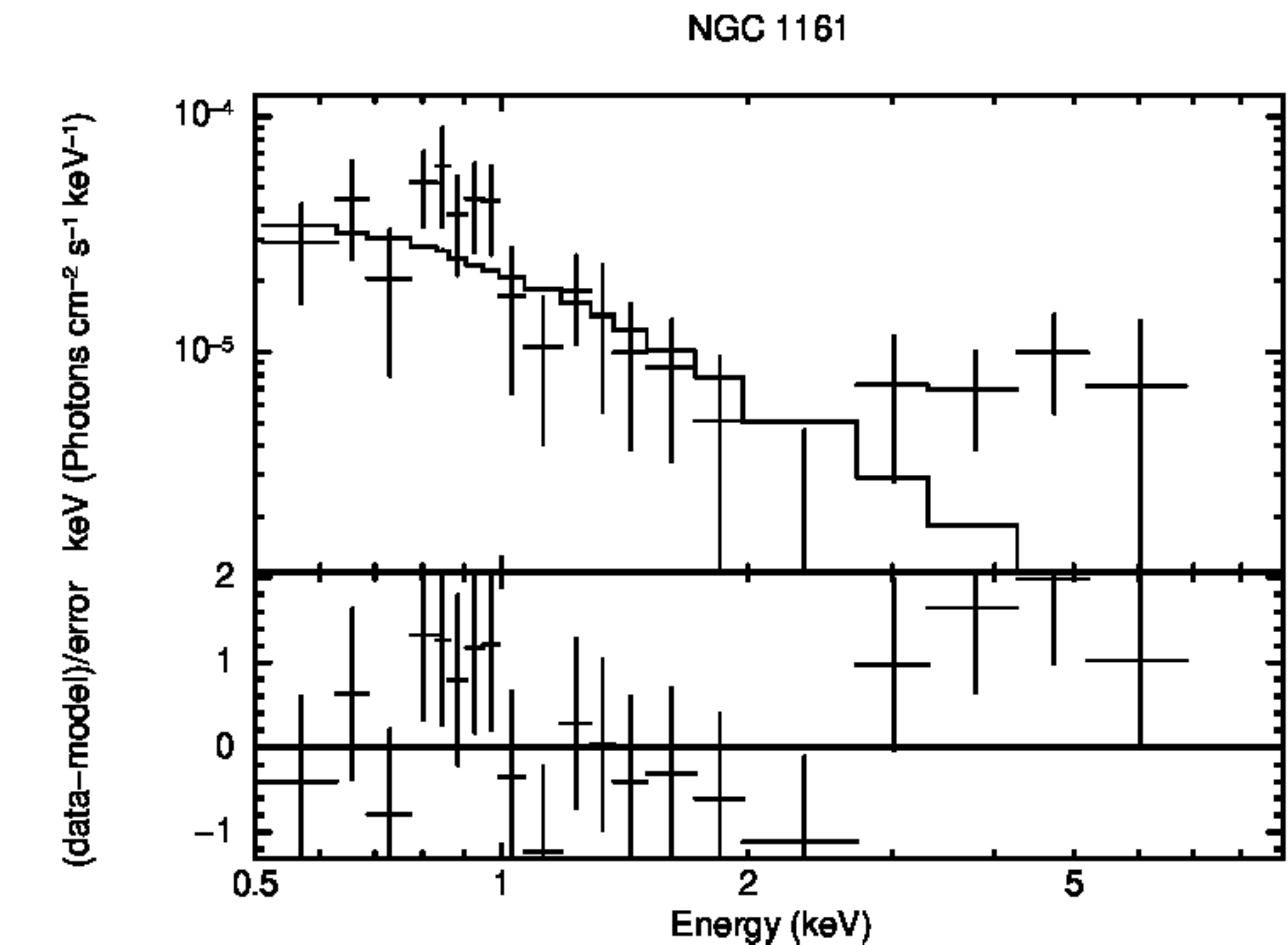}

\end{figure}
\end{center}

%
	 

\begin{center}
 \begin{figure}
	\includegraphics[width=0.89\columnwidth]{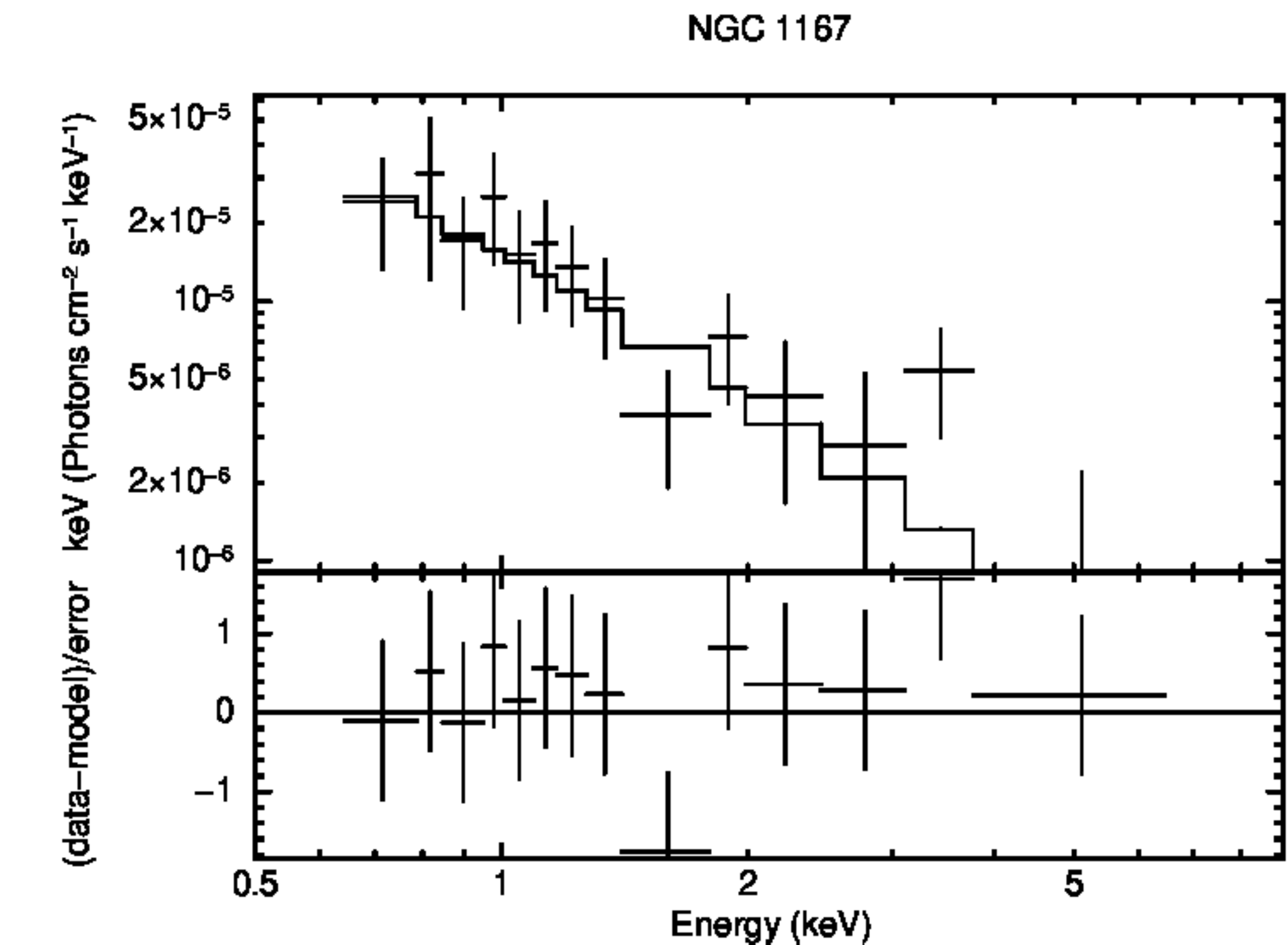}

\end{figure}
\end{center}

%
	 

\begin{center}
 \begin{figure}
	\includegraphics[width=0.89\columnwidth]{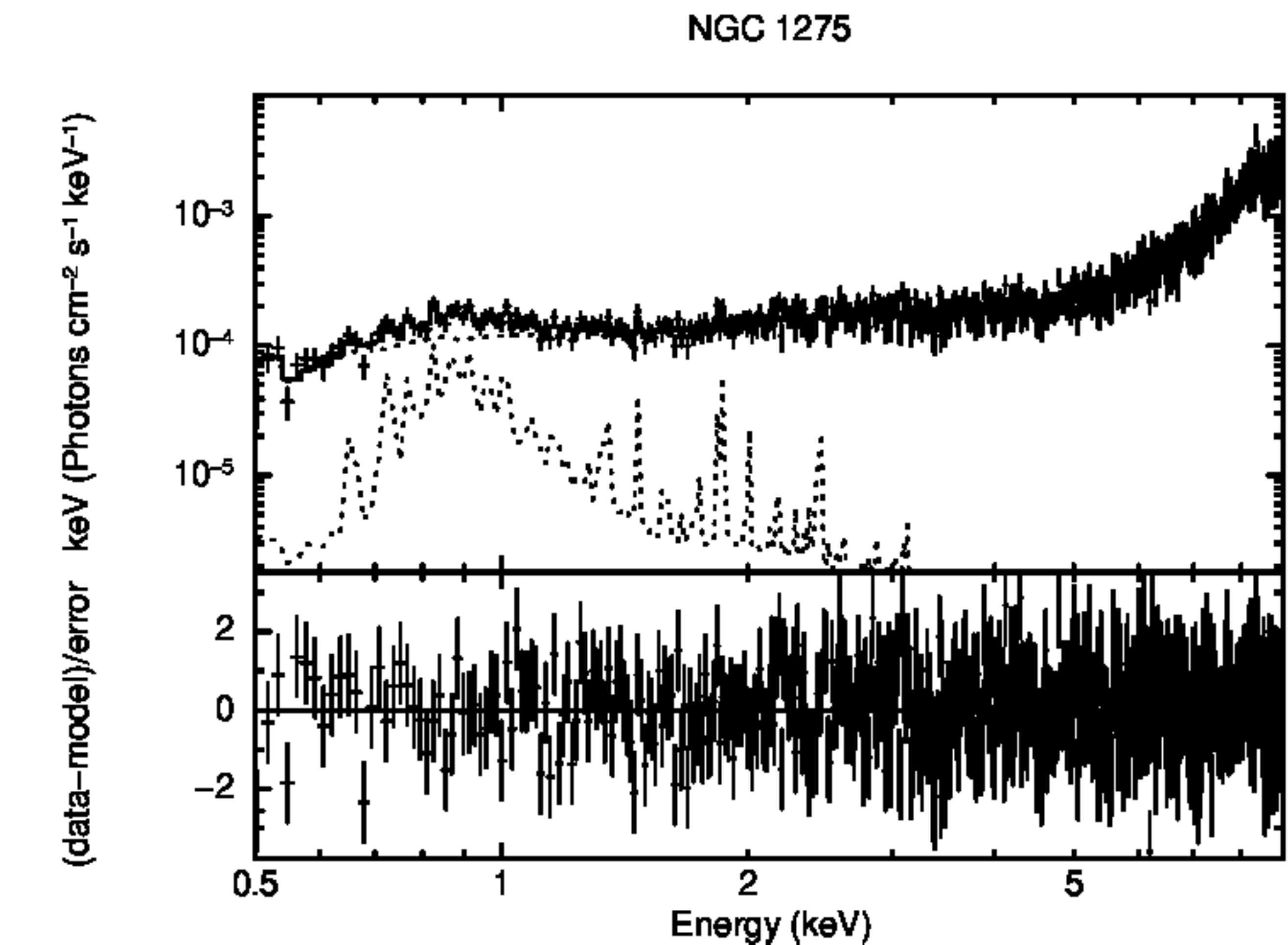}

\end{figure}
\end{center}

%
	 

\begin{center}
 \begin{figure}
	\includegraphics[width=0.89\columnwidth]{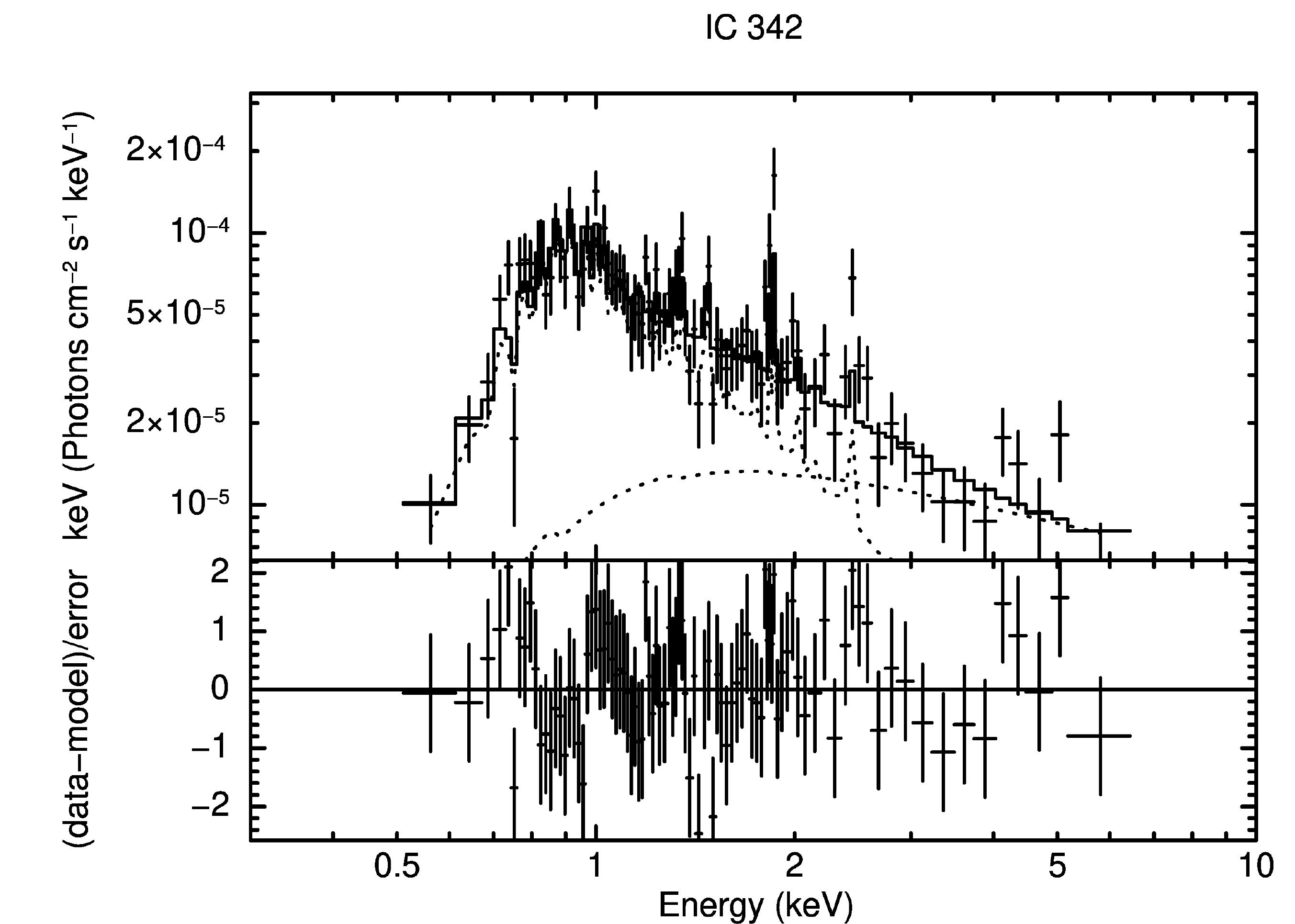}

\end{figure}
\end{center}

\begin{center}
 \begin{figure}
	\includegraphics[width=0.89\columnwidth]{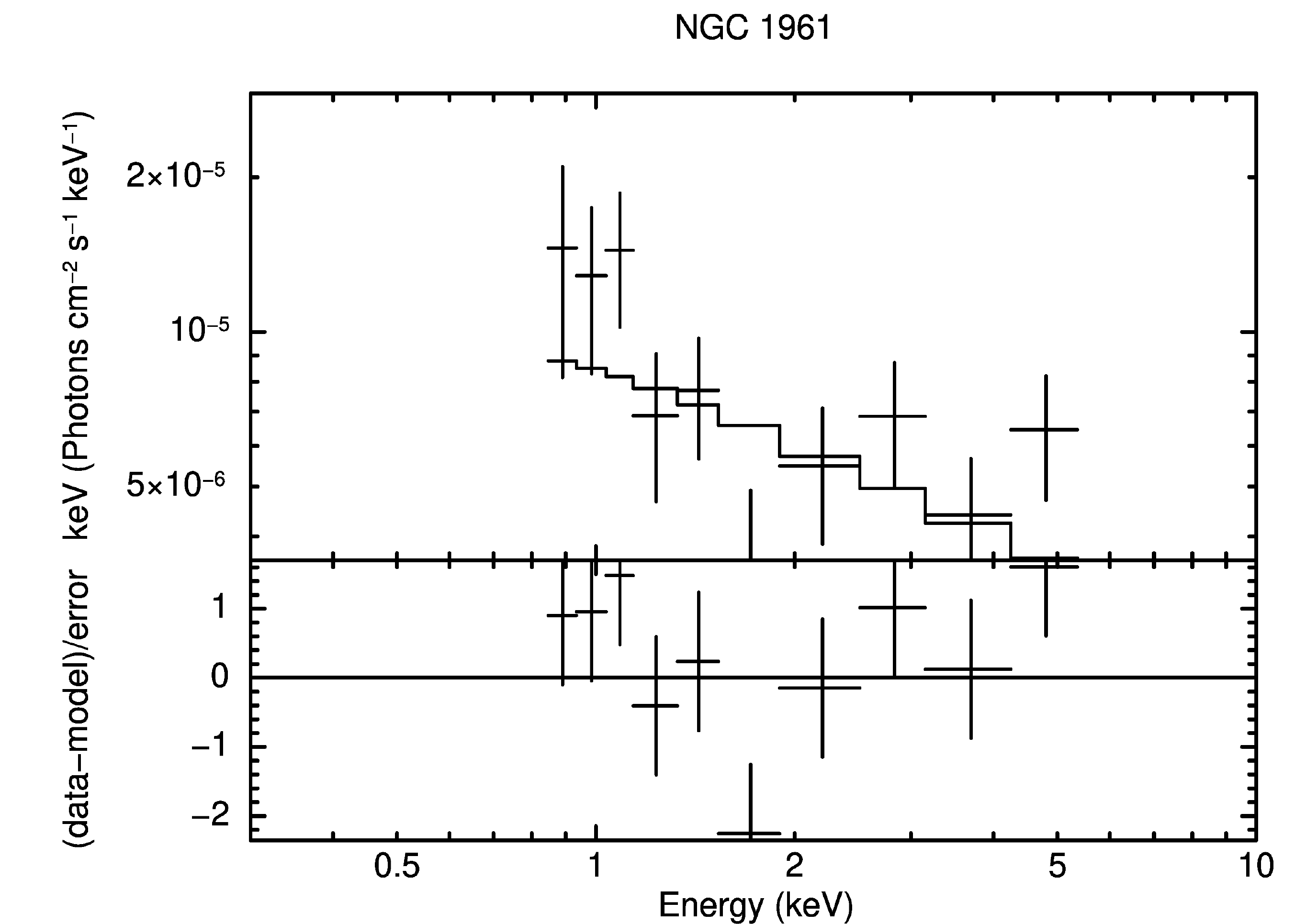}

\end{figure}
\end{center}

\begin{center}
 \begin{figure}
	\includegraphics[width=0.89\columnwidth]{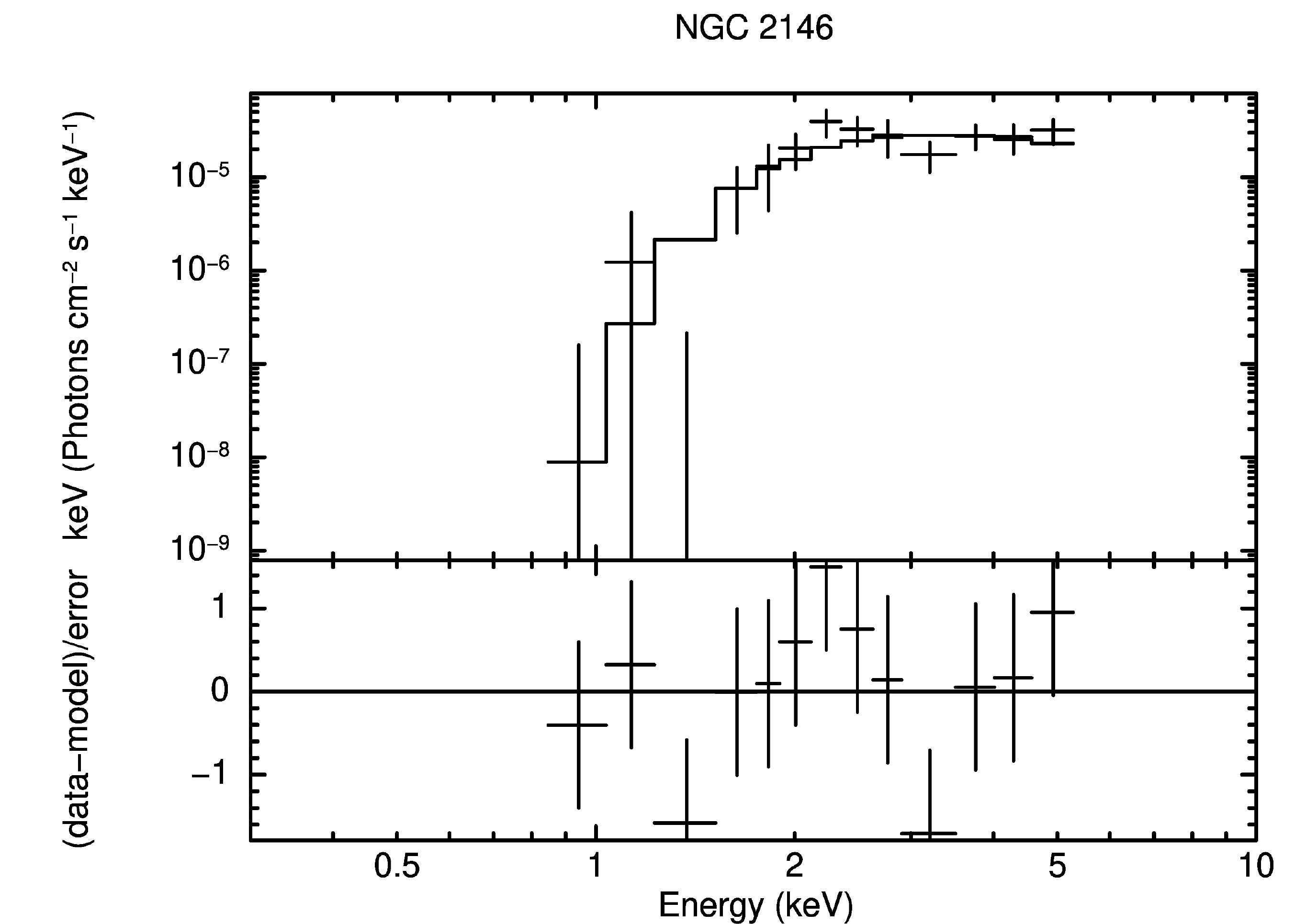}

\end{figure}
\end{center}

\begin{center}
 \begin{figure}
	\includegraphics[width=0.89\columnwidth]{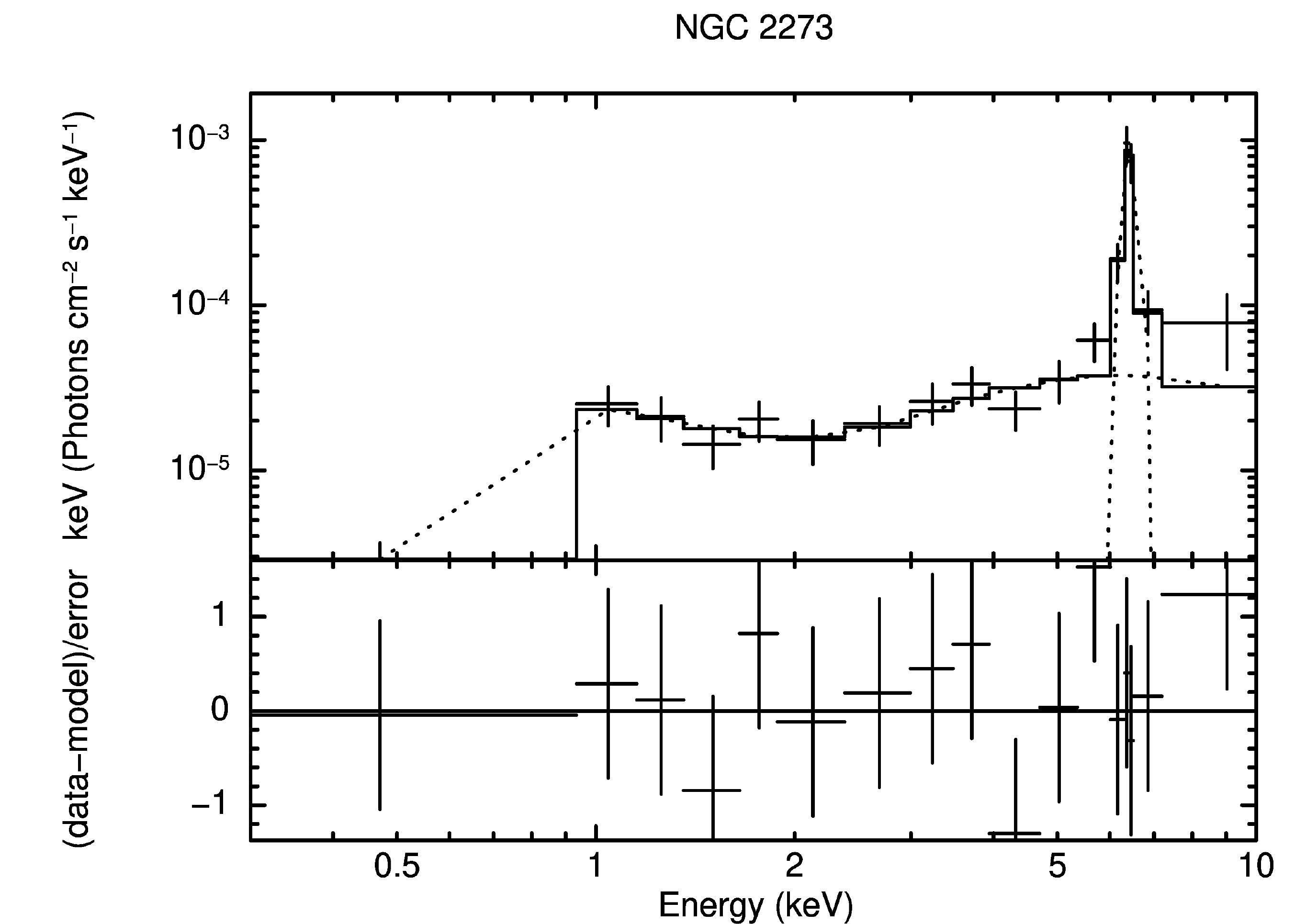}

\end{figure}
\end{center}

\begin{center}
 \begin{figure}
	\includegraphics[width=0.89\columnwidth]{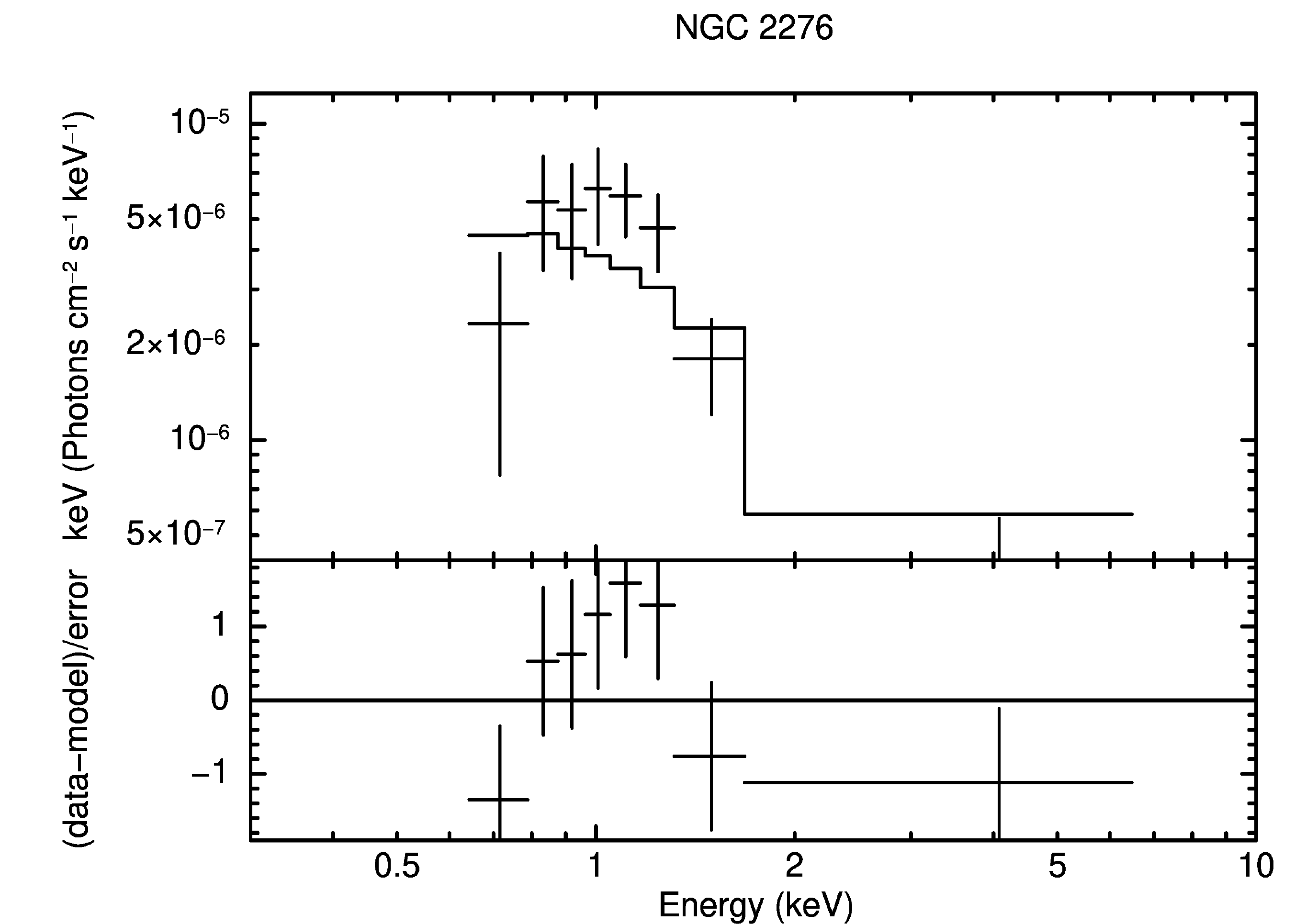}

\end{figure}
\end{center}

\begin{center}
 \begin{figure}
	\includegraphics[width=0.89\columnwidth]{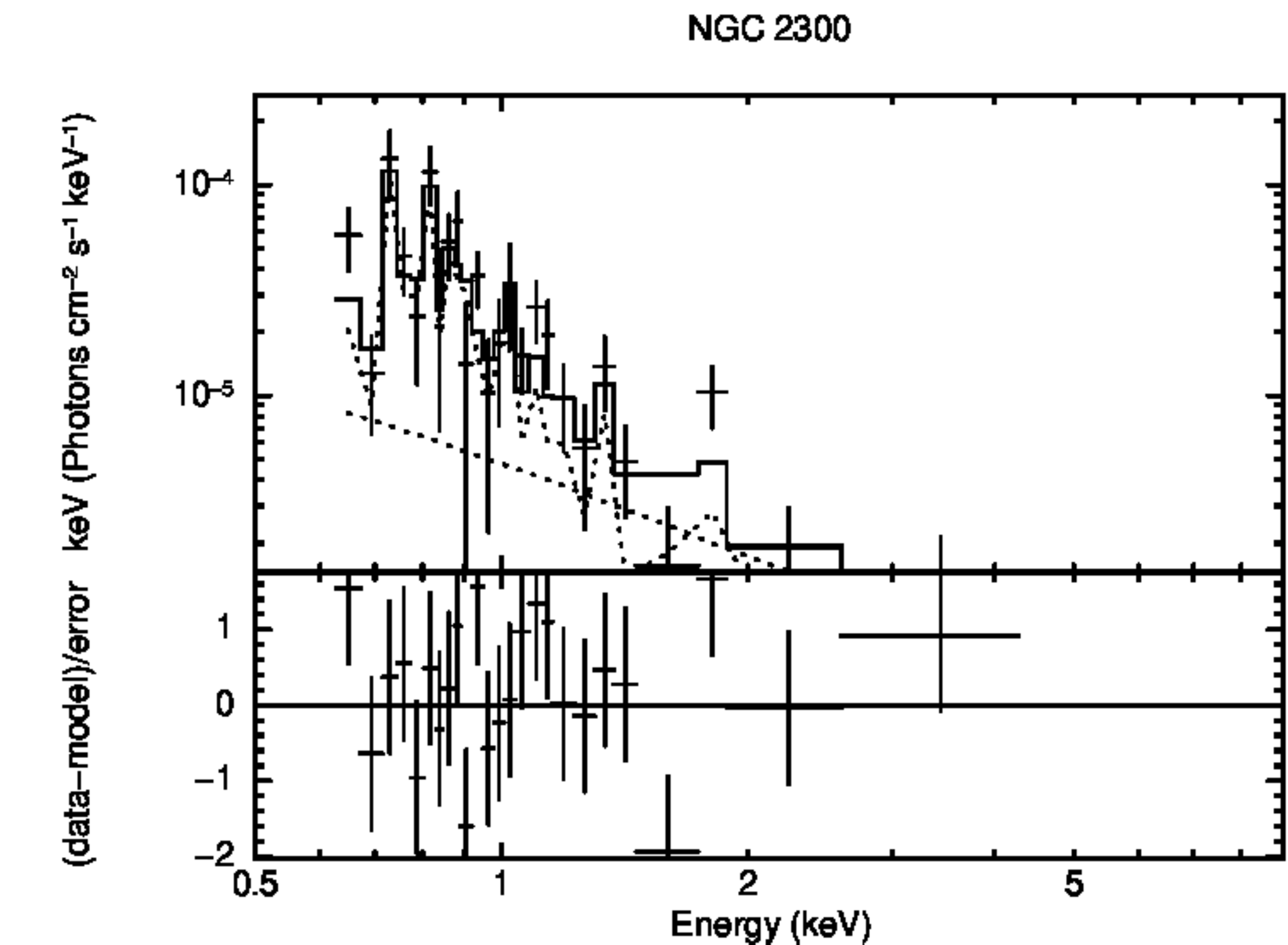}

\end{figure}
\end{center}

\begin{center}
 \begin{figure}
	\includegraphics[width=0.89\columnwidth]{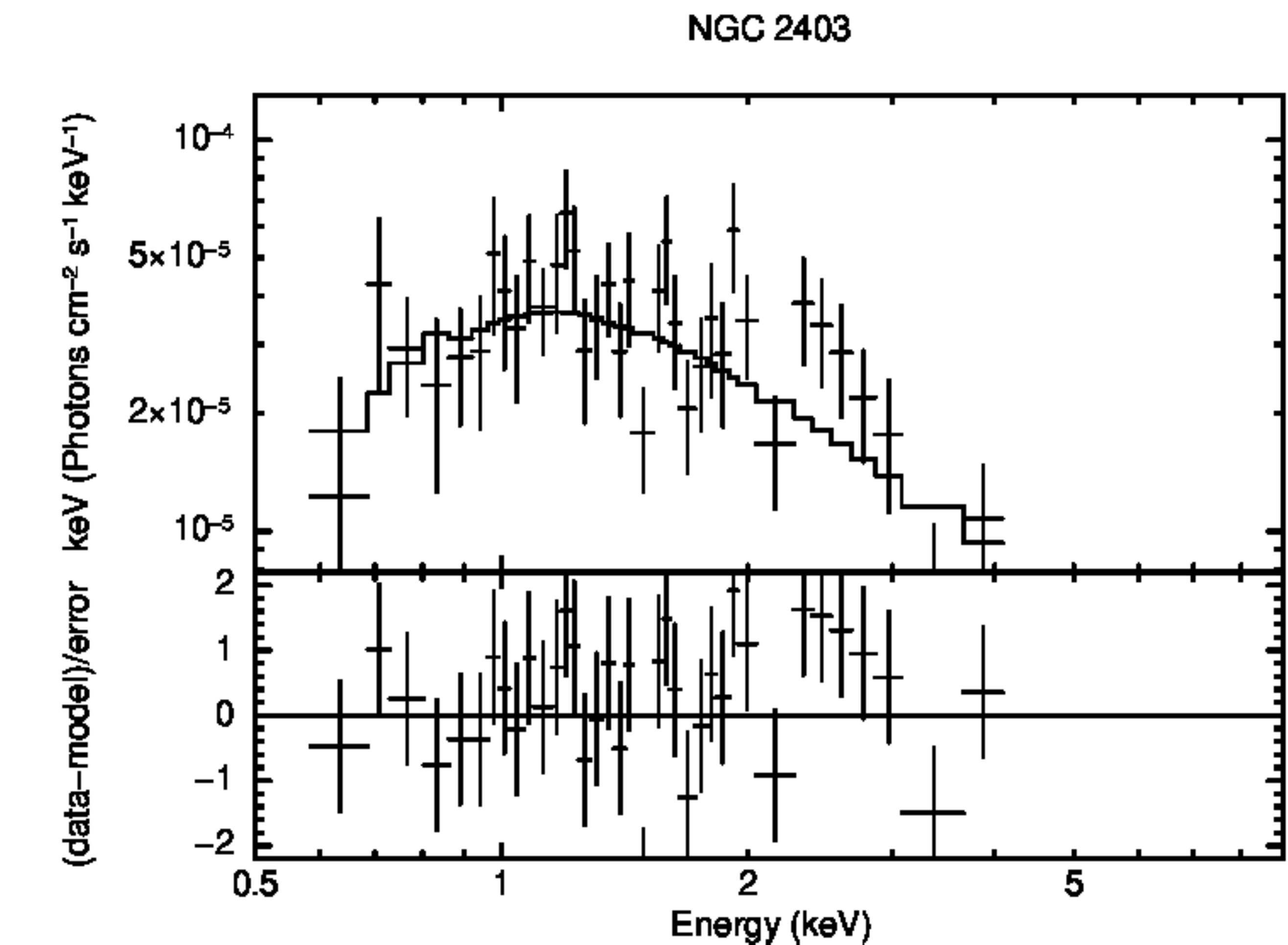}

\end{figure}
\end{center}

%
	 

\begin{center}
 \begin{figure}
	\includegraphics[width=0.89\columnwidth]{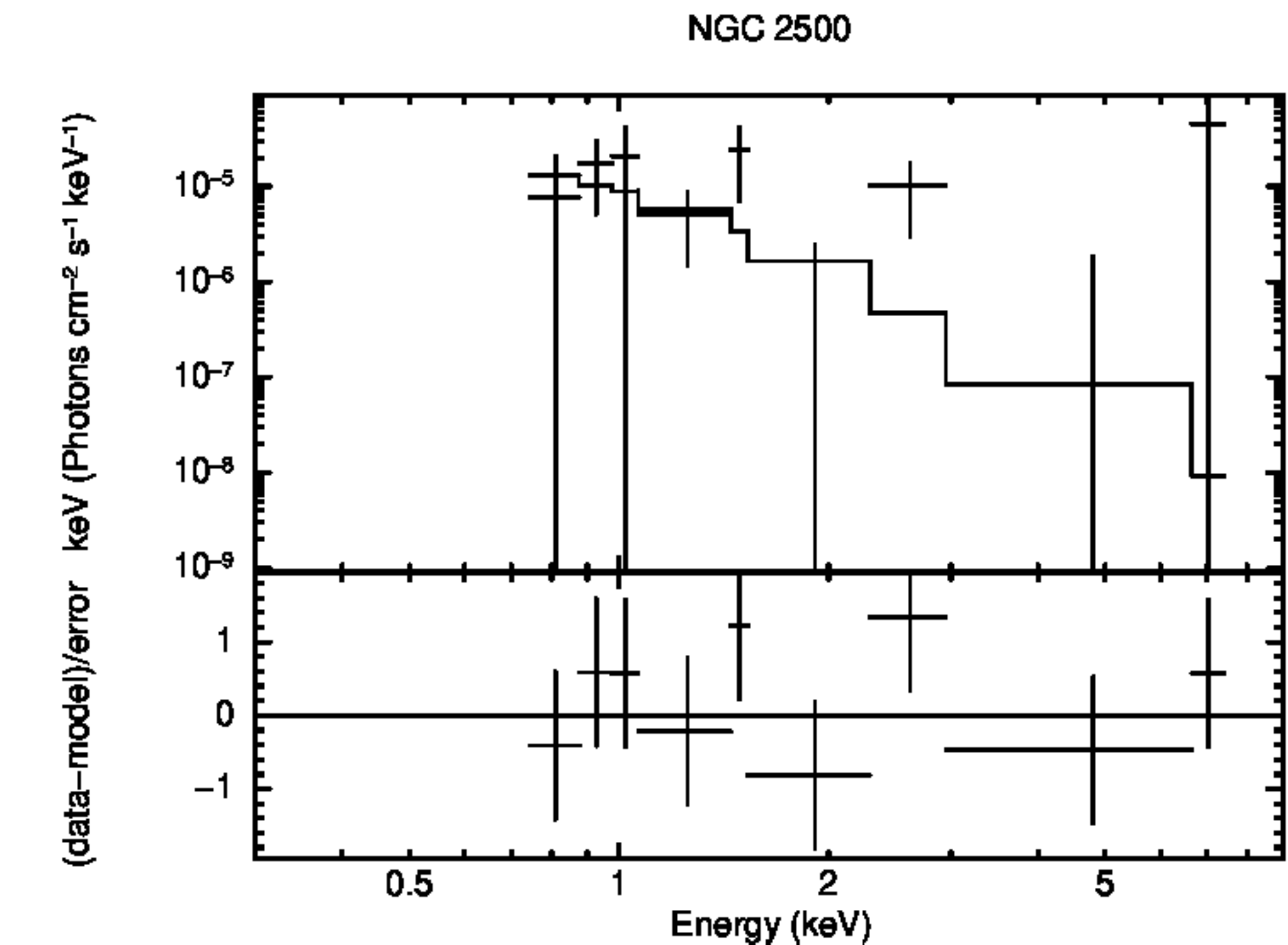}

\end{figure}
\end{center}

%
	 

\begin{center}
 \begin{figure}
	\includegraphics[width=0.89\columnwidth]{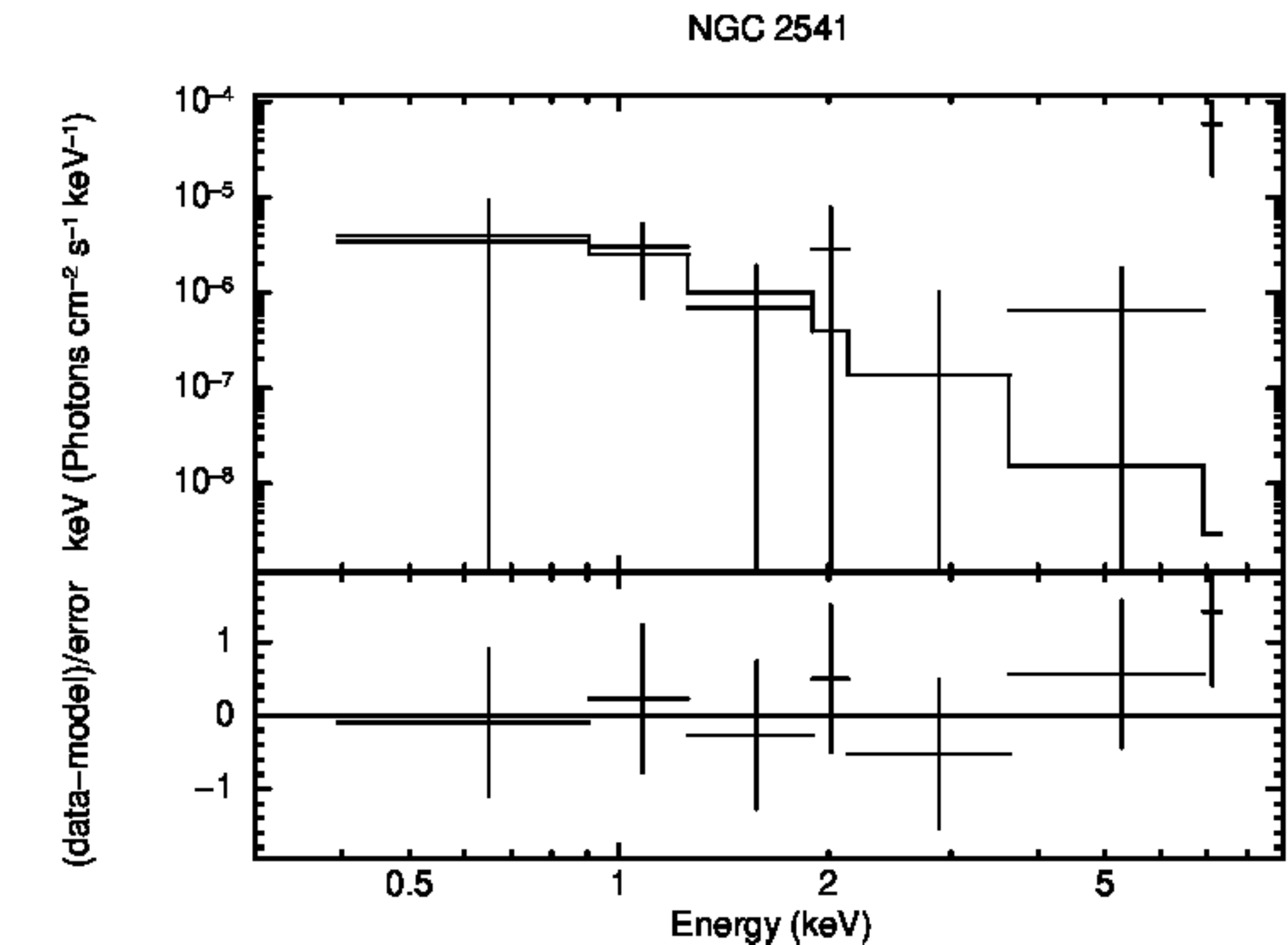}

\end{figure}
\end{center}

%
	 

\begin{center}
 \begin{figure}
	\includegraphics[width=0.89\columnwidth]{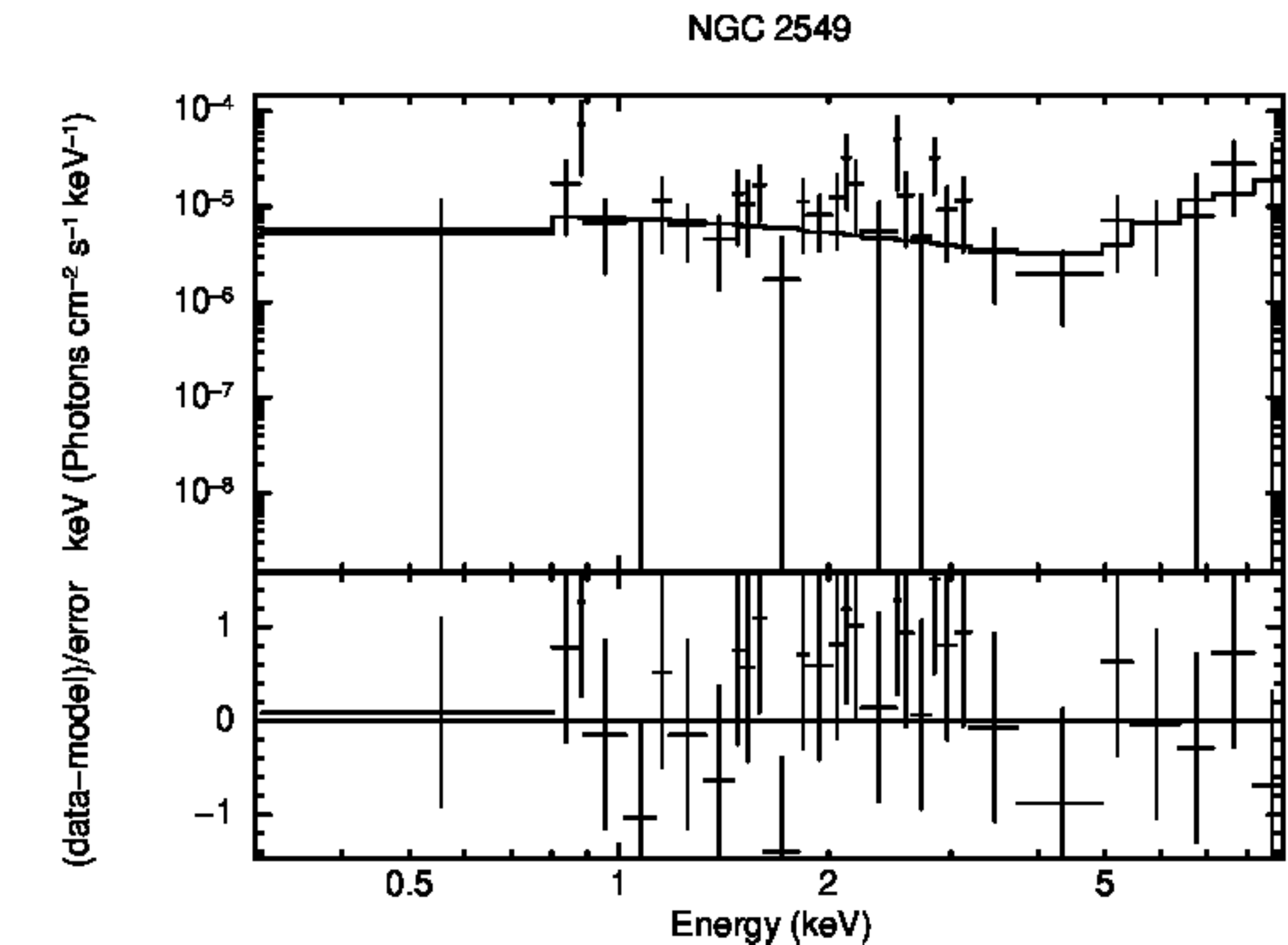}

\end{figure}
\end{center}

%
	 

\begin{center}
 \begin{figure}
	\includegraphics[width=0.89\columnwidth]{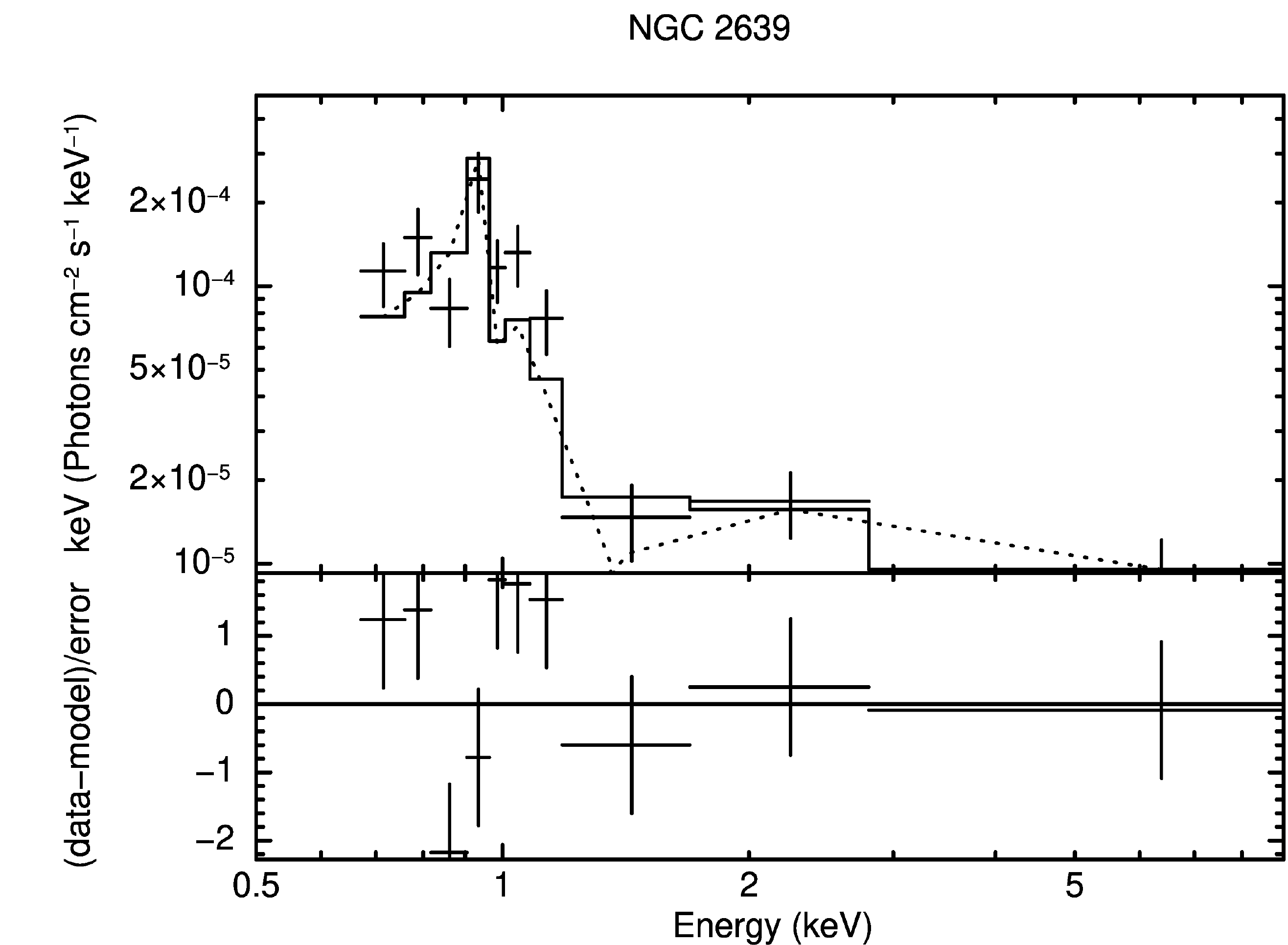}

\end{figure}
\end{center}

\begin{center}
 \begin{figure}
	\includegraphics[width=0.89\columnwidth]{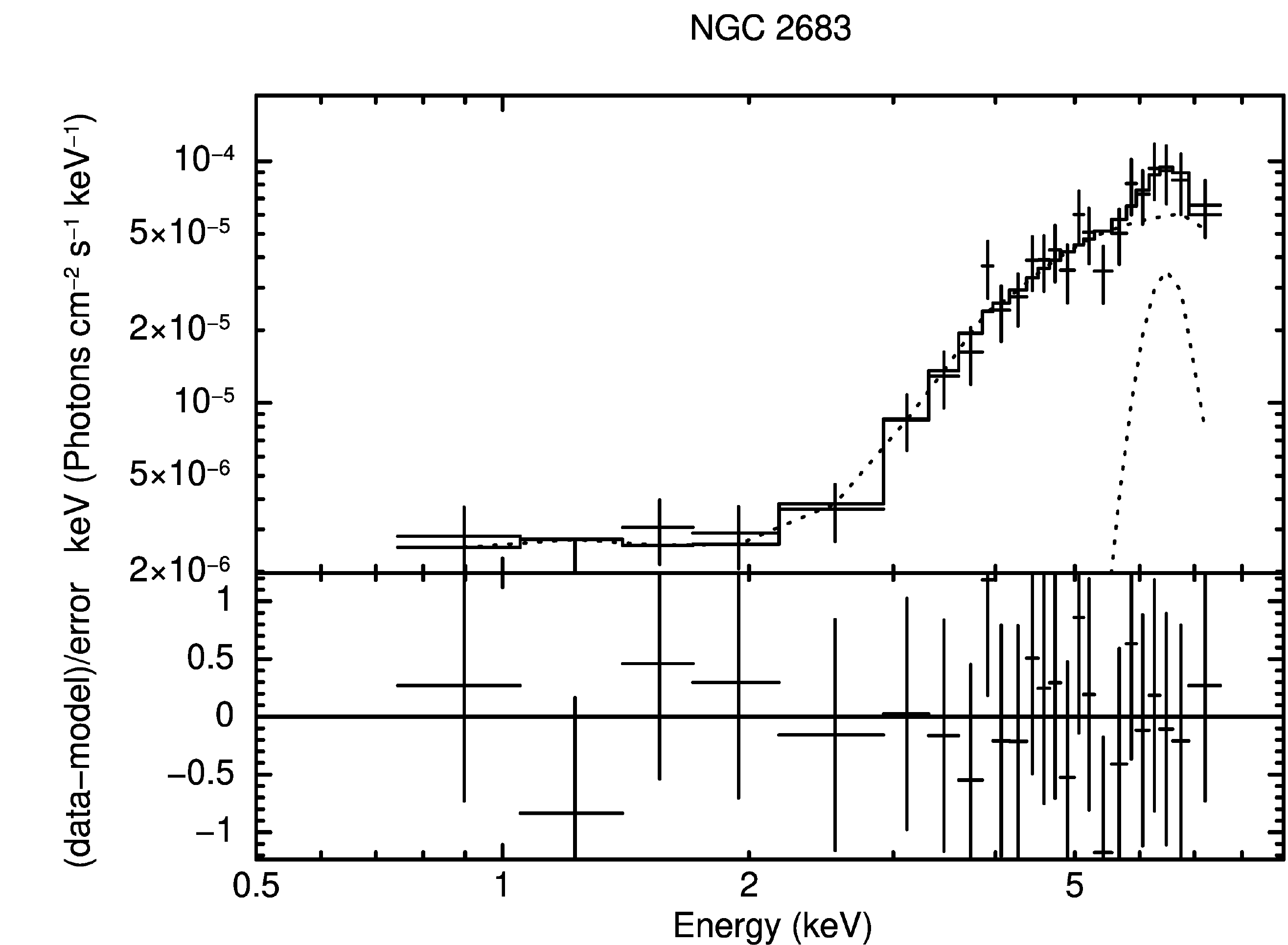}

\end{figure}
\end{center}

\begin{center}
 \begin{figure}
	\includegraphics[width=0.89\columnwidth]{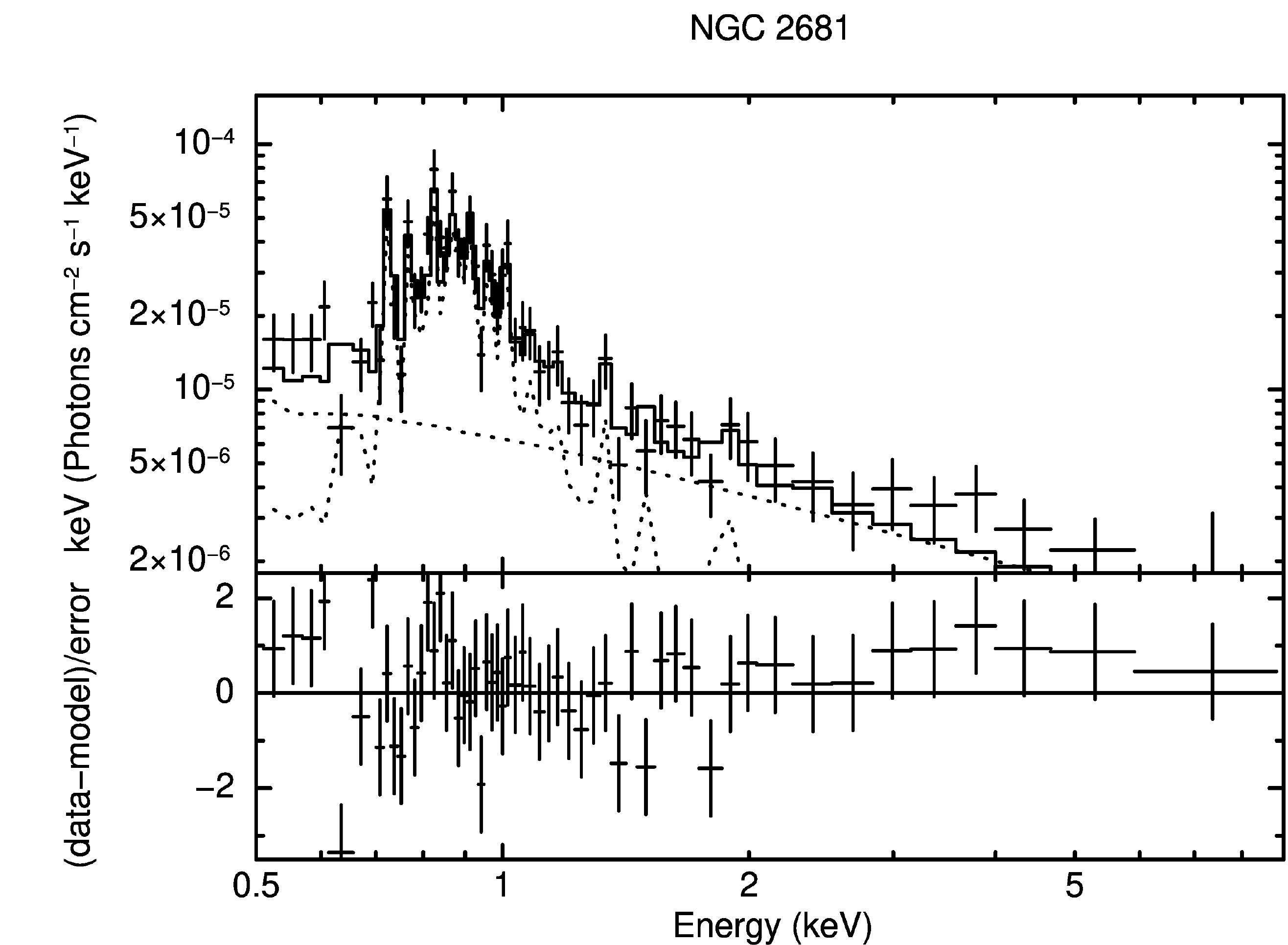}

\end{figure}
\end{center}

\begin{center}
 \begin{figure}
	\includegraphics[width=0.89\columnwidth]{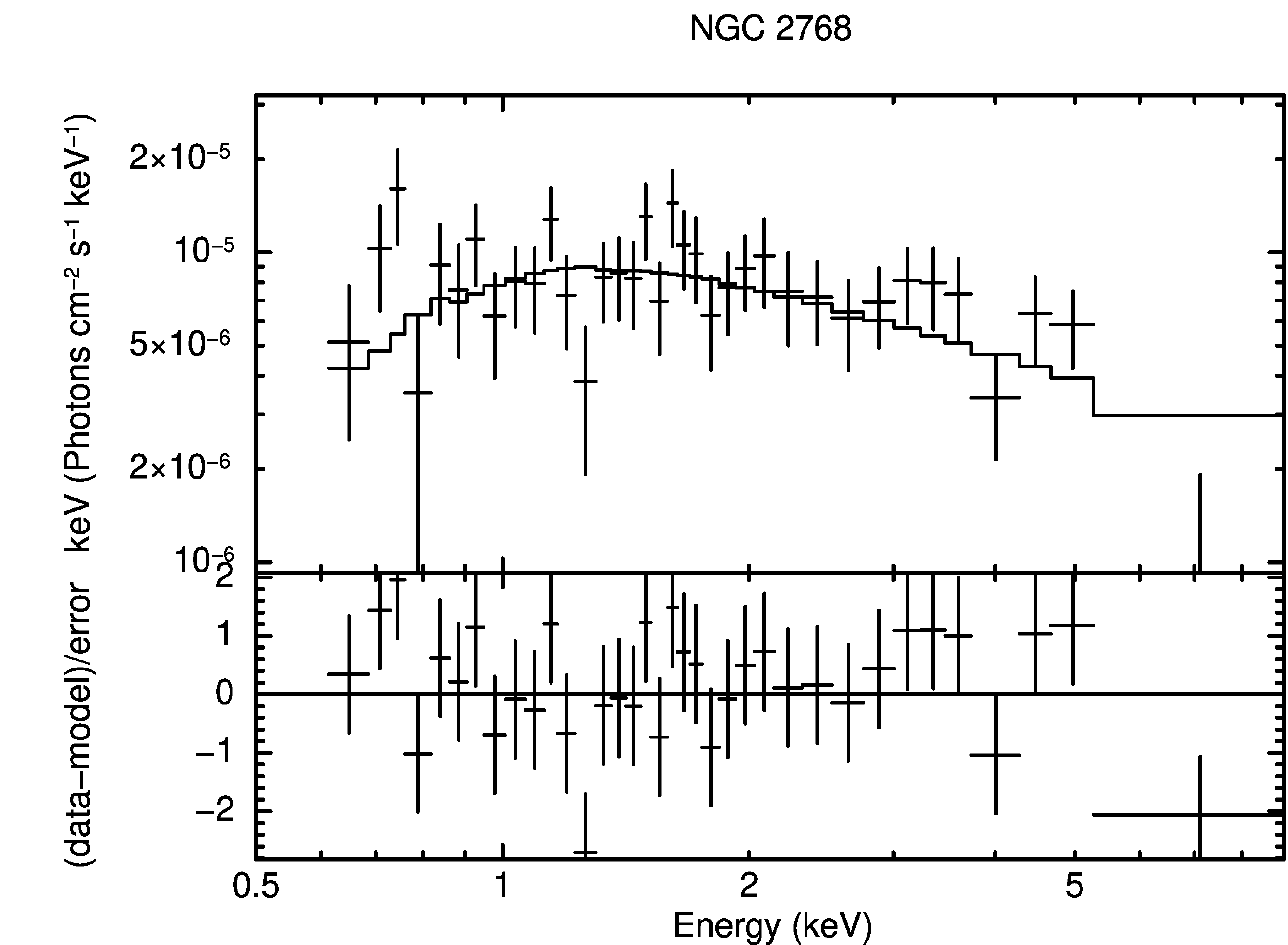}

\end{figure}
\end{center}

\begin{center}
 \begin{figure}
	\includegraphics[width=0.89\columnwidth]{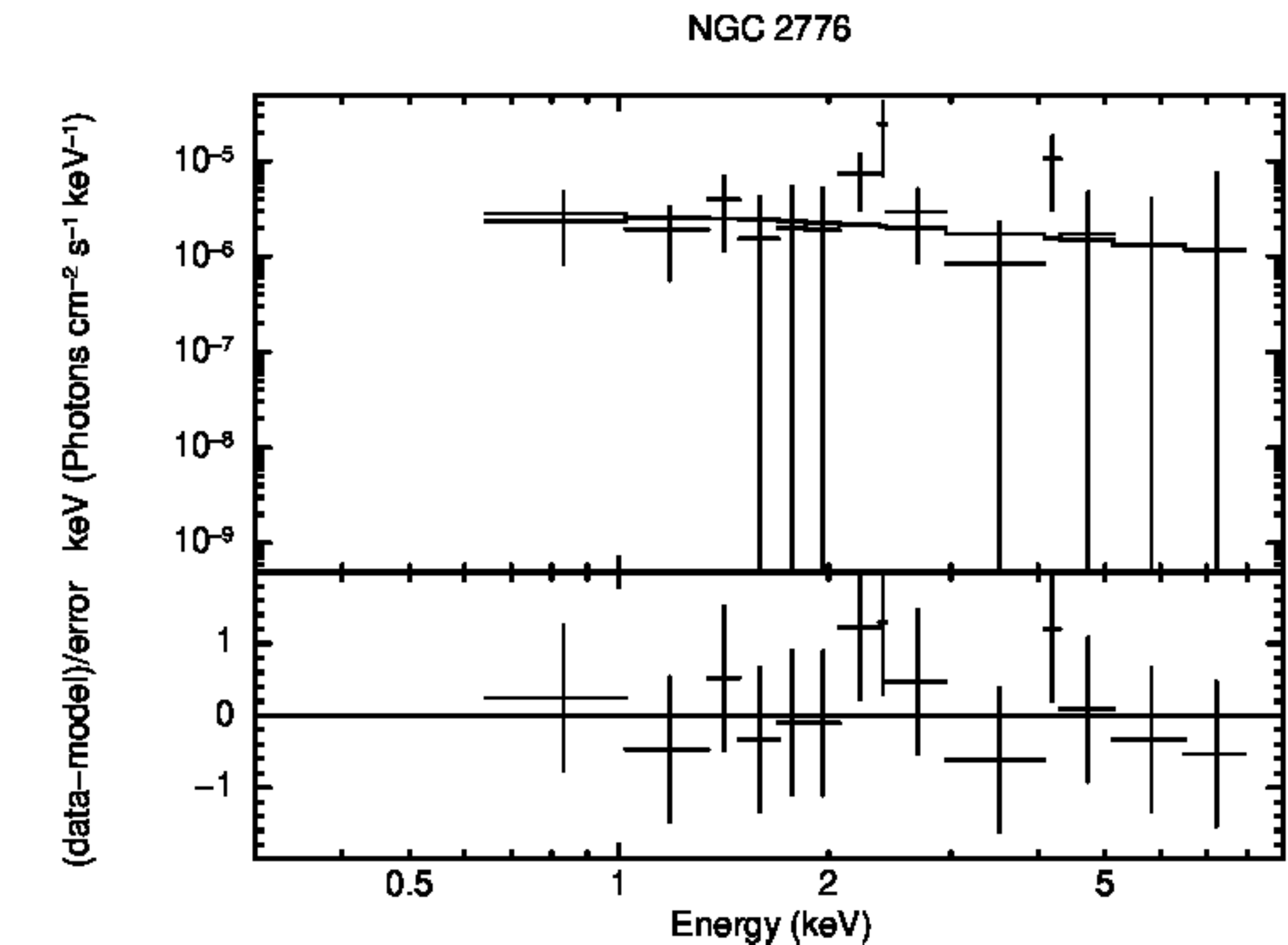}

\end{figure}
\end{center}

%
	 

\begin{center}
 \begin{figure}
	\includegraphics[width=0.89\columnwidth]{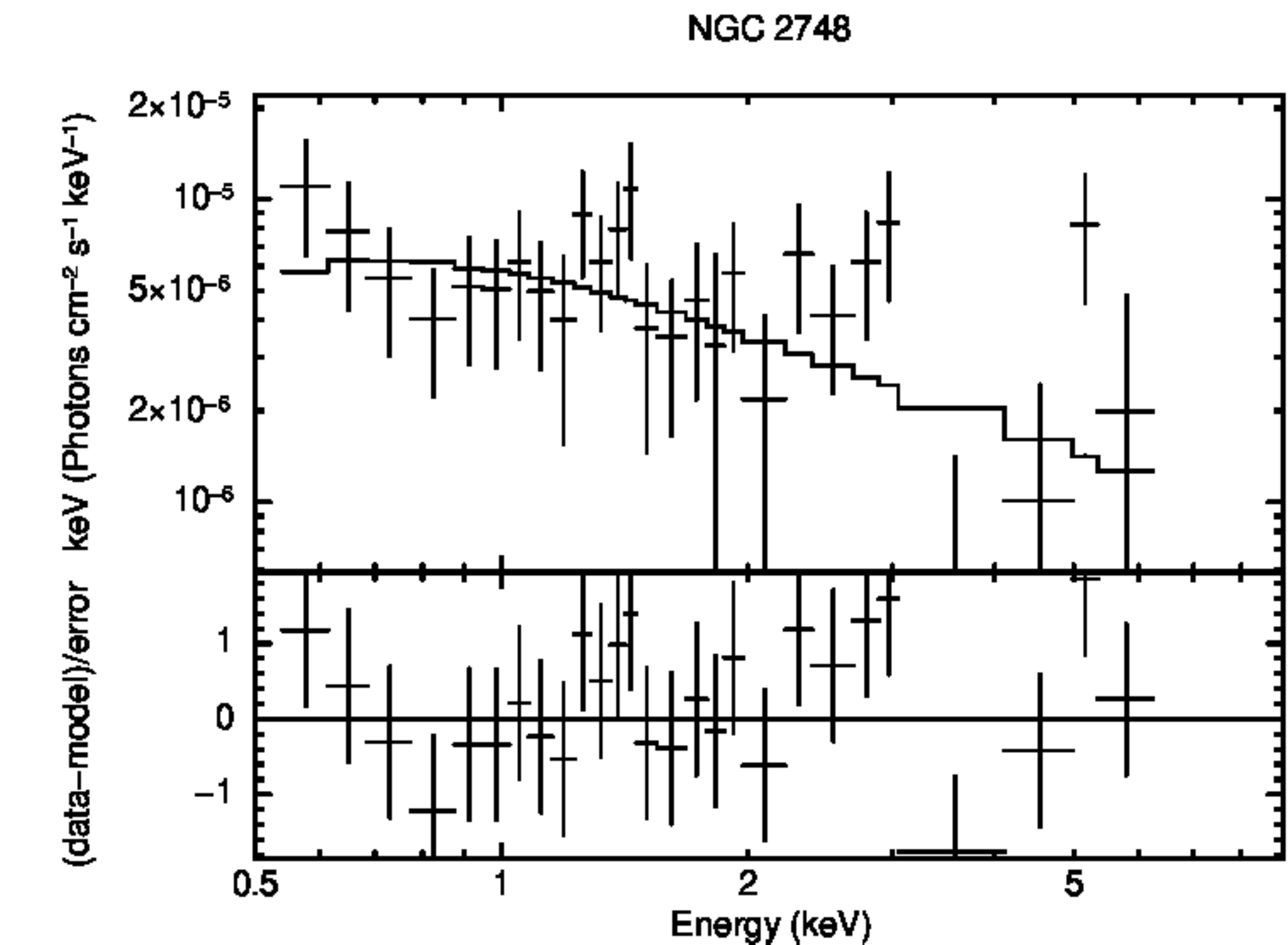}

\end{figure}
\end{center}

%
	 

\begin{center}
 \begin{figure}
	\includegraphics[width=0.89\columnwidth]{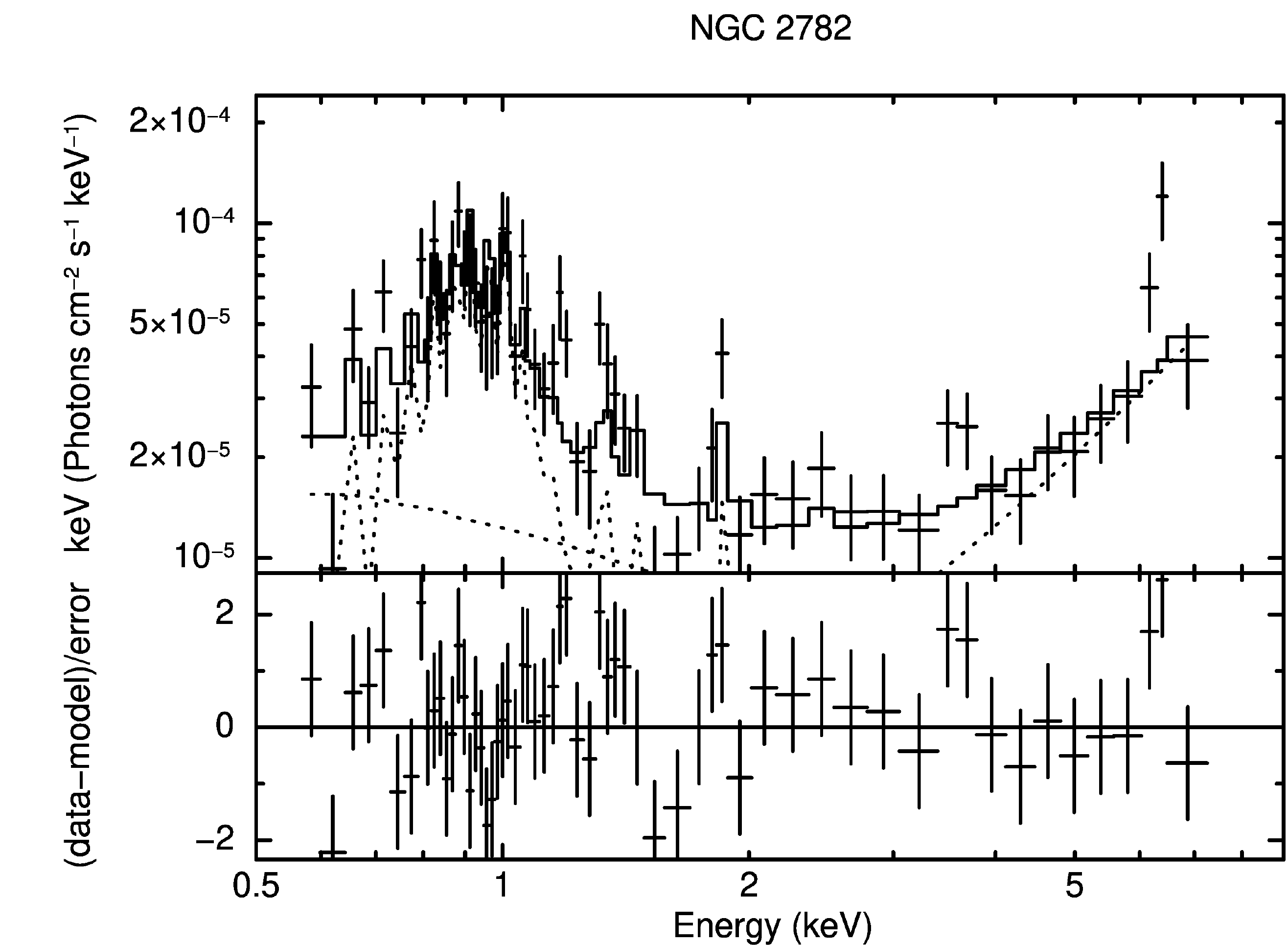}

\end{figure}
\end{center}

\begin{center}
 \begin{figure}
	\includegraphics[width=0.89\columnwidth]{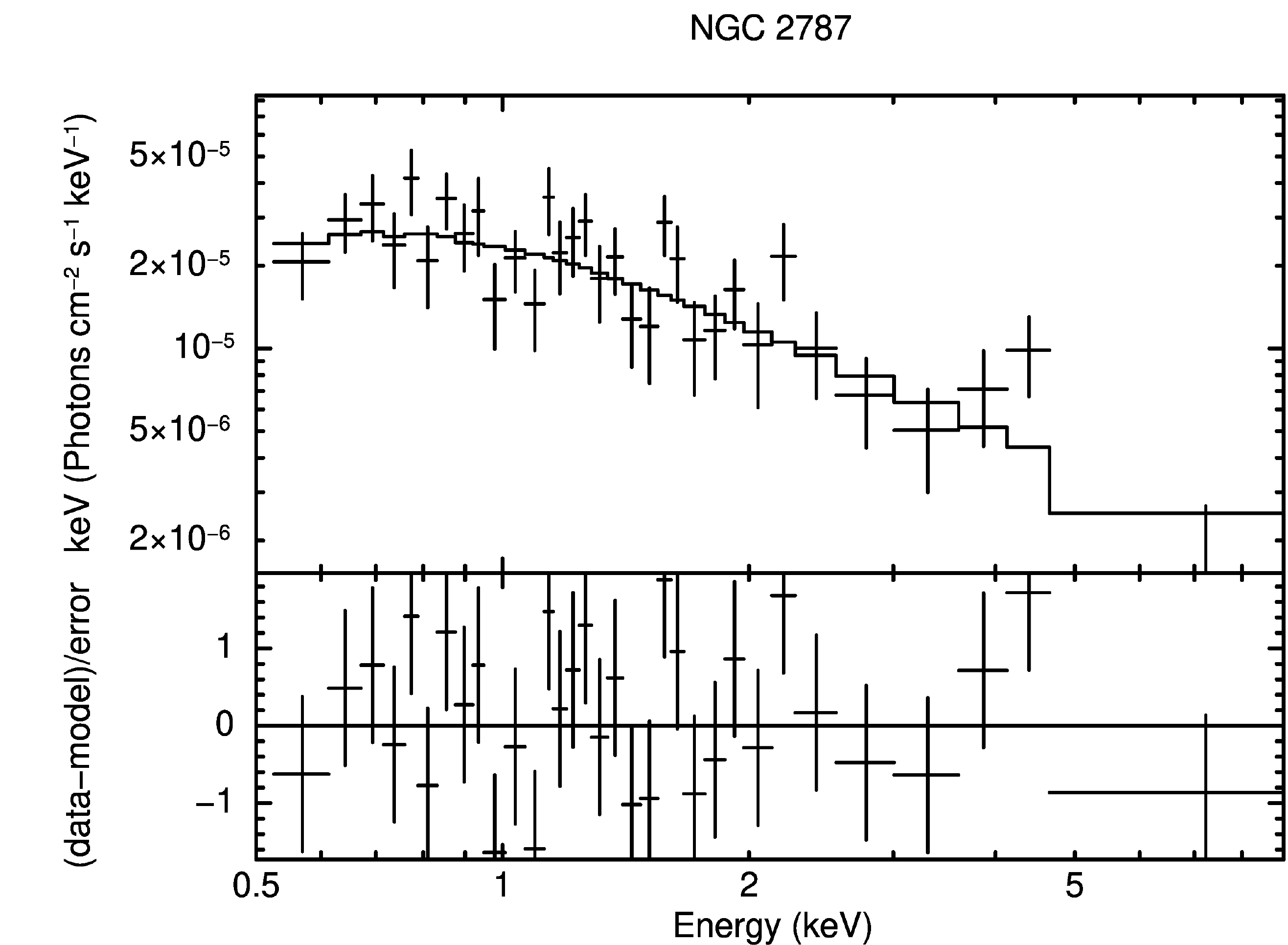}

\end{figure}
\end{center}

\begin{center}
 \begin{figure}
	\includegraphics[width=0.89\columnwidth]{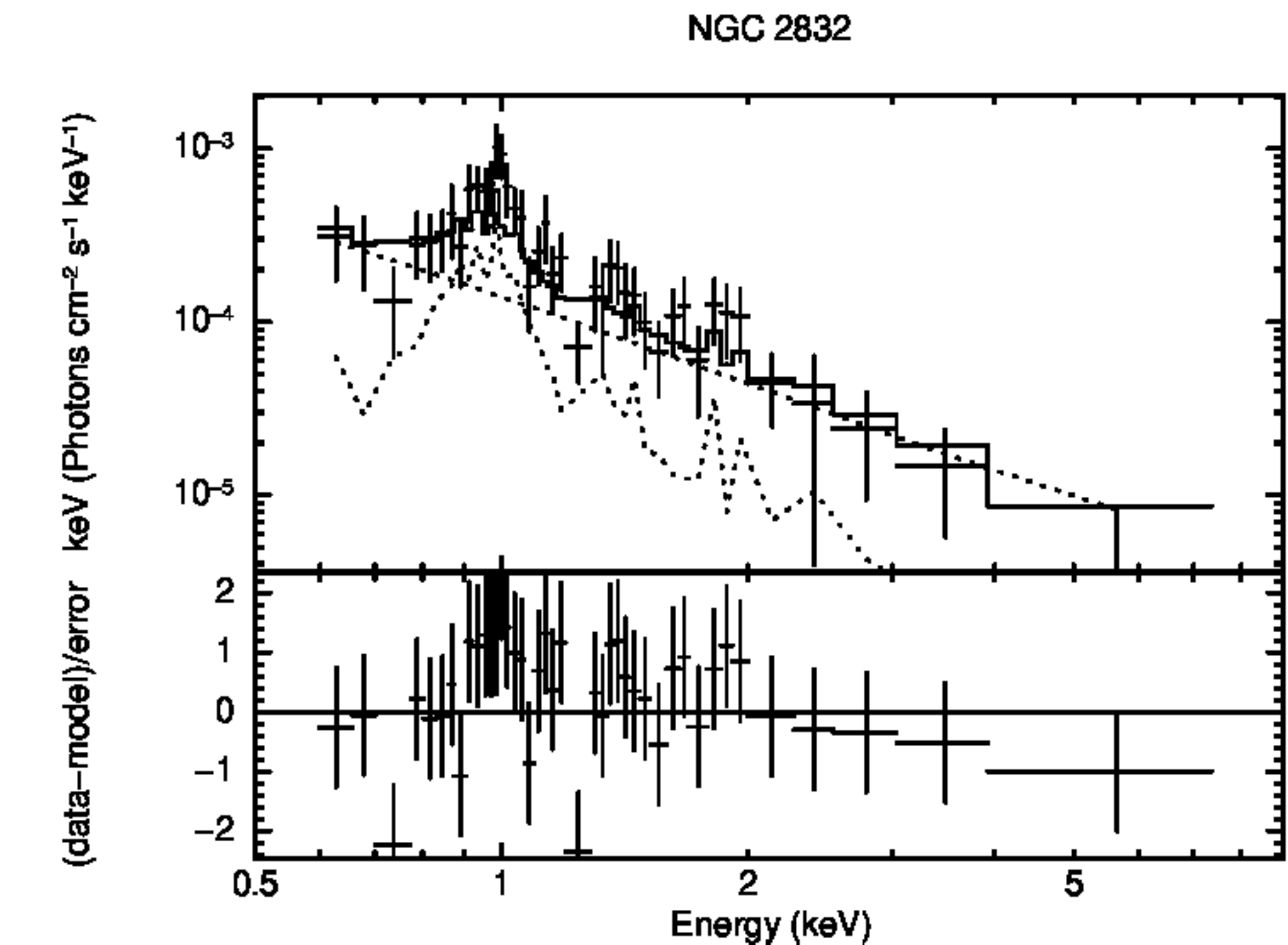}

\end{figure}
\end{center}

%
	 

\begin{center}
 \begin{figure}
	\includegraphics[width=0.89\columnwidth]{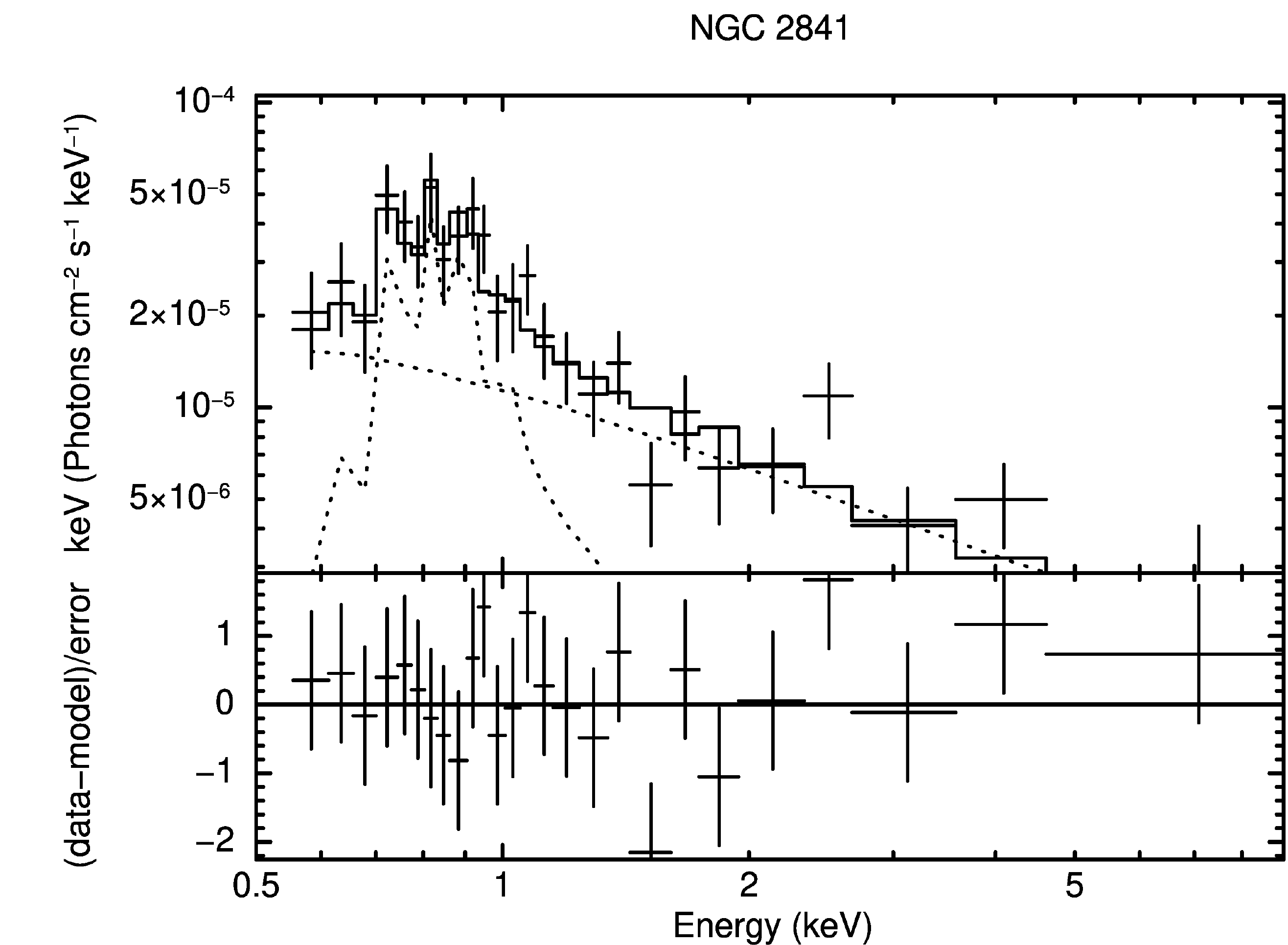}

\end{figure}
\end{center}

\begin{center}
 \begin{figure}
	\includegraphics[width=0.89\columnwidth]{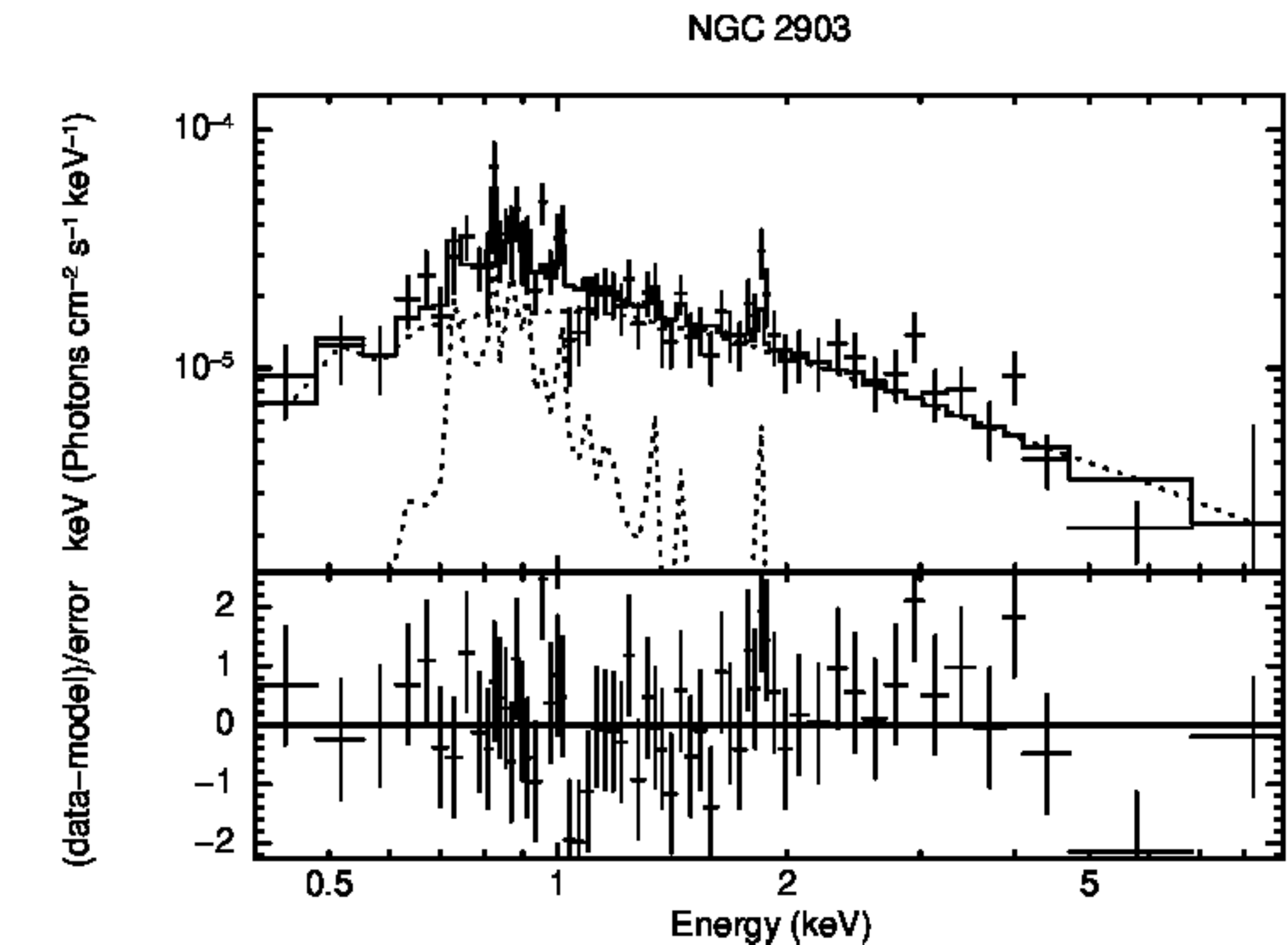}

\end{figure}
\end{center}

%
	 

\begin{center}
 \begin{figure}
	\includegraphics[width=0.89\columnwidth]{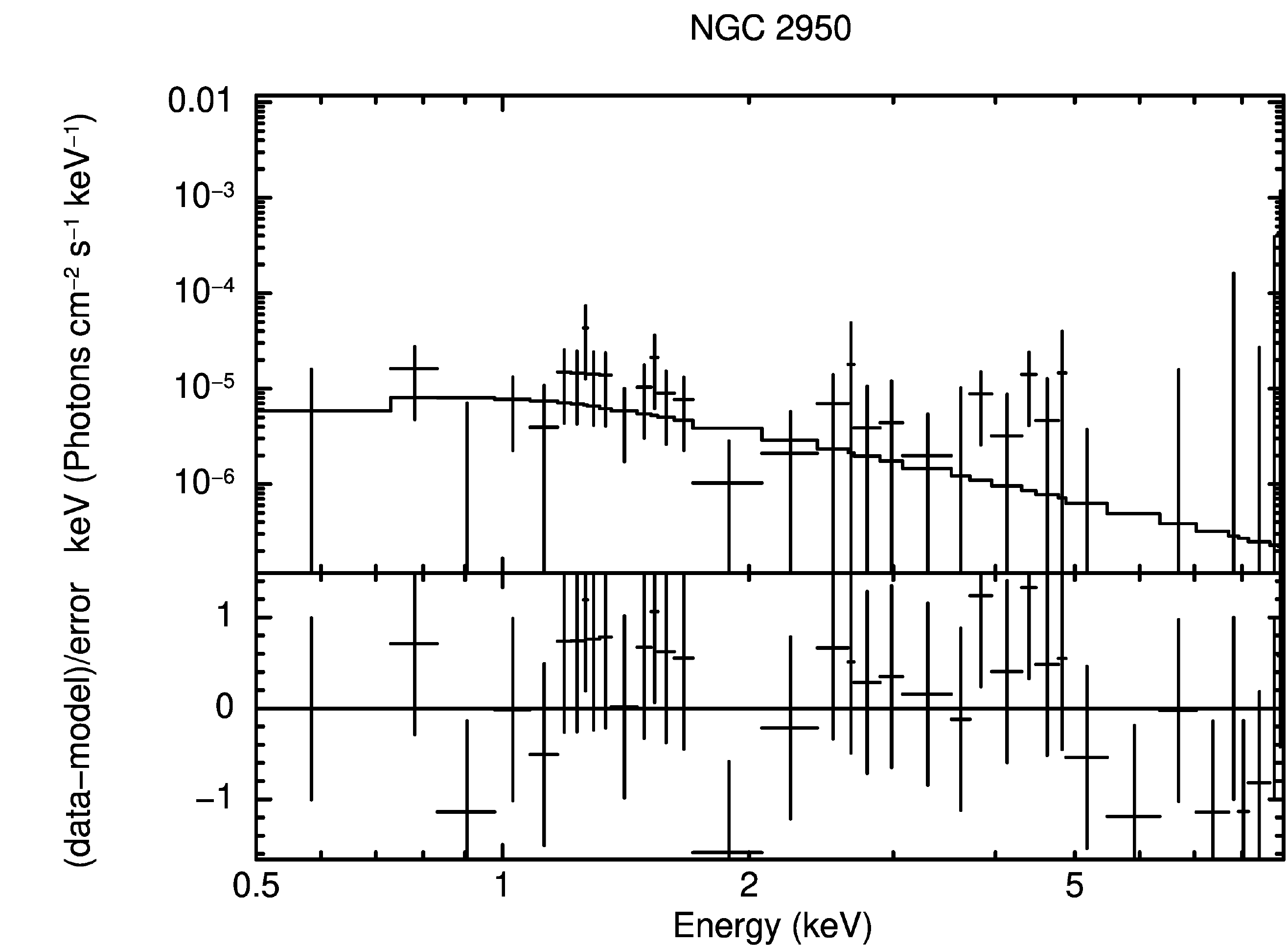}

\end{figure}
\end{center}

\begin{center}
 \begin{figure}
	\includegraphics[width=0.89\columnwidth]{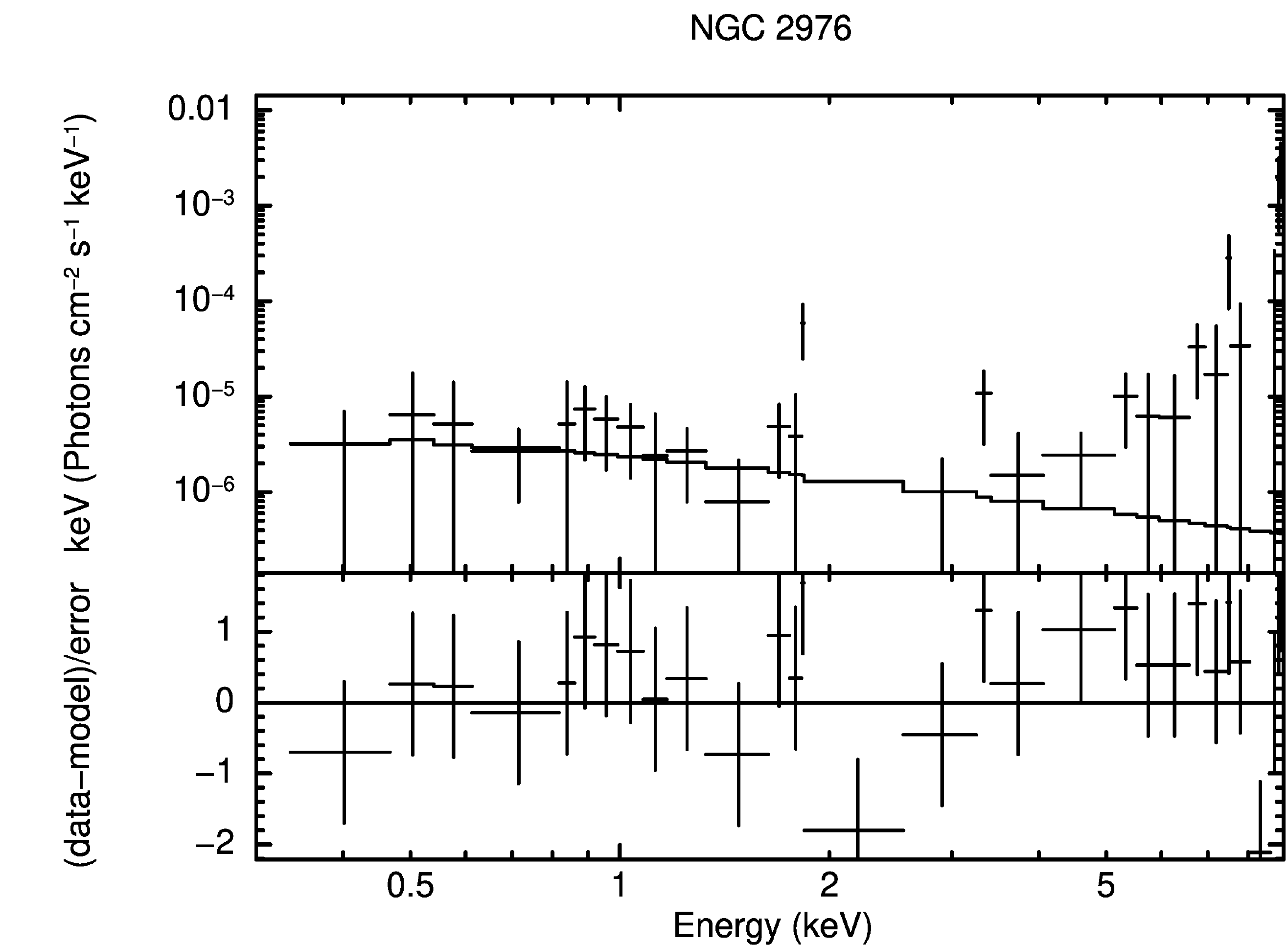}

\end{figure}
\end{center}

\begin{center}
 \begin{figure}
	\includegraphics[width=0.89\columnwidth]{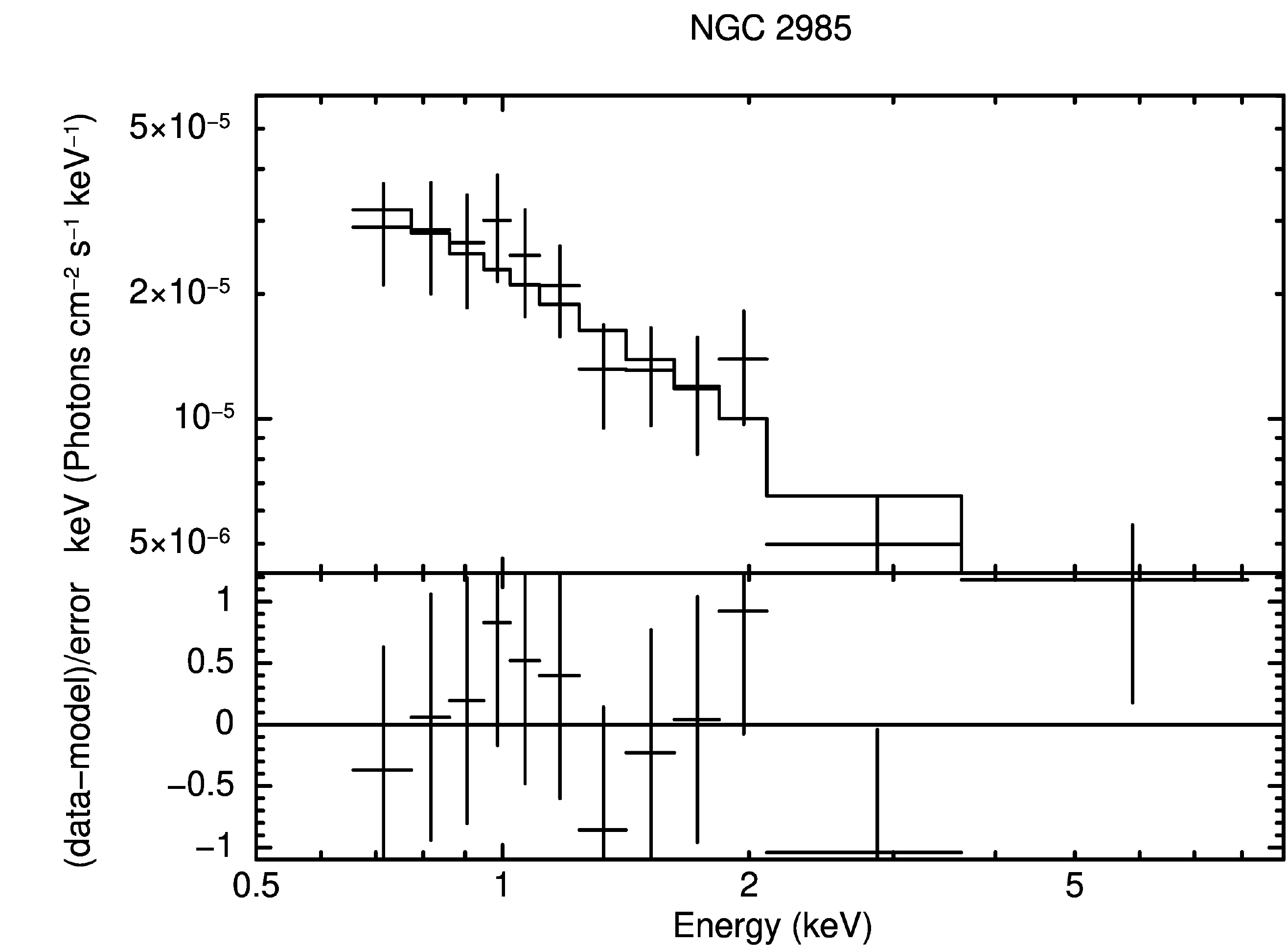}

\end{figure}
\end{center}

\begin{center}
 \begin{figure}
	\includegraphics[width=0.89\columnwidth]{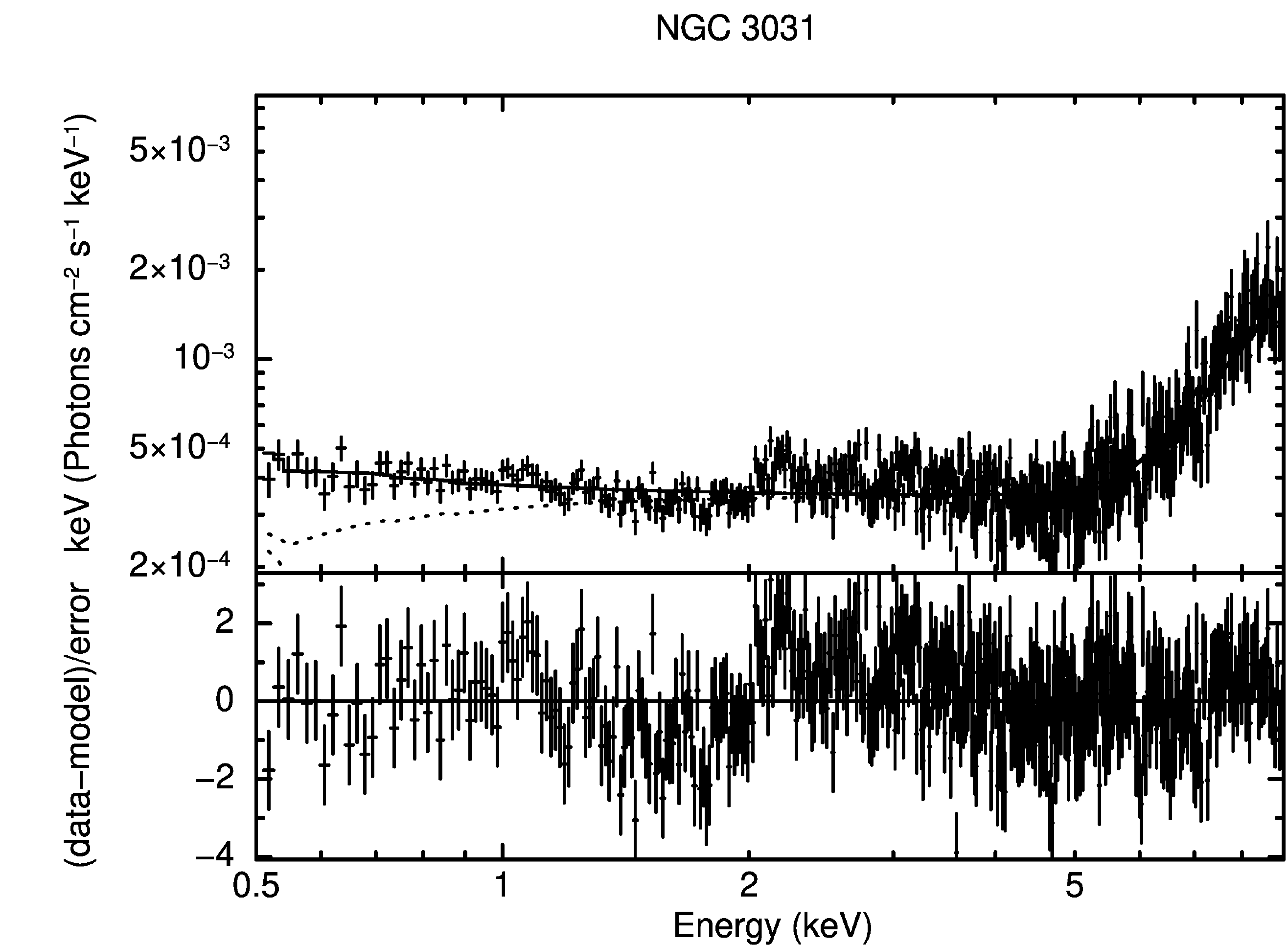}

\end{figure}
\end{center}

\begin{center}
 \begin{figure}
	\includegraphics[width=0.89\columnwidth]{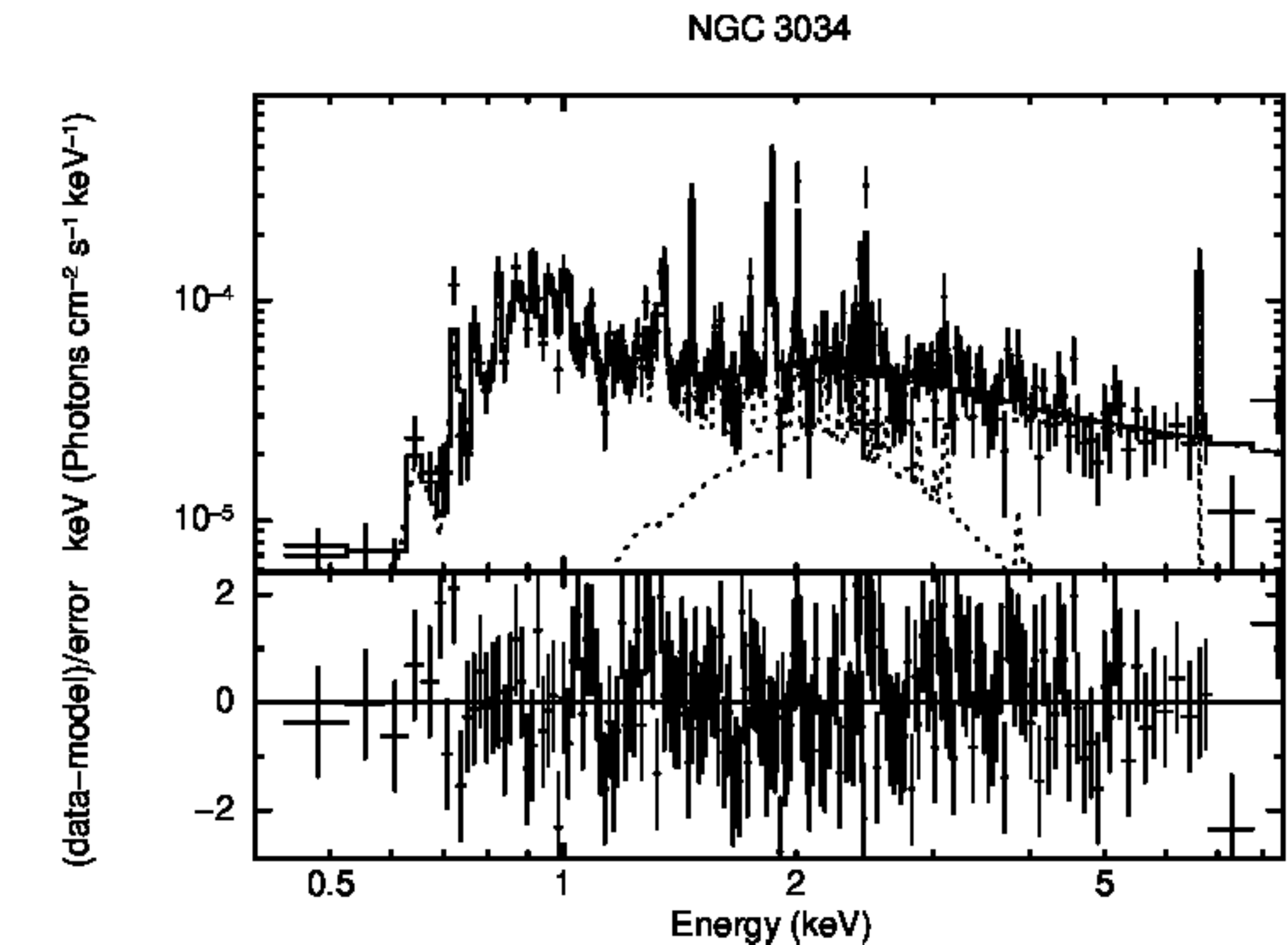}

\end{figure}
\end{center}

%
	 

\begin{center}
 \begin{figure}
	\includegraphics[width=0.89\columnwidth]{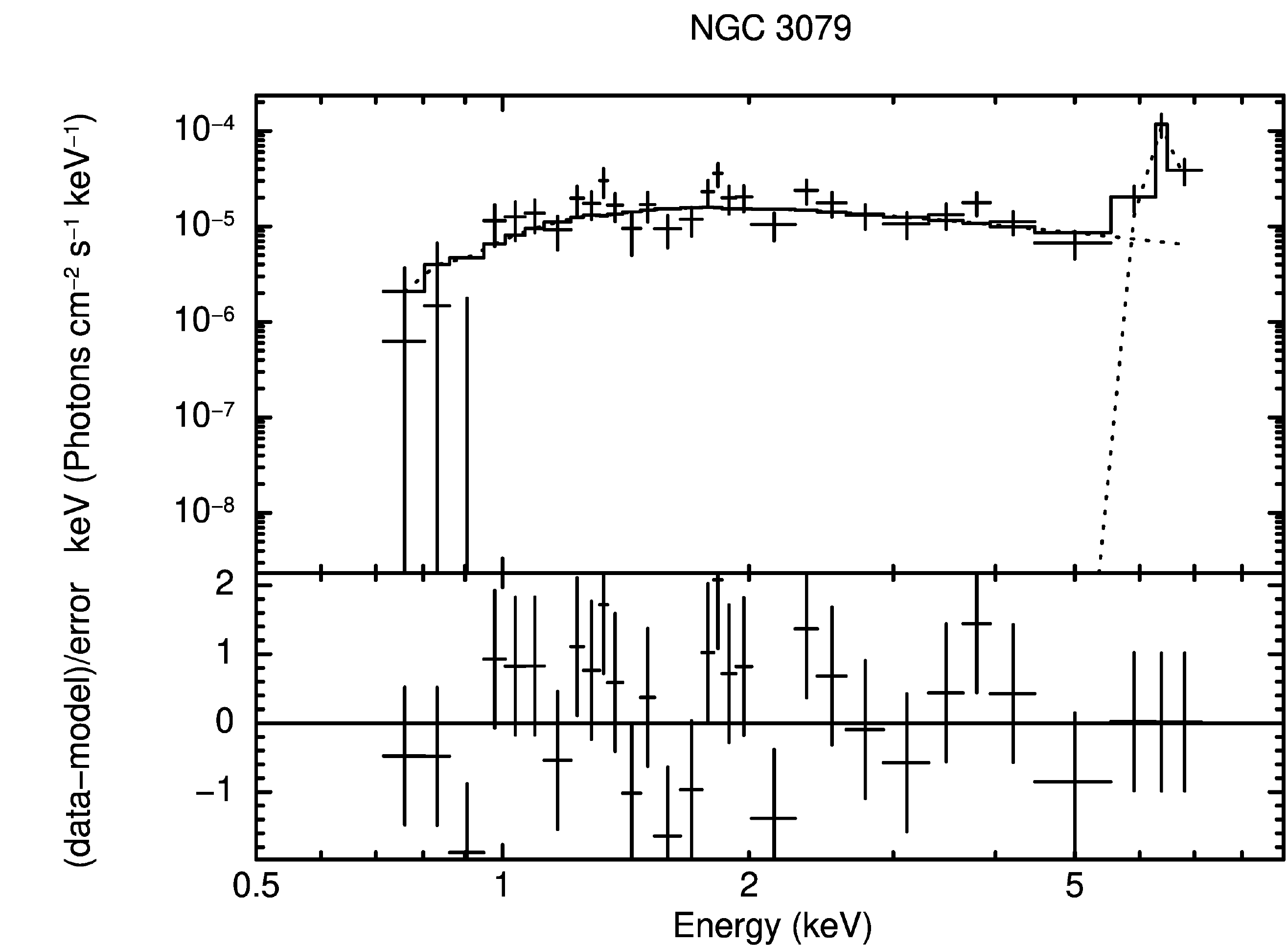}

\end{figure}
\end{center}

\begin{center}
 \begin{figure}
	\includegraphics[width=0.89\columnwidth]{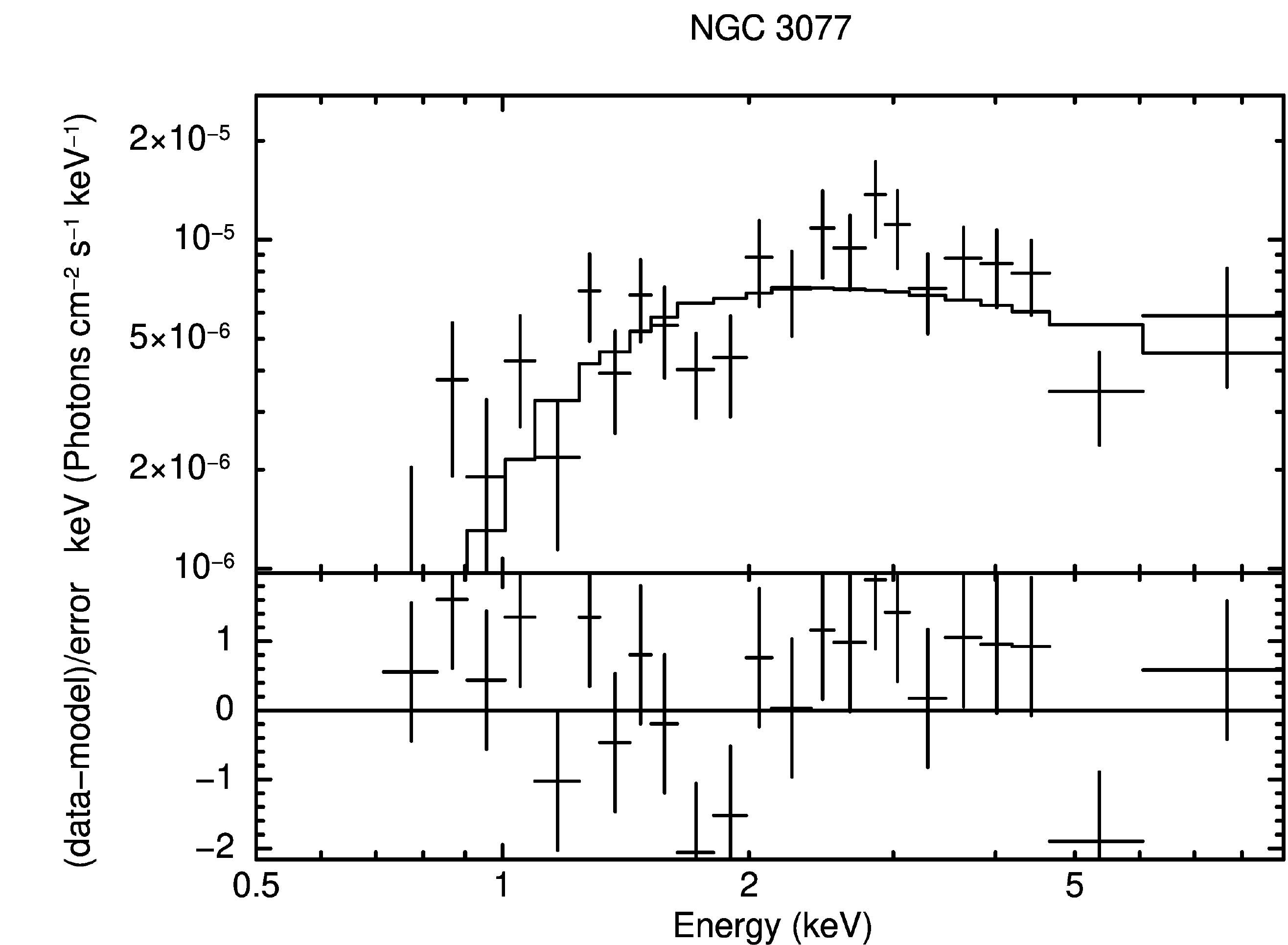}

\end{figure}
\end{center}

\begin{center}
 \begin{figure}
	\includegraphics[width=0.89\columnwidth]{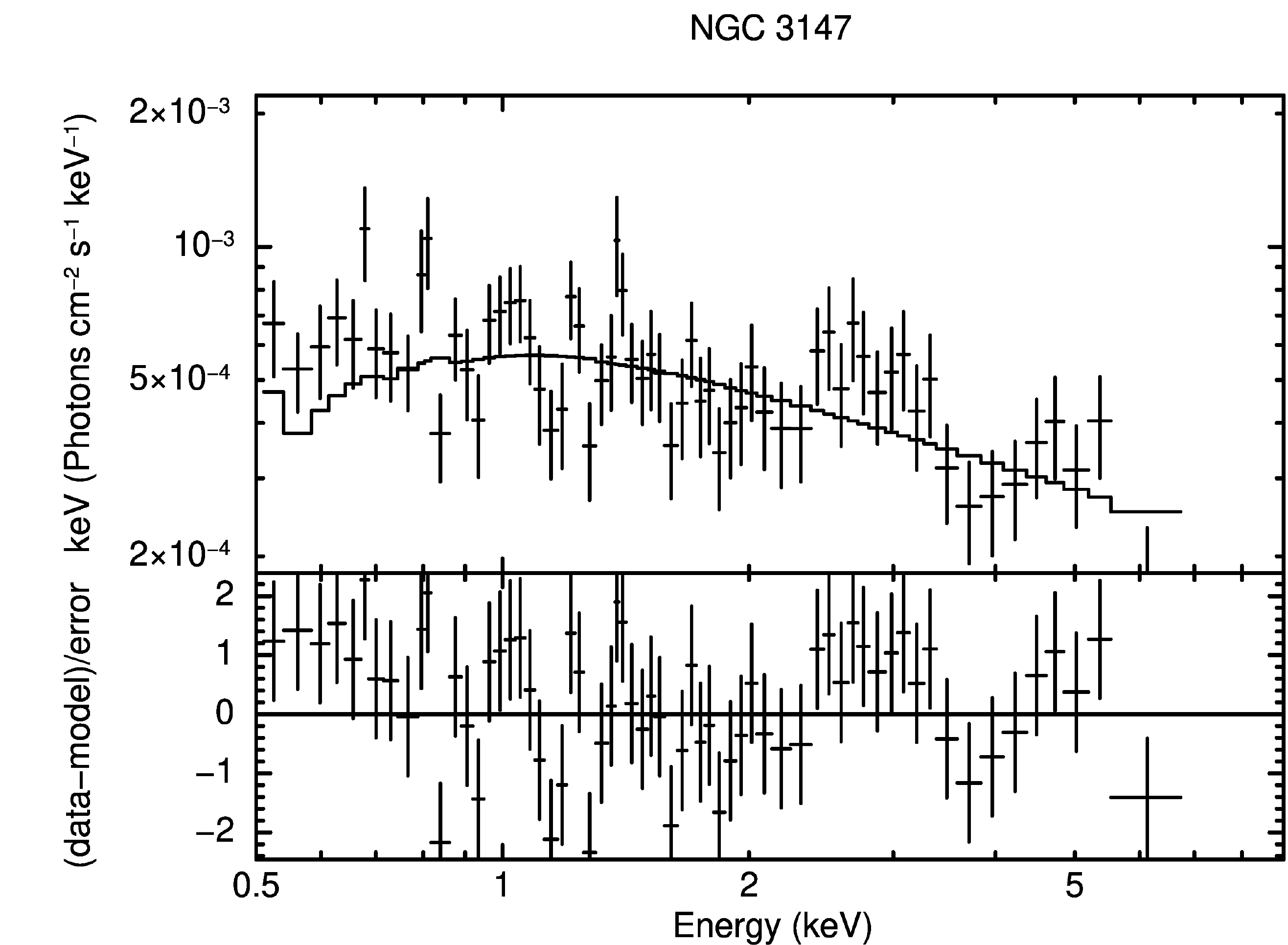}

\end{figure}
\end{center}

\begin{center}
 \begin{figure}
	\includegraphics[width=0.89\columnwidth]{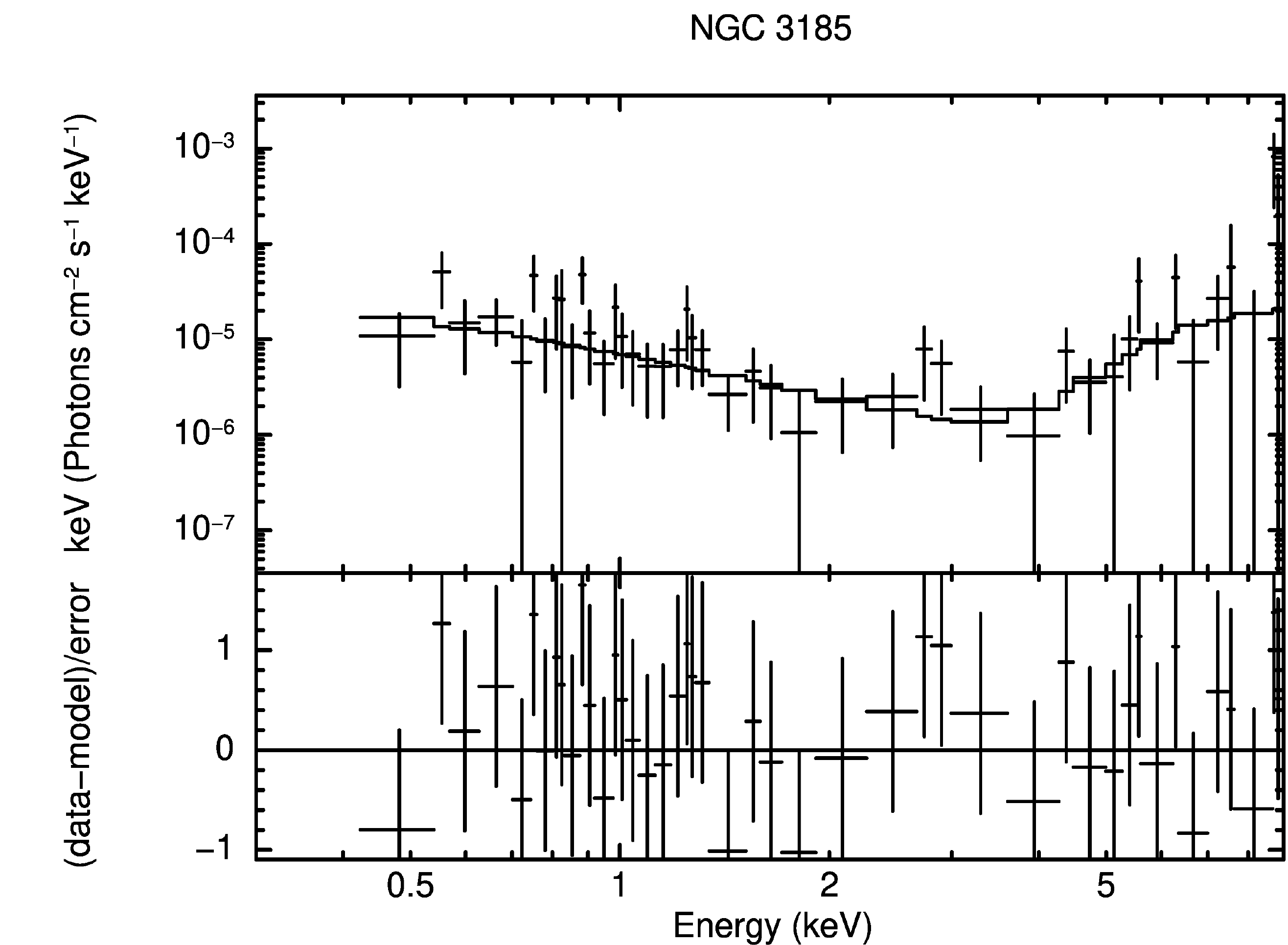}

\end{figure}
\end{center}

\begin{center}
 \begin{figure}
	\includegraphics[width=0.89\columnwidth]{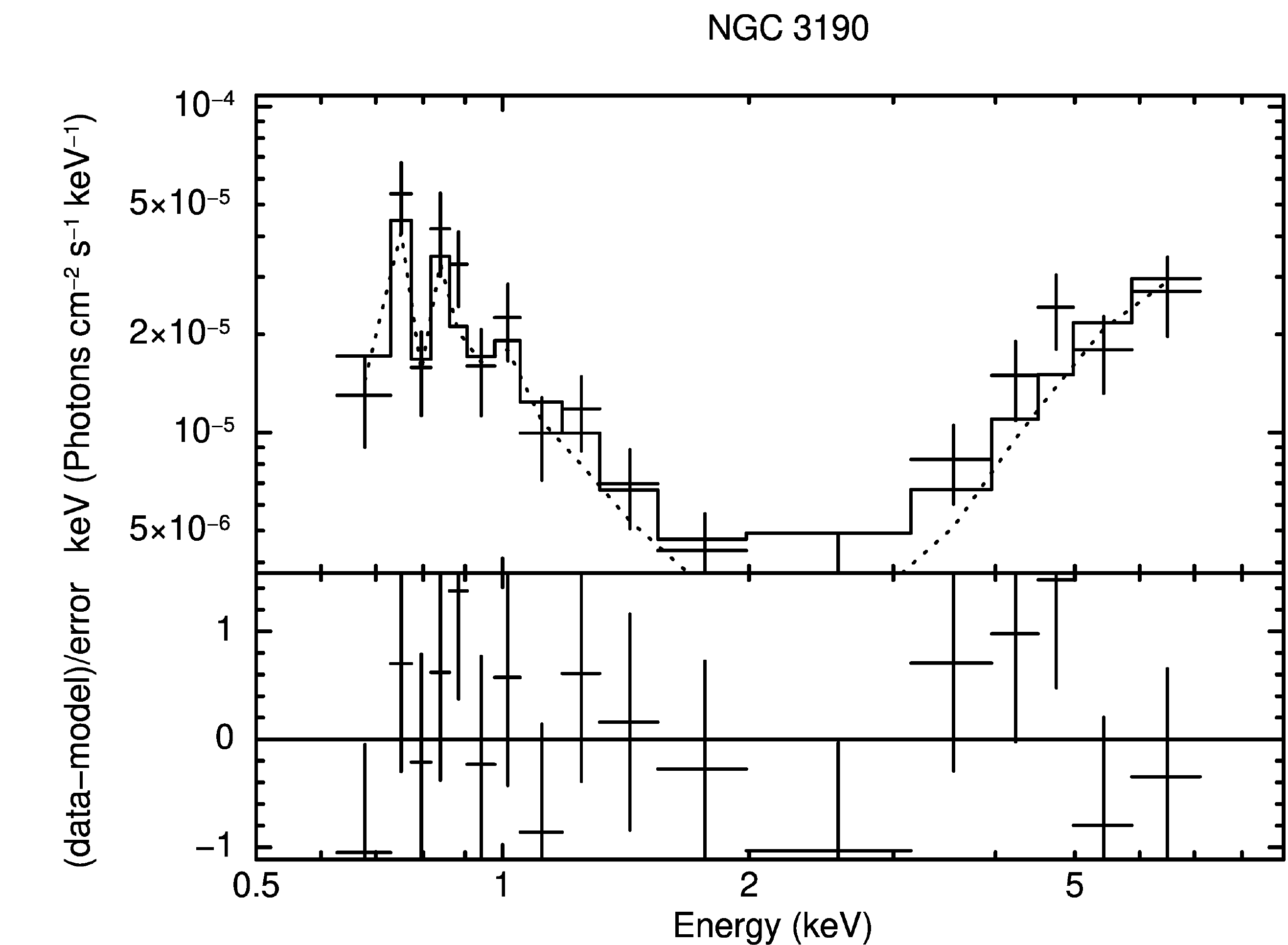}

\end{figure}
\end{center}

\begin{center}
 \begin{figure}
	\includegraphics[width=0.89\columnwidth]{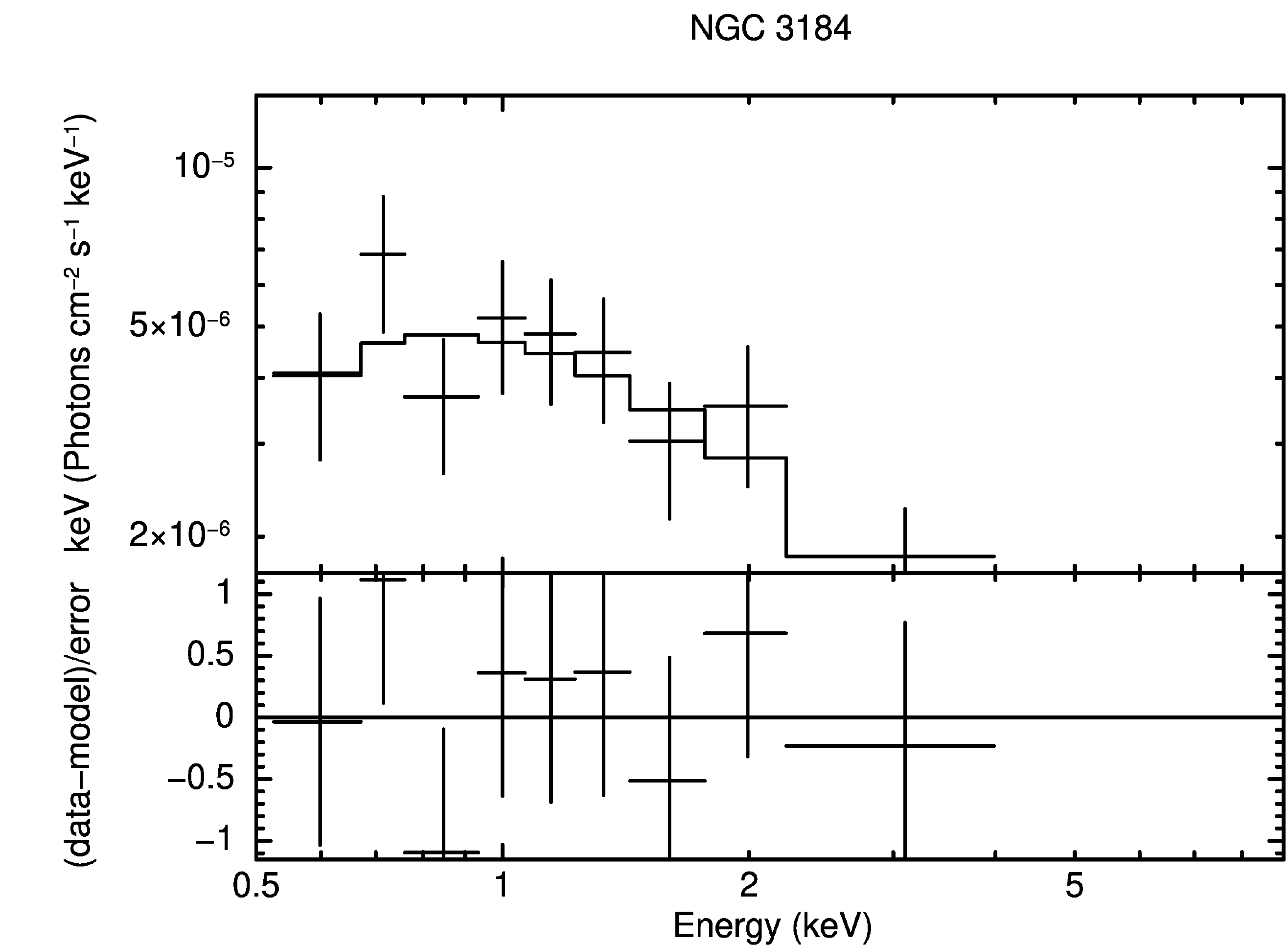}

\end{figure}
\end{center}

\begin{center}
 \begin{figure}
	\includegraphics[width=0.89\columnwidth]{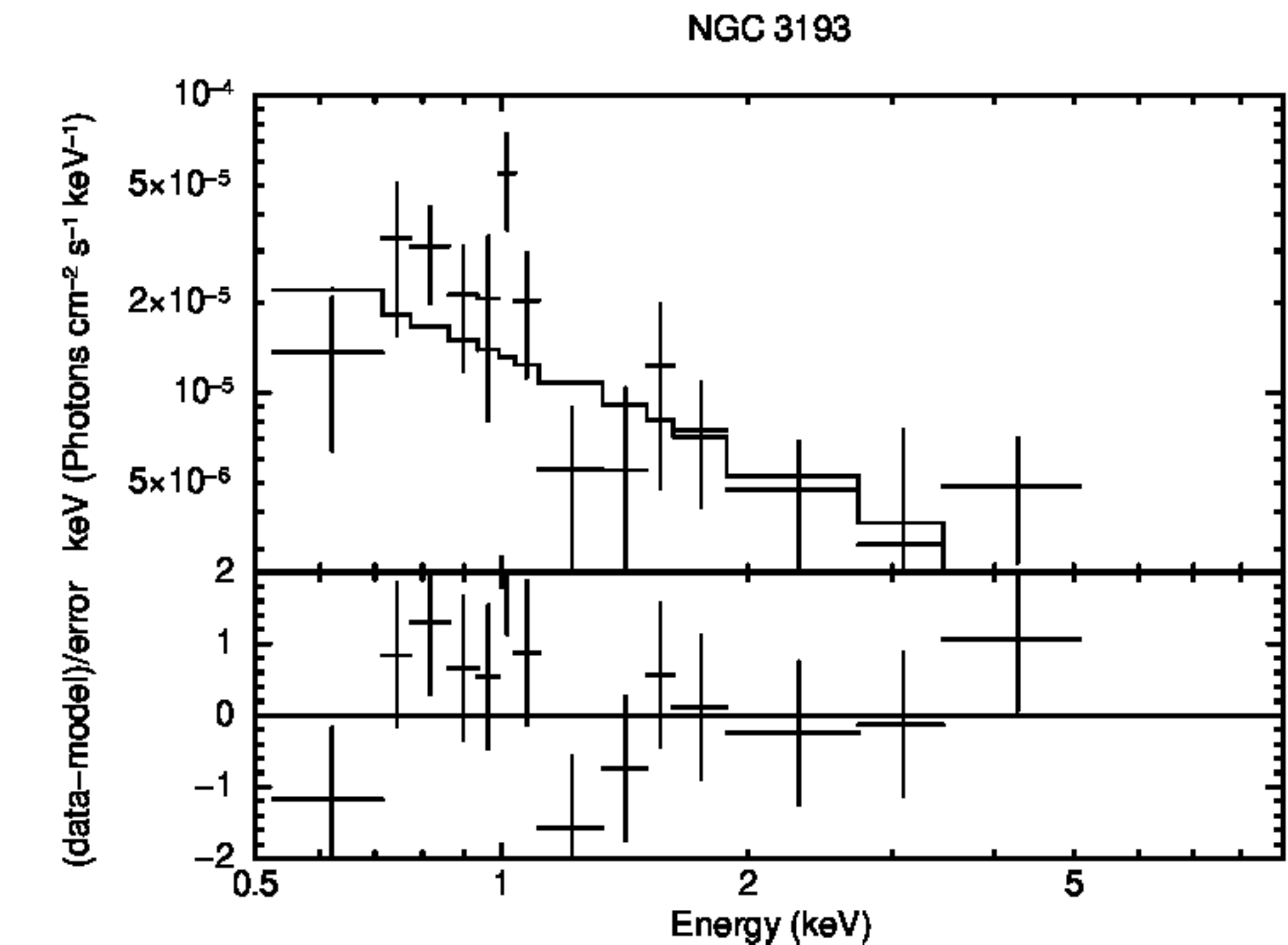}

\end{figure}
\end{center}

%
	 

\begin{center}
 \begin{figure}
	\includegraphics[width=0.89\columnwidth]{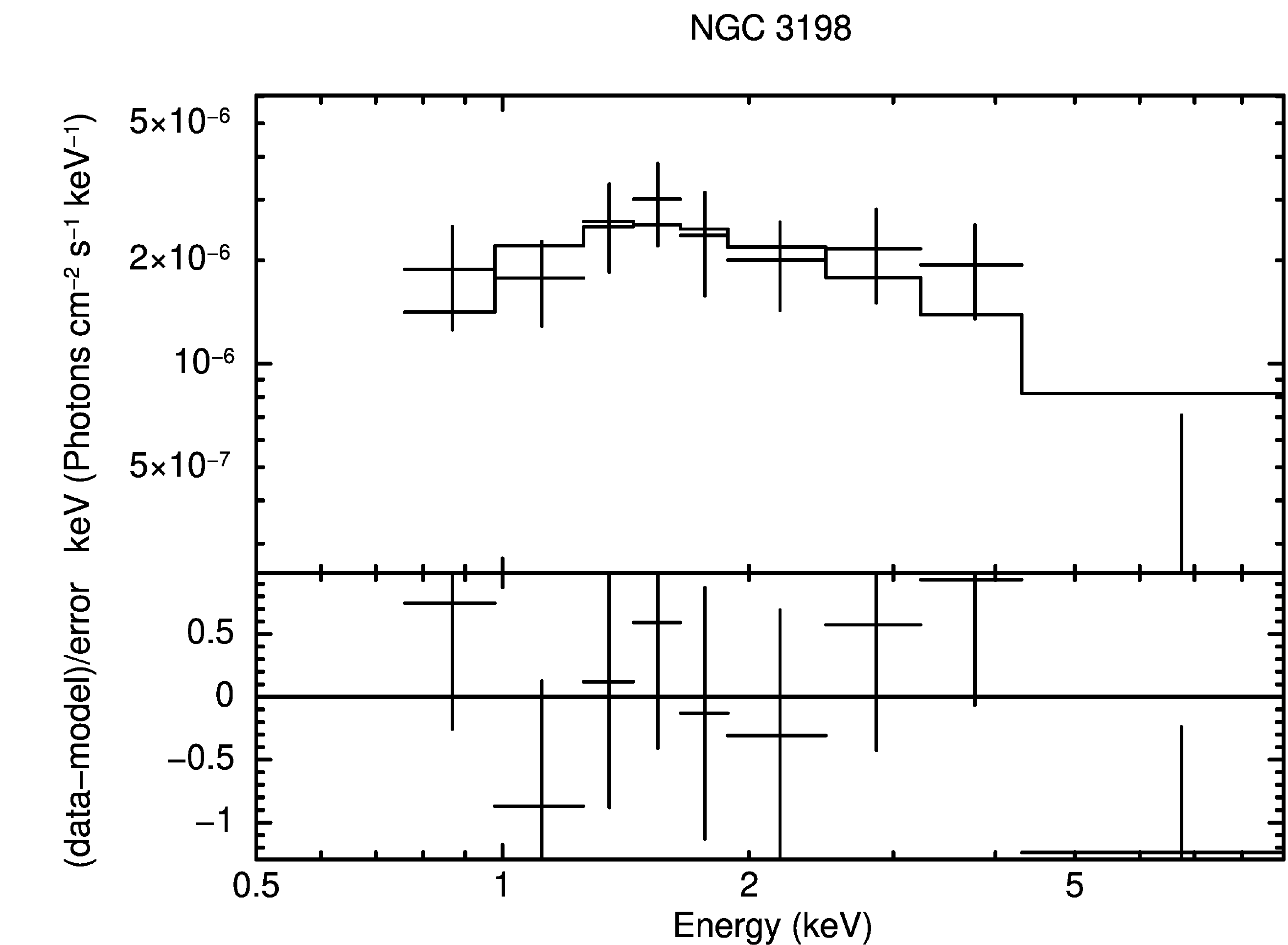}

\end{figure}
\end{center}

\begin{center}
 \begin{figure}
	\includegraphics[width=0.89\columnwidth]{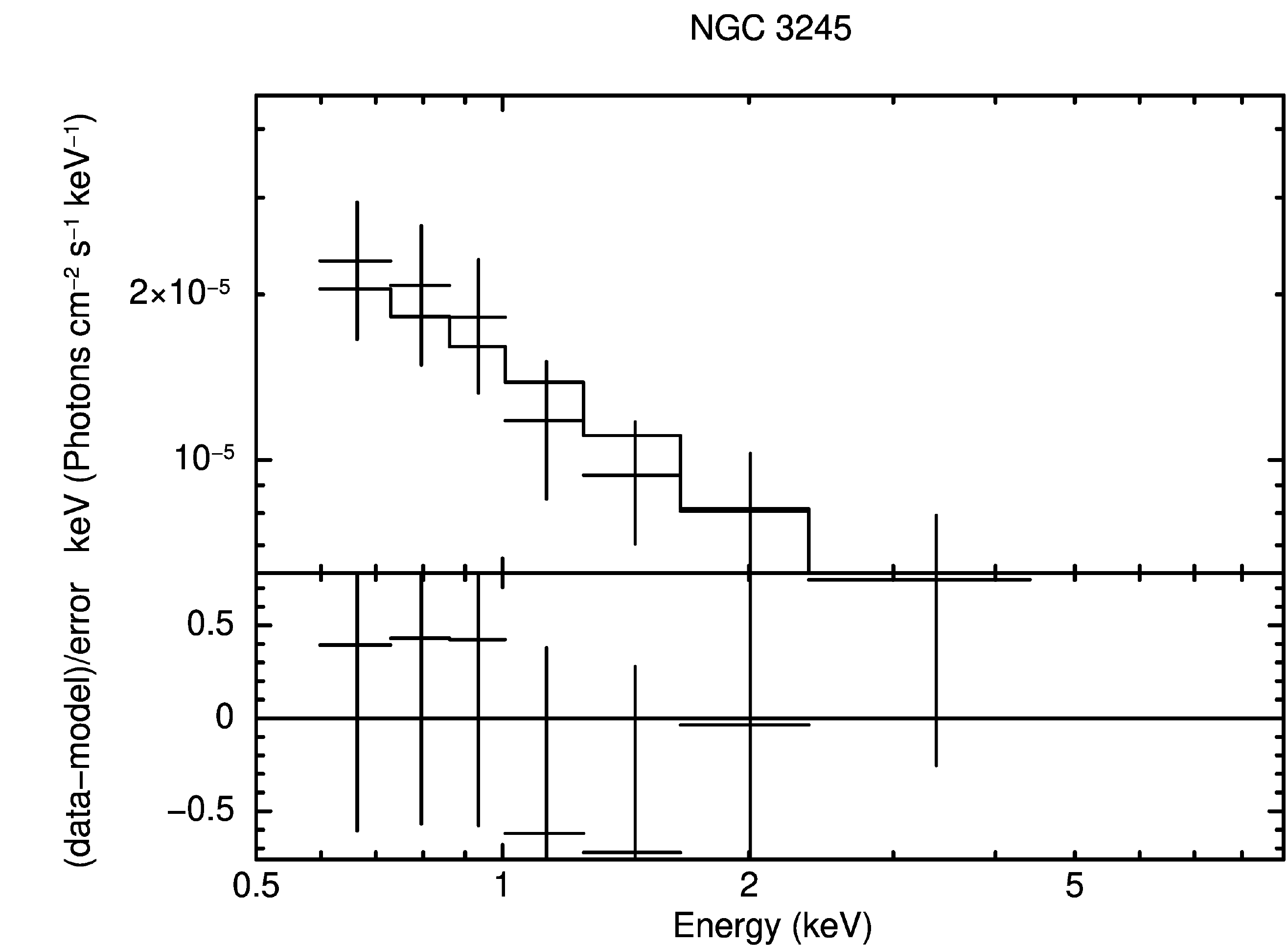}

\end{figure}
\end{center}

\begin{center}
 \begin{figure}
	\includegraphics[width=0.89\columnwidth]{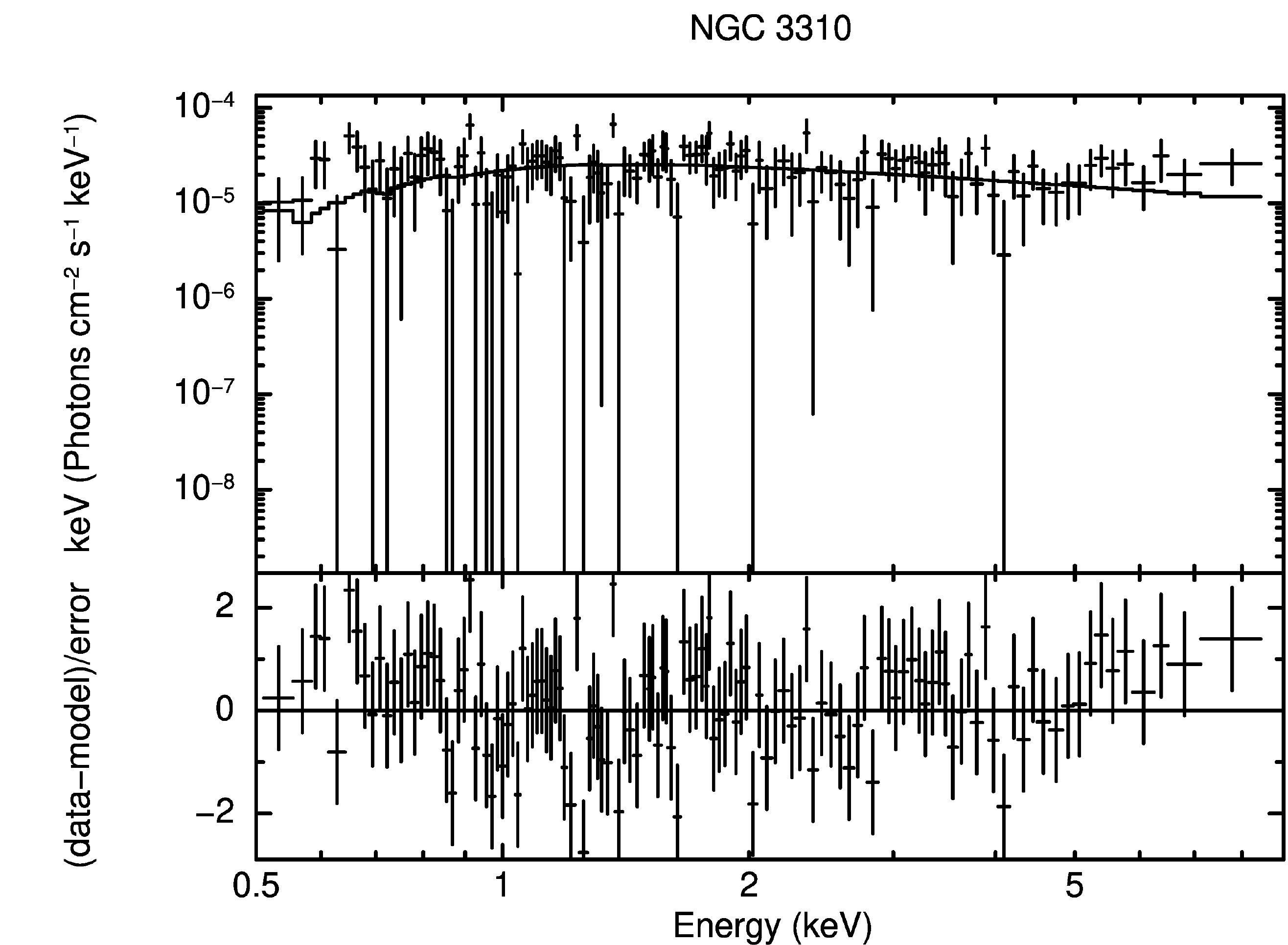}

\end{figure}
\end{center}

\begin{center}
 \begin{figure}
	\includegraphics[width=0.89\columnwidth]{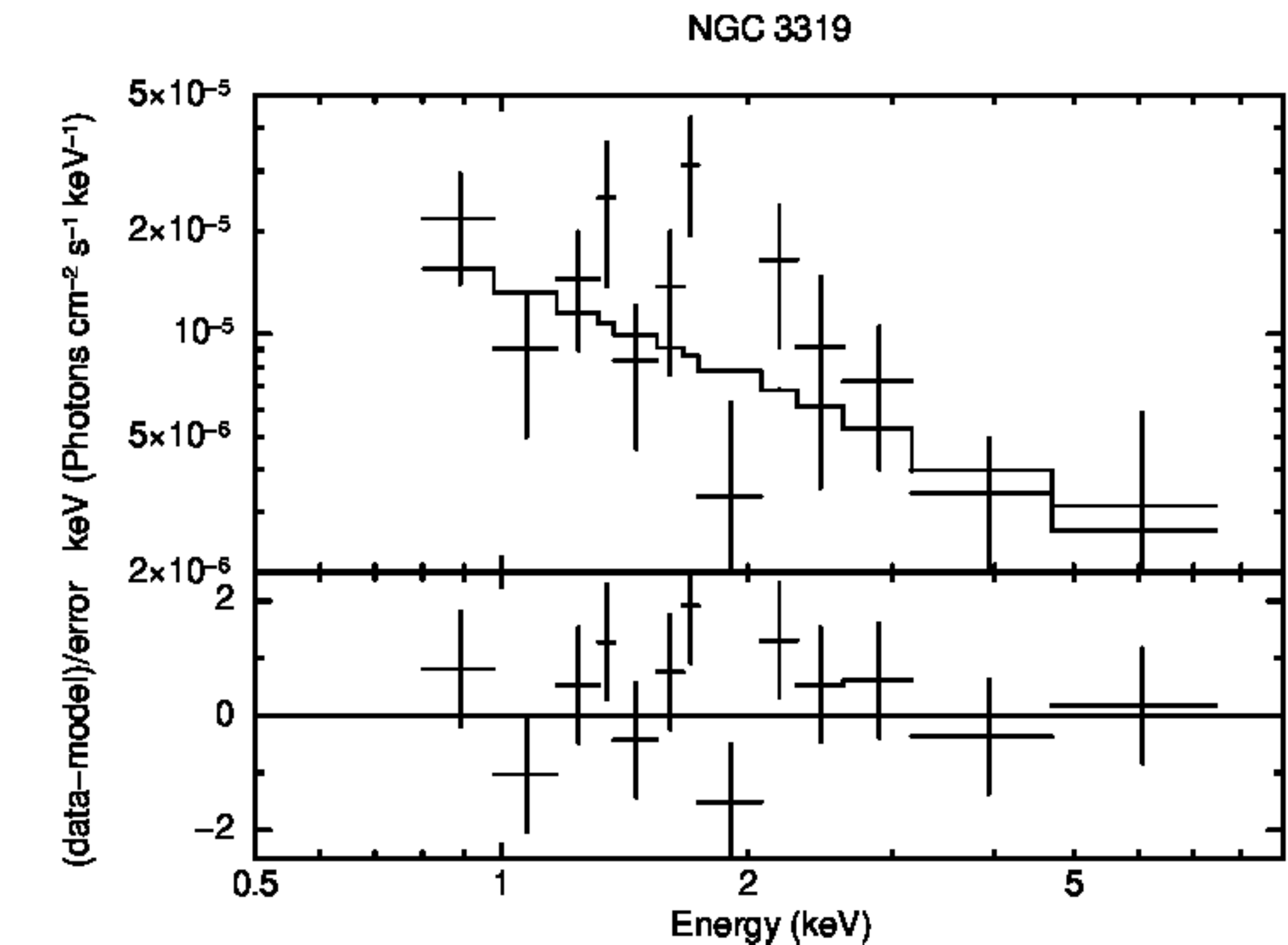}

\end{figure}
\end{center}

%
	 

\begin{center}
 \begin{figure}
	\includegraphics[width=0.89\columnwidth]{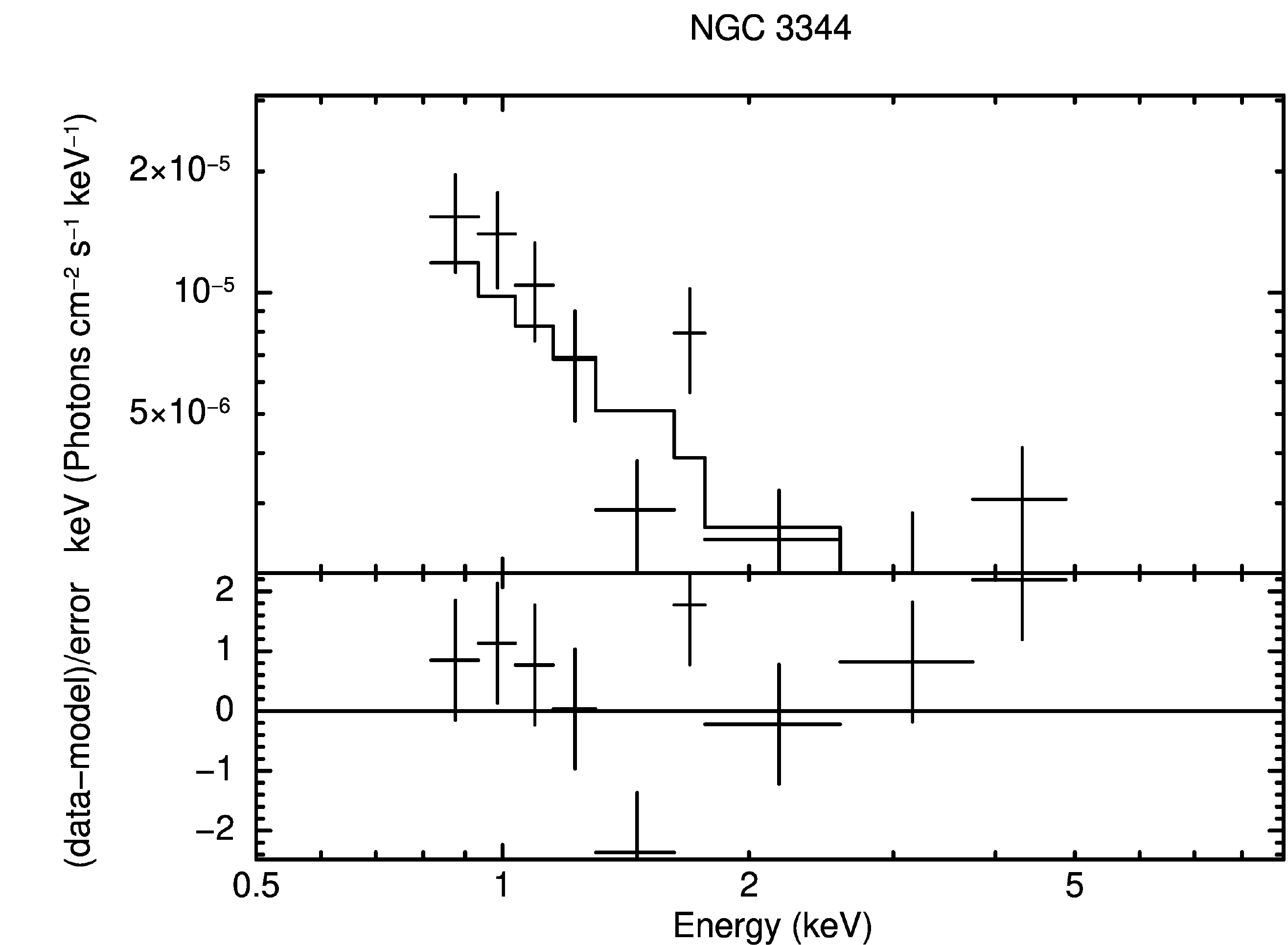}

\end{figure}
\end{center}

\begin{center}
 \begin{figure}
	\includegraphics[width=0.89\columnwidth]{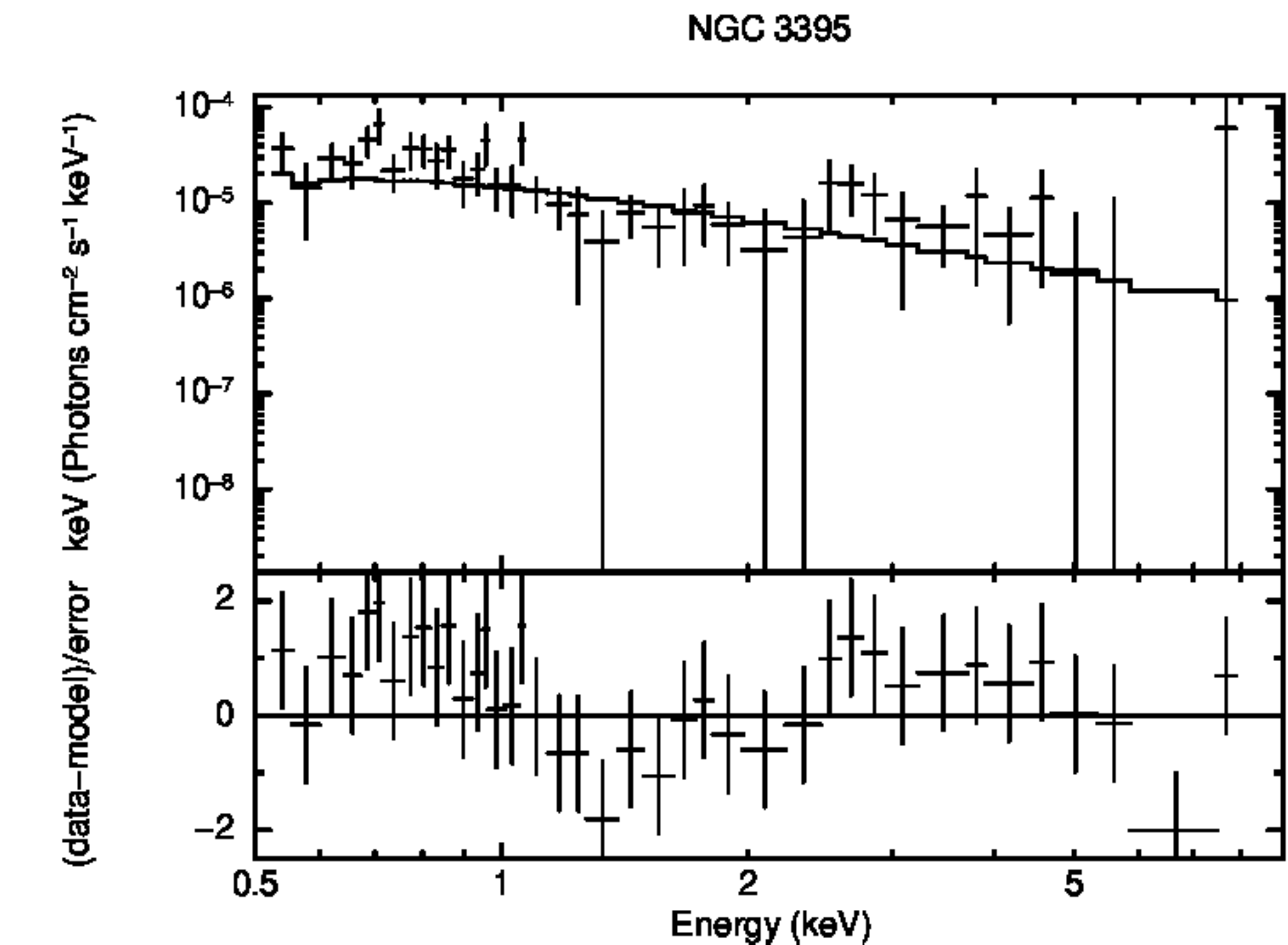}

\end{figure}
\end{center}

%
	 

\begin{center}
 \begin{figure}
	\includegraphics[width=0.89\columnwidth]{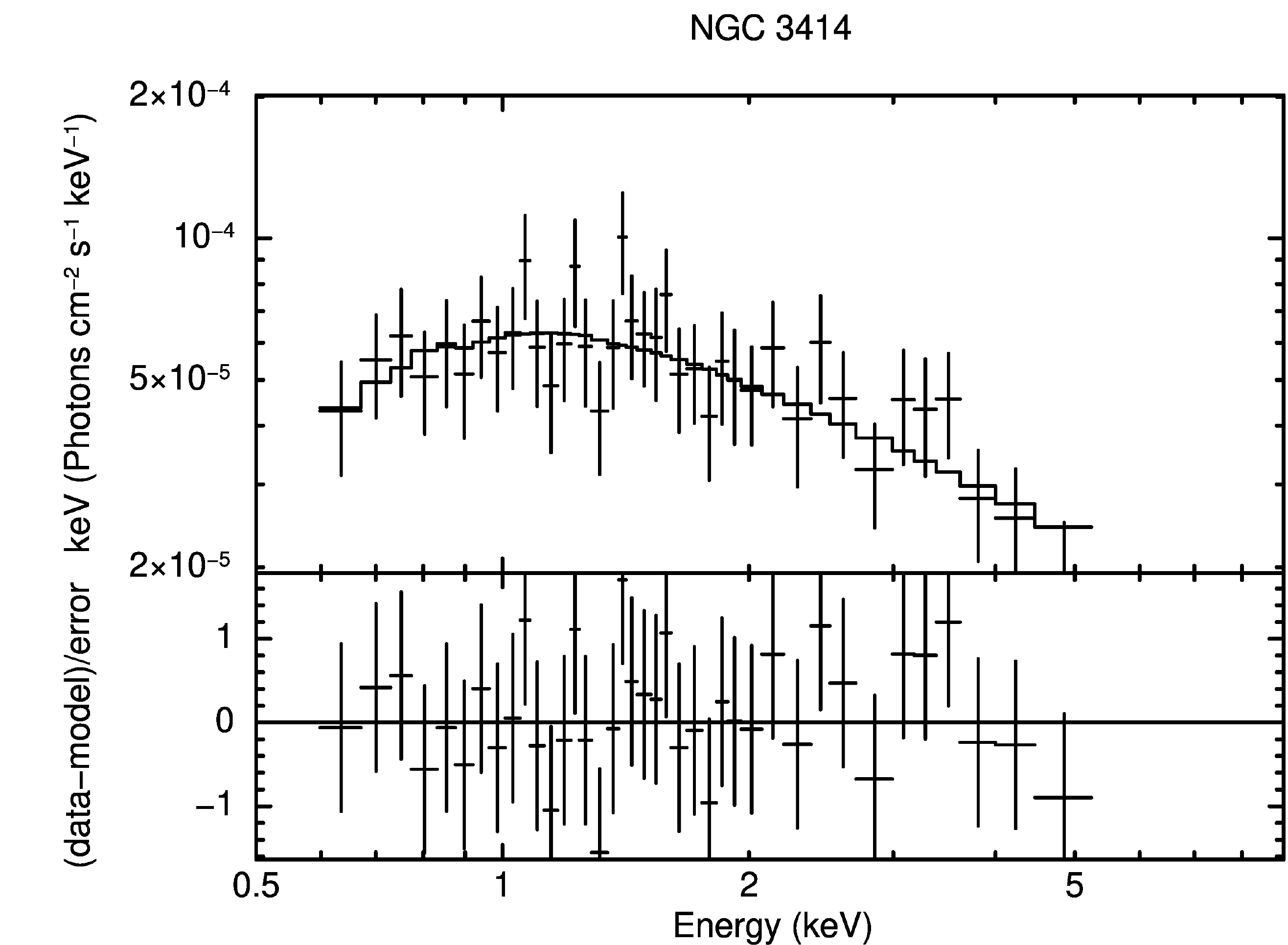}

\end{figure}
\end{center}

\begin{center}
 \begin{figure}
	\includegraphics[width=0.89\columnwidth]{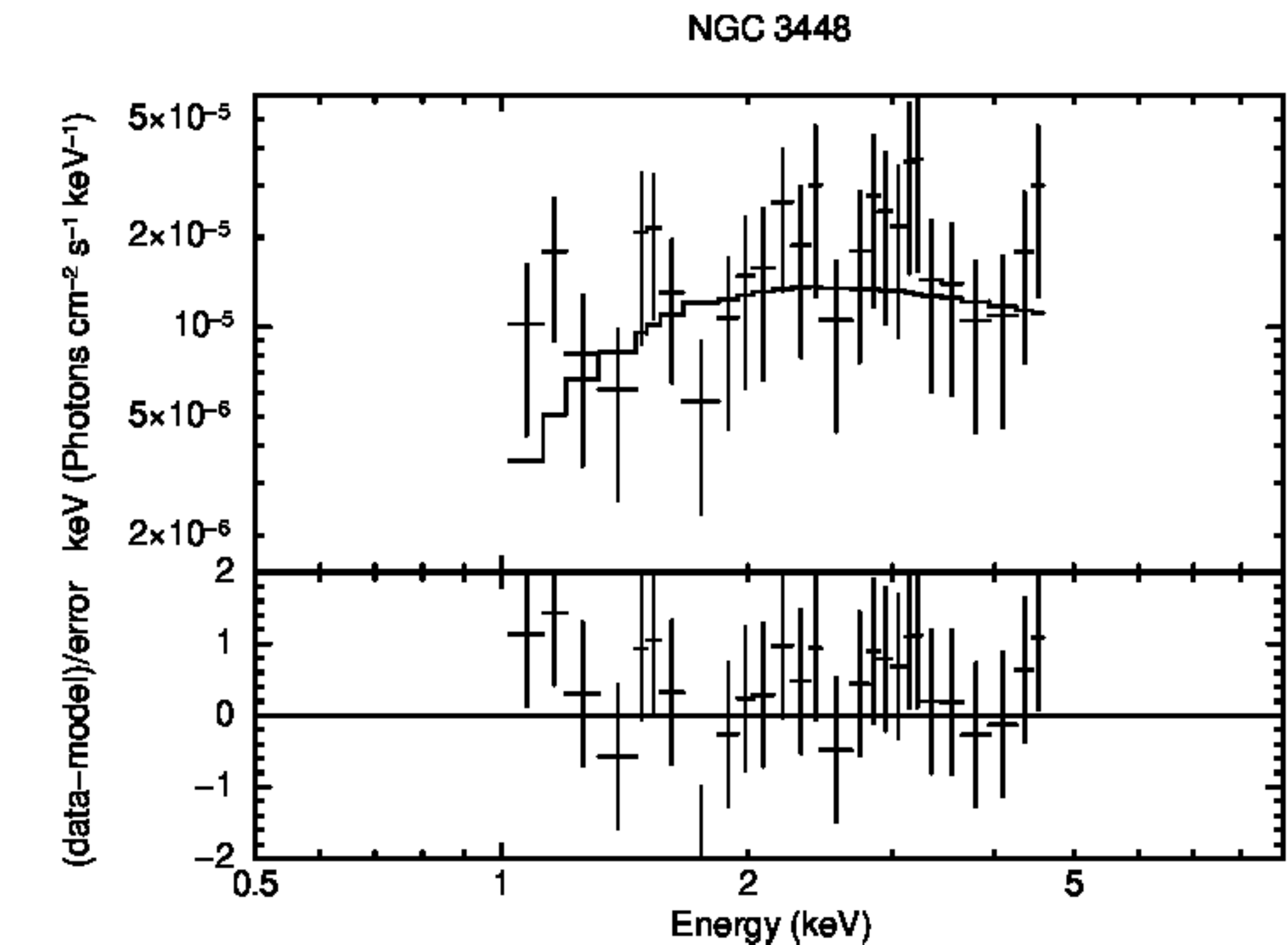}

\end{figure}
\end{center}

%
	 

\begin{center}
 \begin{figure}
	\includegraphics[width=0.89\columnwidth]{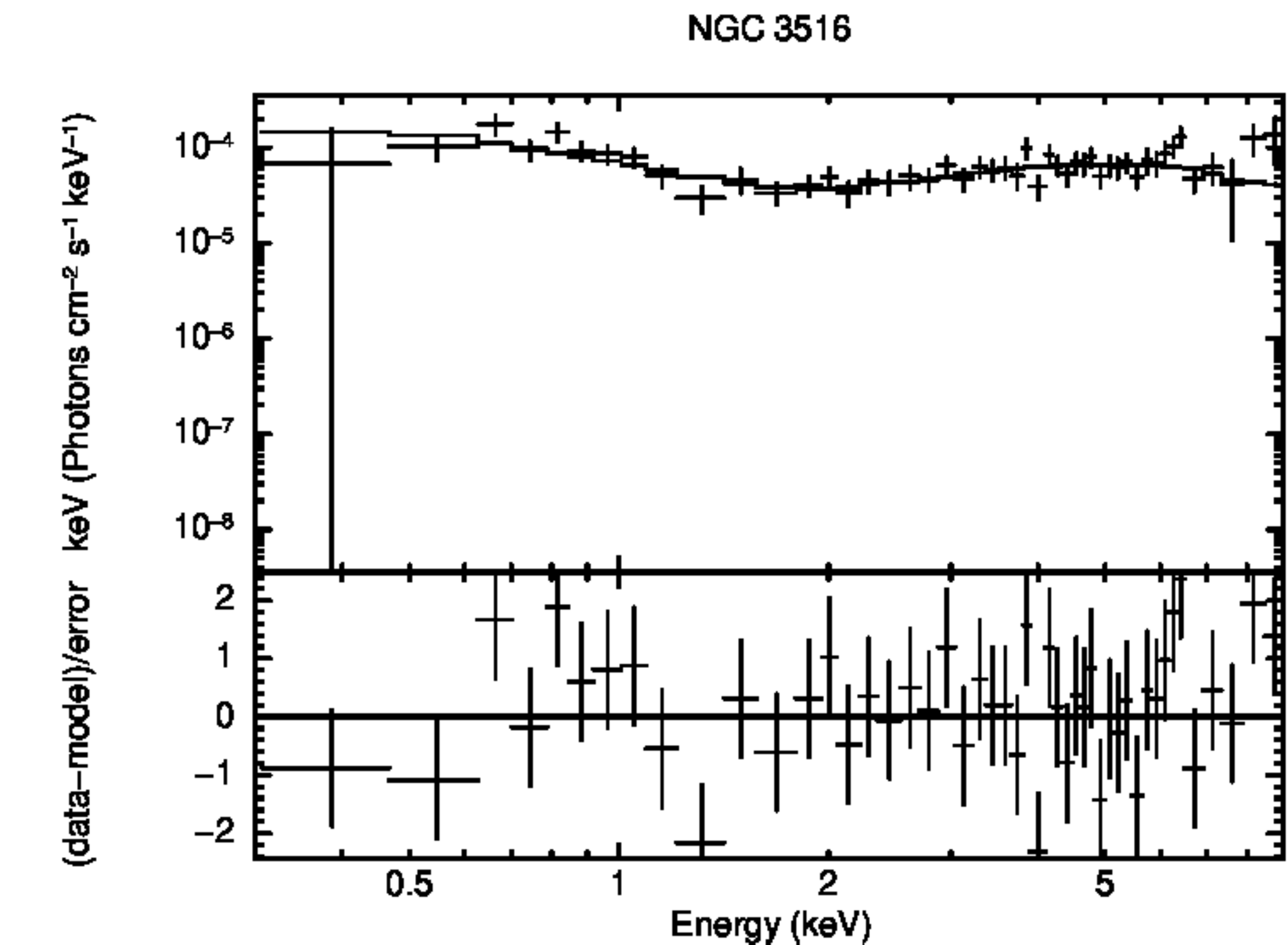}

\end{figure}
\end{center}

%
	 

\begin{center}
 \begin{figure}
	\includegraphics[width=0.89\columnwidth]{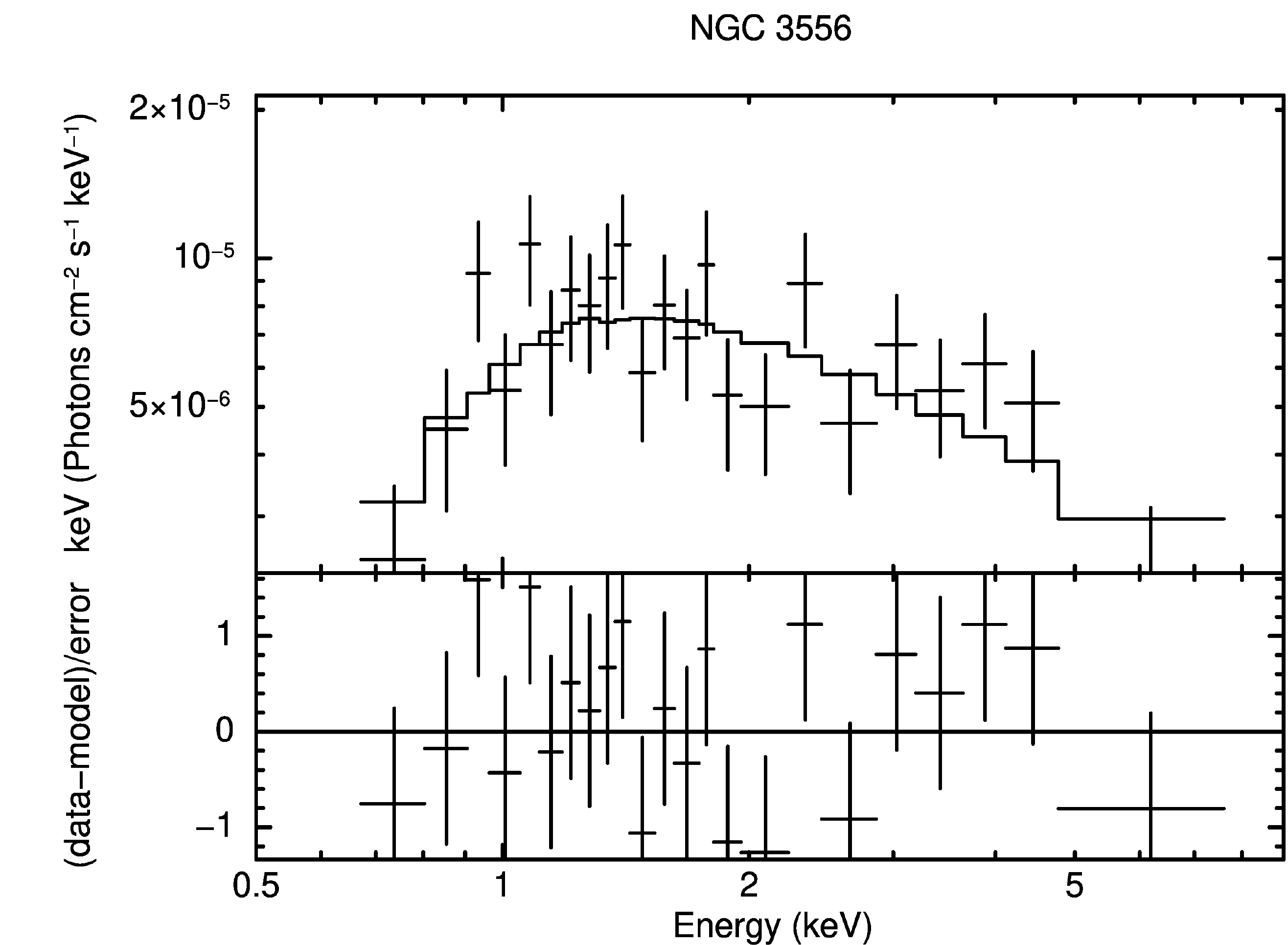}

\end{figure}
\end{center}

\begin{center}
 \begin{figure}
	\includegraphics[width=0.89\columnwidth]{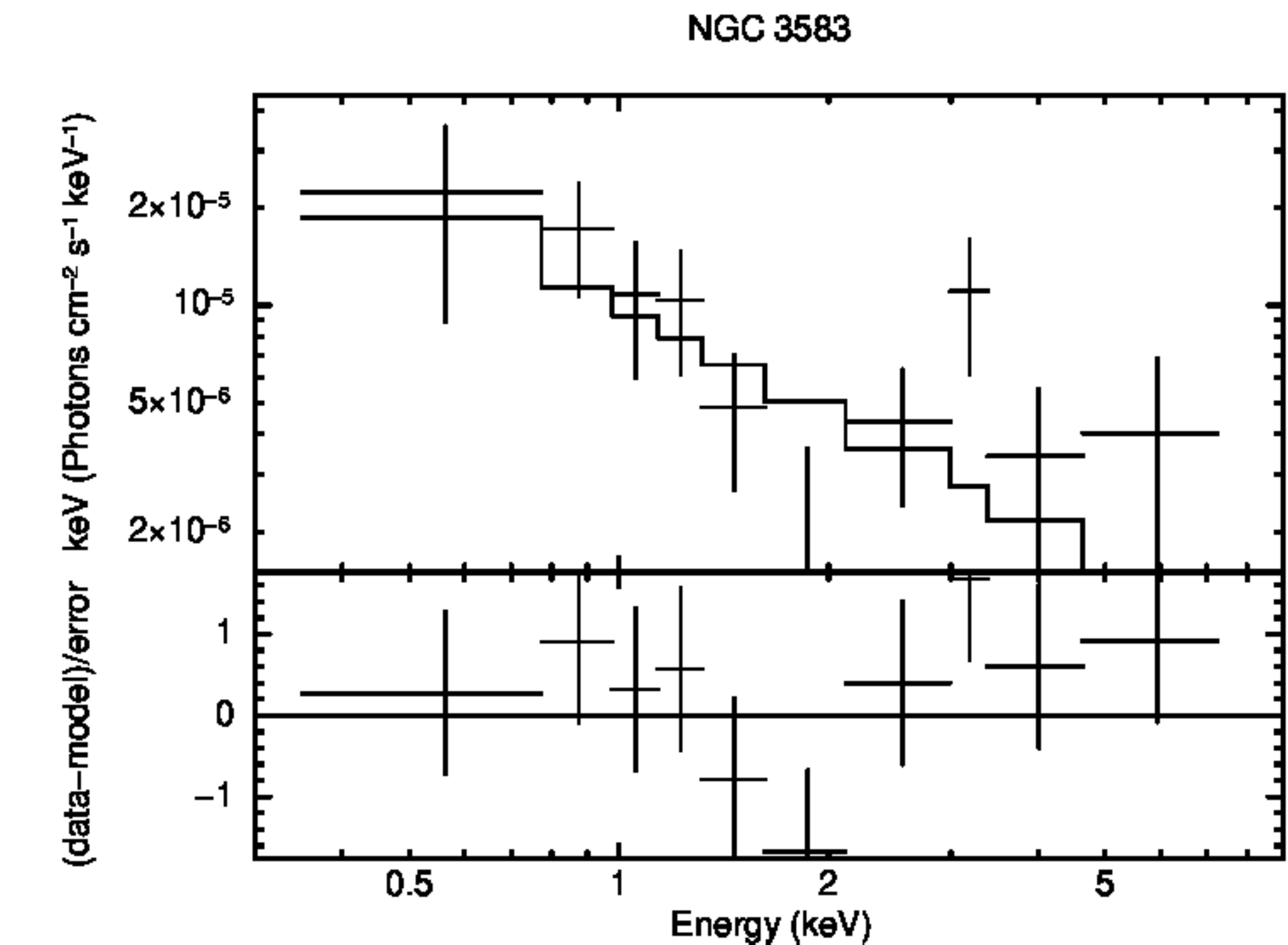}

\end{figure}
\end{center}

%
	 

\begin{center}
 \begin{figure}
	\includegraphics[width=0.89\columnwidth]{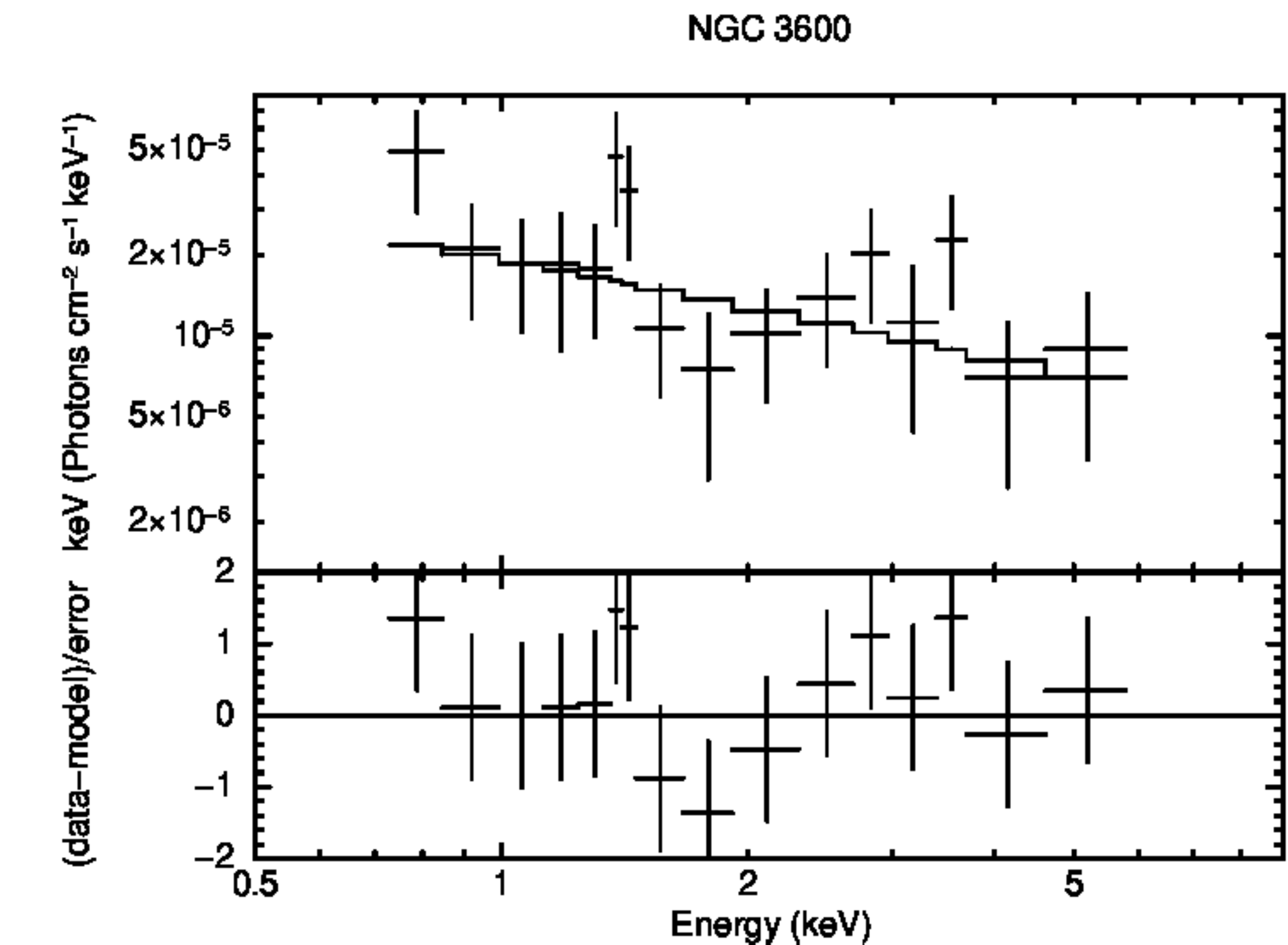}

\end{figure}
\end{center}

%
	 

\begin{center}
 \begin{figure}
	\includegraphics[width=0.89\columnwidth]{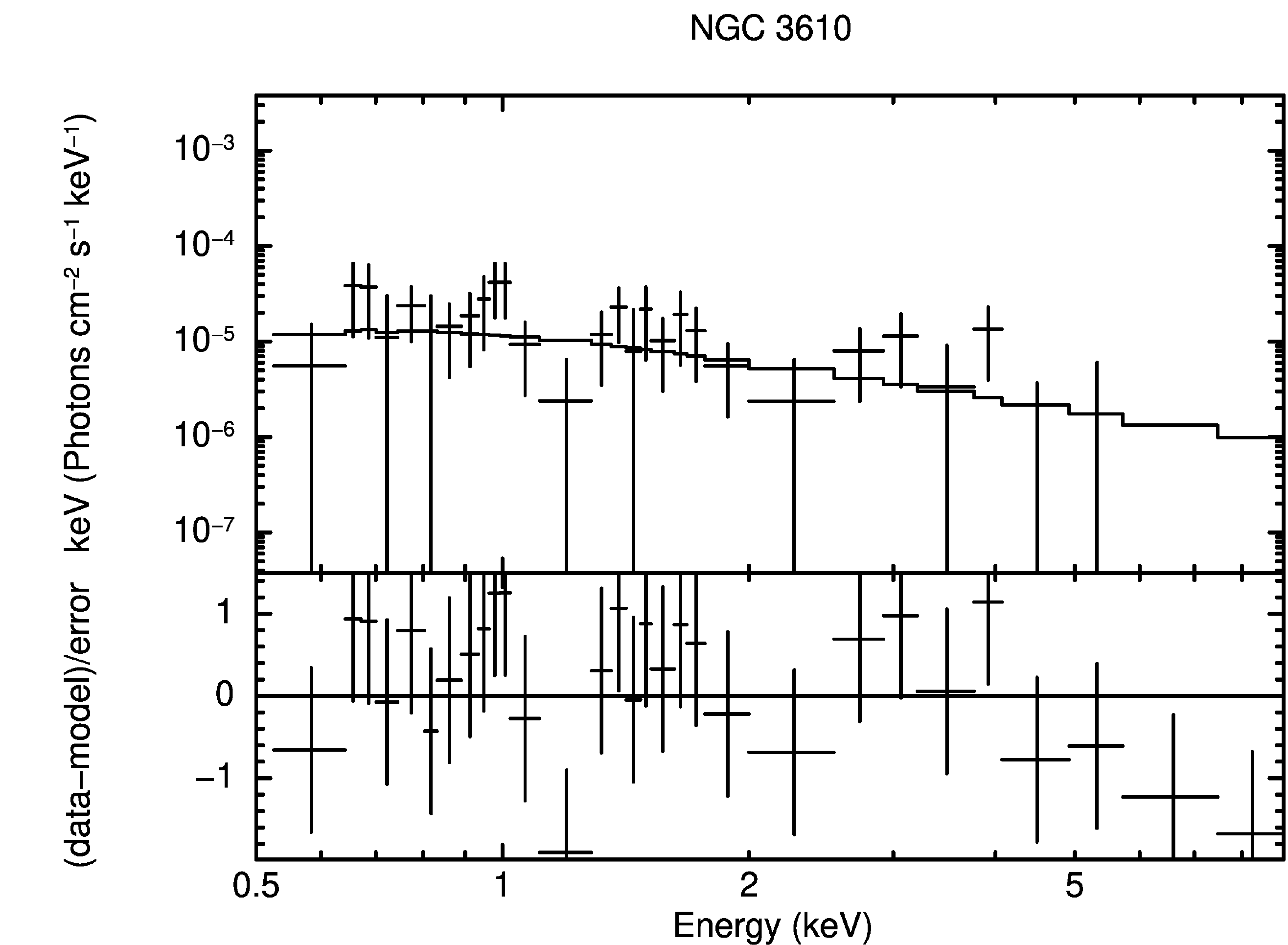}

\end{figure}
\end{center}

\begin{center}
 \begin{figure}
	\includegraphics[width=0.89\columnwidth]{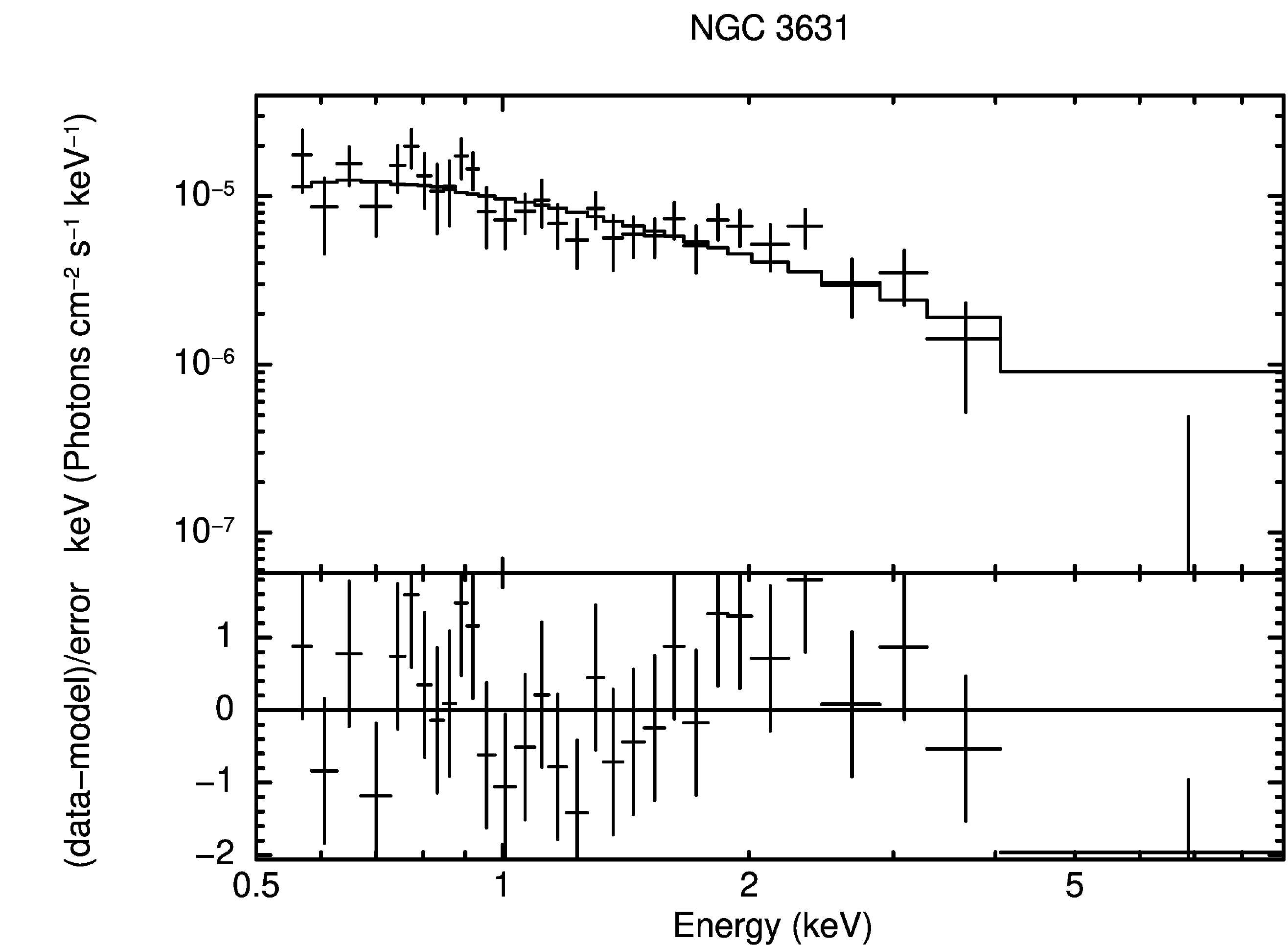}

\end{figure}
\end{center}

\begin{center}
 \begin{figure}
	\includegraphics[width=0.89\columnwidth]{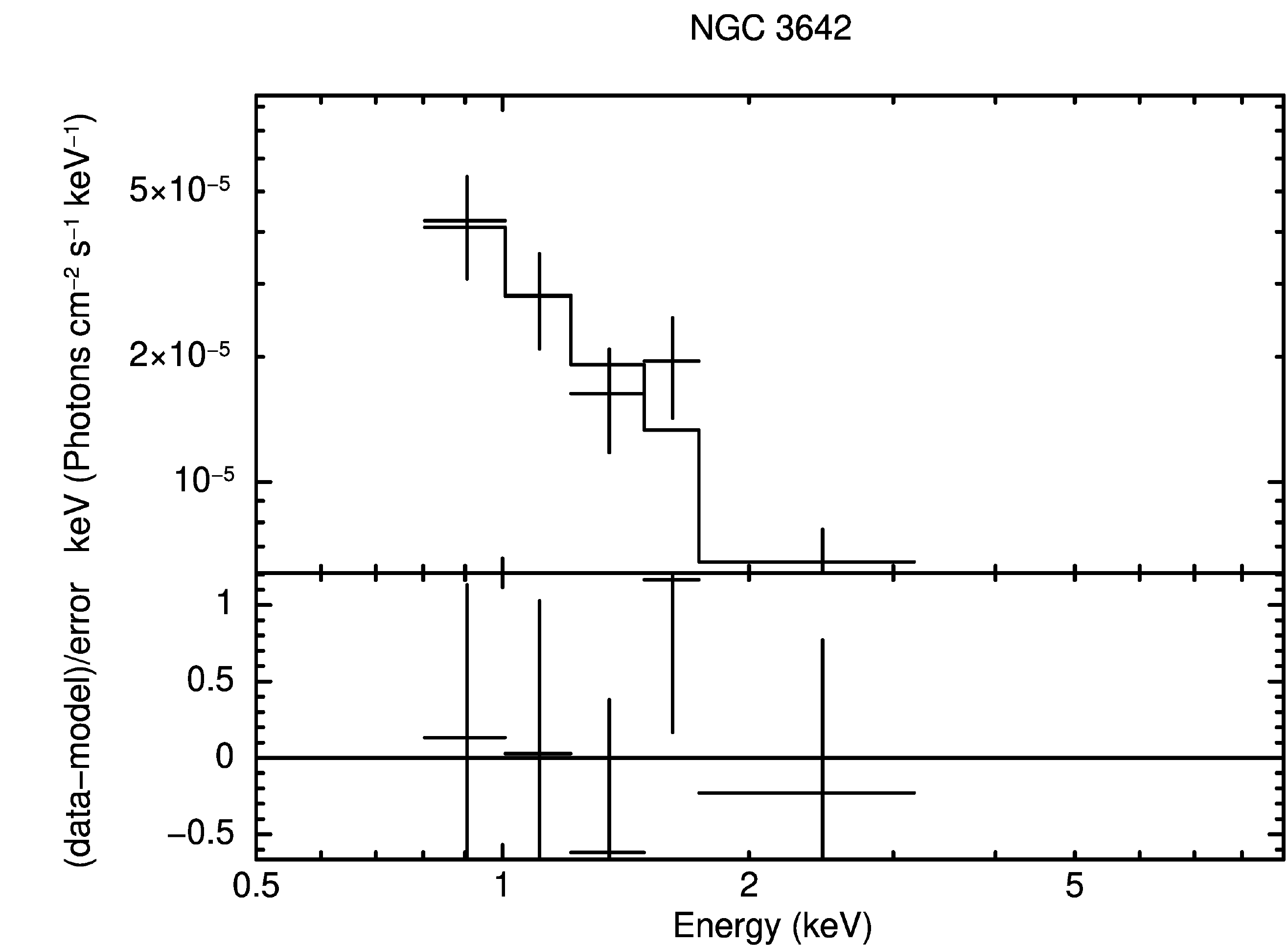}

\end{figure}
\end{center}

\begin{center}
 \begin{figure}
	\includegraphics[width=0.89\columnwidth]{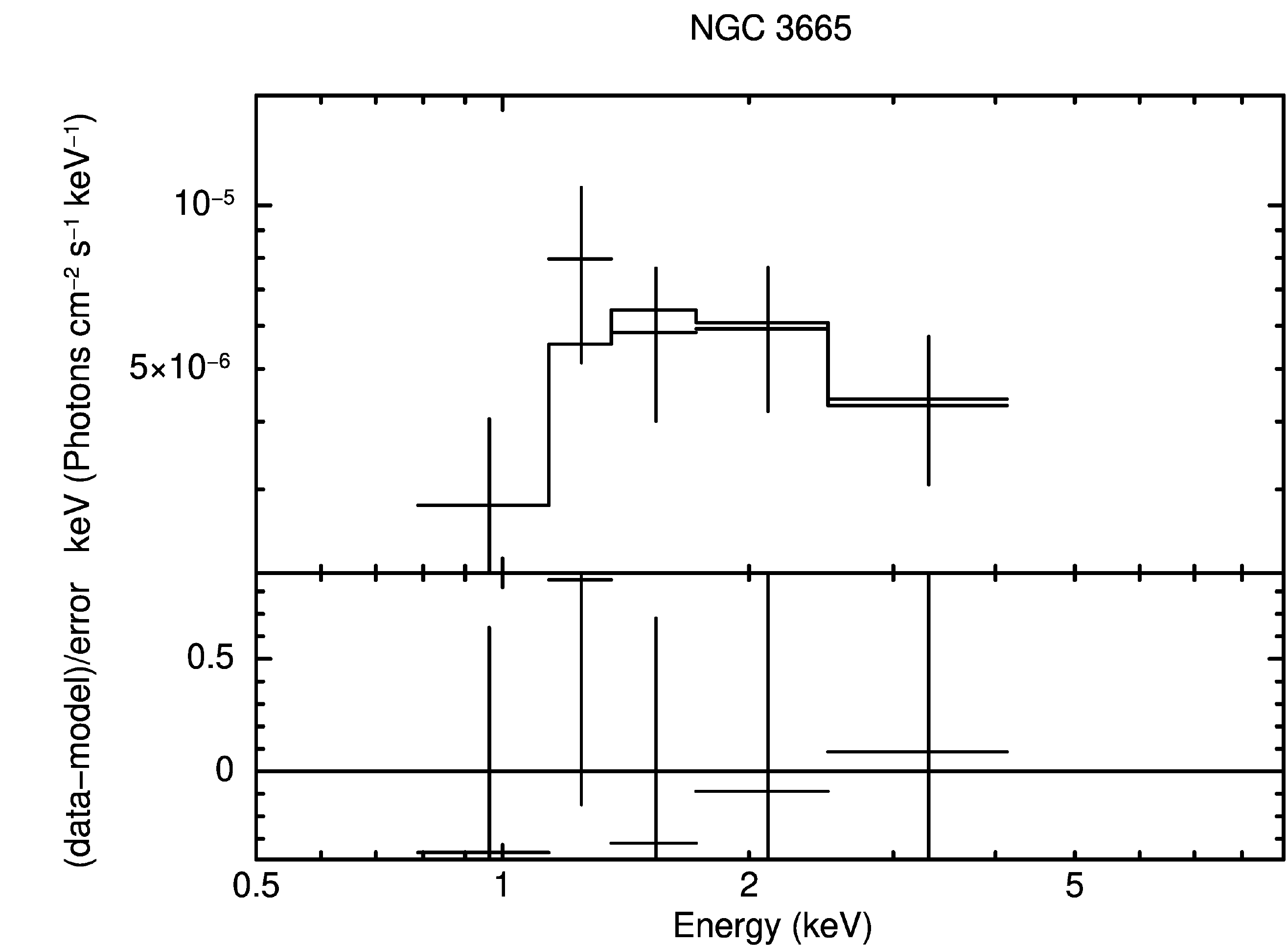}

\end{figure}
\end{center}

\begin{center}
 \begin{figure}
	\includegraphics[width=0.89\columnwidth]{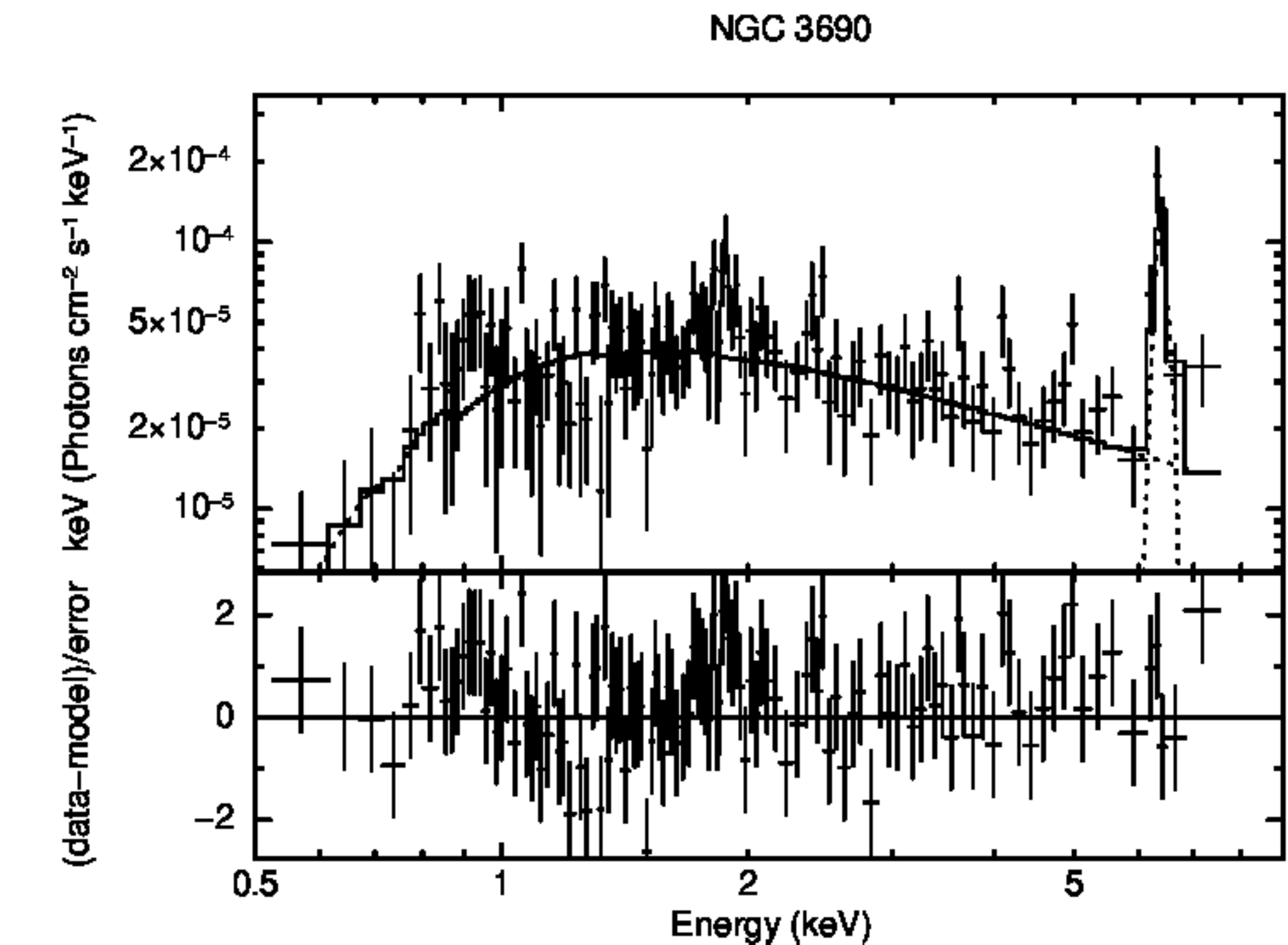}

\end{figure}
\end{center}

\begin{center}
 \begin{figure}
	\includegraphics[width=0.89\columnwidth]{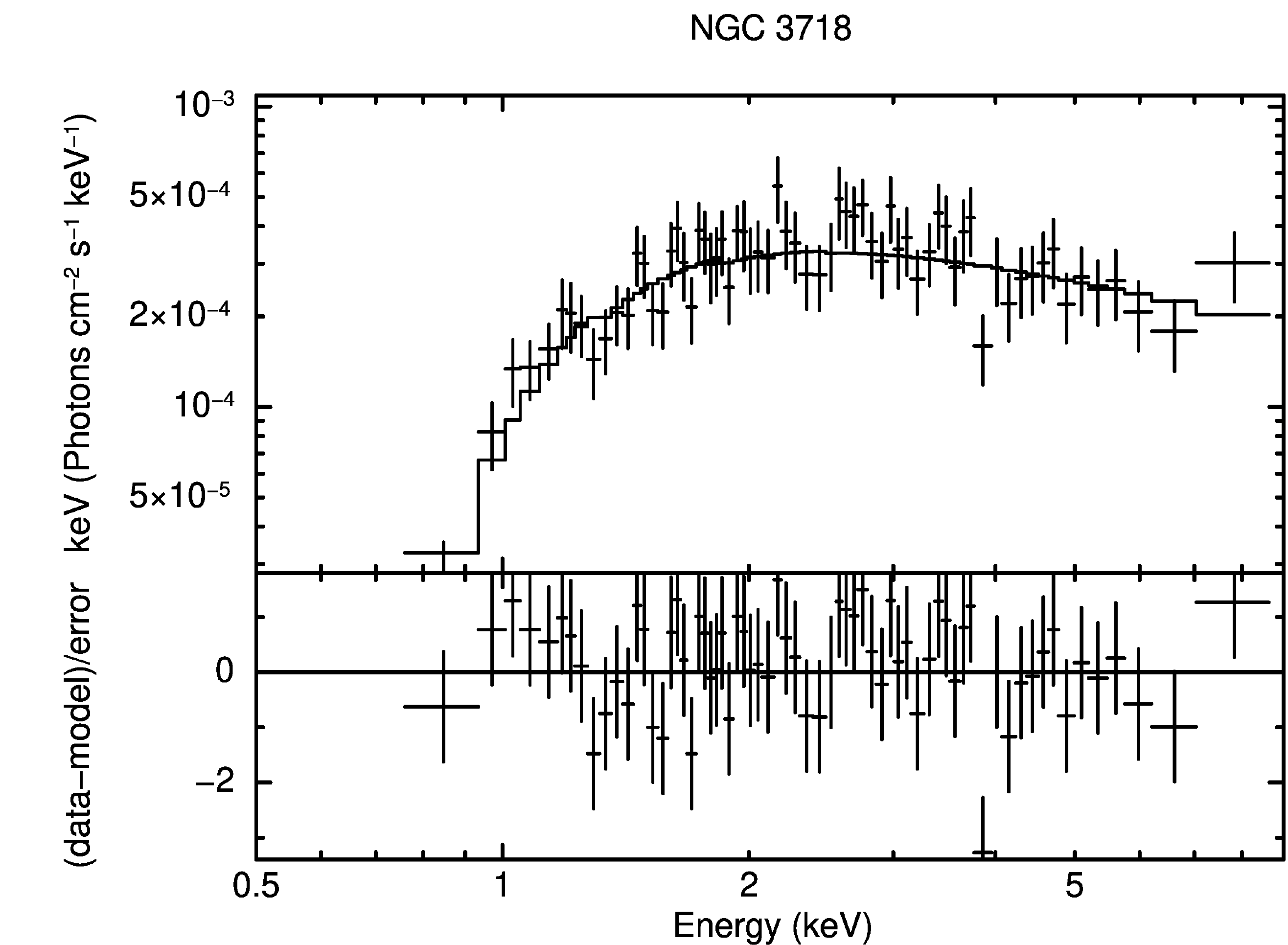}

\end{figure}
\end{center}

\begin{center}
 \begin{figure}
	\includegraphics[width=0.9\columnwidth]{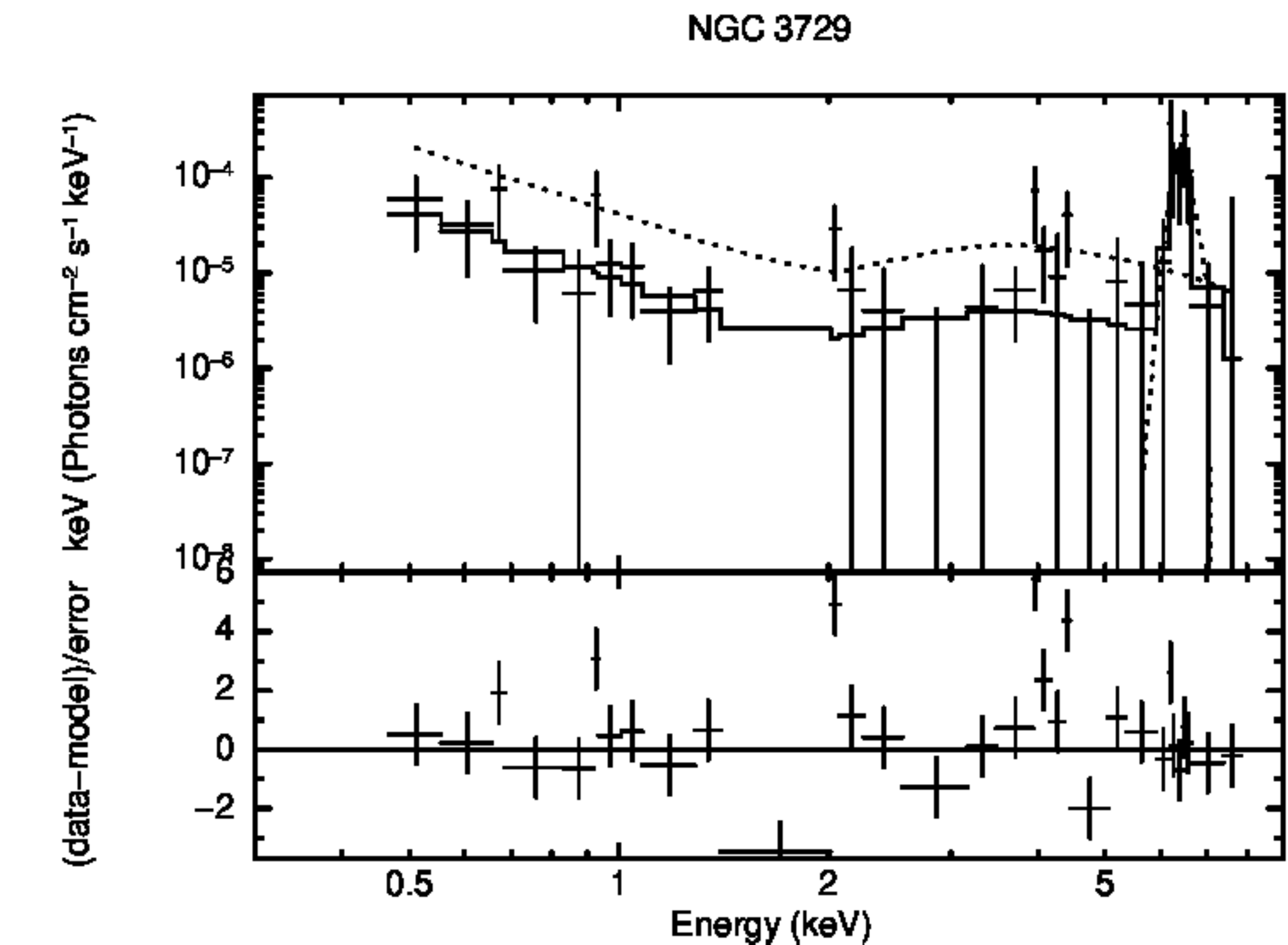}

\end{figure}
\end{center}

%
	 

\begin{center}
 \begin{figure}
	\includegraphics[width=0.89\columnwidth]{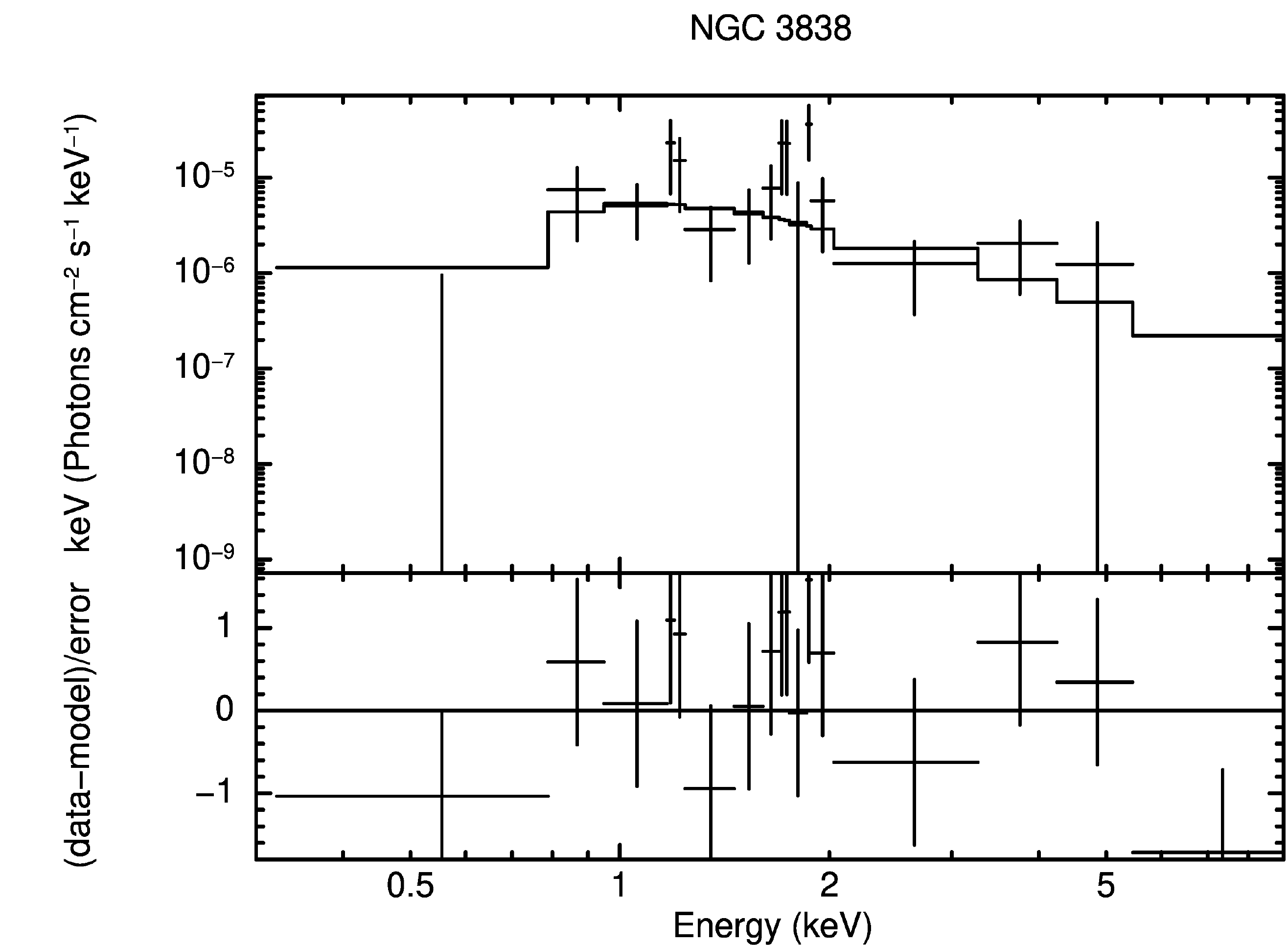}

\end{figure}
\end{center}

\begin{center}
 \begin{figure}
	\includegraphics[width=0.89\columnwidth]{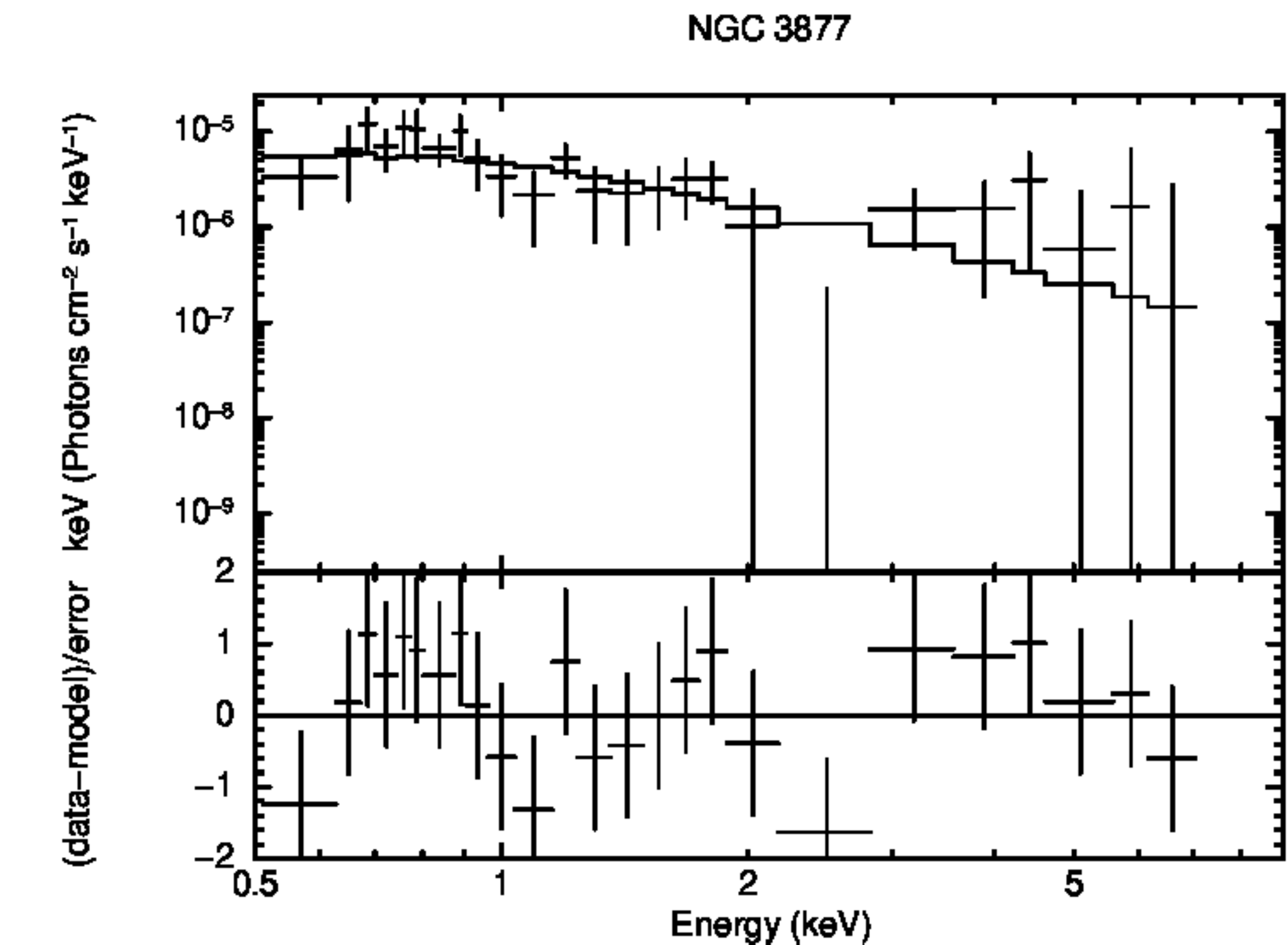}

\end{figure}
\end{center}

%
	 

\begin{center}
 \begin{figure}
	\includegraphics[width=0.89\columnwidth]{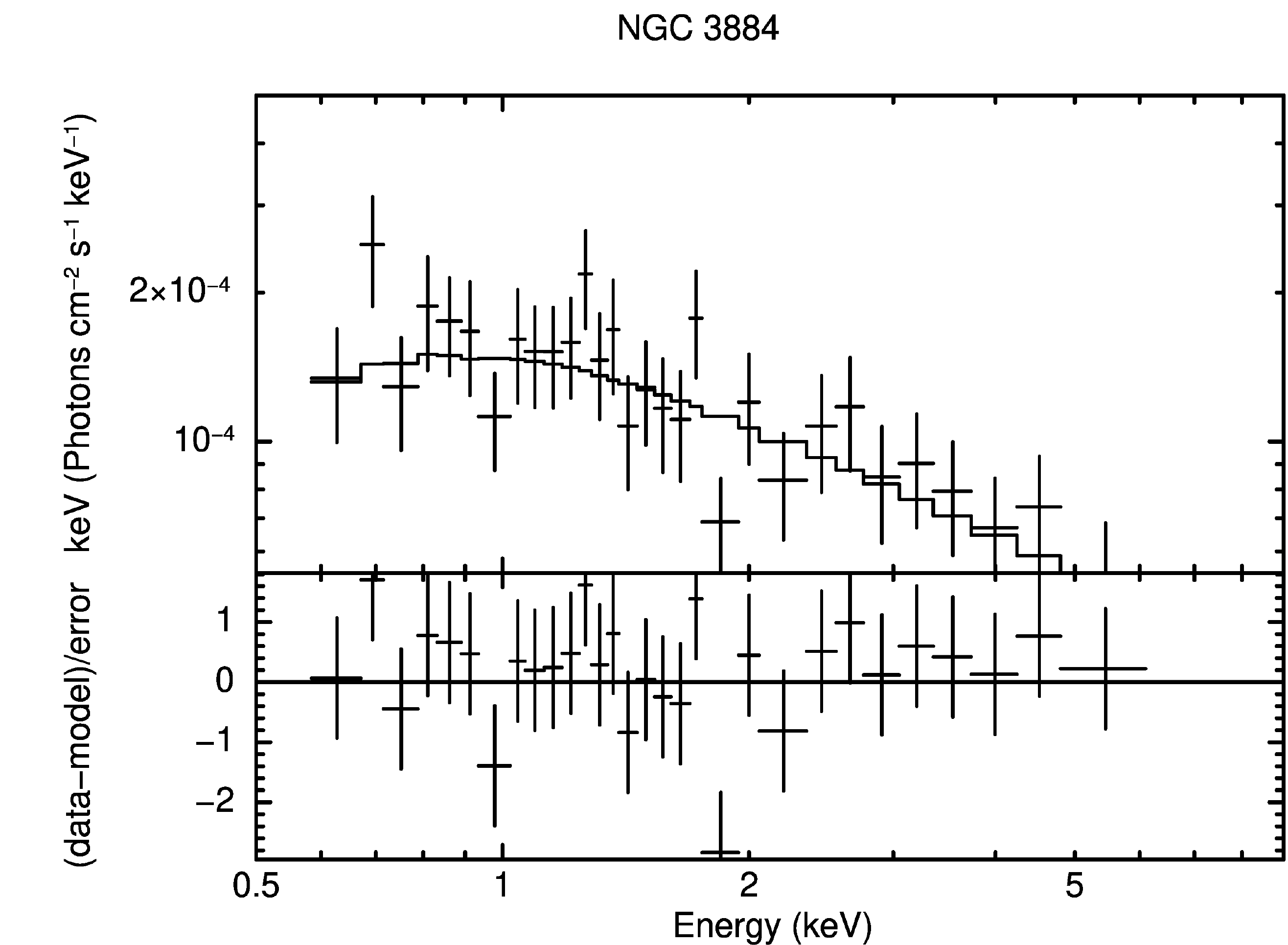}

\end{figure}
\end{center}

\begin{center}
 \begin{figure}
	\includegraphics[width=0.89\columnwidth]{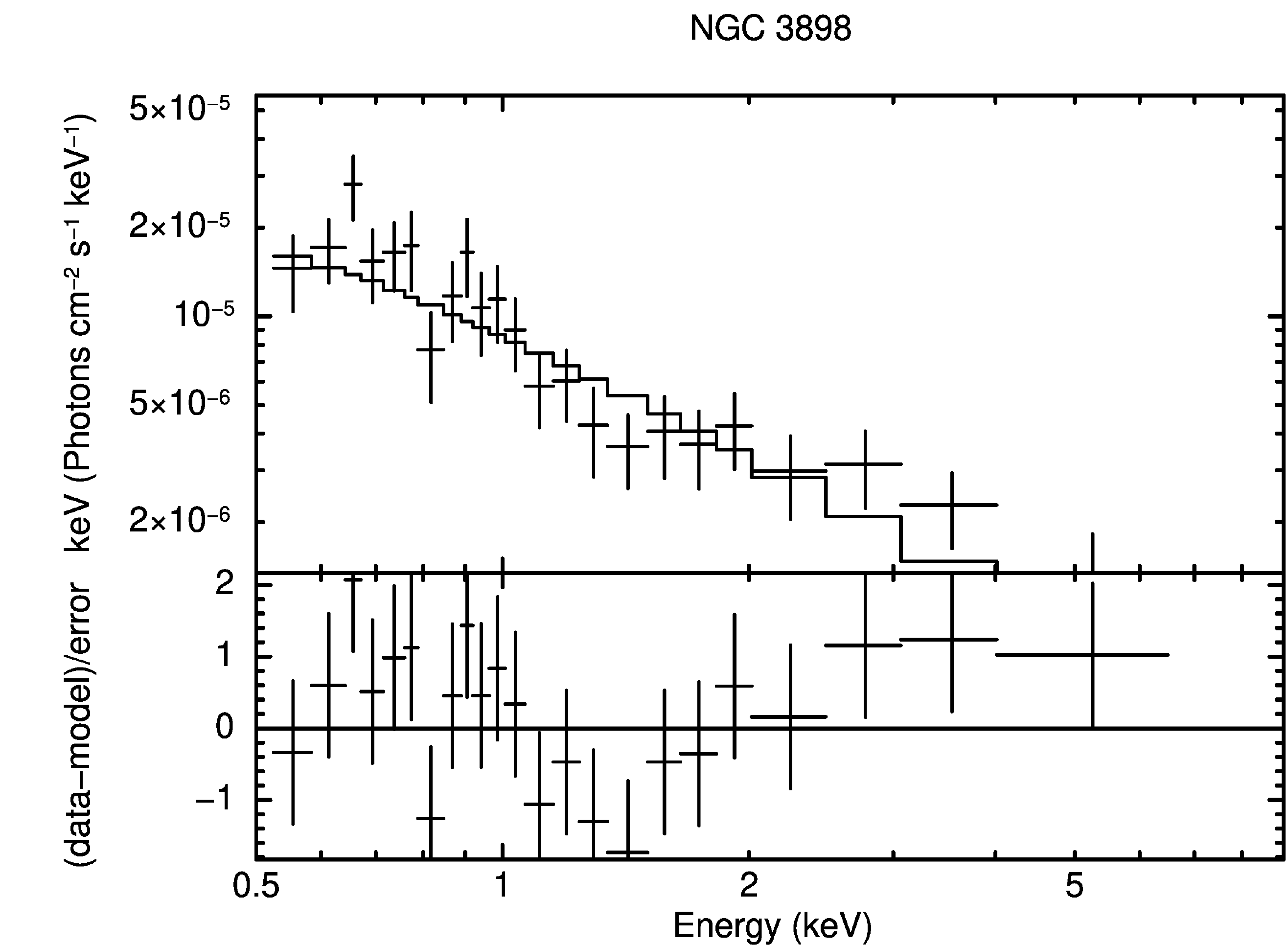}

\end{figure}
\end{center}

\begin{center}
 \begin{figure}
	\includegraphics[width=0.89\columnwidth]{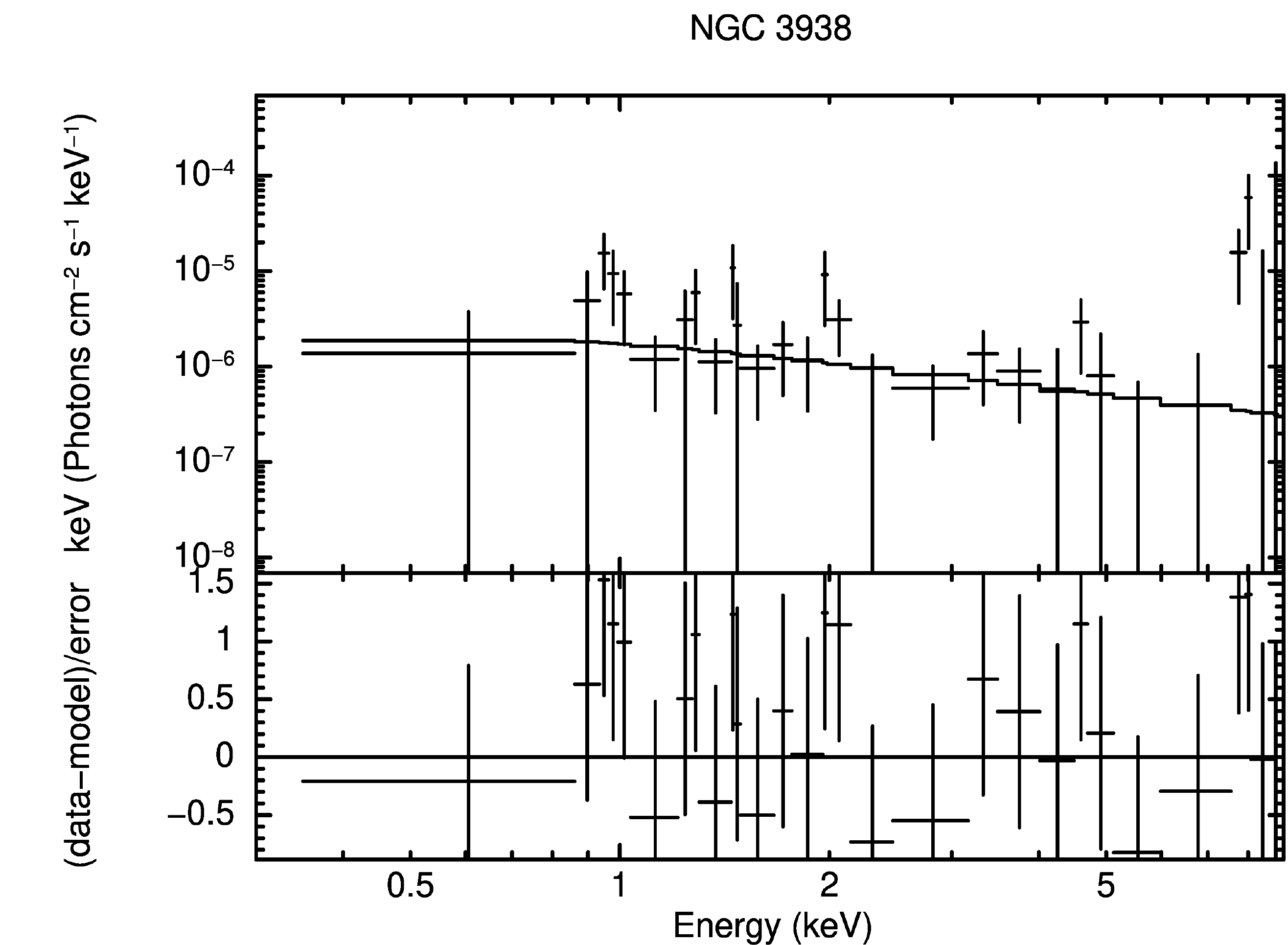}

\end{figure}
\end{center}

\begin{center}
 \begin{figure}
	\includegraphics[width=0.89\columnwidth]{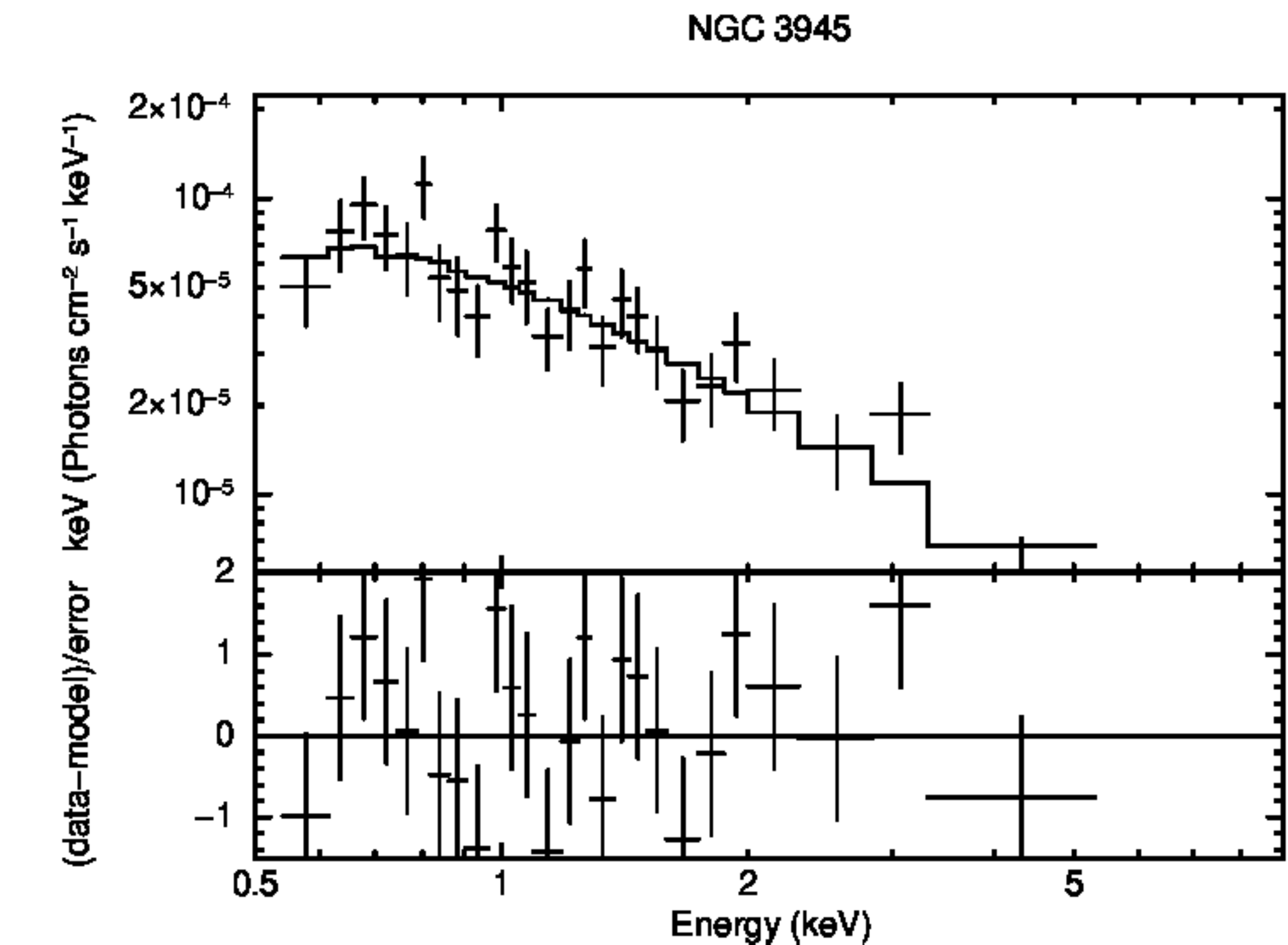}

\end{figure}
\end{center}

\begin{center}
 \begin{figure}
	\includegraphics[width=0.89\columnwidth]{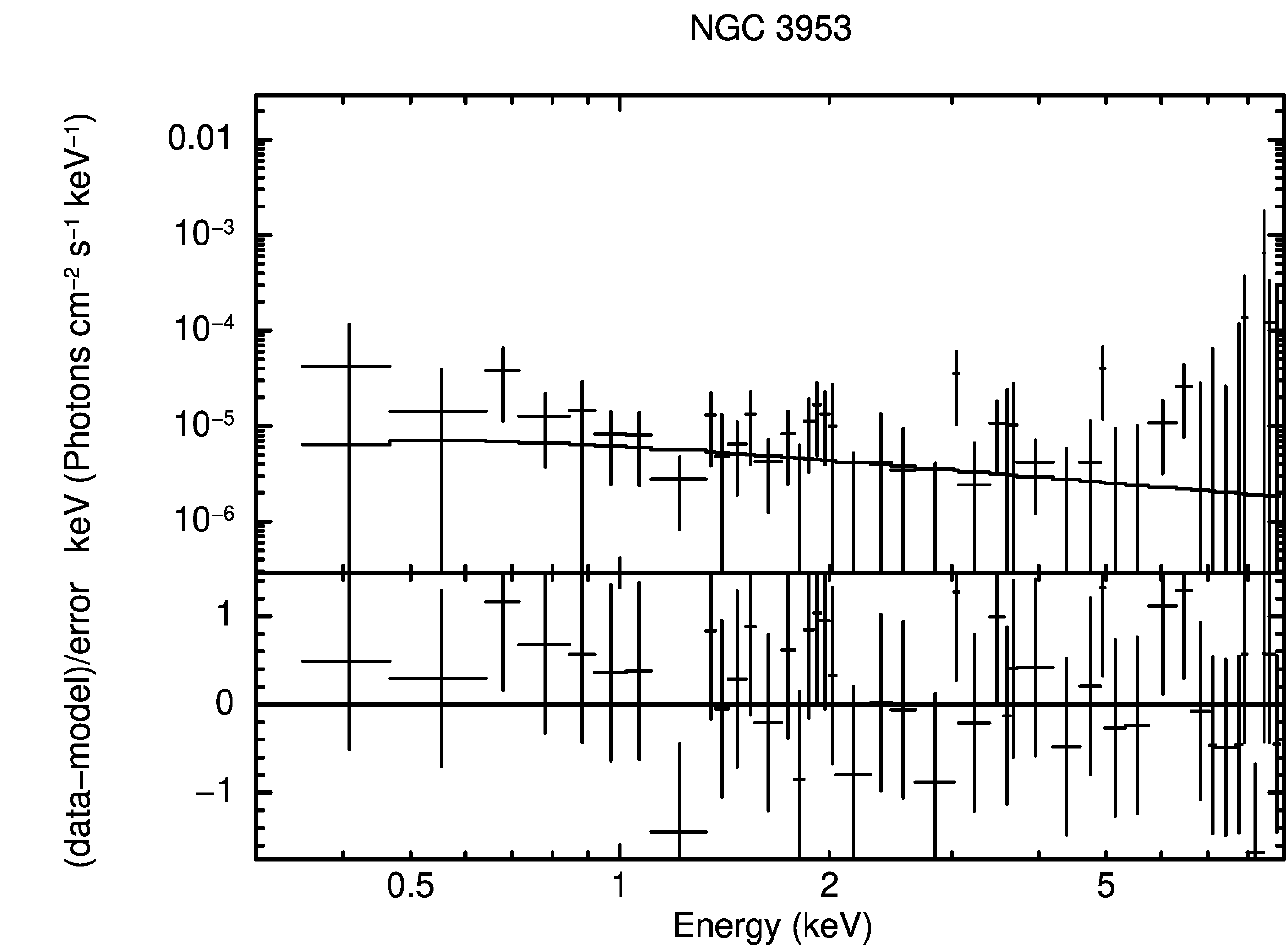}

\end{figure}
\end{center}

\begin{center}
 \begin{figure}
	\includegraphics[width=0.89\columnwidth]{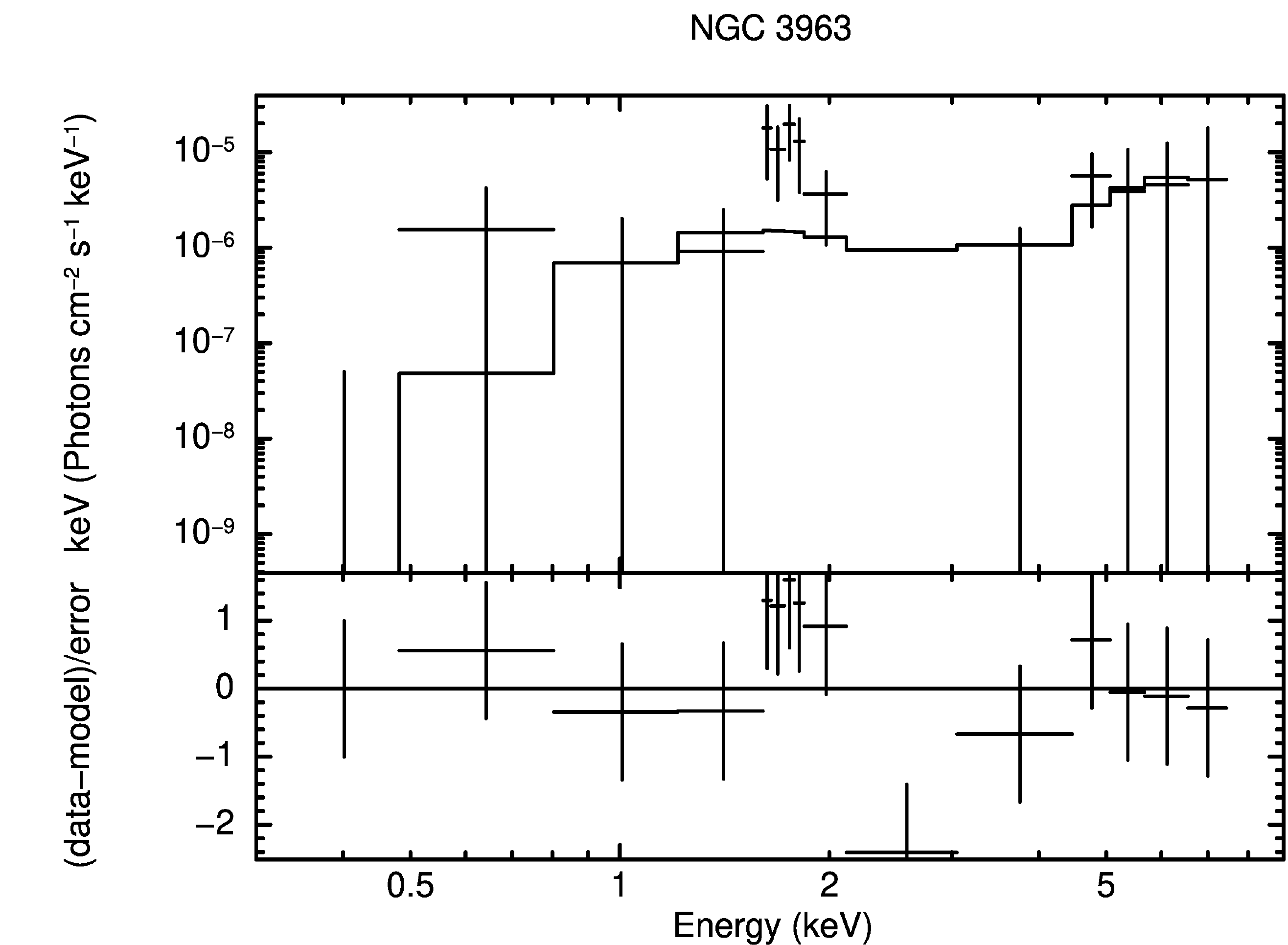}

\end{figure}
\end{center}

\begin{center}
 \begin{figure}
	\includegraphics[width=0.89\columnwidth]{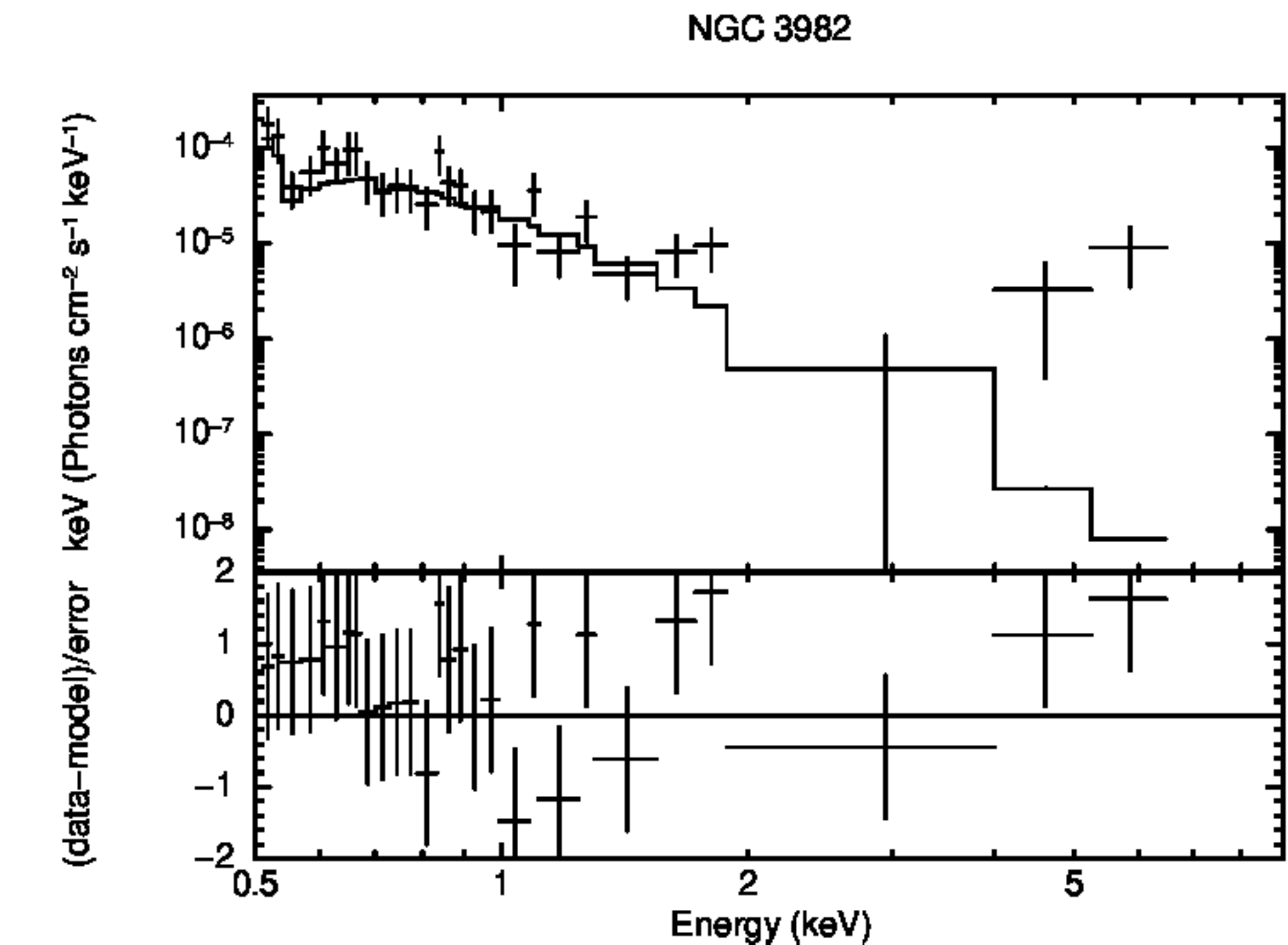}

\end{figure}
\end{center}

%
	 

\begin{center}
 \begin{figure}
	\includegraphics[width=0.89\columnwidth]{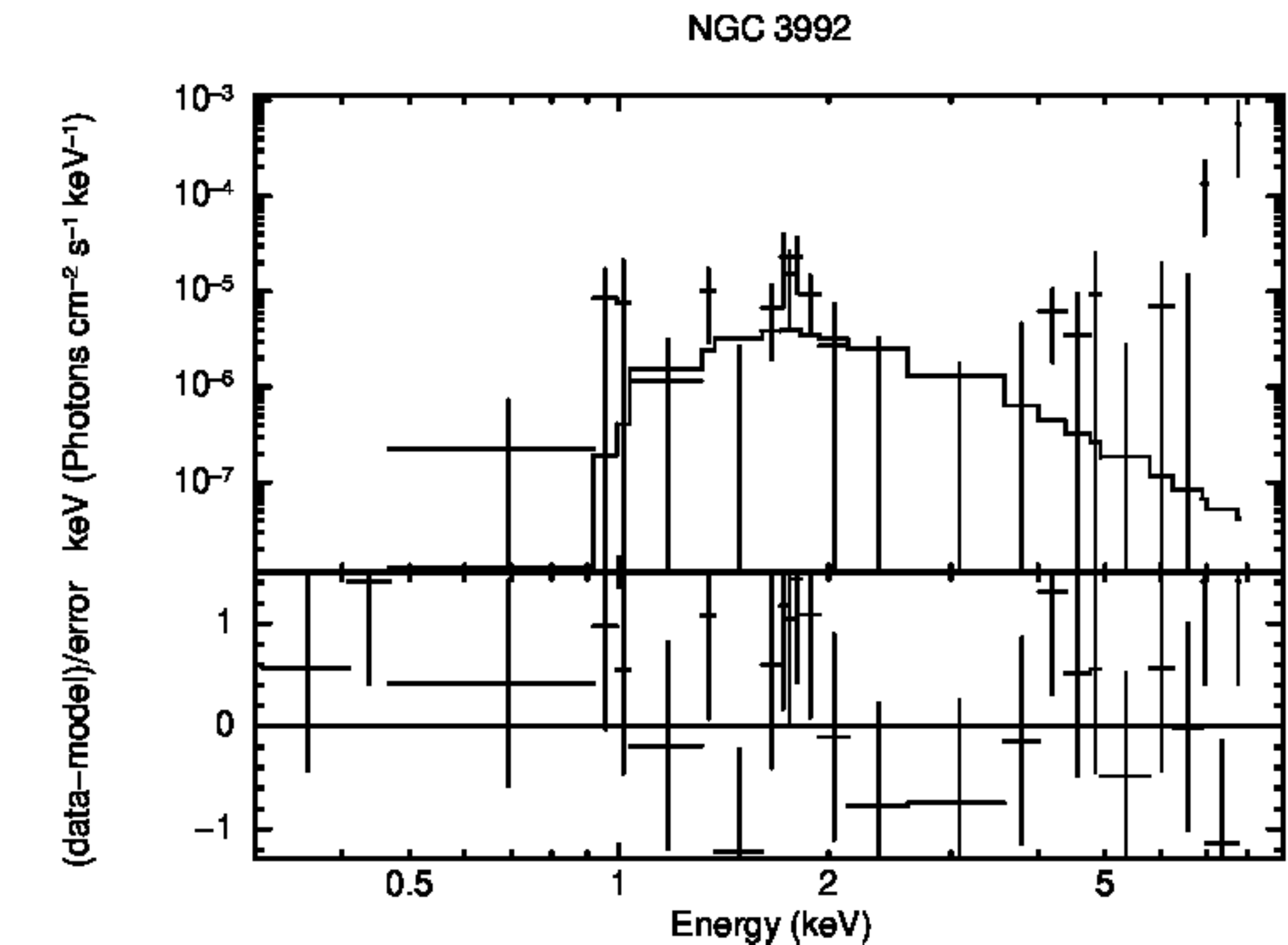}

\end{figure}
\end{center}

%
	 

\begin{center}
 \begin{figure}
	\includegraphics[width=0.89\columnwidth]{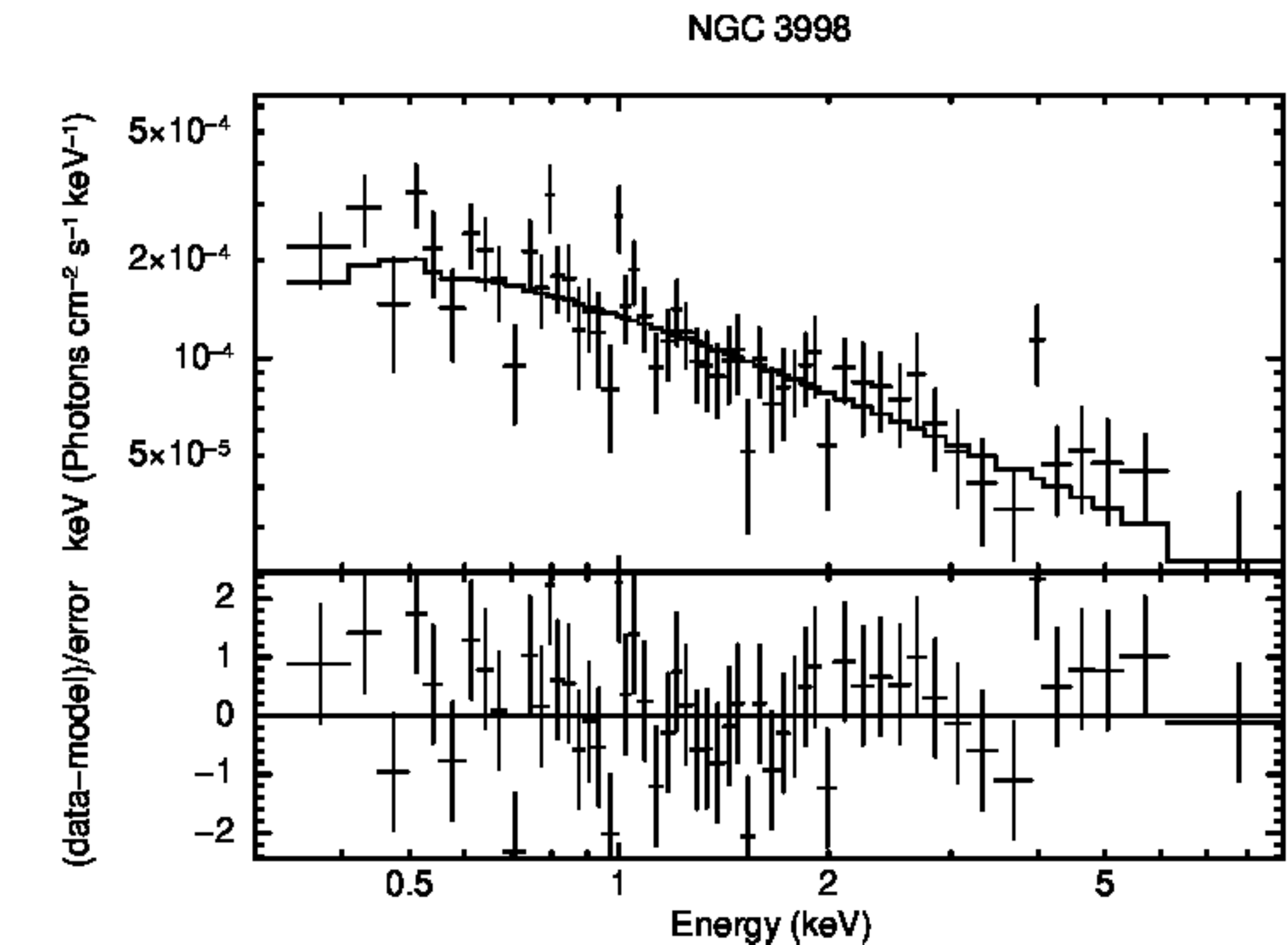}

\end{figure}
\end{center}

%
	 

\begin{center}
 \begin{figure}
	\includegraphics[width=0.89\columnwidth]{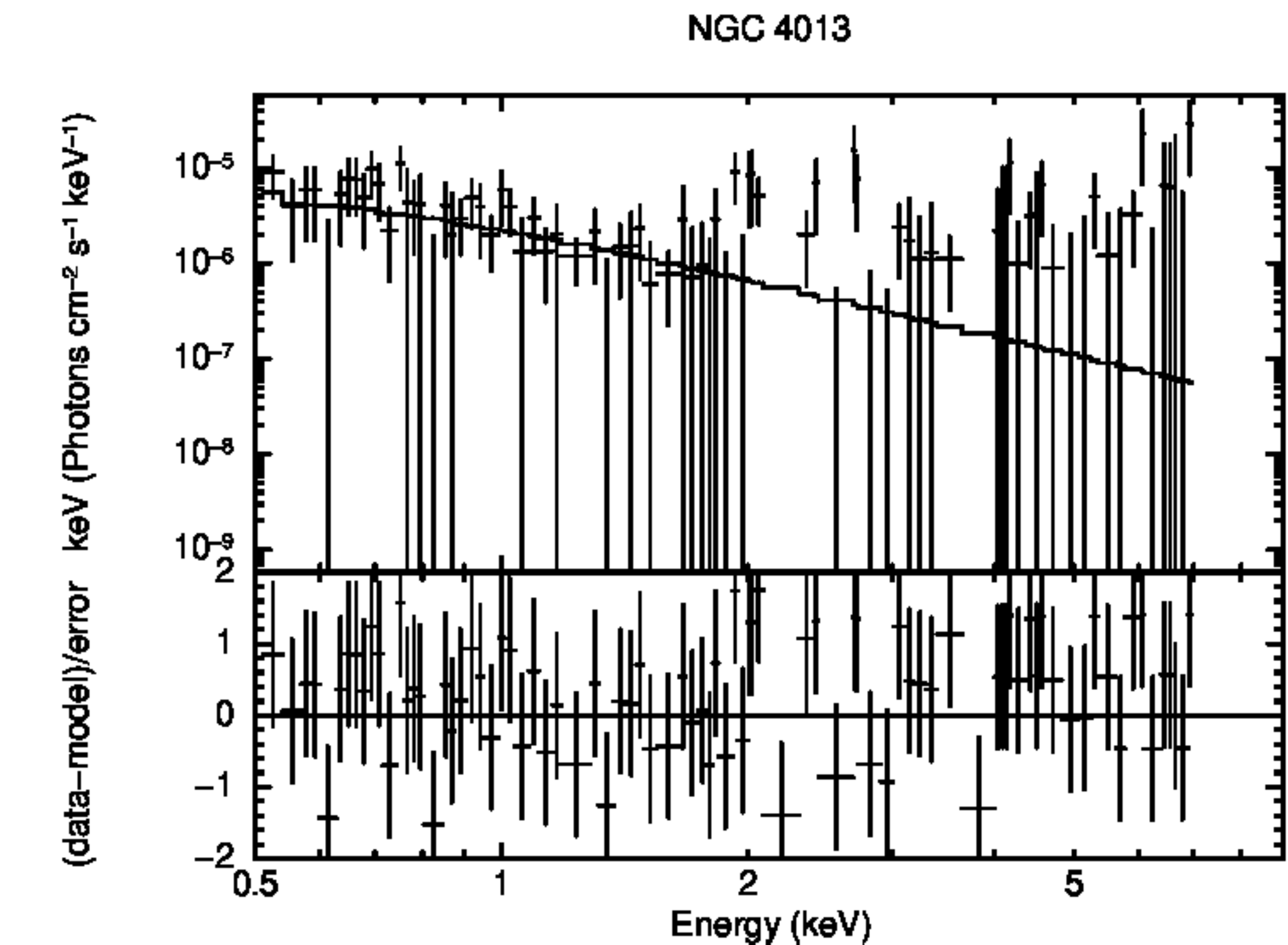}

\end{figure}
\end{center}

%
	 

\begin{center}
 \begin{figure}
	\includegraphics[width=0.89\columnwidth]{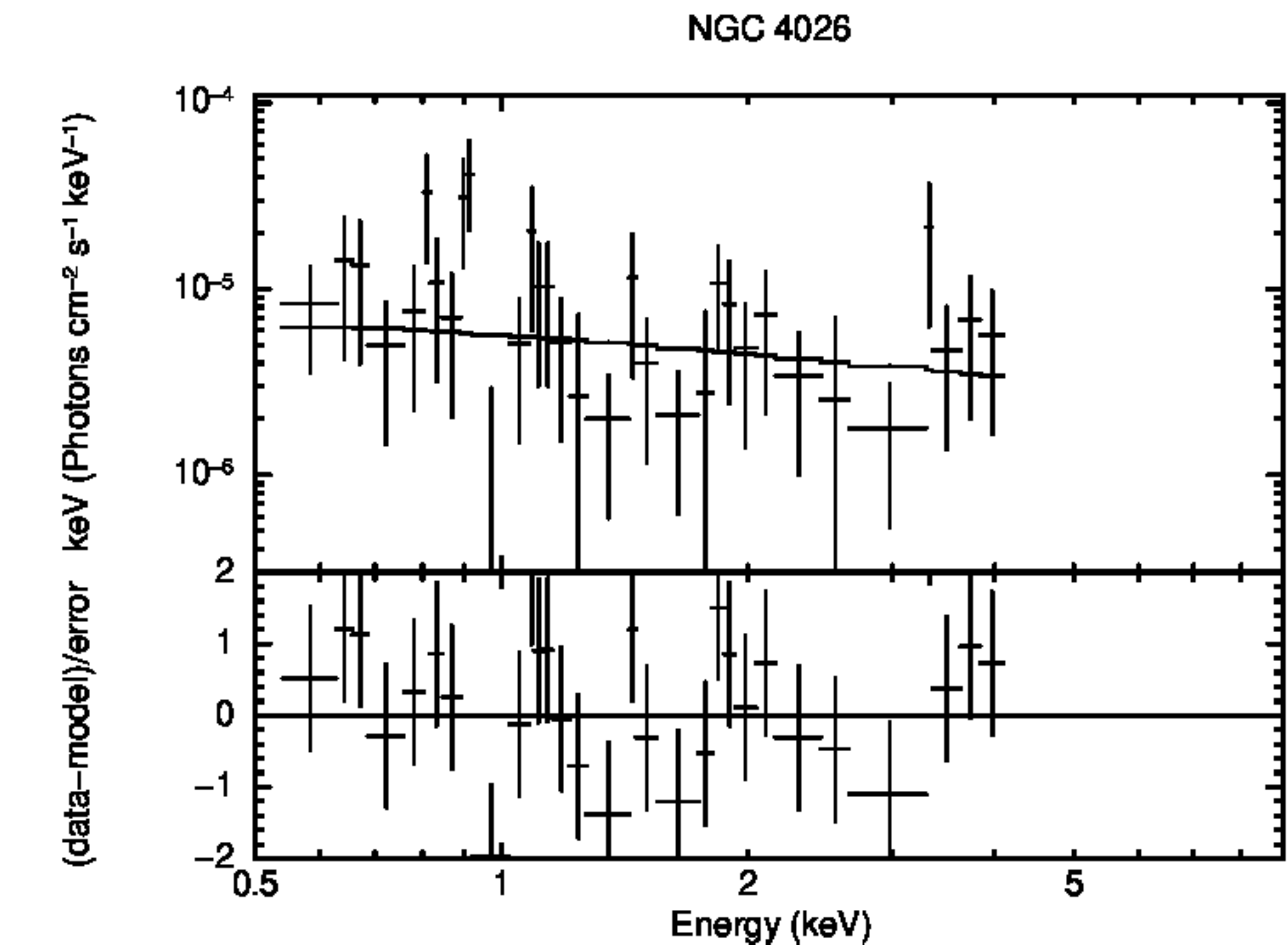}

\end{figure}
\end{center}

%
	 

\begin{center}
 \begin{figure}
	\includegraphics[width=0.89\columnwidth]{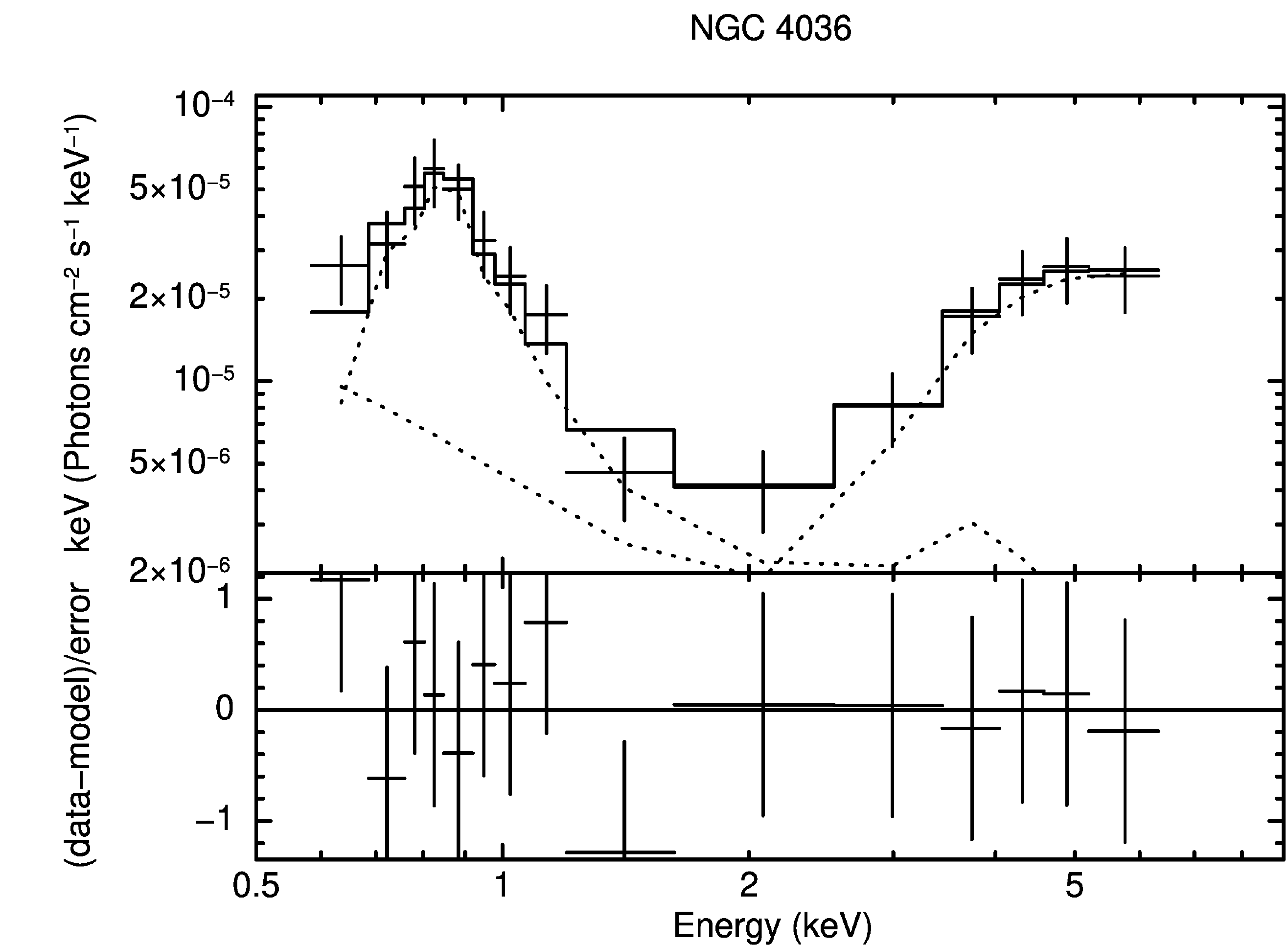}

\end{figure}
\end{center}

\begin{center}
 \begin{figure}
	\includegraphics[width=0.89\columnwidth]{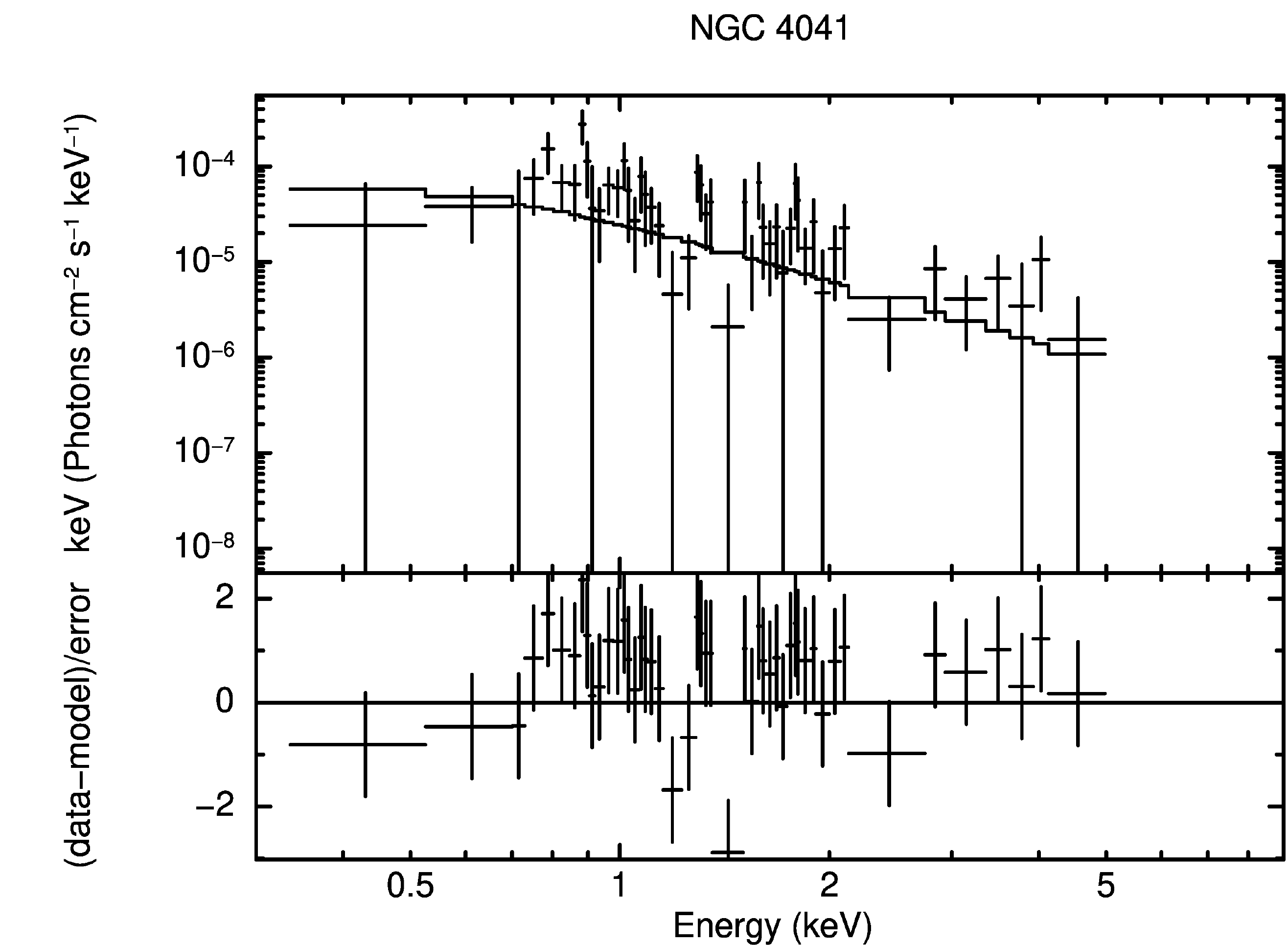}

\end{figure}
\end{center}

\begin{center}
 \begin{figure}
	\includegraphics[width=0.89\columnwidth]{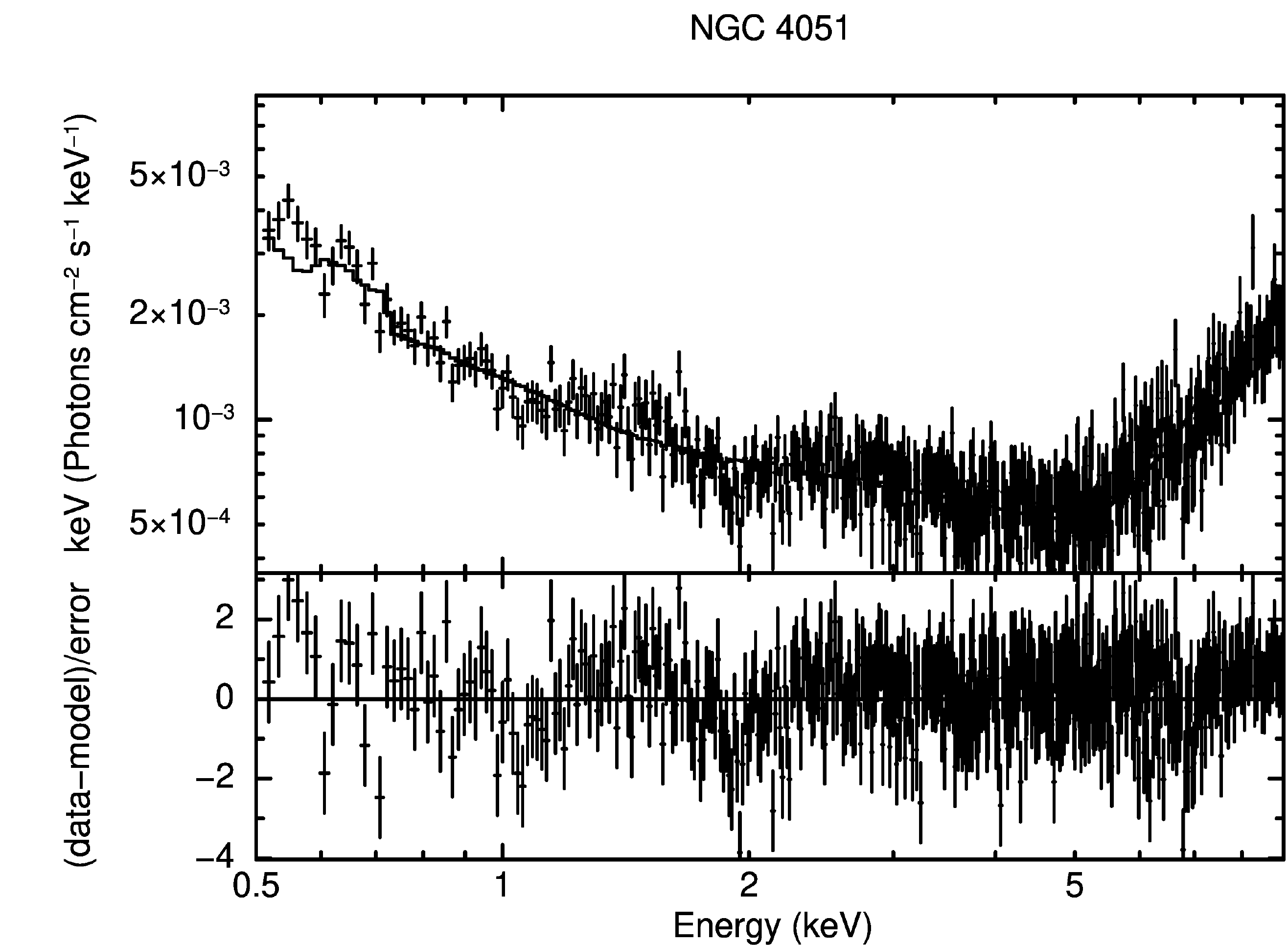}

\end{figure}
\end{center}

\begin{center}
 \begin{figure}
	\includegraphics[width=0.89\columnwidth]{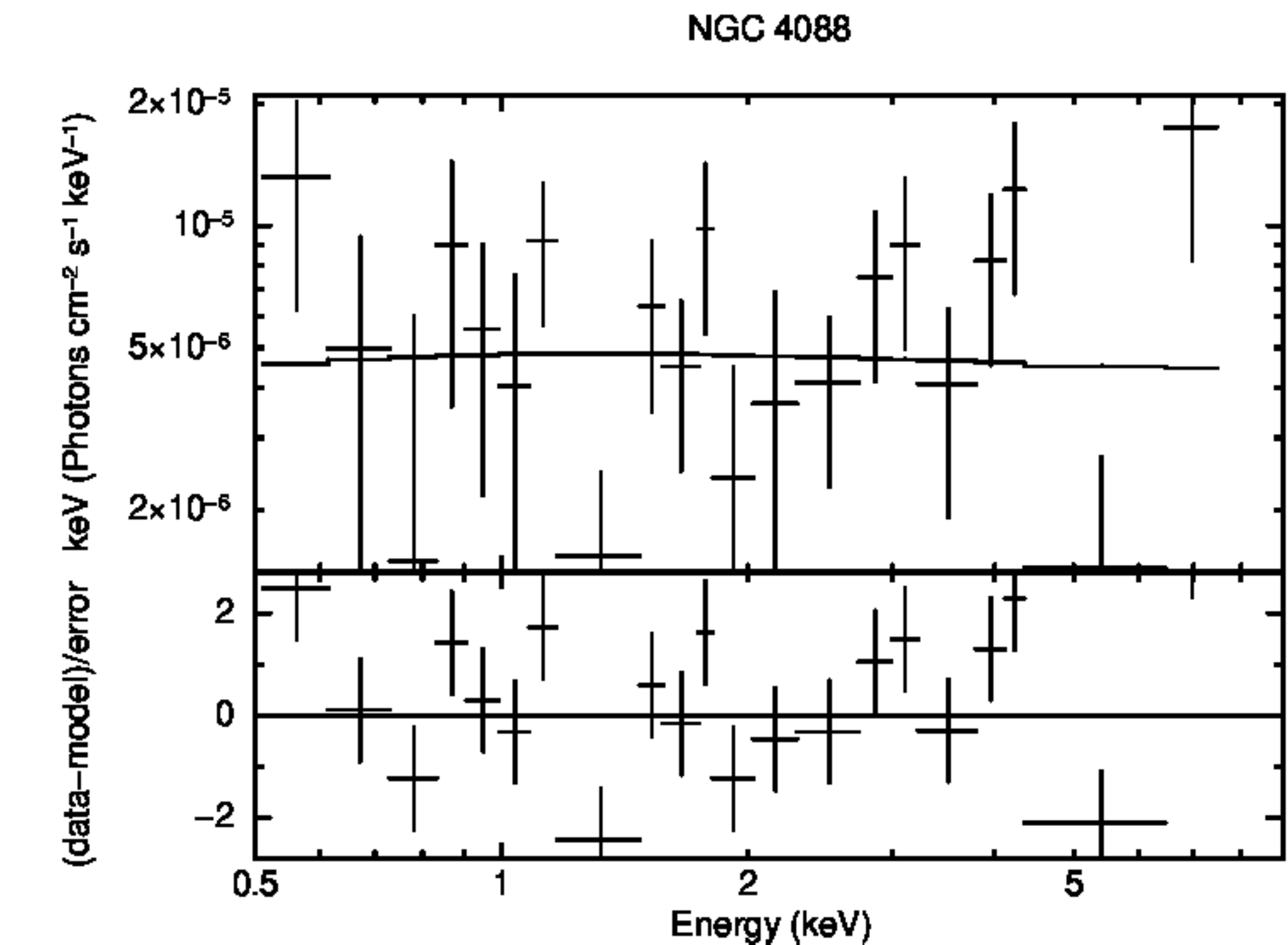}

\end{figure}
\end{center}

%
	 

\begin{center}
 \begin{figure}
	\includegraphics[width=0.89\columnwidth]{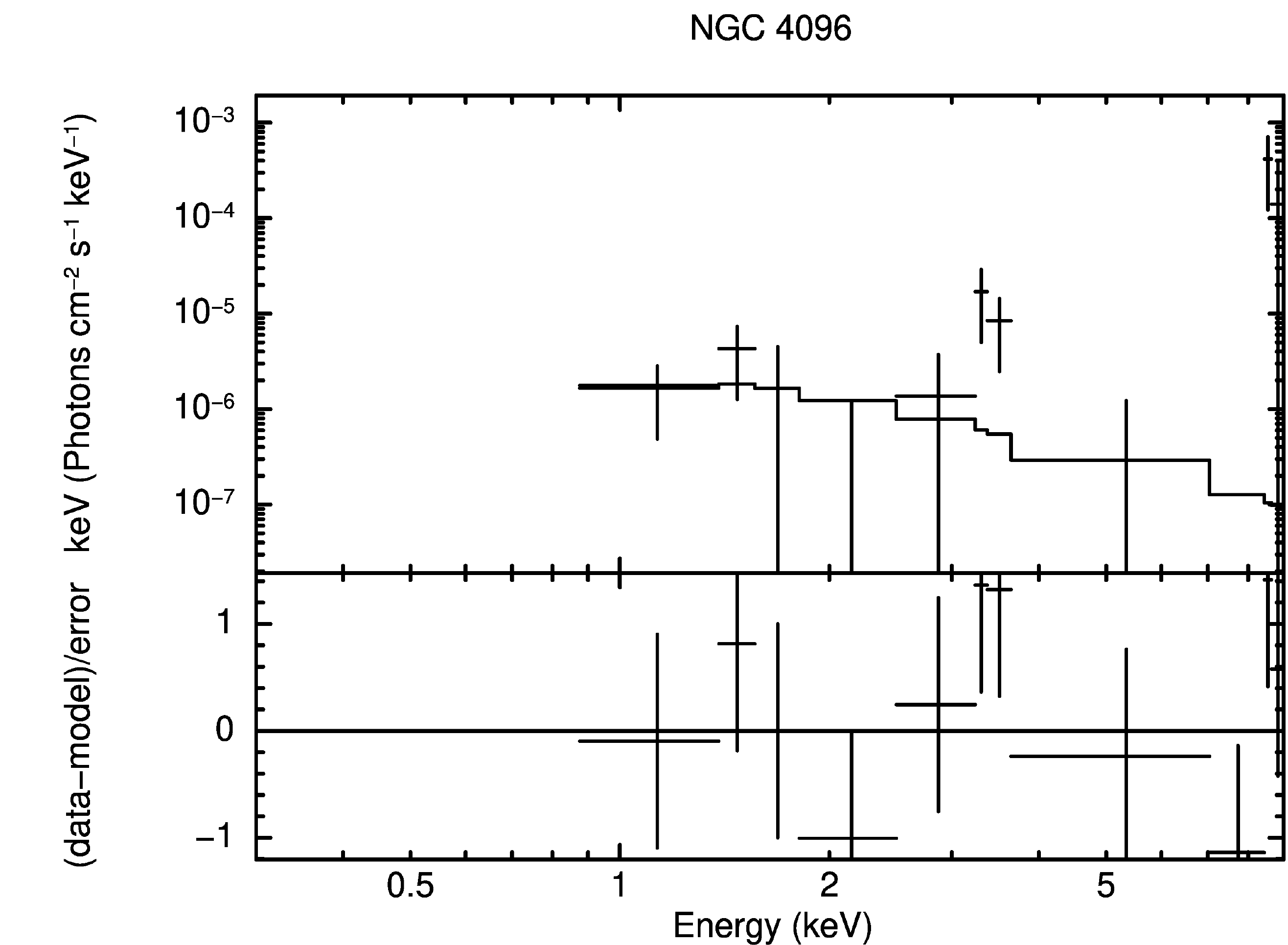}

\end{figure}
\end{center}

\begin{center}
 \begin{figure}
	\includegraphics[width=0.89\columnwidth]{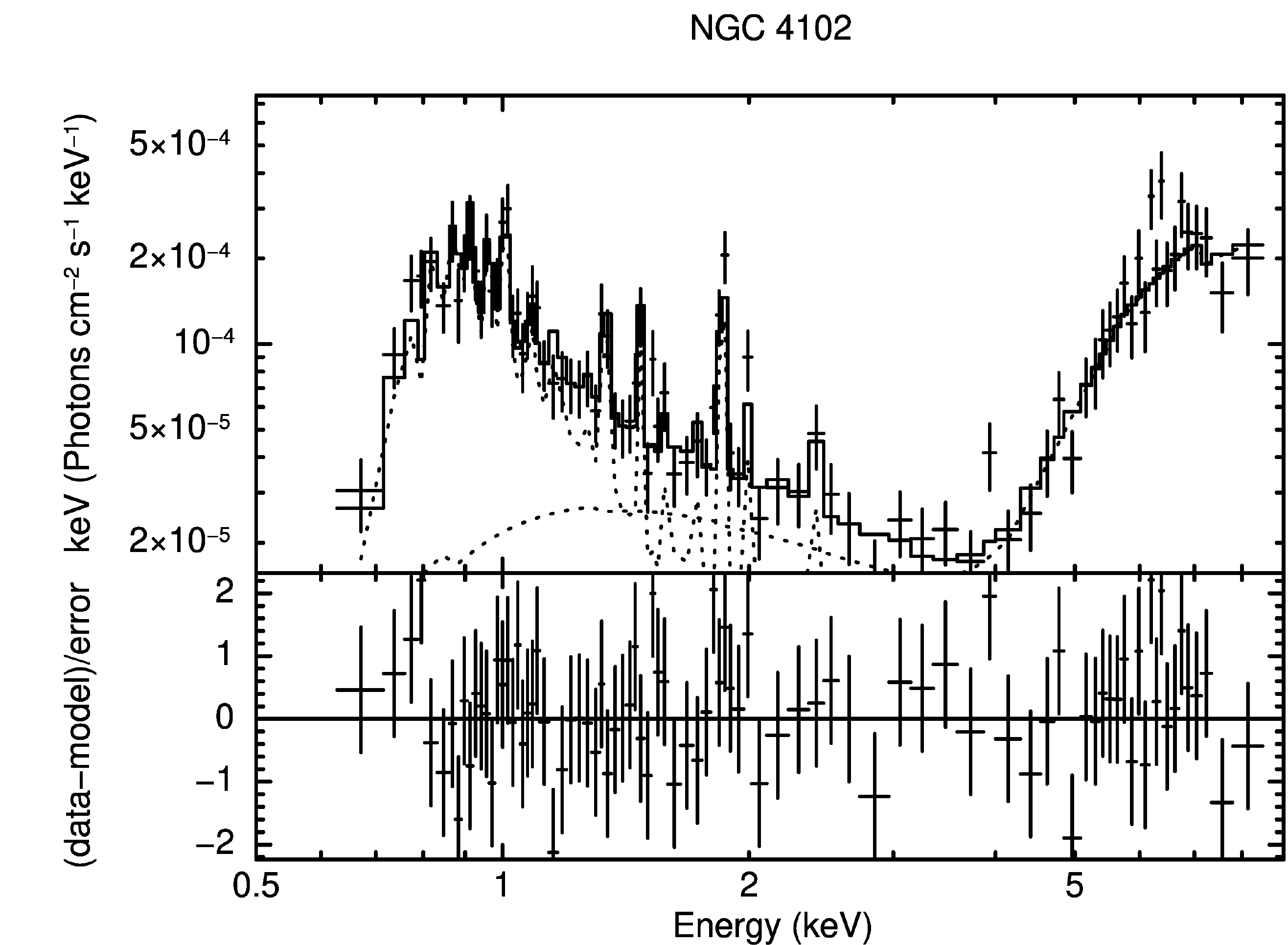}

\end{figure}
\end{center}

\begin{center}
 \begin{figure}
	\includegraphics[width=0.89\columnwidth]{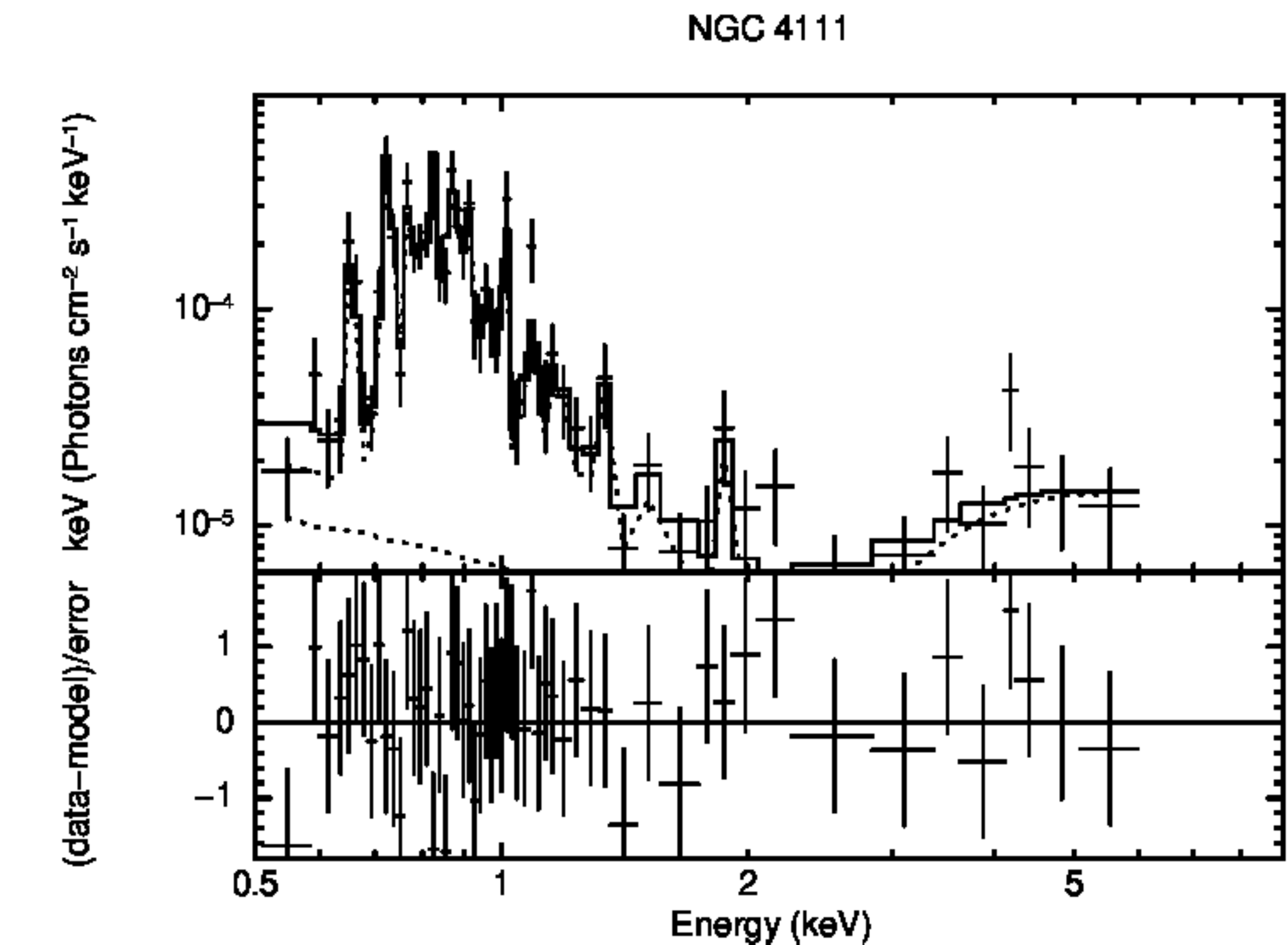}

\end{figure}
\end{center}

%


\begin{center}
 \begin{figure}
	\includegraphics[width=0.89\columnwidth]{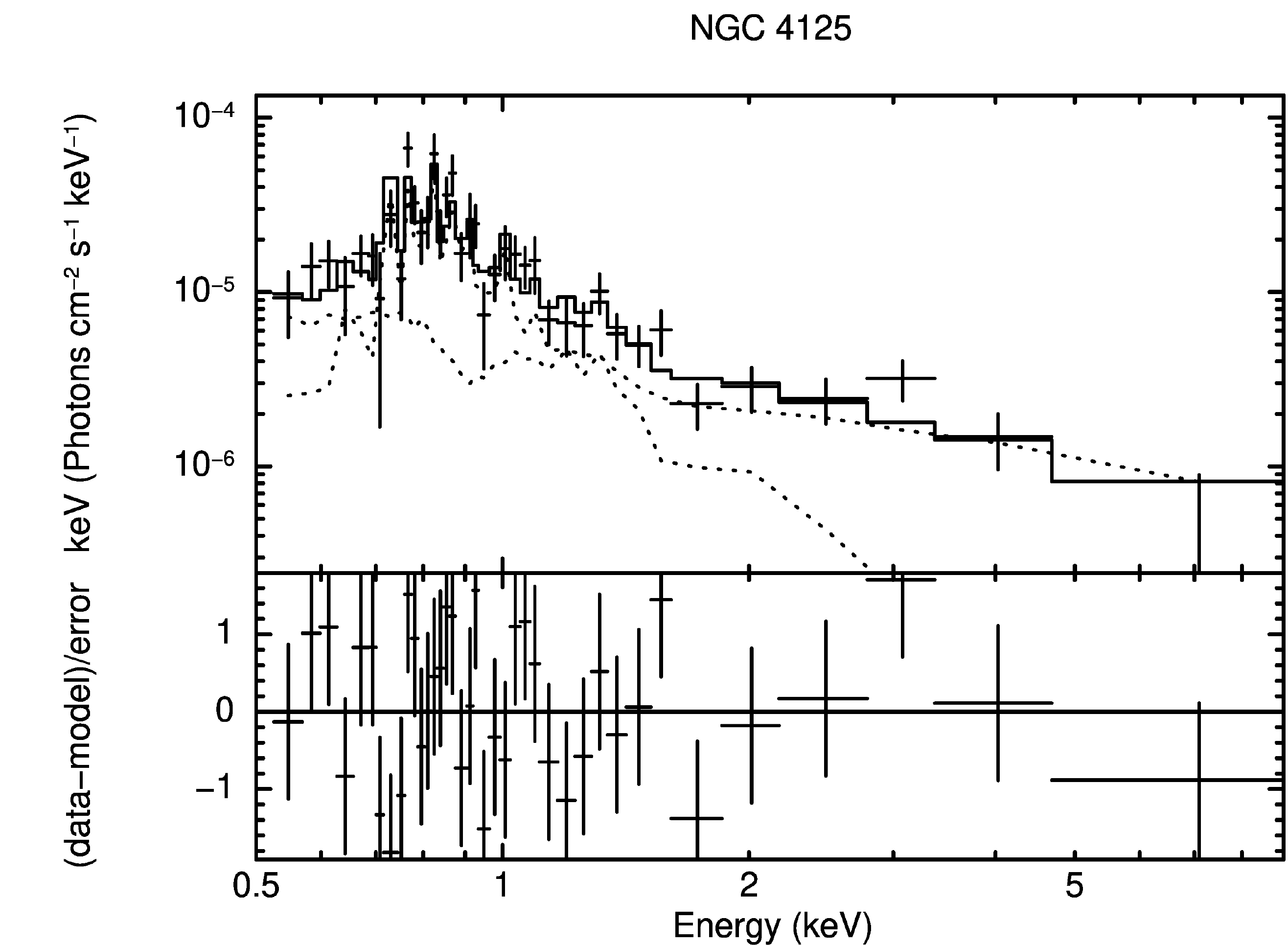}

\end{figure}
\end{center}

\begin{center}
 \begin{figure}
	\includegraphics[width=0.89\columnwidth]{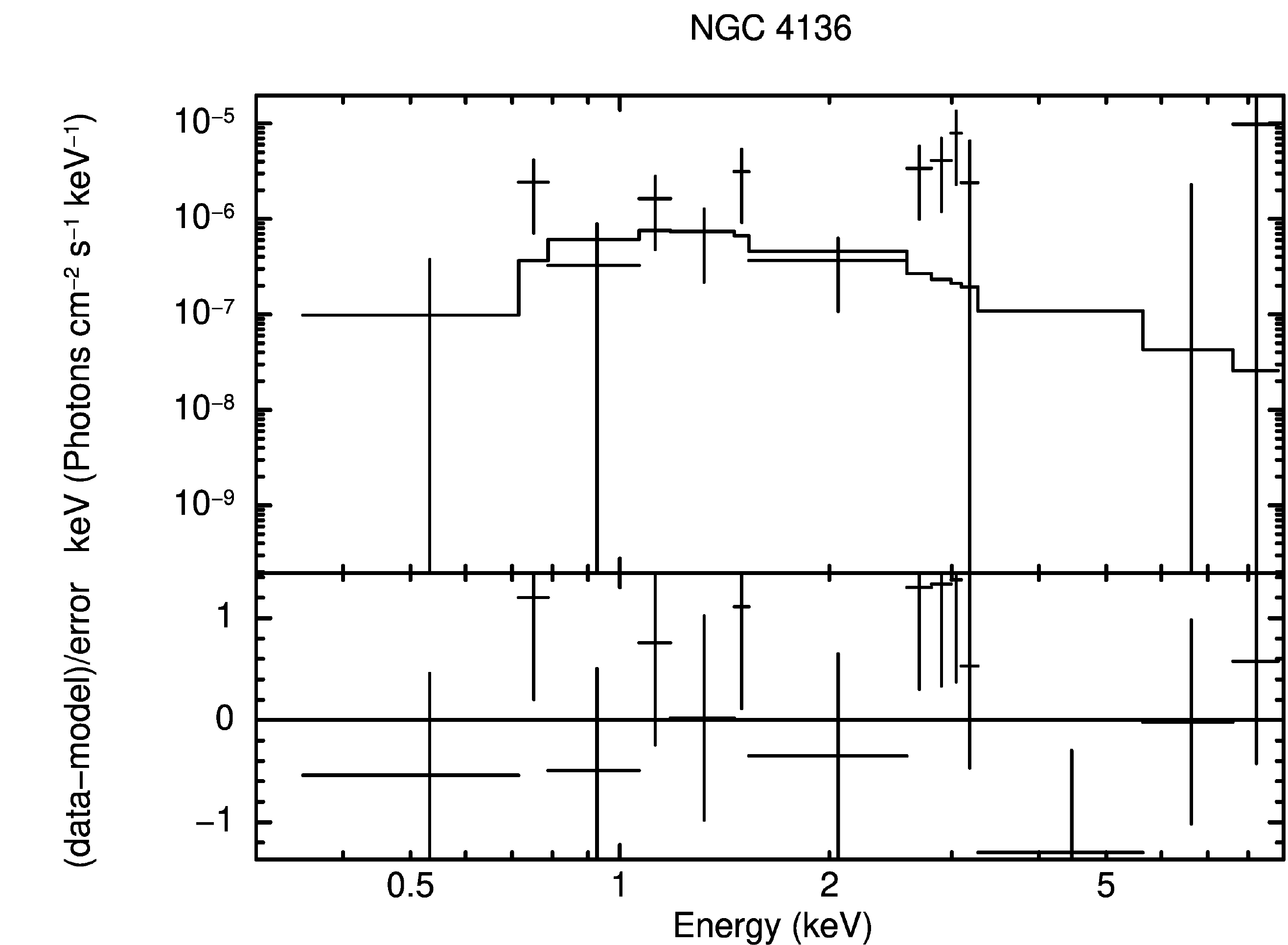}

\end{figure}
\end{center}

\begin{center}
 \begin{figure}
	\includegraphics[width=0.89\columnwidth]{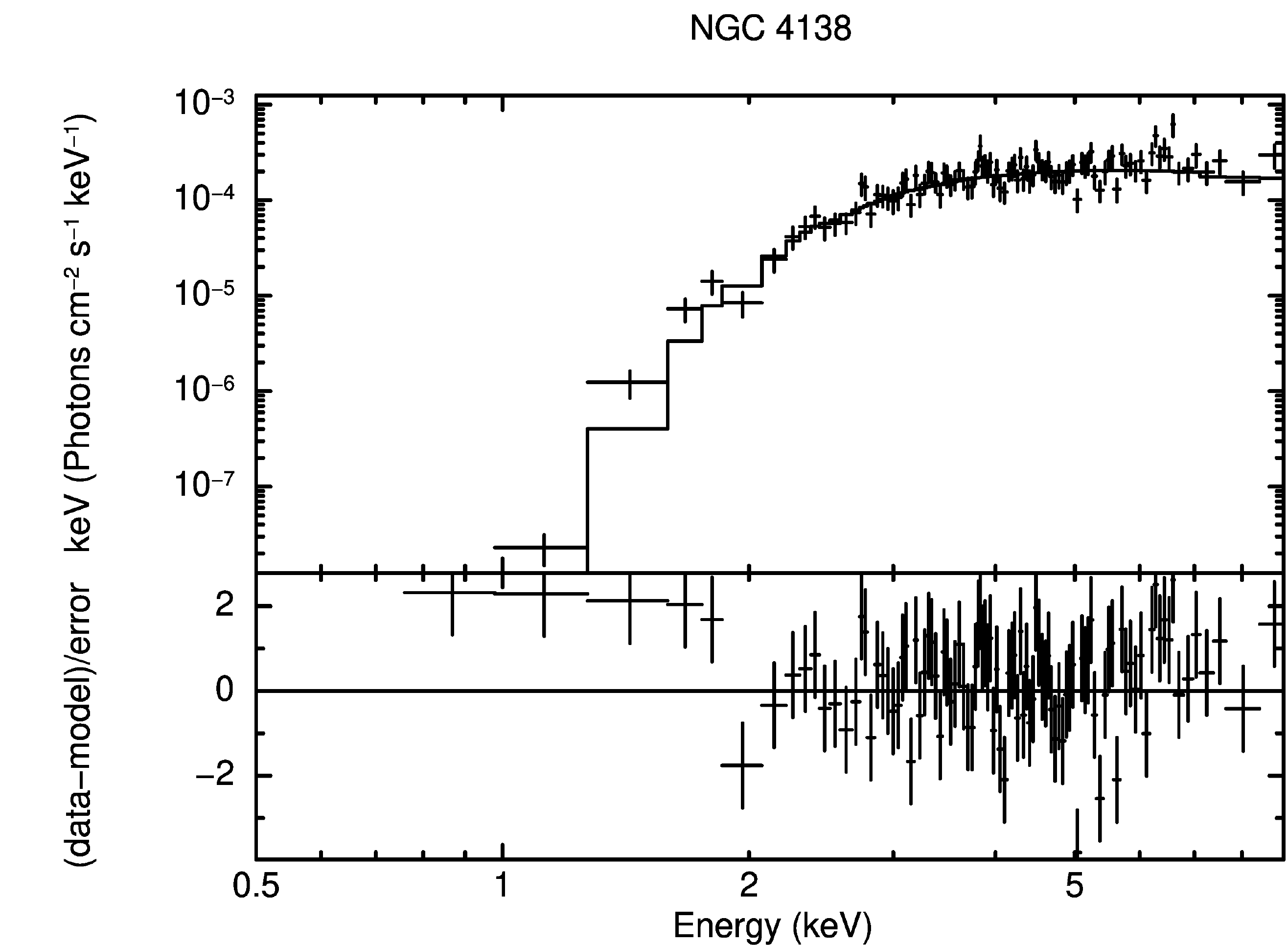}

\end{figure}
\end{center}

\begin{center}
 \begin{figure}
	\includegraphics[width=0.89\columnwidth]{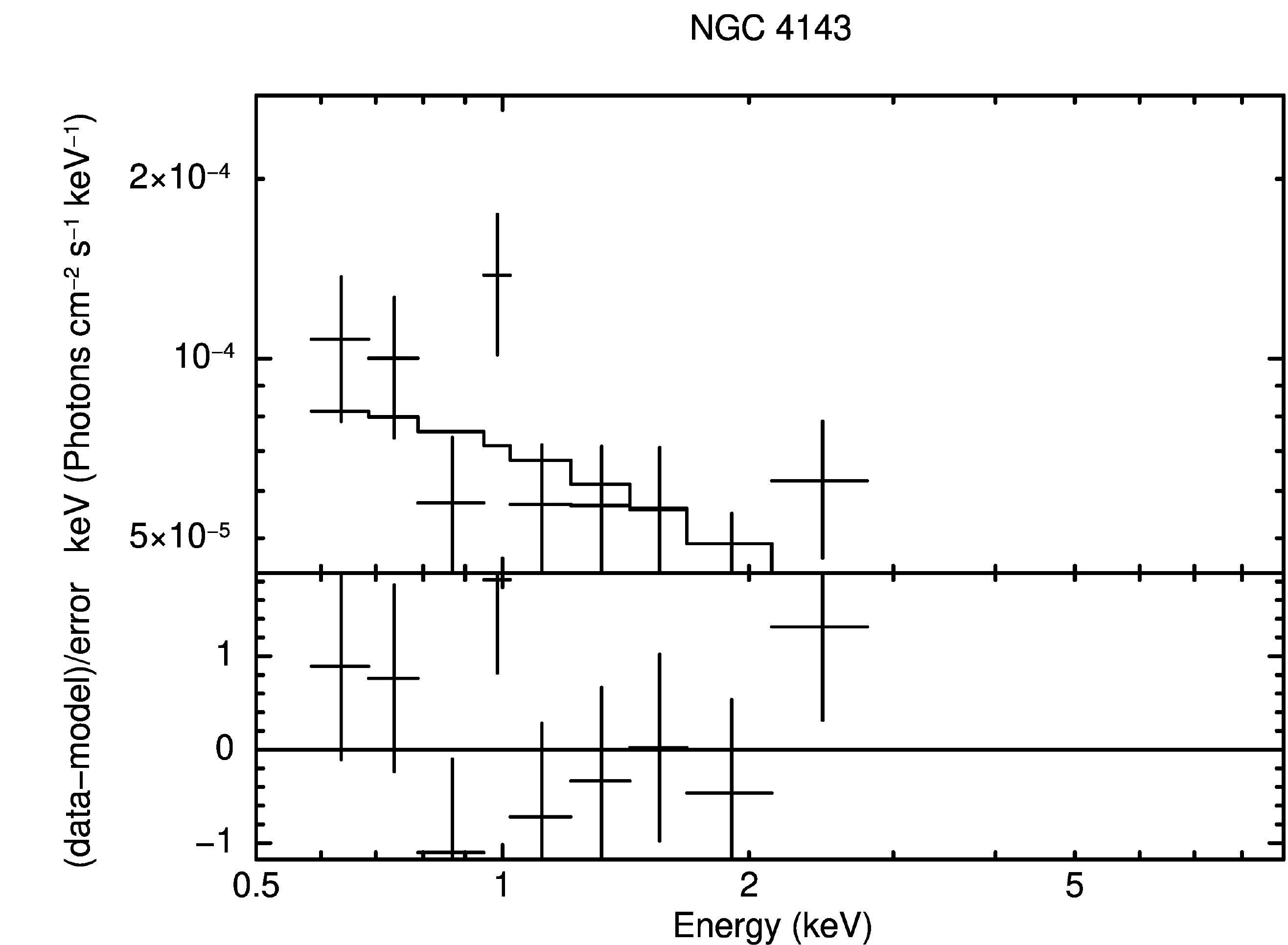}

\end{figure}
\end{center}

\begin{center}
 \begin{figure}
	\includegraphics[width=0.89\columnwidth]{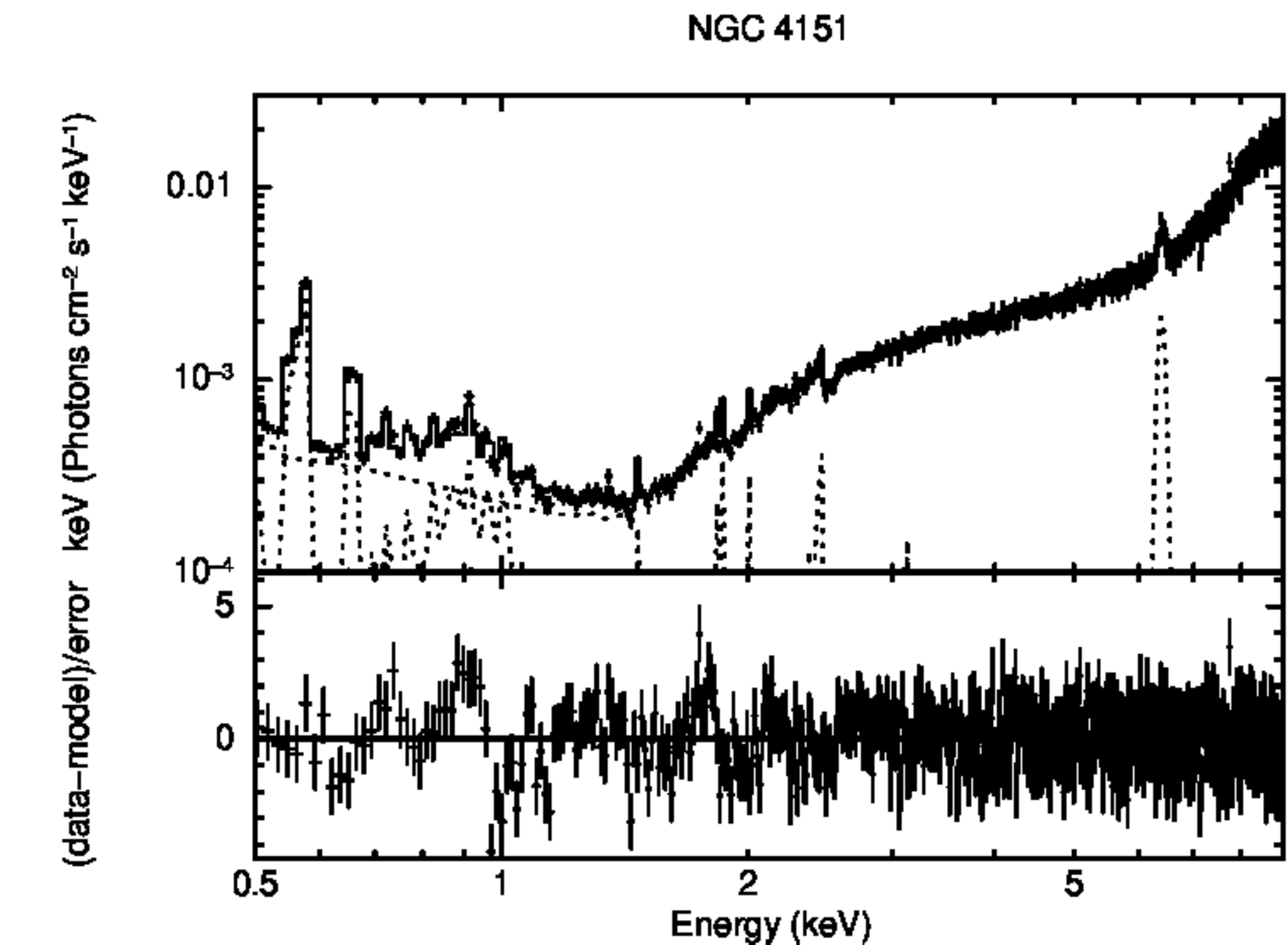}

\end{figure}
\end{center}

%
	 

\begin{center}
 \begin{figure}
	\includegraphics[width=0.89\columnwidth]{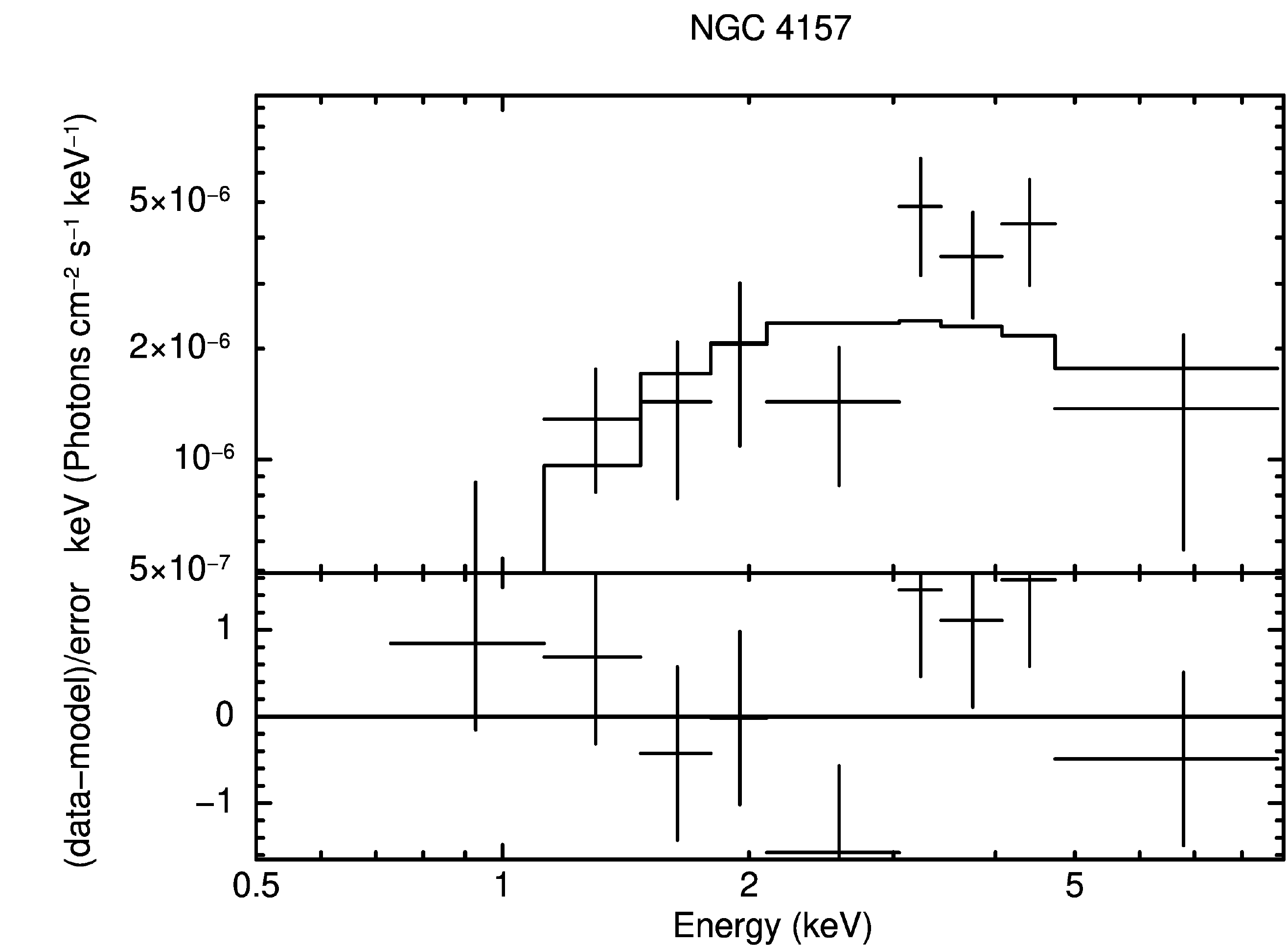}

\end{figure}
\end{center}\begin{center}
 \begin{figure}
	\includegraphics[width=0.89\columnwidth]{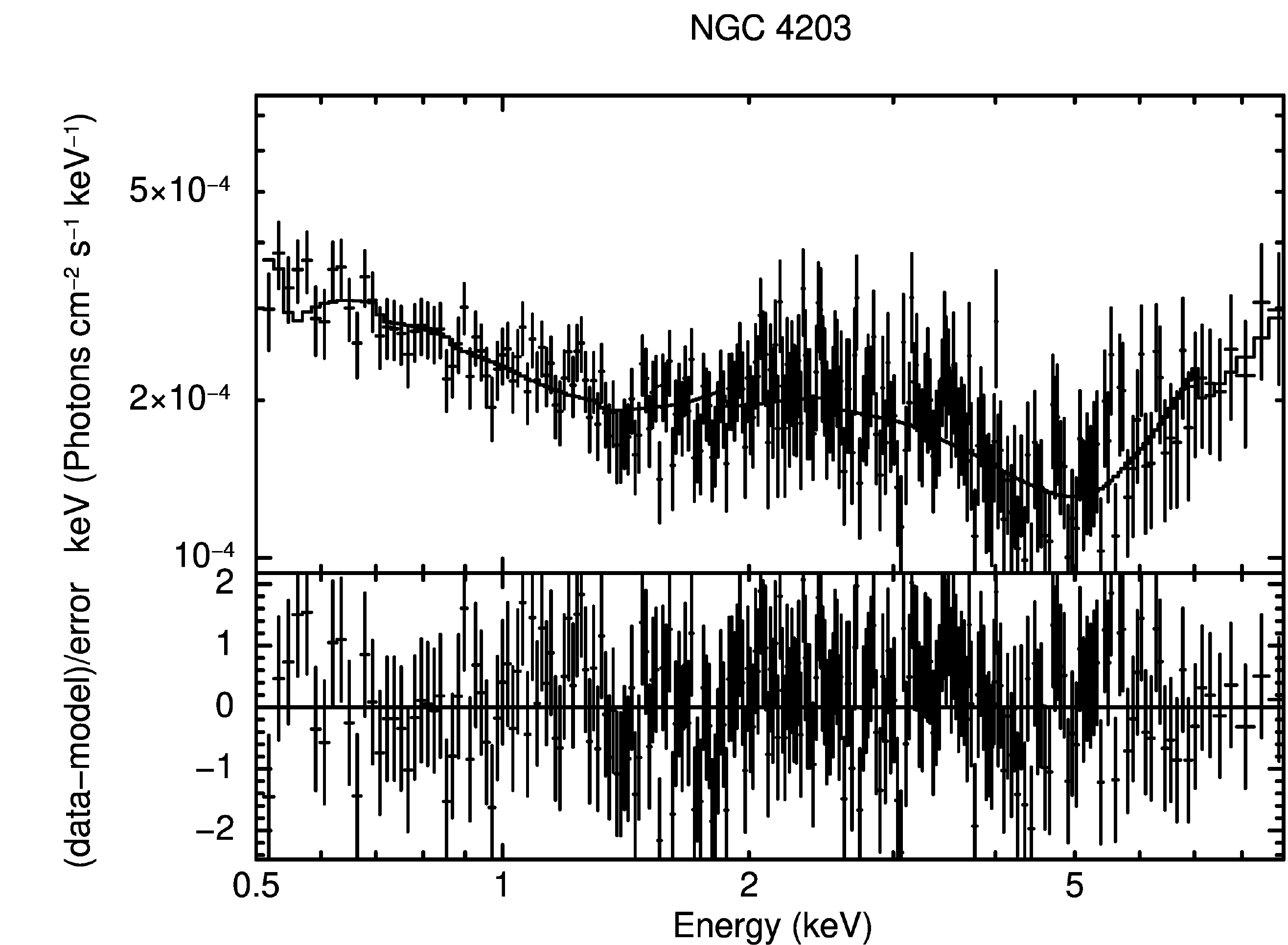}

\end{figure}
\end{center}

\begin{center}
 \begin{figure}
	\includegraphics[width=0.89\columnwidth]{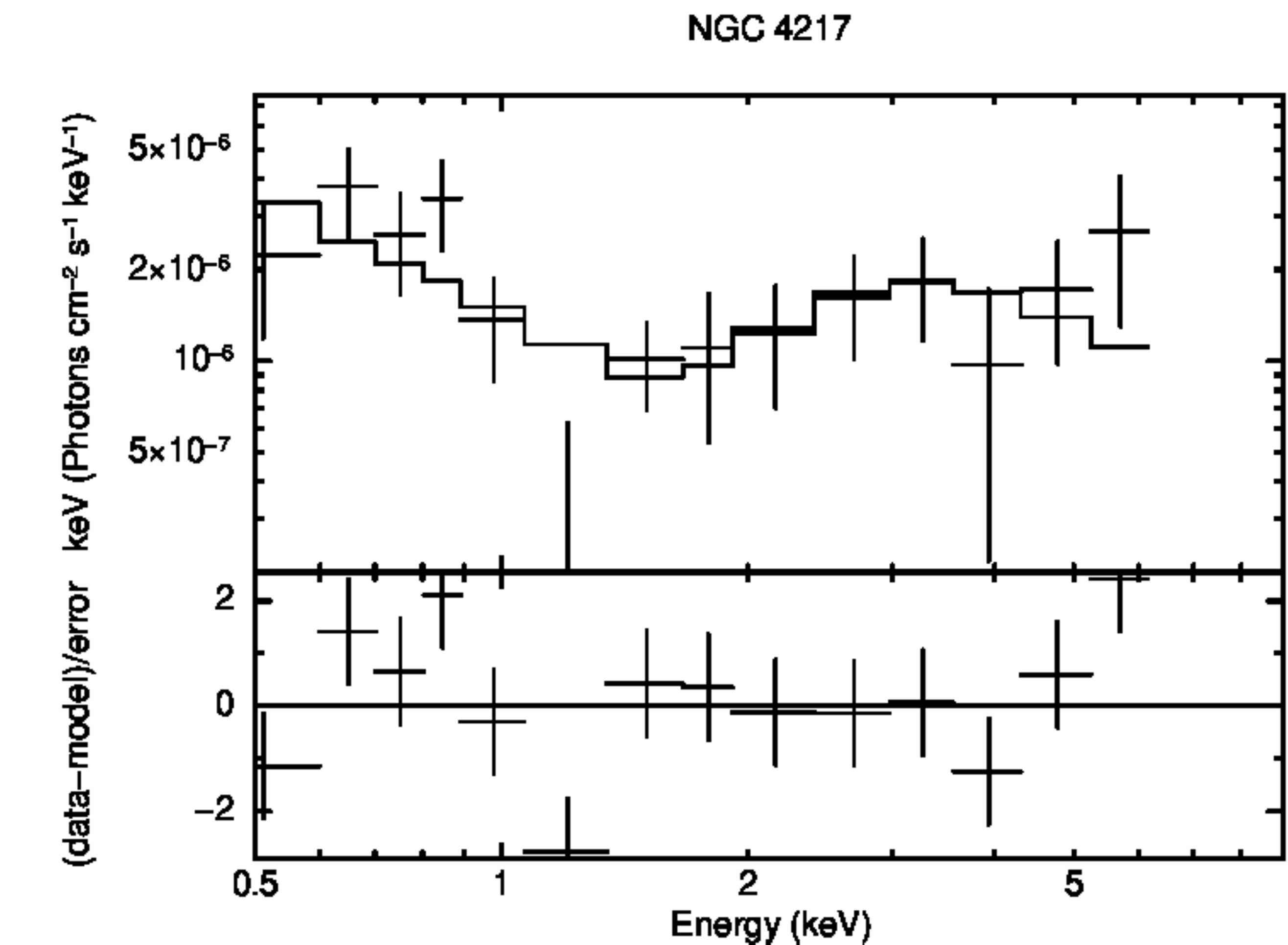}

\end{figure}
\end{center}

%
	 

\begin{center}
 \begin{figure}
	\includegraphics[width=0.89\columnwidth]{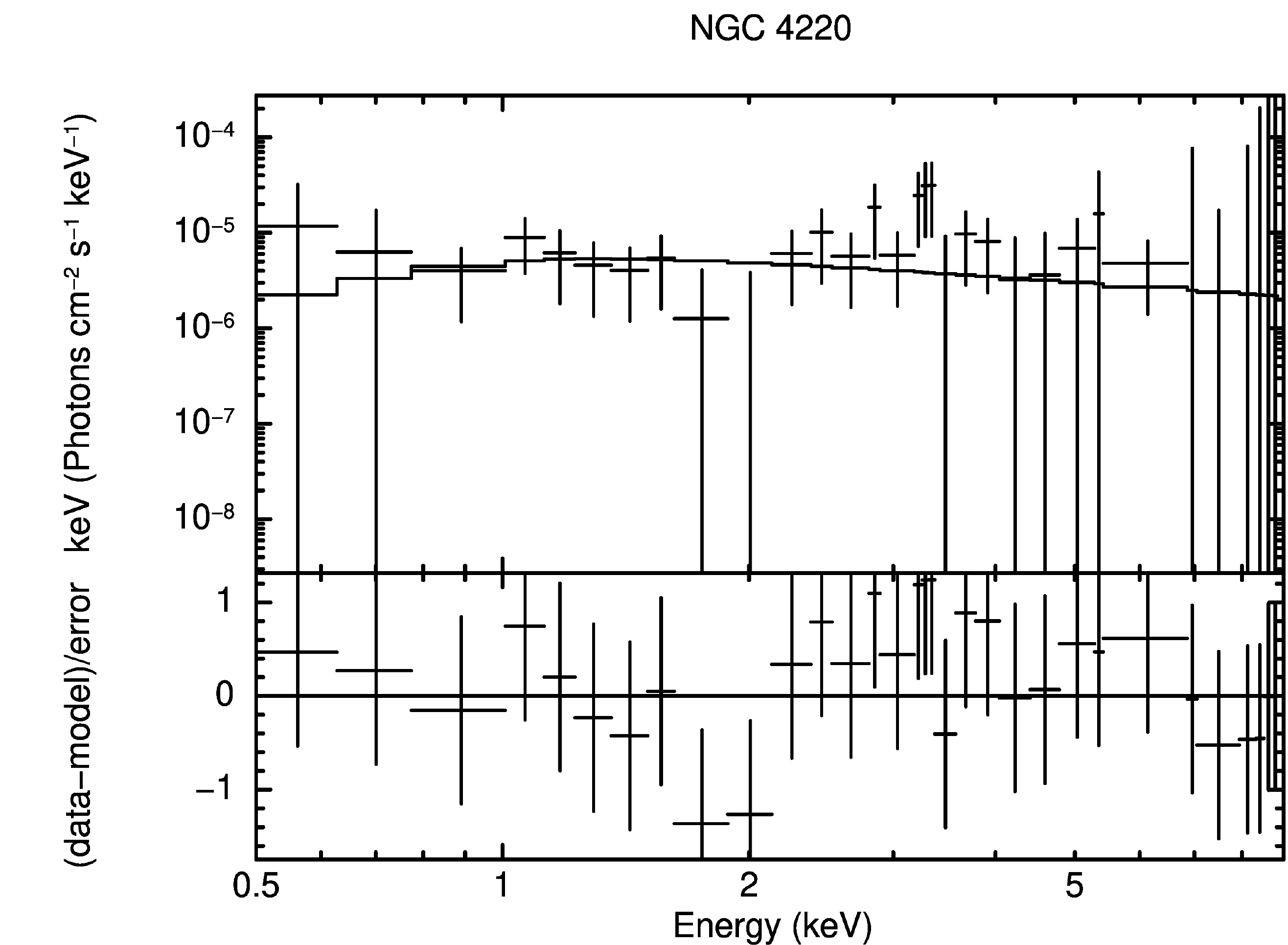}

\end{figure}
\end{center}

\begin{center}
 \begin{figure}
	\includegraphics[width=0.89\columnwidth]{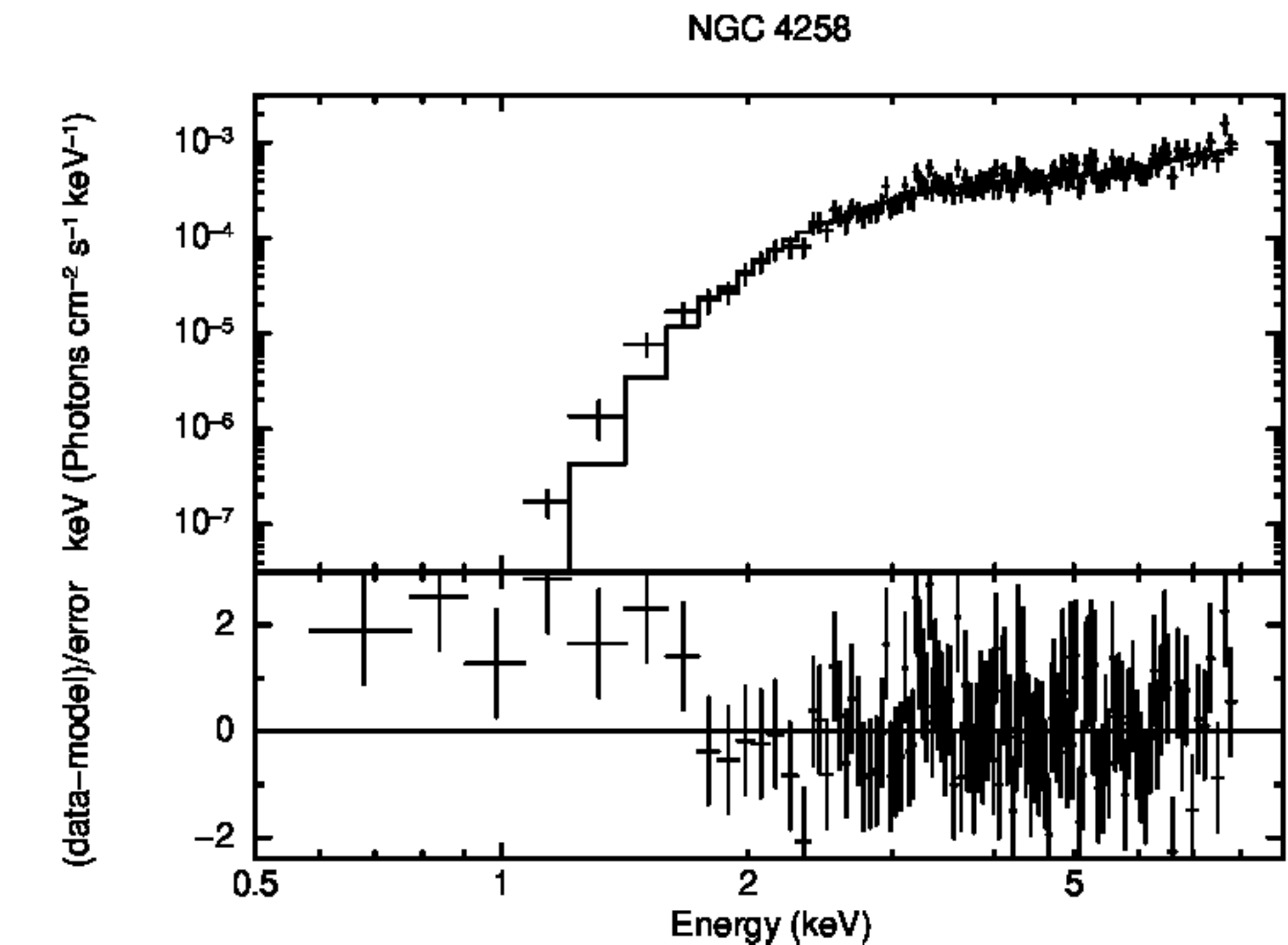}

\end{figure}
\end{center}

%
	 

\begin{center}
 \begin{figure}
	\includegraphics[width=0.89\columnwidth]{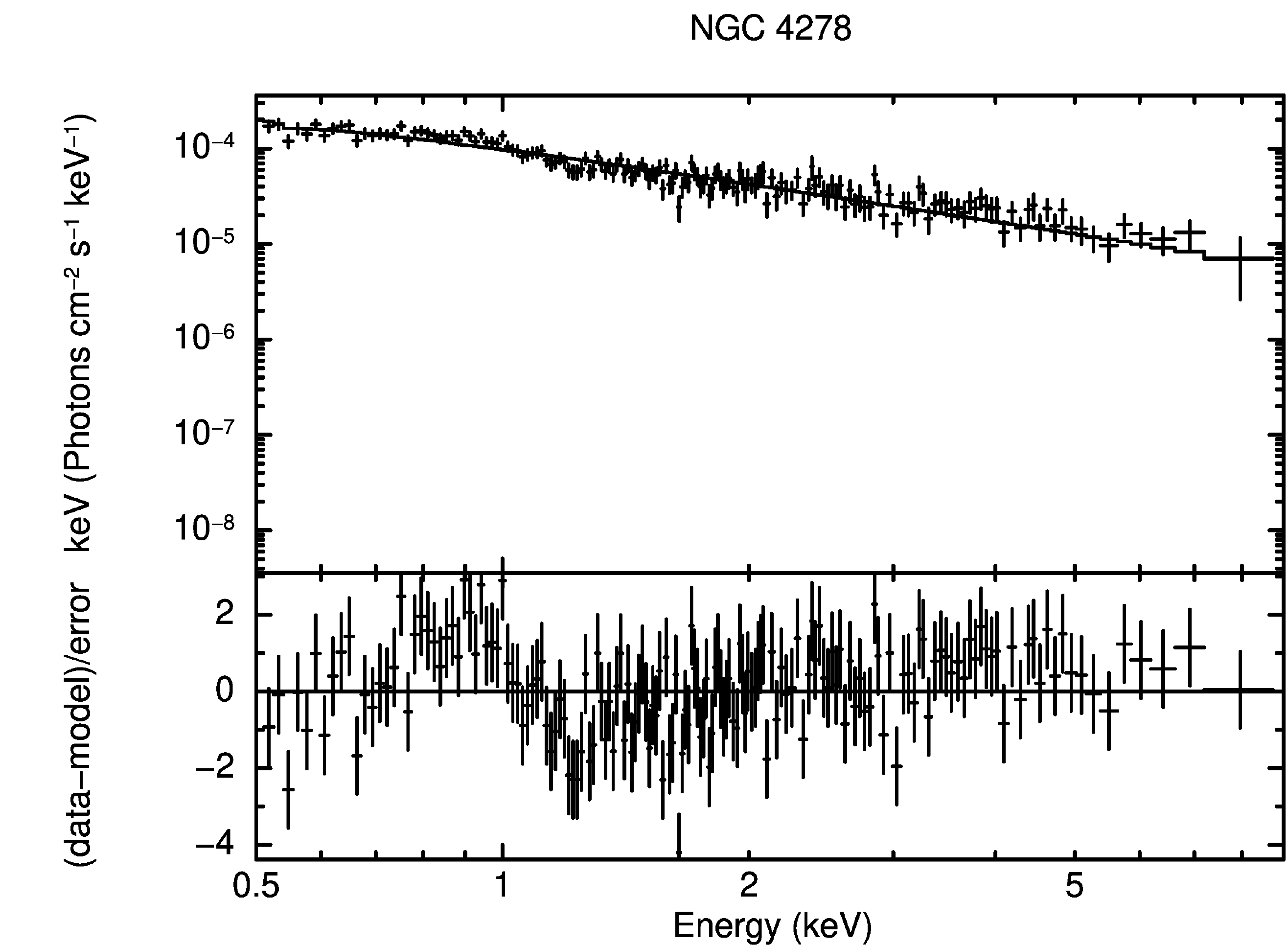}

\end{figure}
\end{center}

\begin{center}
 \begin{figure}
	\includegraphics[width=0.89\columnwidth]{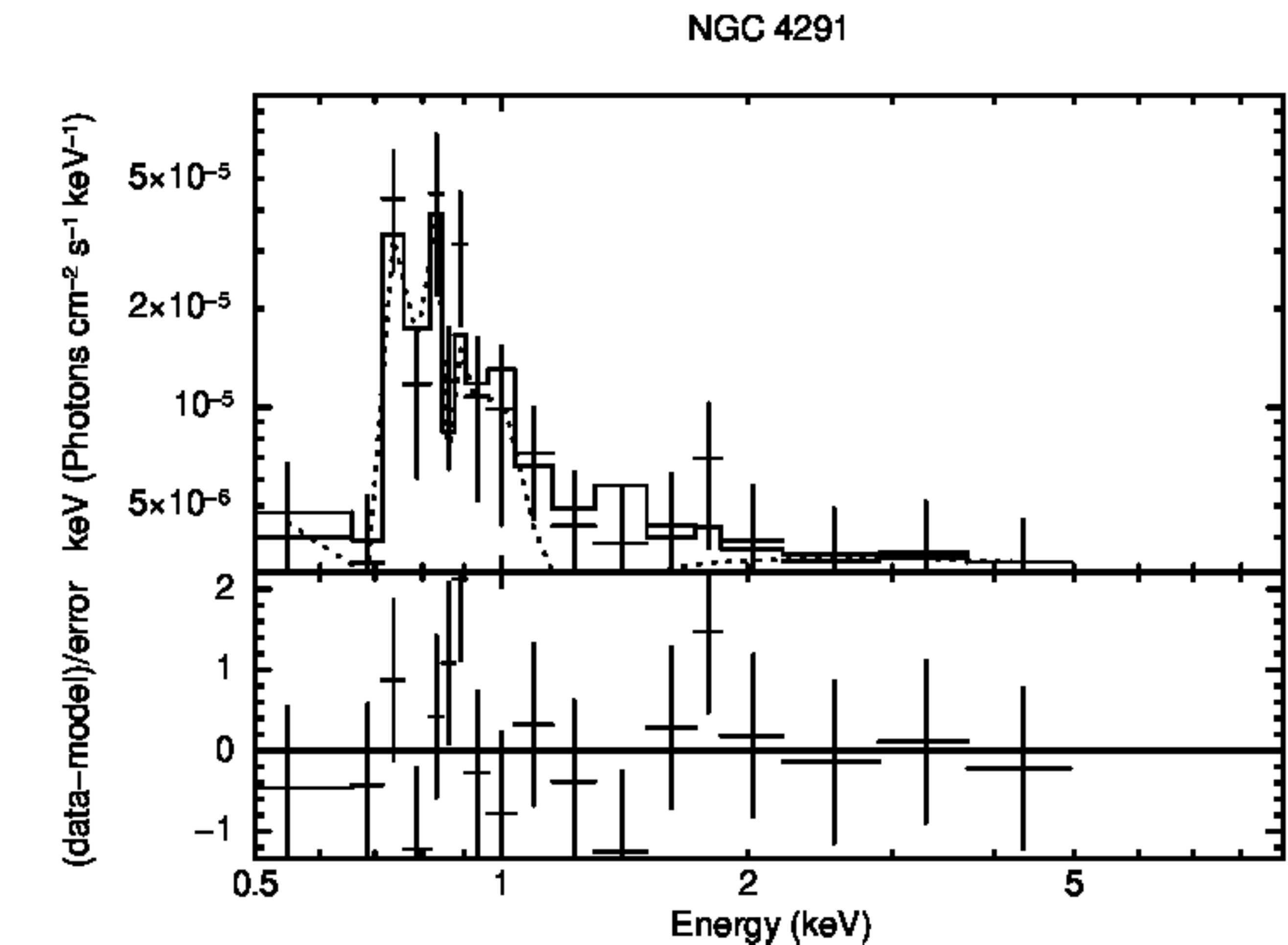}

\end{figure}
\end{center}

%
	 

\begin{center}
 \begin{figure}
	\includegraphics[width=0.89\columnwidth]{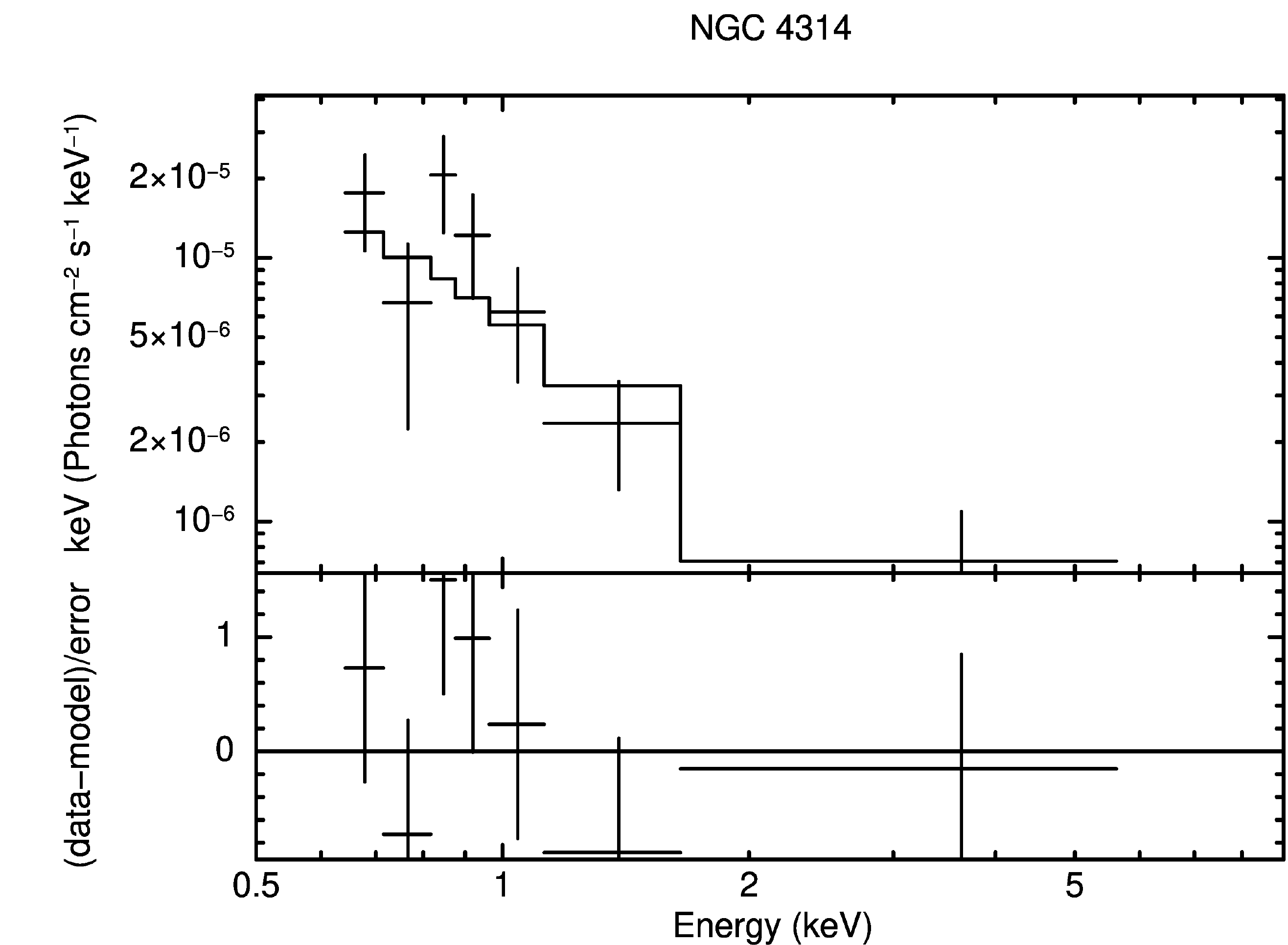}

\end{figure}
\end{center}

\begin{center}
 \begin{figure}
	\includegraphics[width=0.89\columnwidth]{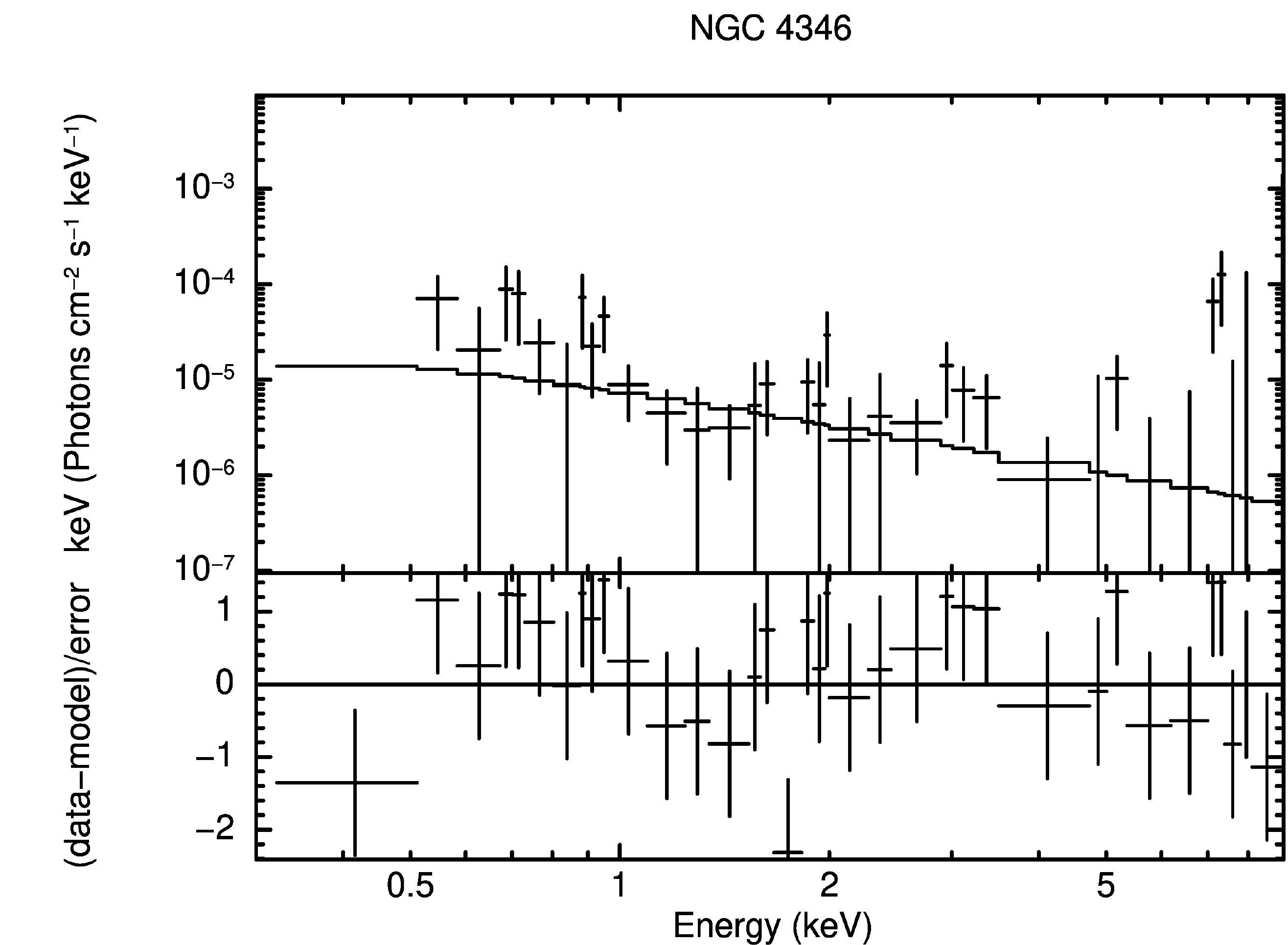}

\end{figure}
\end{center}

\begin{center}
 \begin{figure}
	\includegraphics[width=0.89\columnwidth]{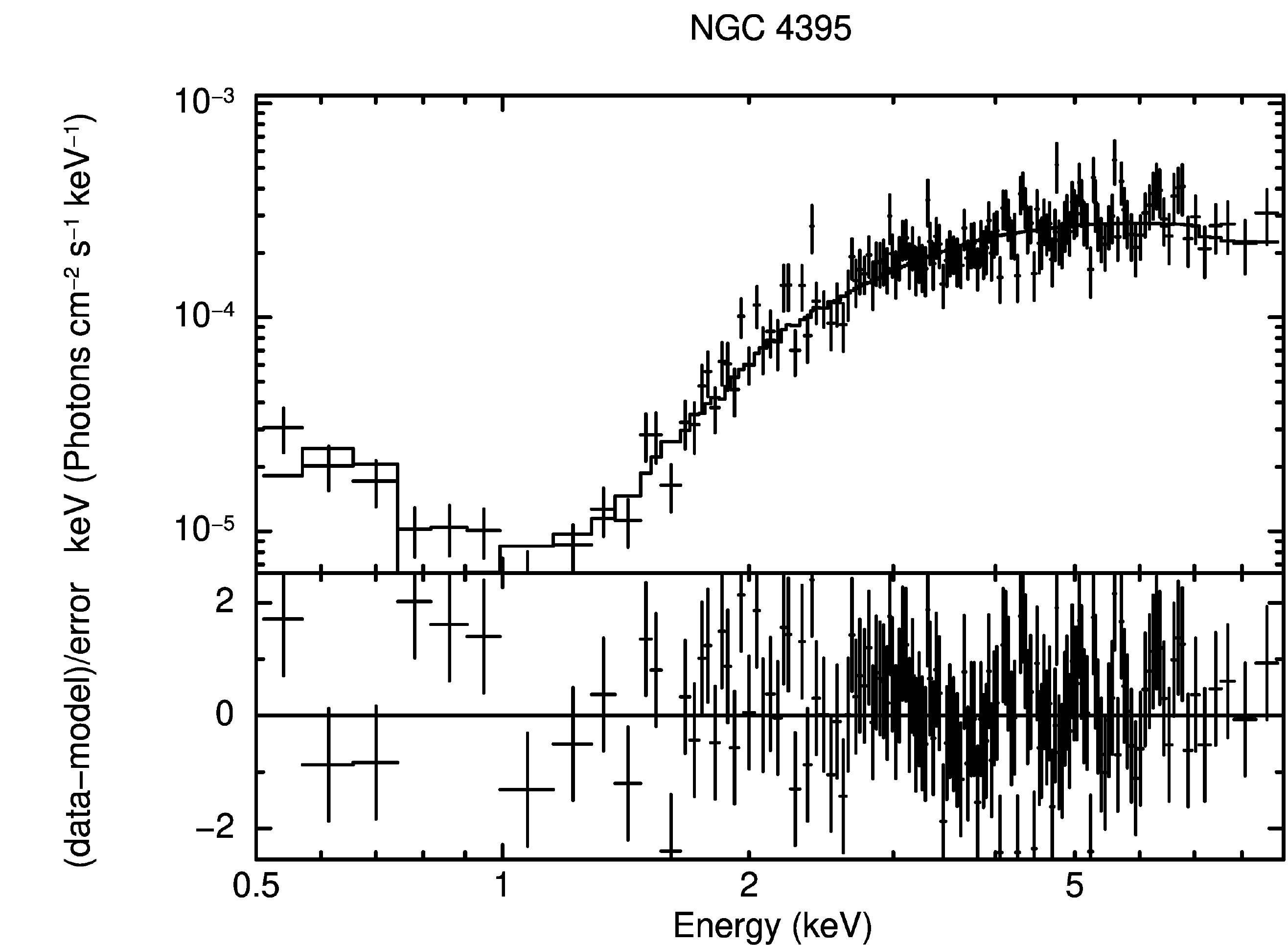}

\end{figure}
\end{center}

\begin{center}
 \begin{figure}
	\includegraphics[width=0.89\columnwidth]{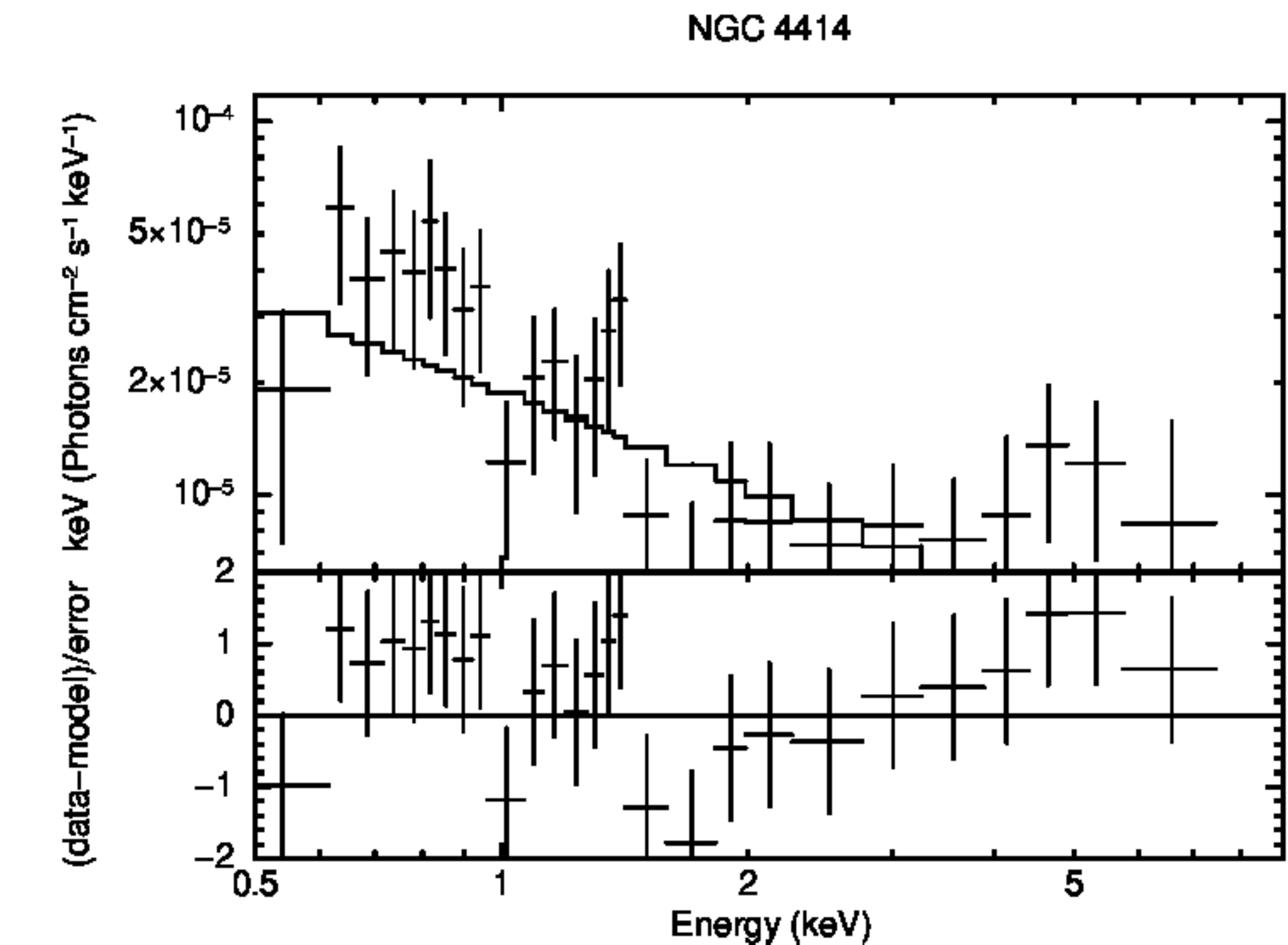}

\end{figure}
\end{center}

%
	 

\begin{center}
 \begin{figure}
	\includegraphics[width=0.89\columnwidth]{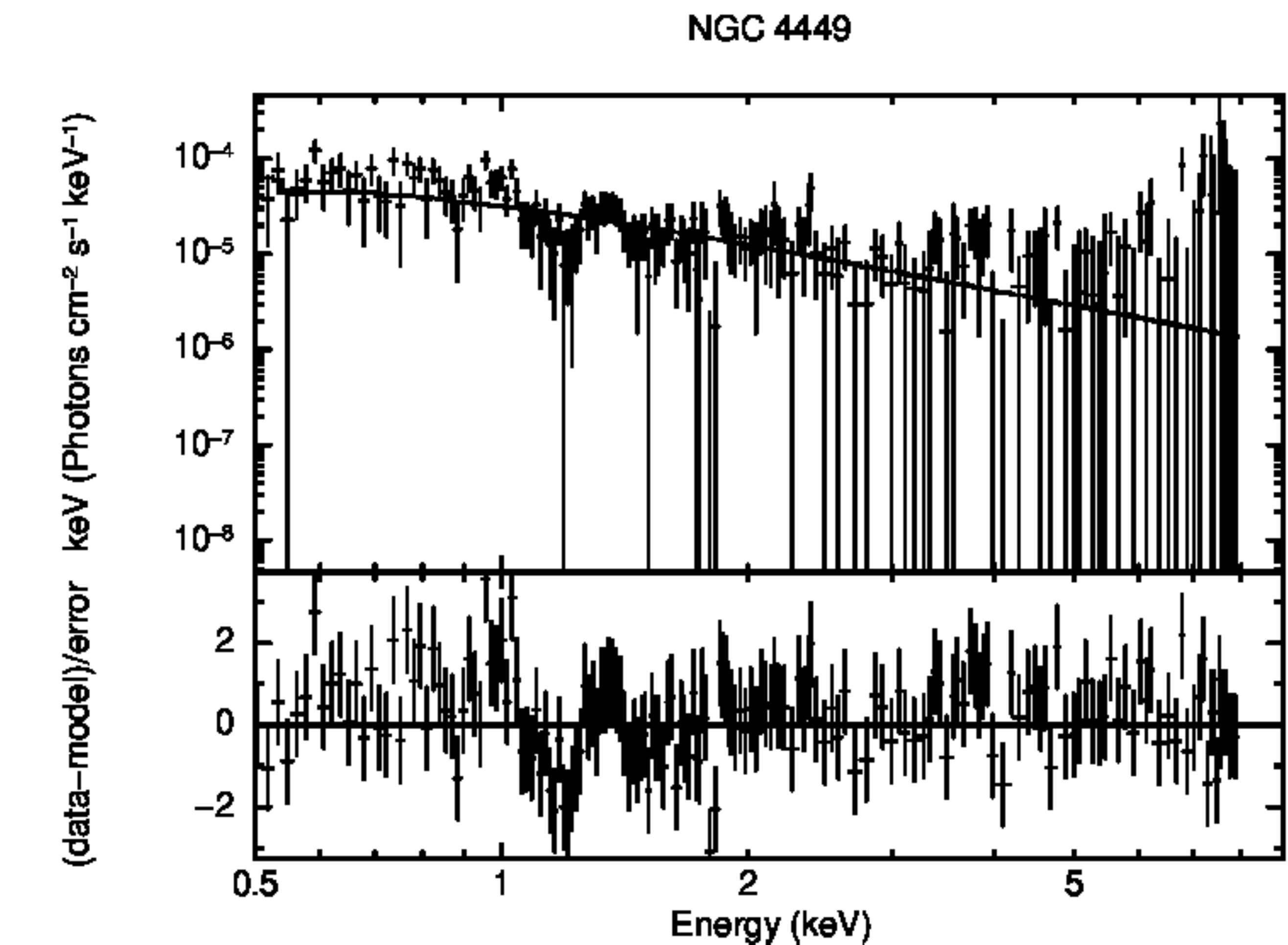}

\end{figure}
\end{center}

%
	 

\begin{center}
 \begin{figure}
	\includegraphics[width=0.89\columnwidth]{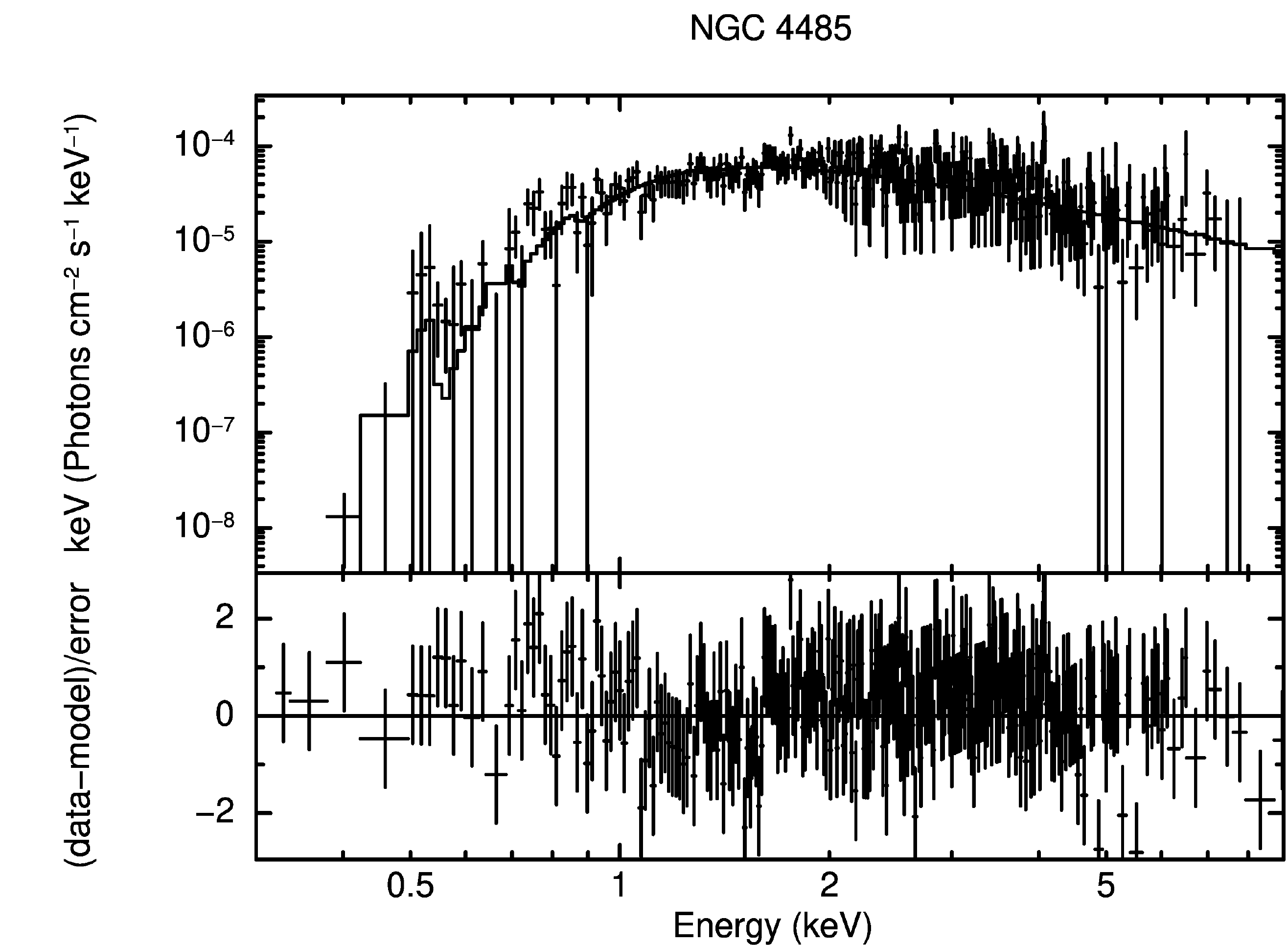}

\end{figure}
\end{center}

\begin{center}
 \begin{figure}
	\includegraphics[width=0.89\columnwidth]{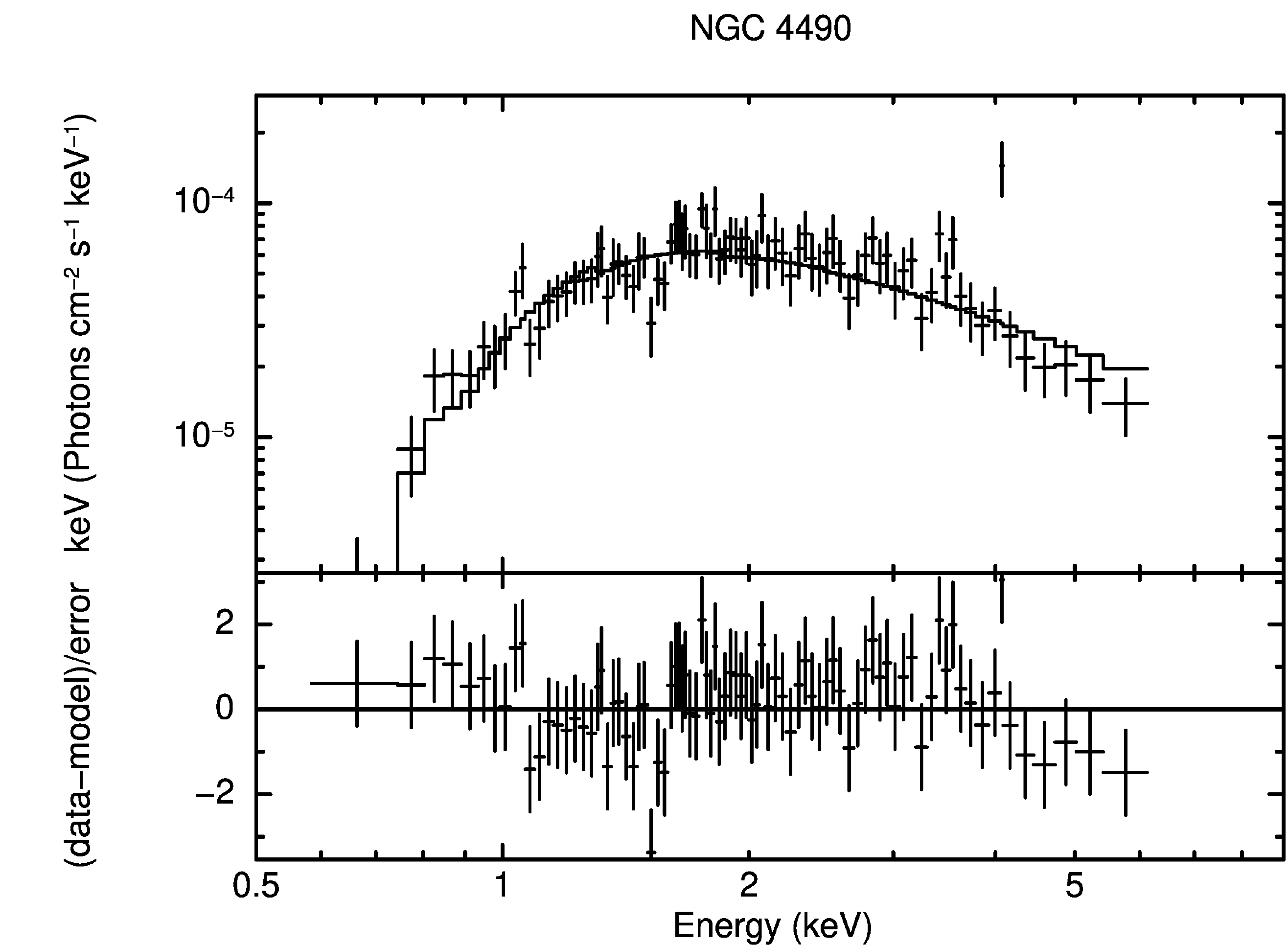}

\end{figure}
\end{center}

\begin{center}
 \begin{figure}
	\includegraphics[width=0.89\columnwidth]{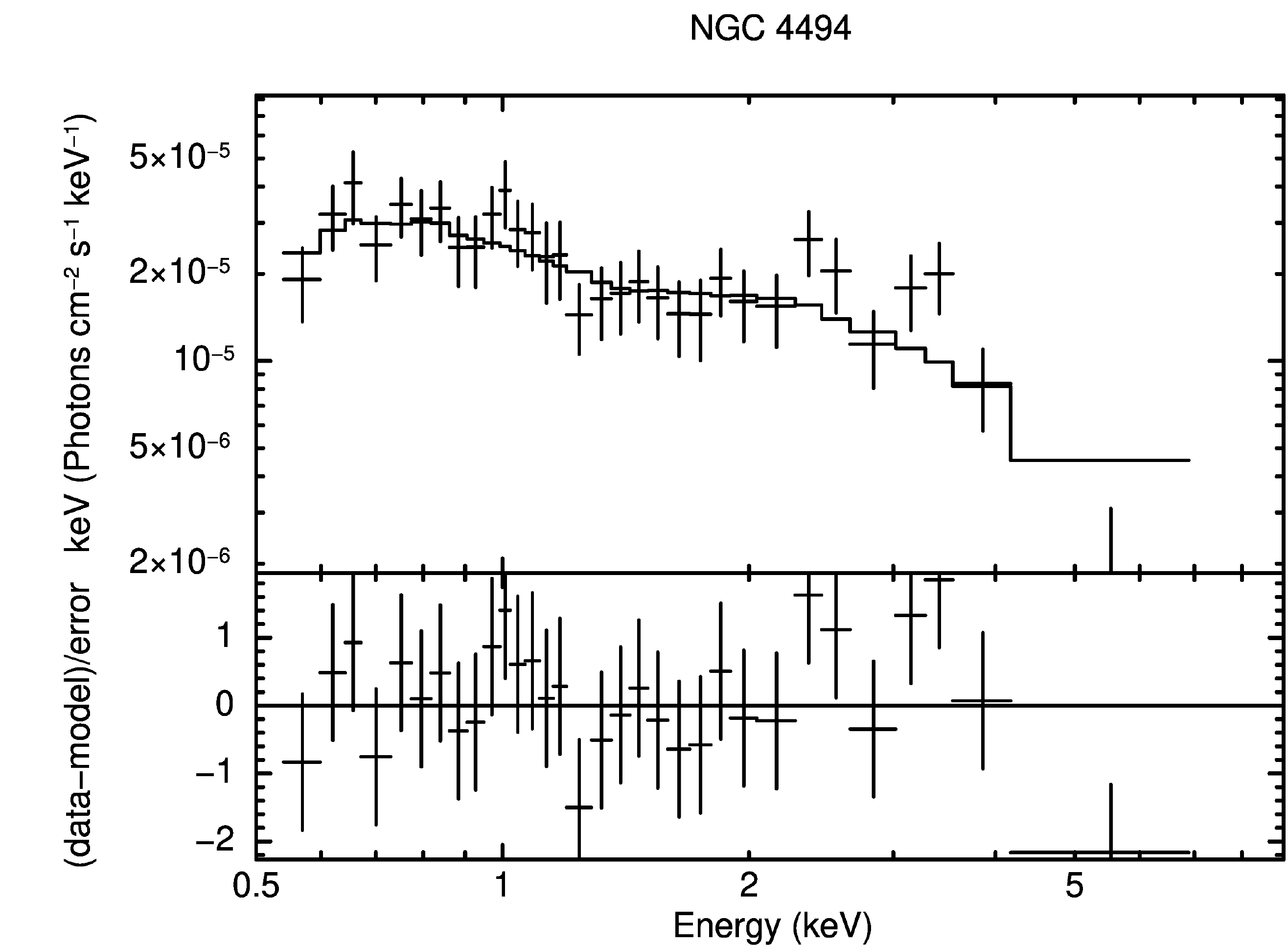}

\end{figure}
\end{center}

\begin{center}
 \begin{figure}
	\includegraphics[width=0.89\columnwidth]{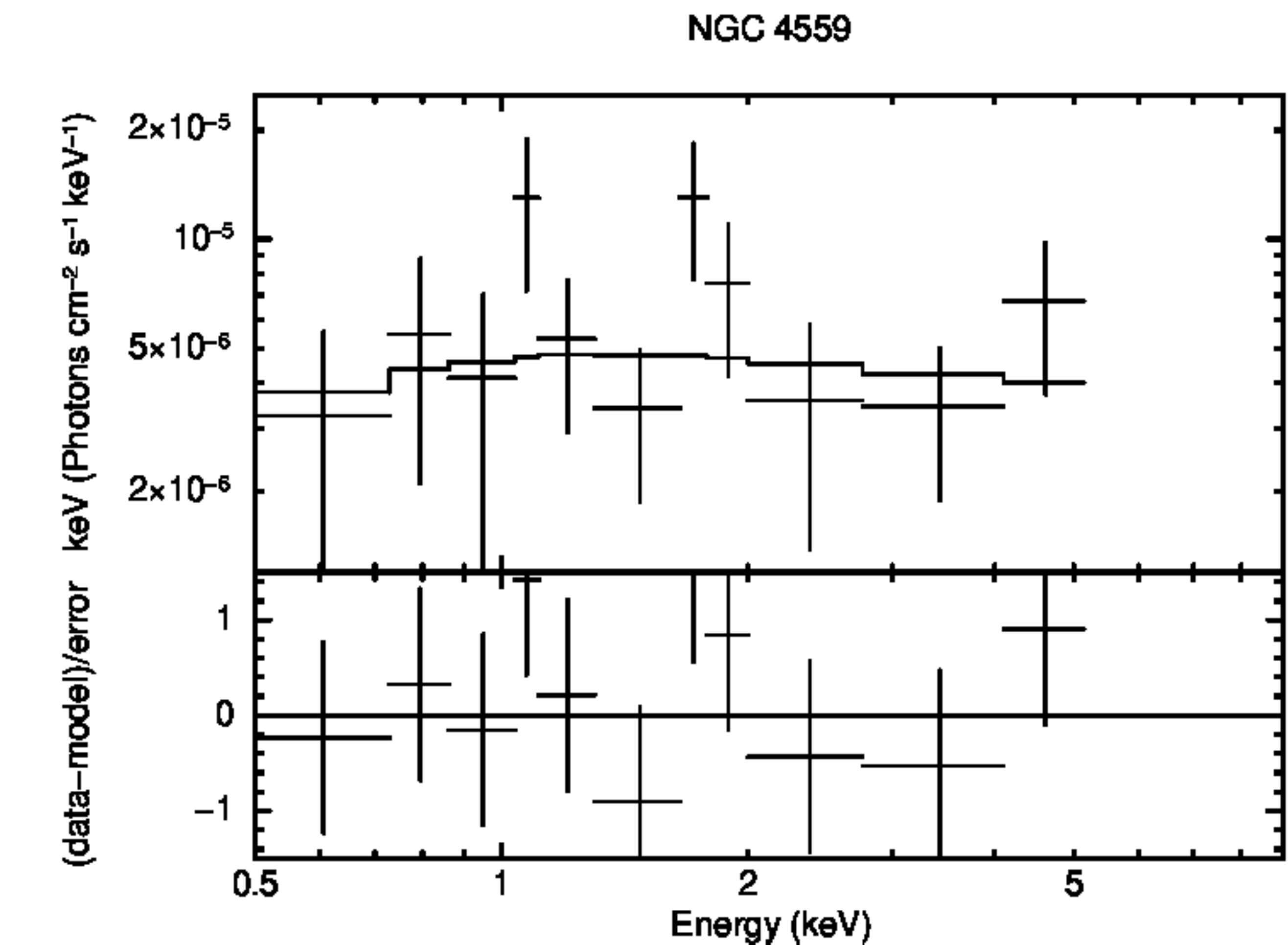}

\end{figure}
\end{center}

%
	 

\begin{center}
 \begin{figure}
	\includegraphics[width=0.89\columnwidth]{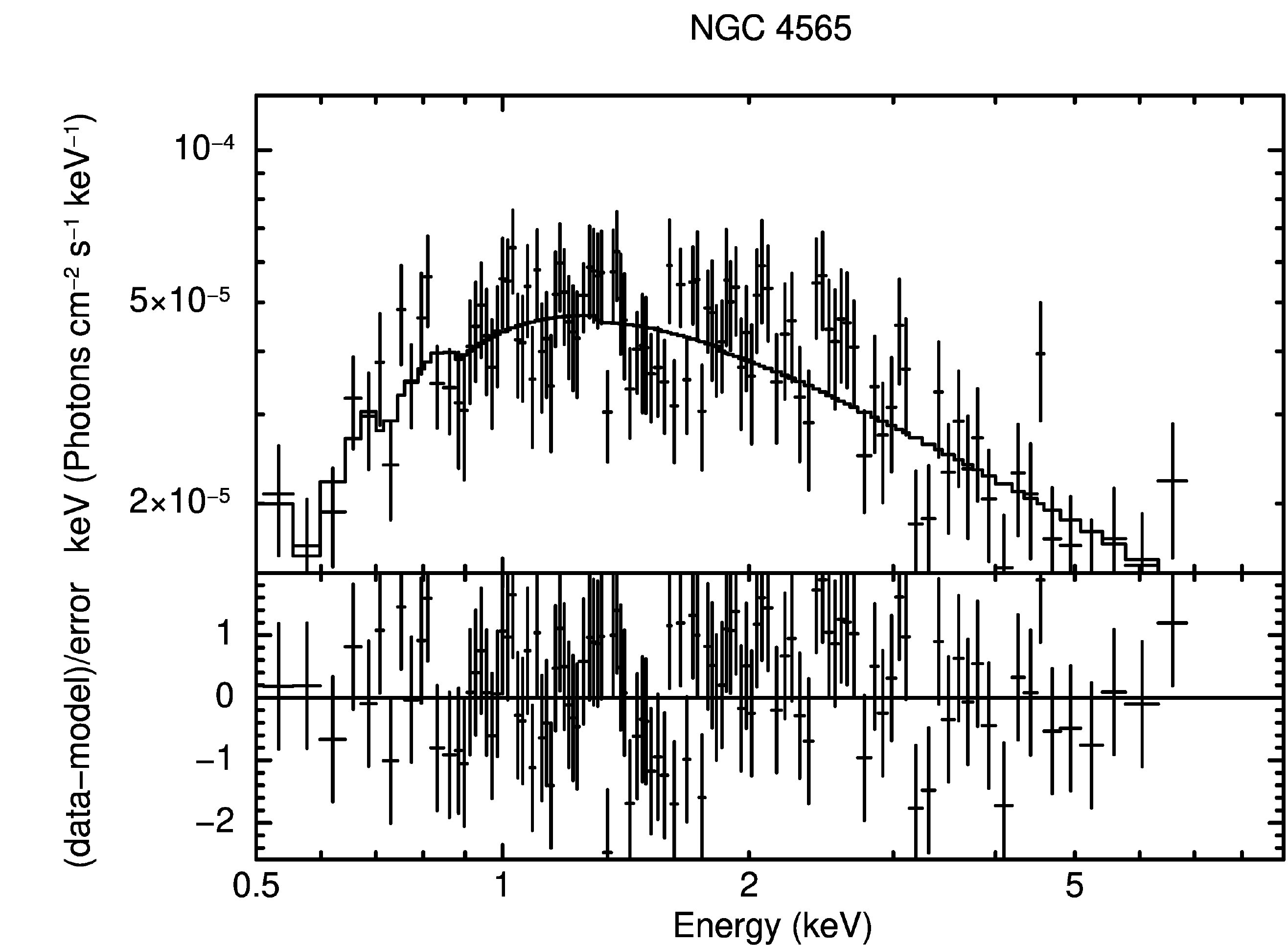}

\end{figure}
\end{center}

\begin{center}
 \begin{figure}
	\includegraphics[width=0.89\columnwidth]{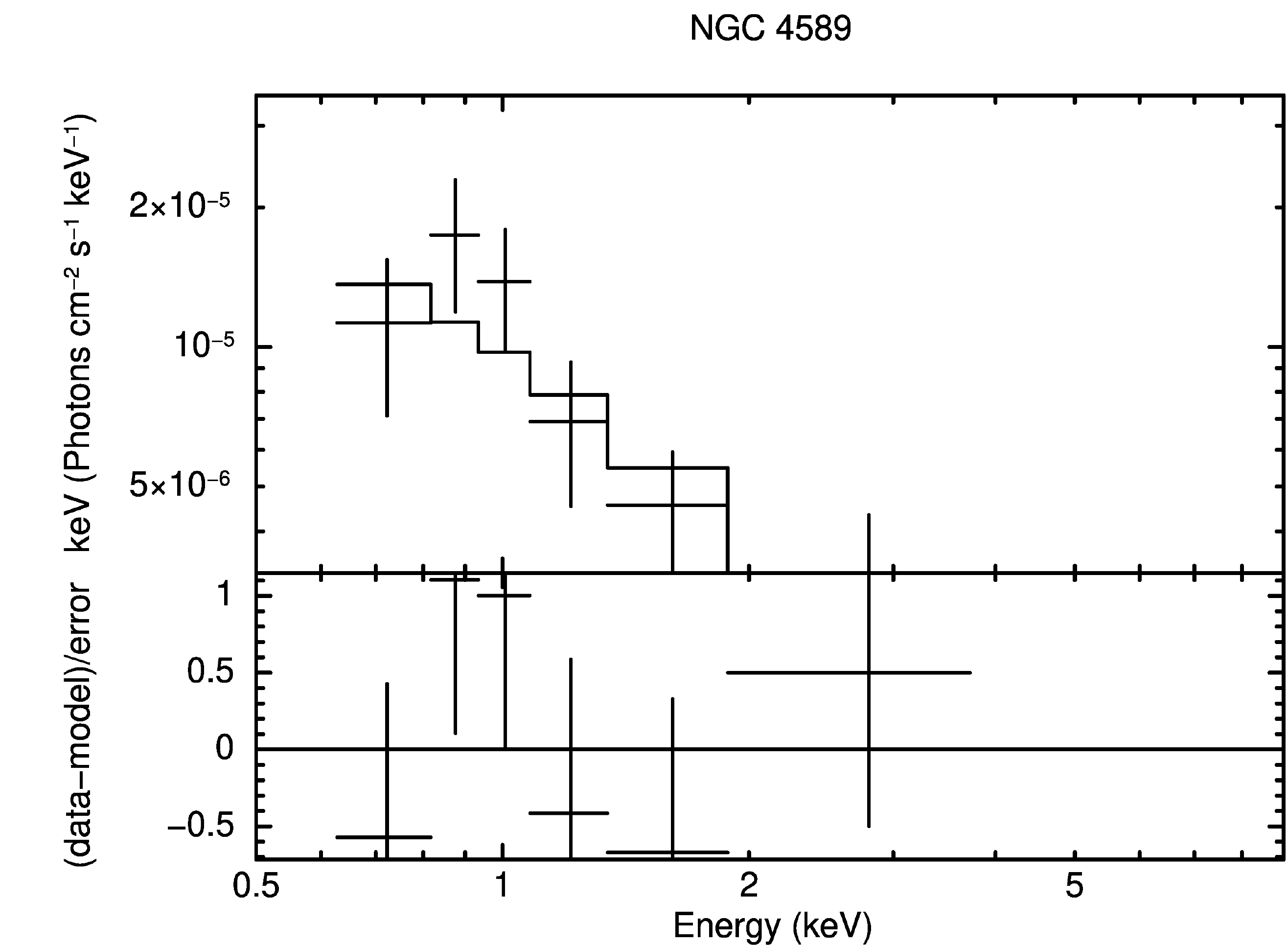}

\end{figure}
\end{center}

\begin{center}
 \begin{figure}
	\includegraphics[width=0.89\columnwidth]{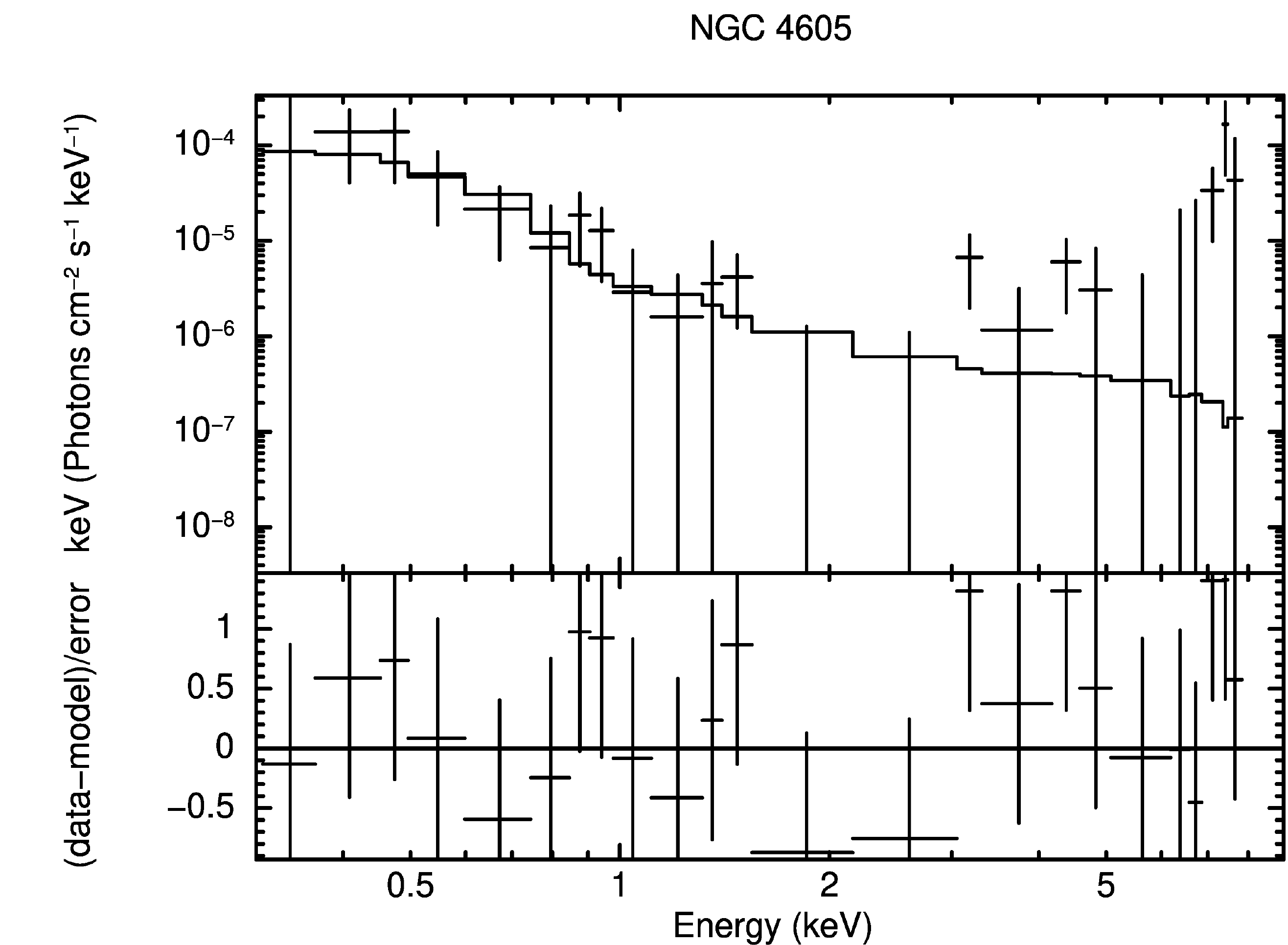}

\end{figure}
\end{center}

\begin{center}
 \begin{figure}
	\includegraphics[width=0.89\columnwidth]{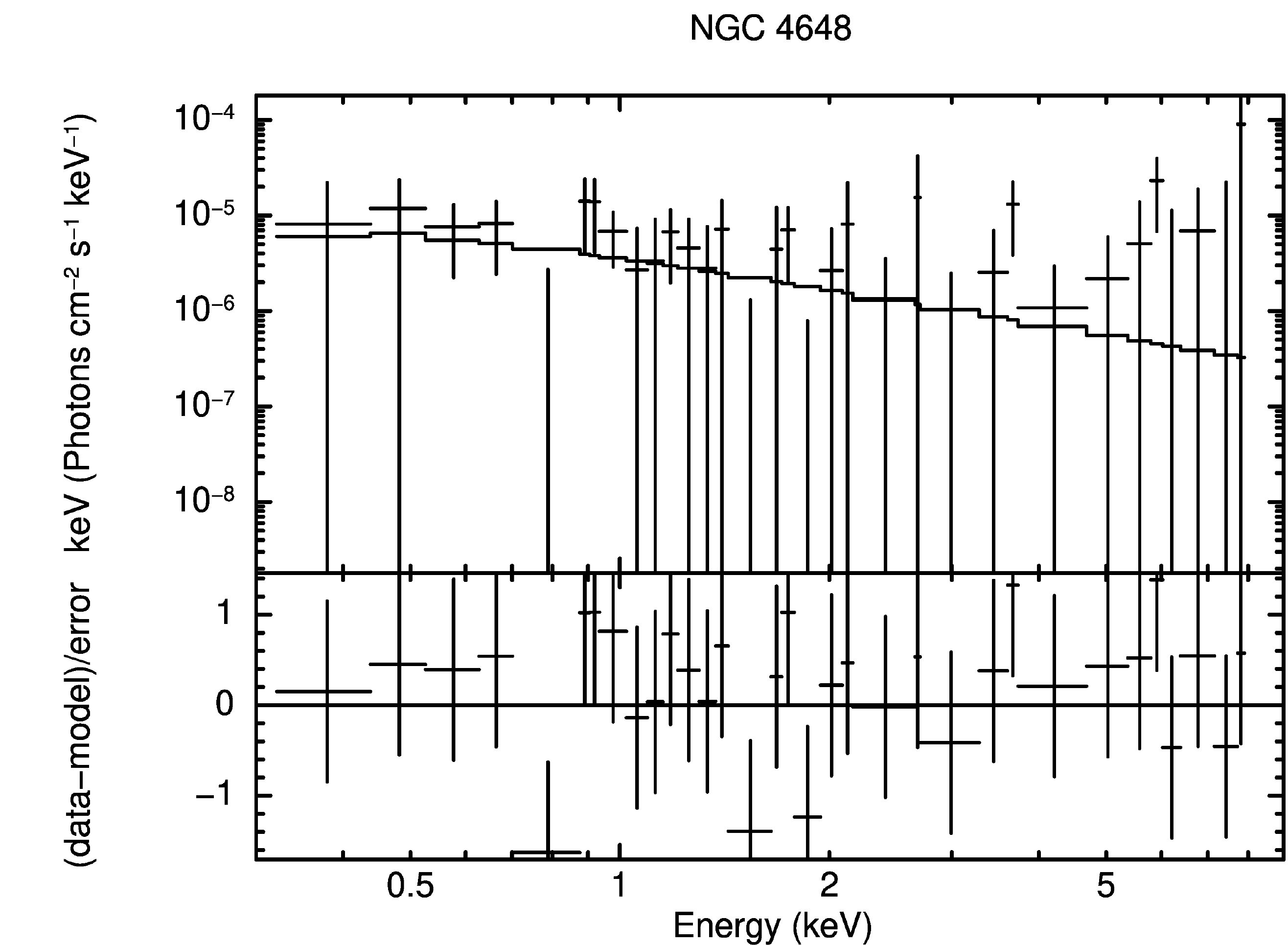}

\end{figure}
\end{center}

\begin{center}
 \begin{figure}
	\includegraphics[width=0.89\columnwidth]{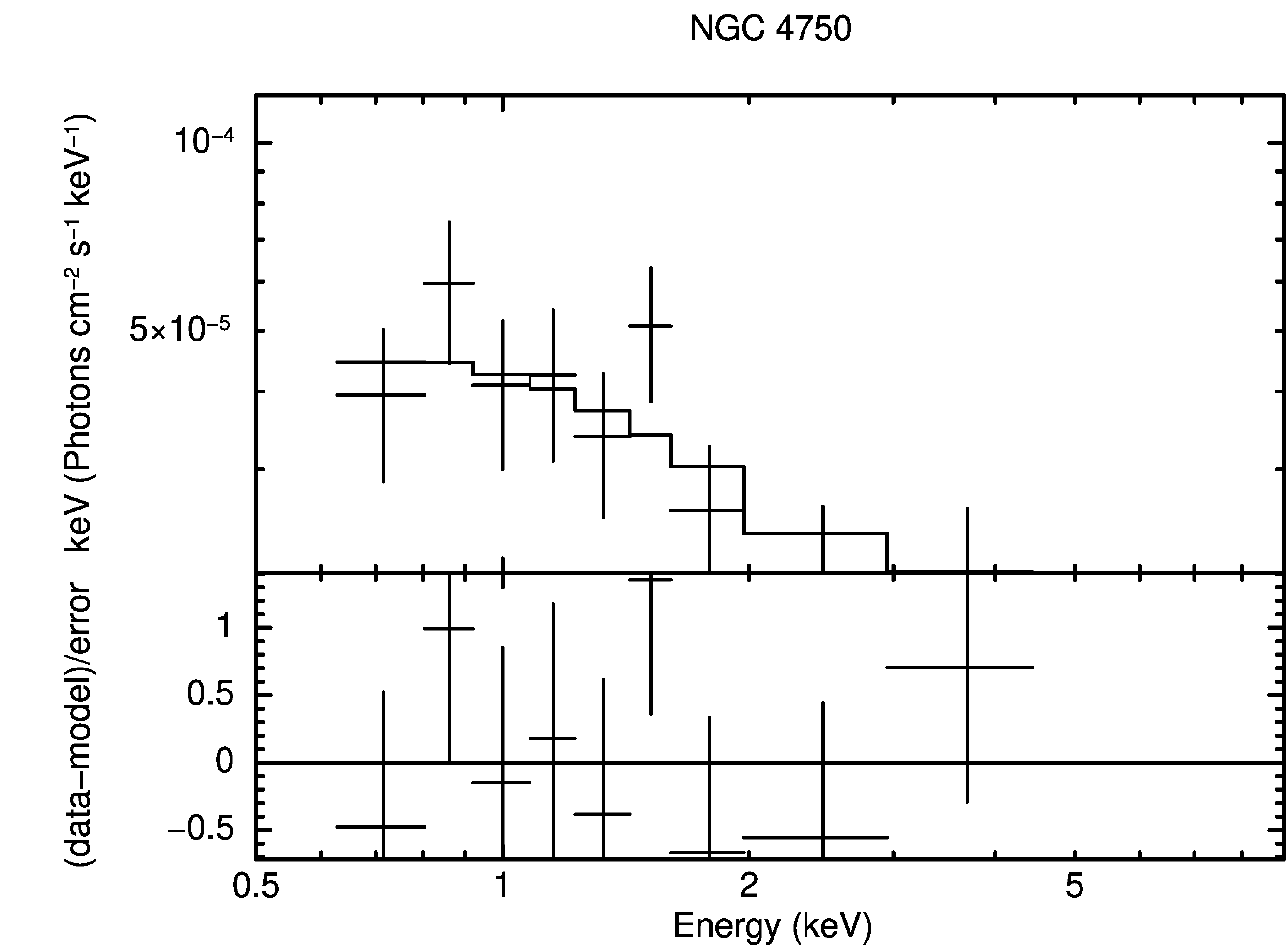}

\end{figure}
\end{center}

\begin{center}
 \begin{figure}
	\includegraphics[width=0.89\columnwidth]{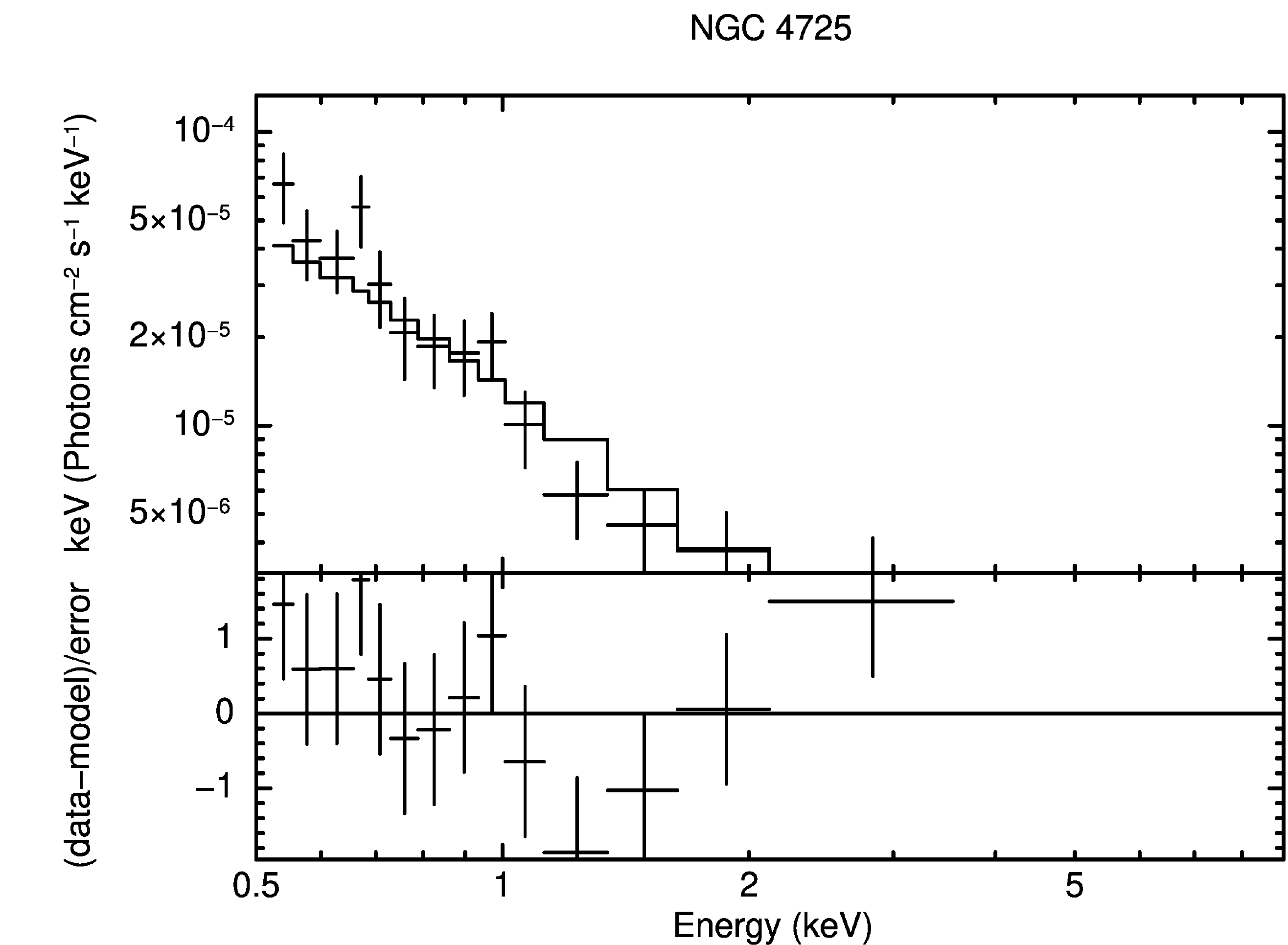}

\end{figure}
\end{center}

\begin{center}
 \begin{figure}
	\includegraphics[width=0.89\columnwidth]{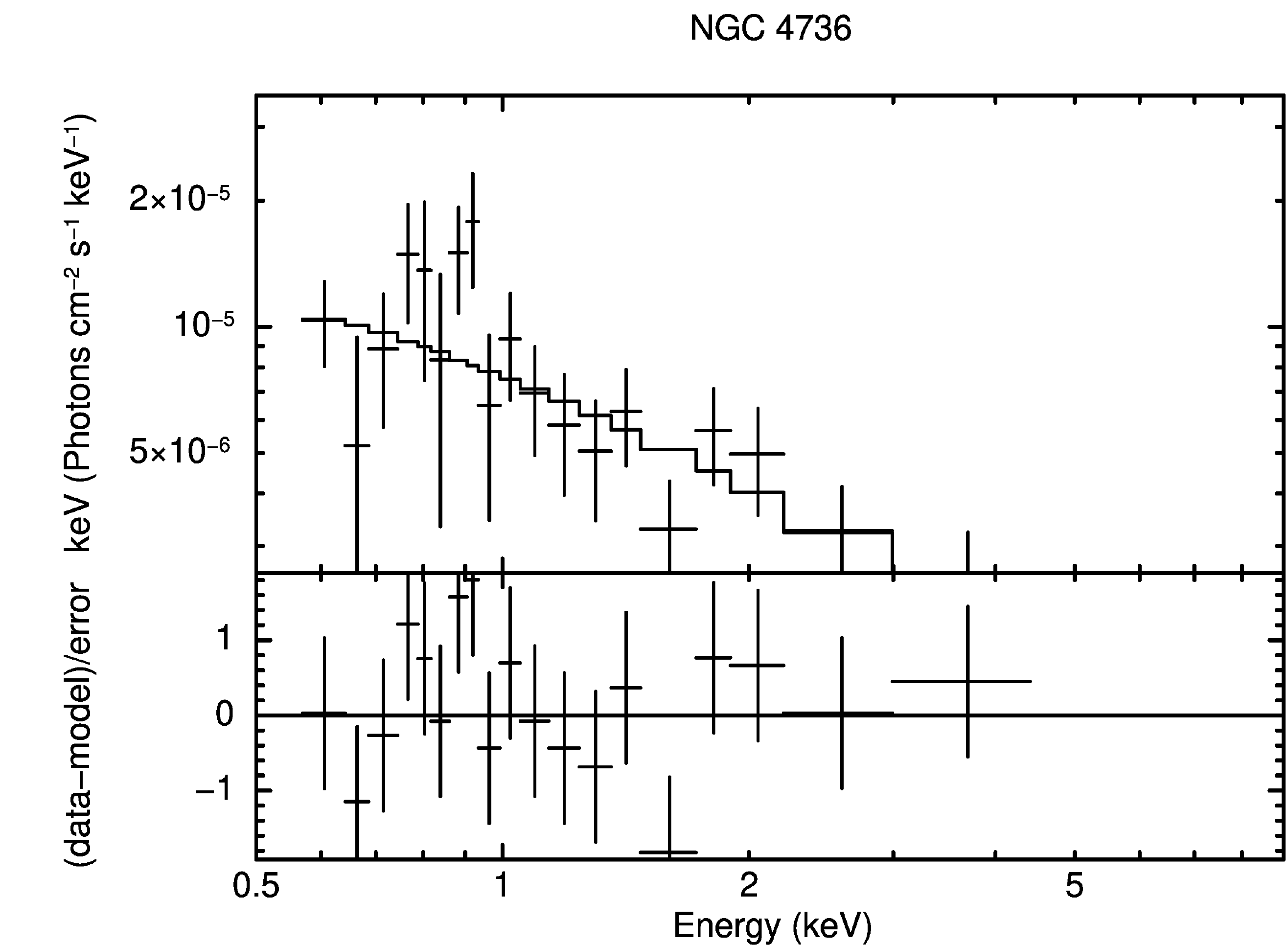}

\end{figure}
\end{center}

\begin{center}
 \begin{figure}
	\includegraphics[width=0.89\columnwidth]{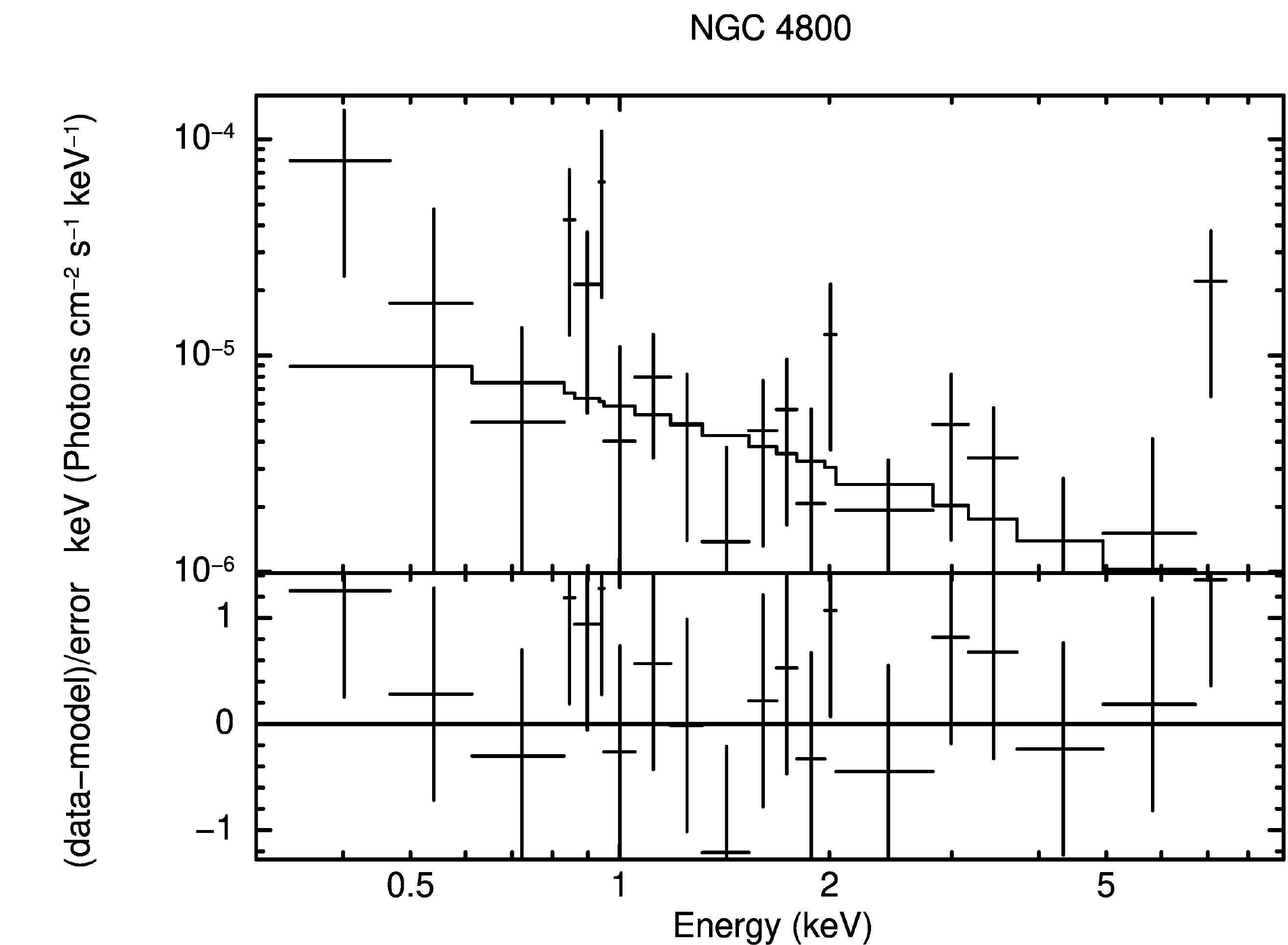}

\end{figure}
\end{center}

\begin{center}
 \begin{figure}
	\includegraphics[width=0.89\columnwidth]{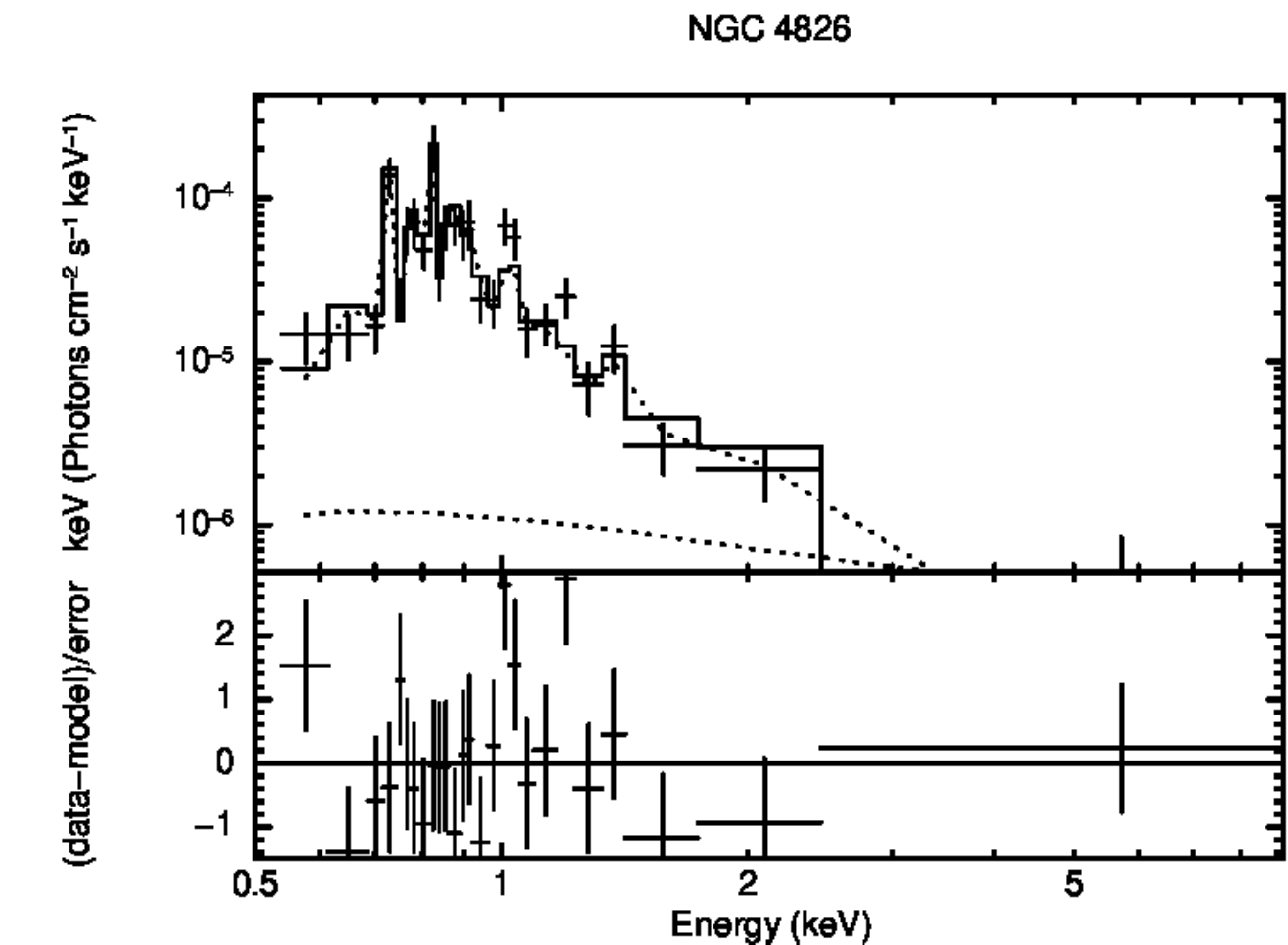}

\end{figure}
\end{center}

%
	 

\begin{center}
 \begin{figure}
	\includegraphics[width=0.89\columnwidth]{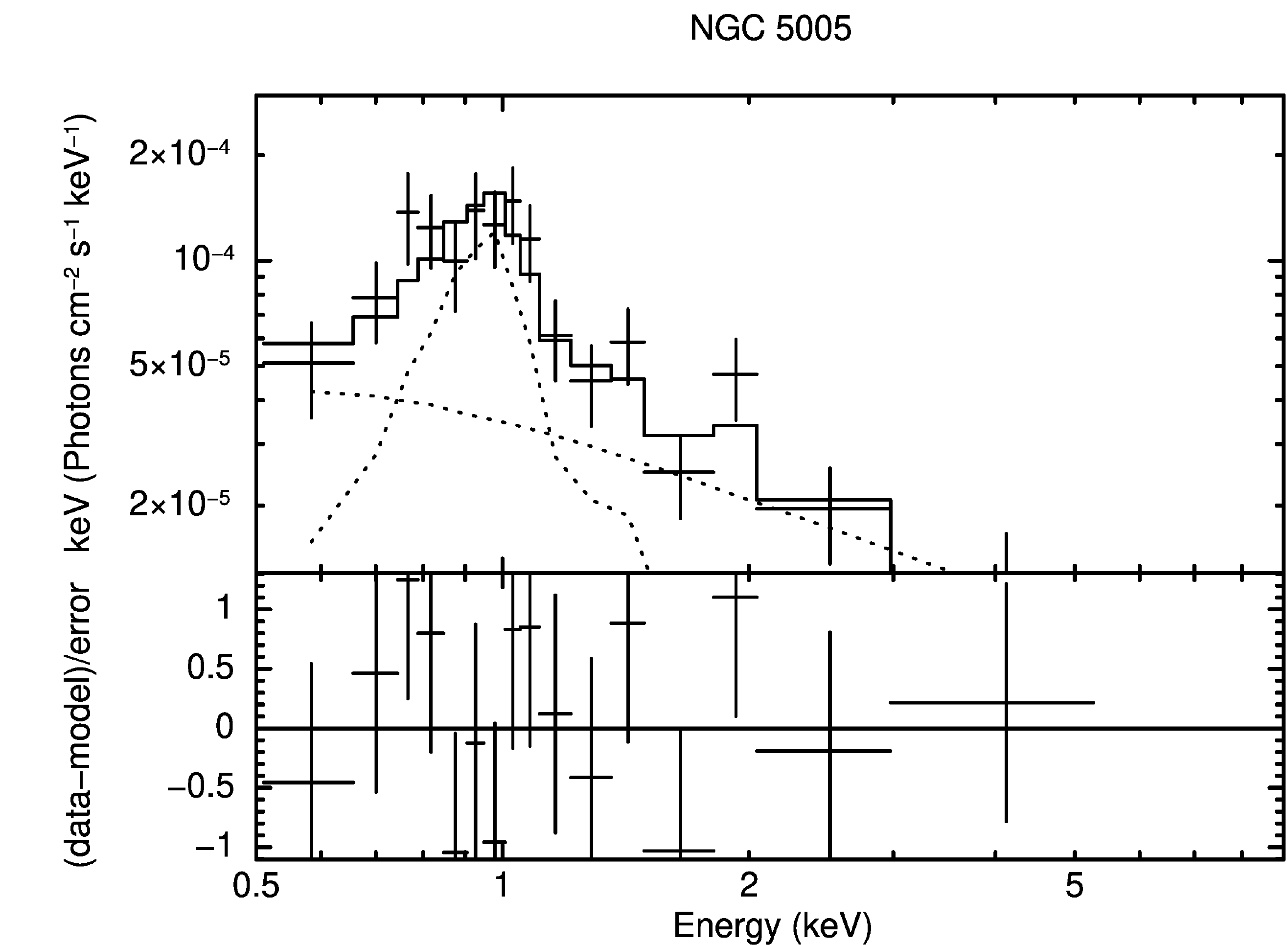}

\end{figure}
\end{center}

\begin{center}
 \begin{figure}
	\includegraphics[width=0.89\columnwidth]{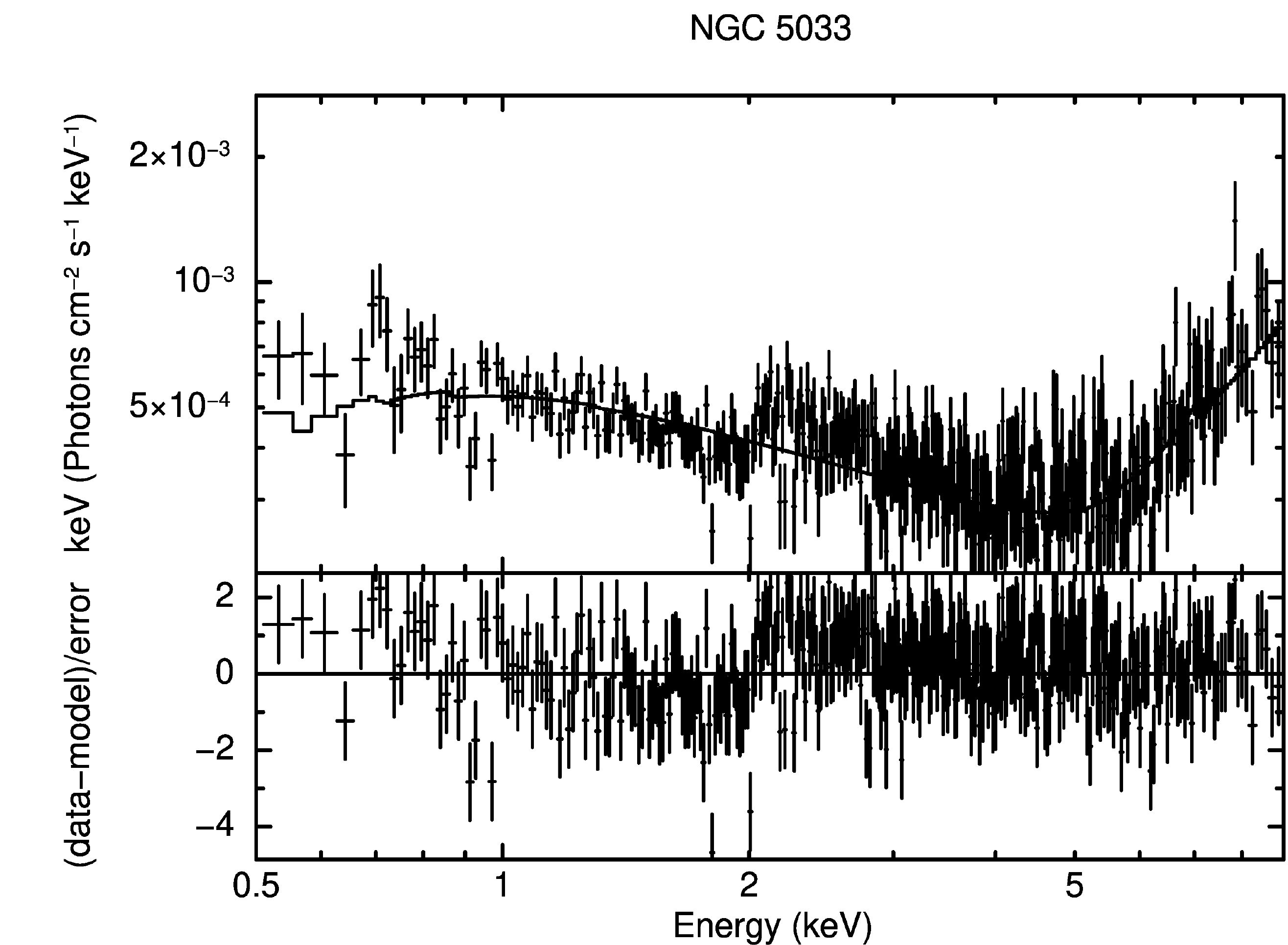}

\end{figure}
\end{center}

\begin{center}
 \begin{figure}
	\includegraphics[width=0.89\columnwidth]{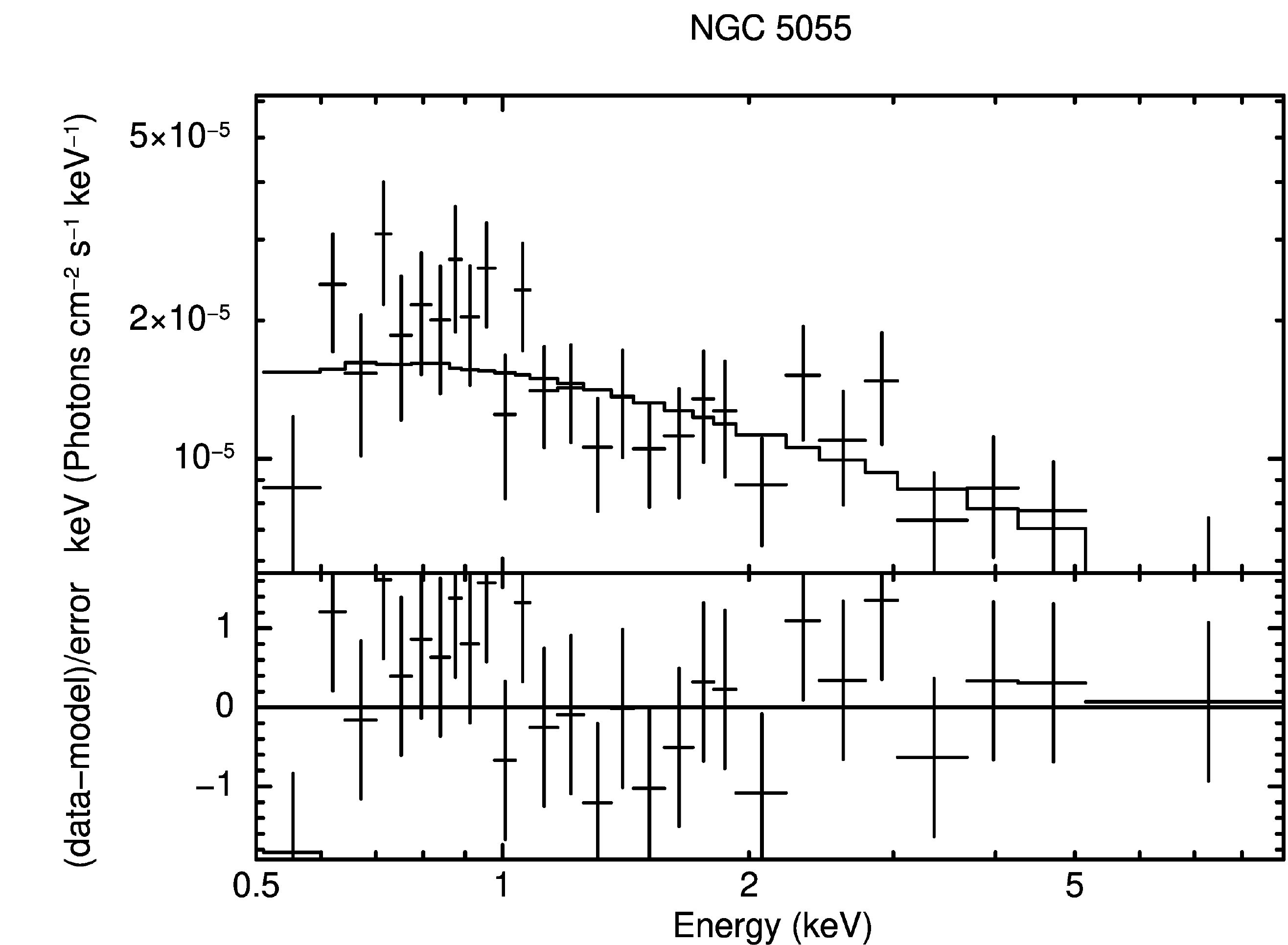}

\end{figure}
\end{center}

\begin{center}
 \begin{figure}
	\includegraphics[width=0.89\columnwidth]{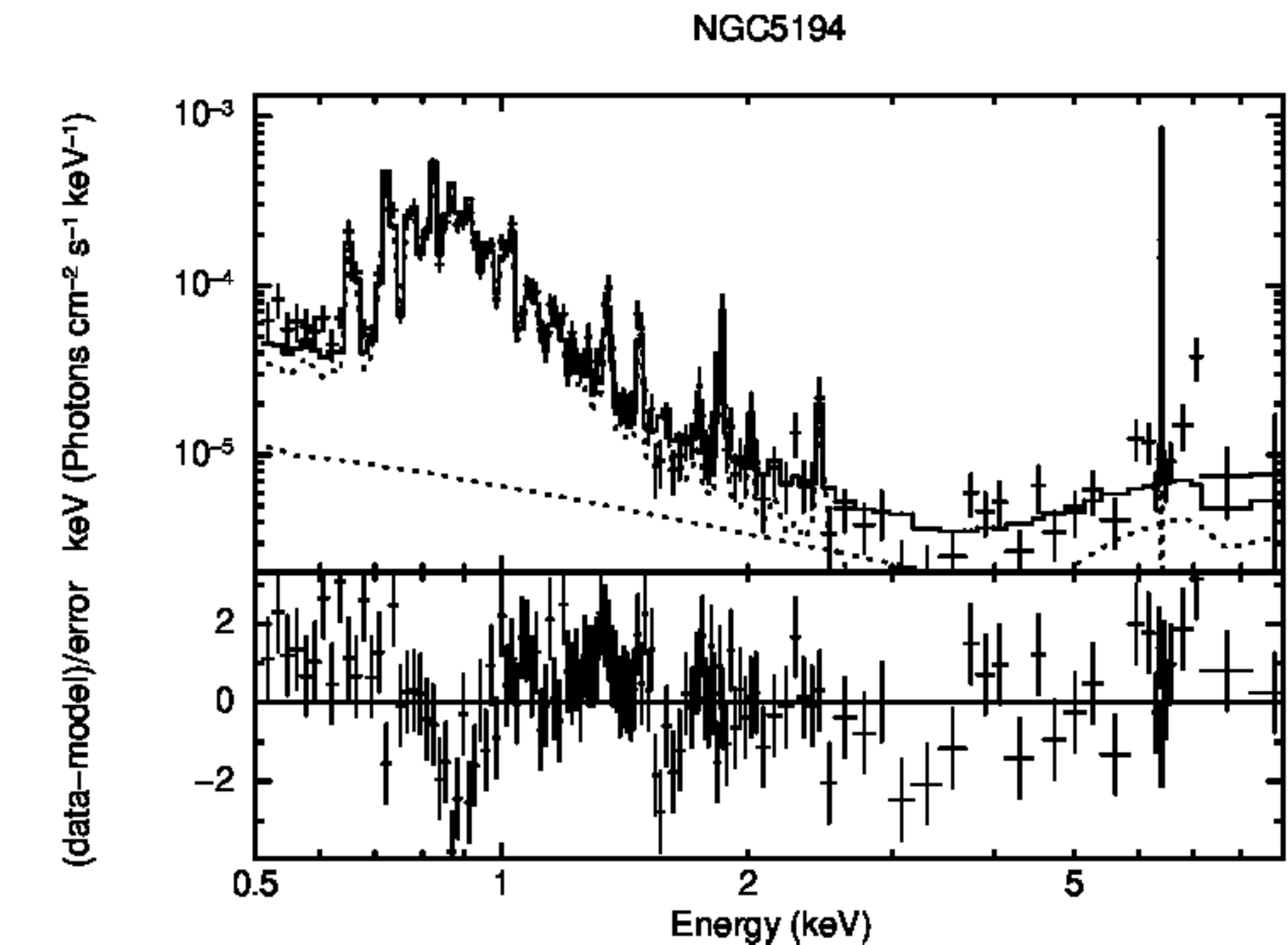}

\end{figure}
\end{center}

\begin{center}
 \begin{figure}
	\includegraphics[width=0.89\columnwidth]{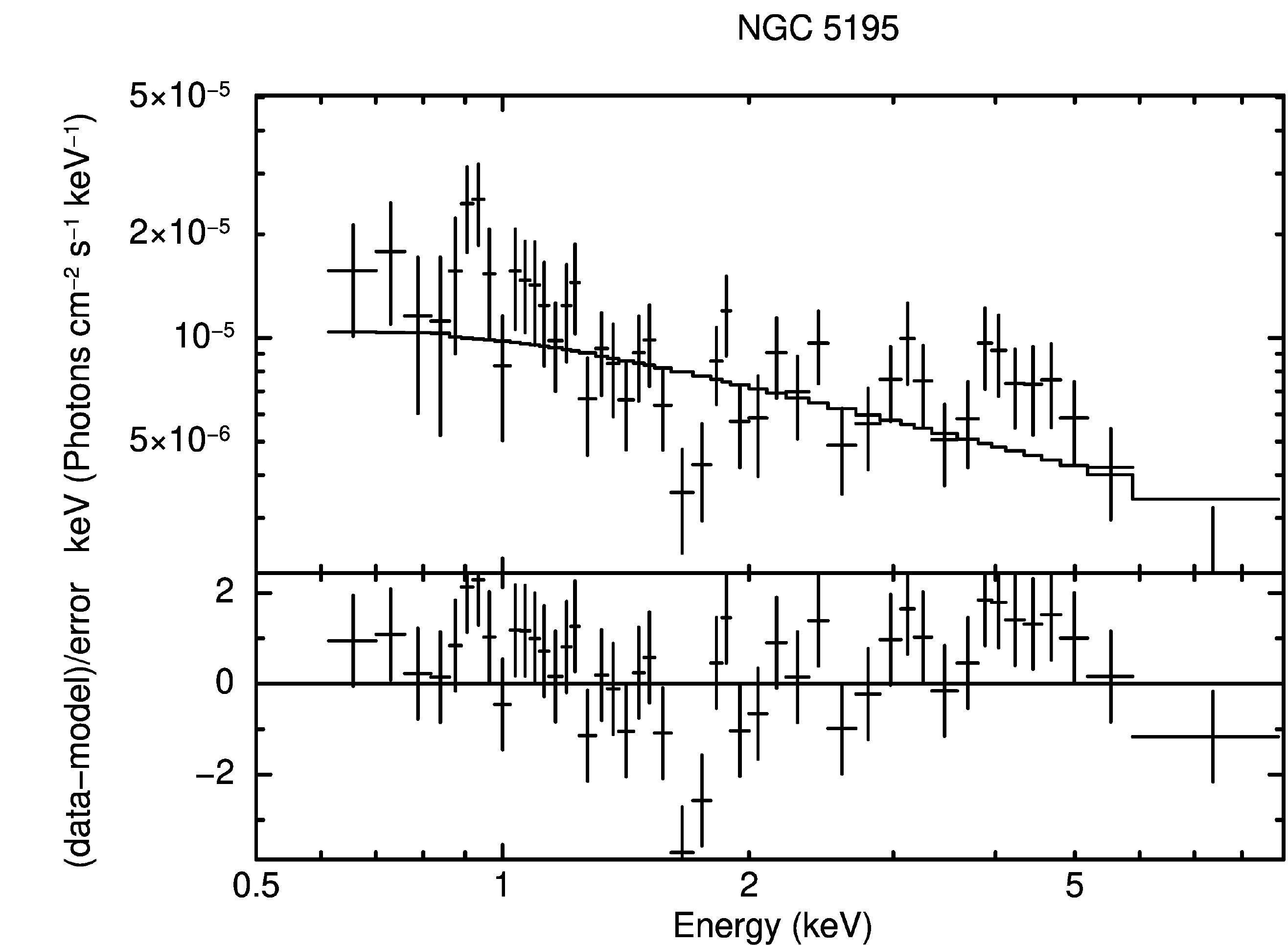}

\end{figure}
\end{center}

\begin{center}
 \begin{figure}
	\includegraphics[width=0.89\columnwidth]{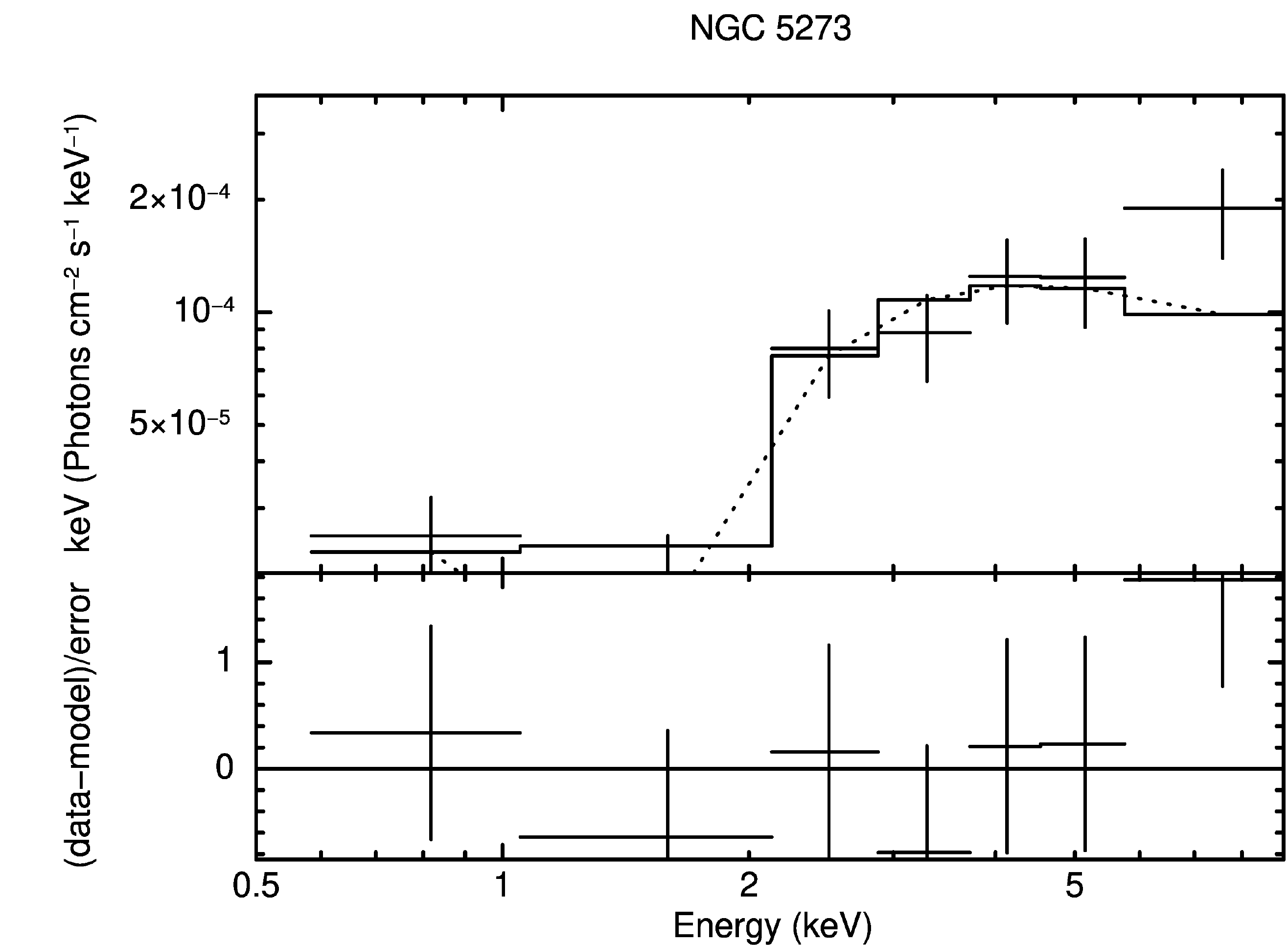}

\end{figure}
\end{center}

\begin{center}
 \begin{figure}
	\includegraphics[width=0.89\columnwidth]{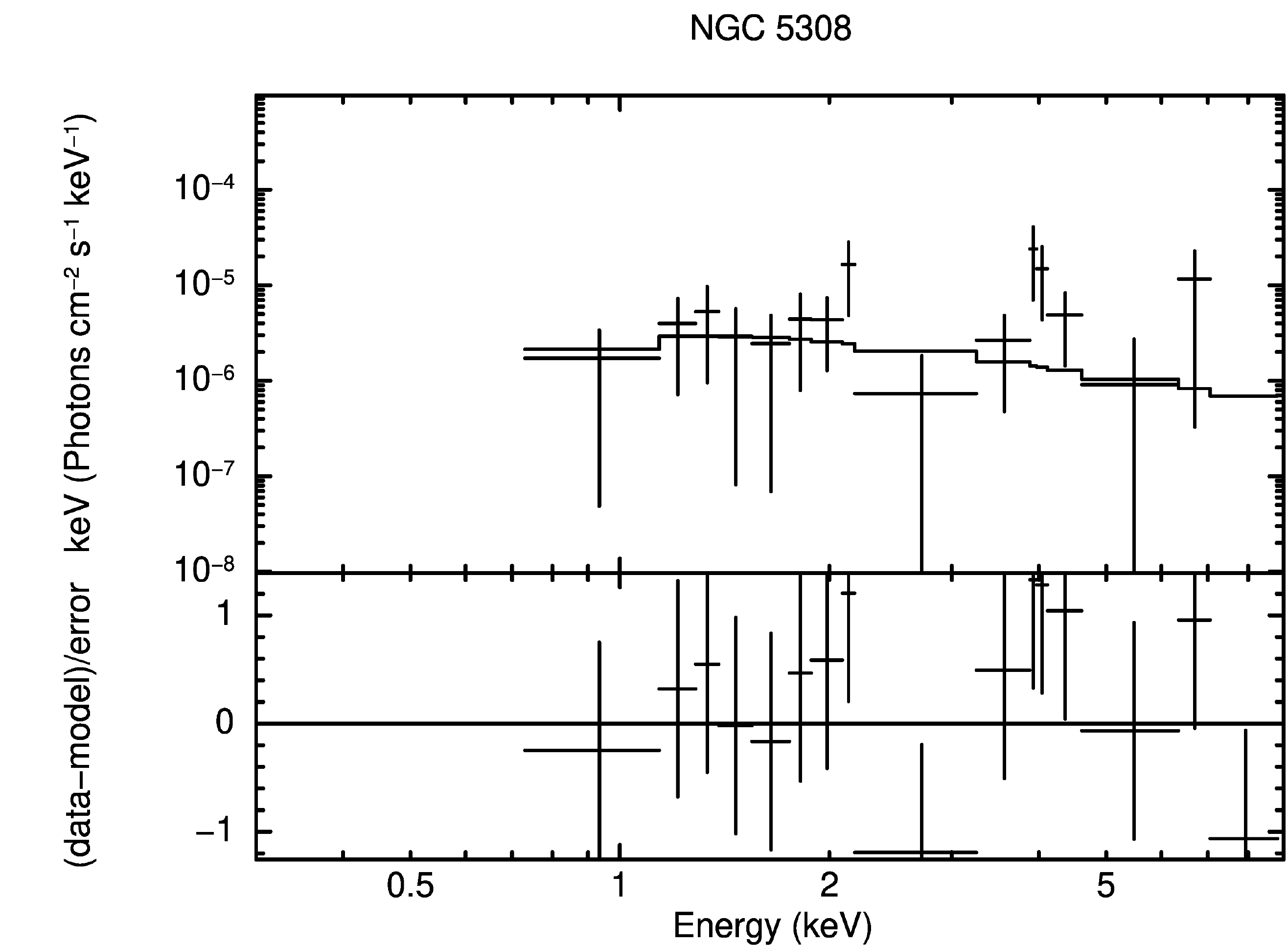}

\end{figure}
\end{center}

\begin{center}
 \begin{figure}
	\includegraphics[width=0.89\columnwidth]{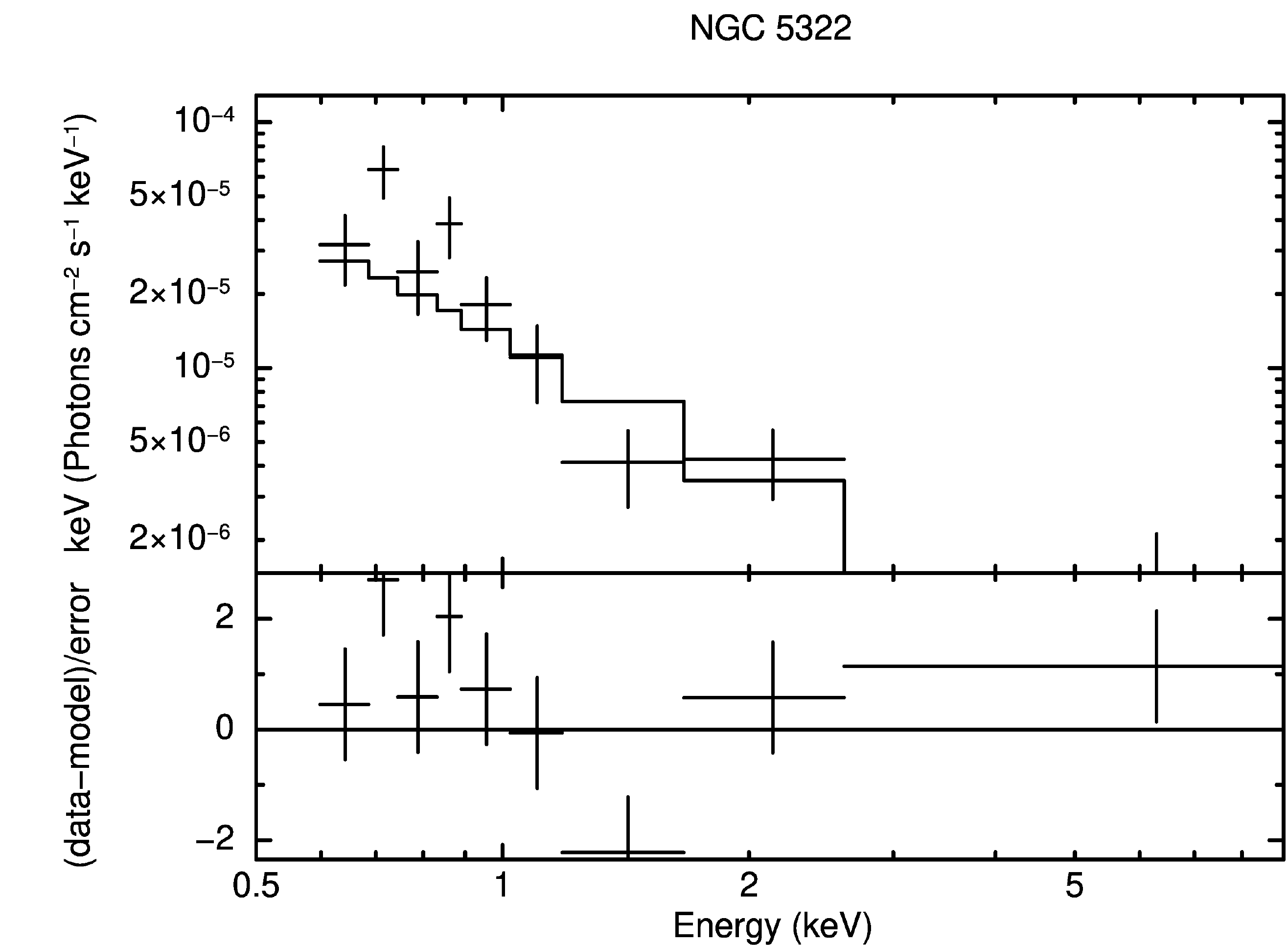}

\end{figure}
\end{center}

\begin{center}
 \begin{figure}
	\includegraphics[width=0.89\columnwidth]{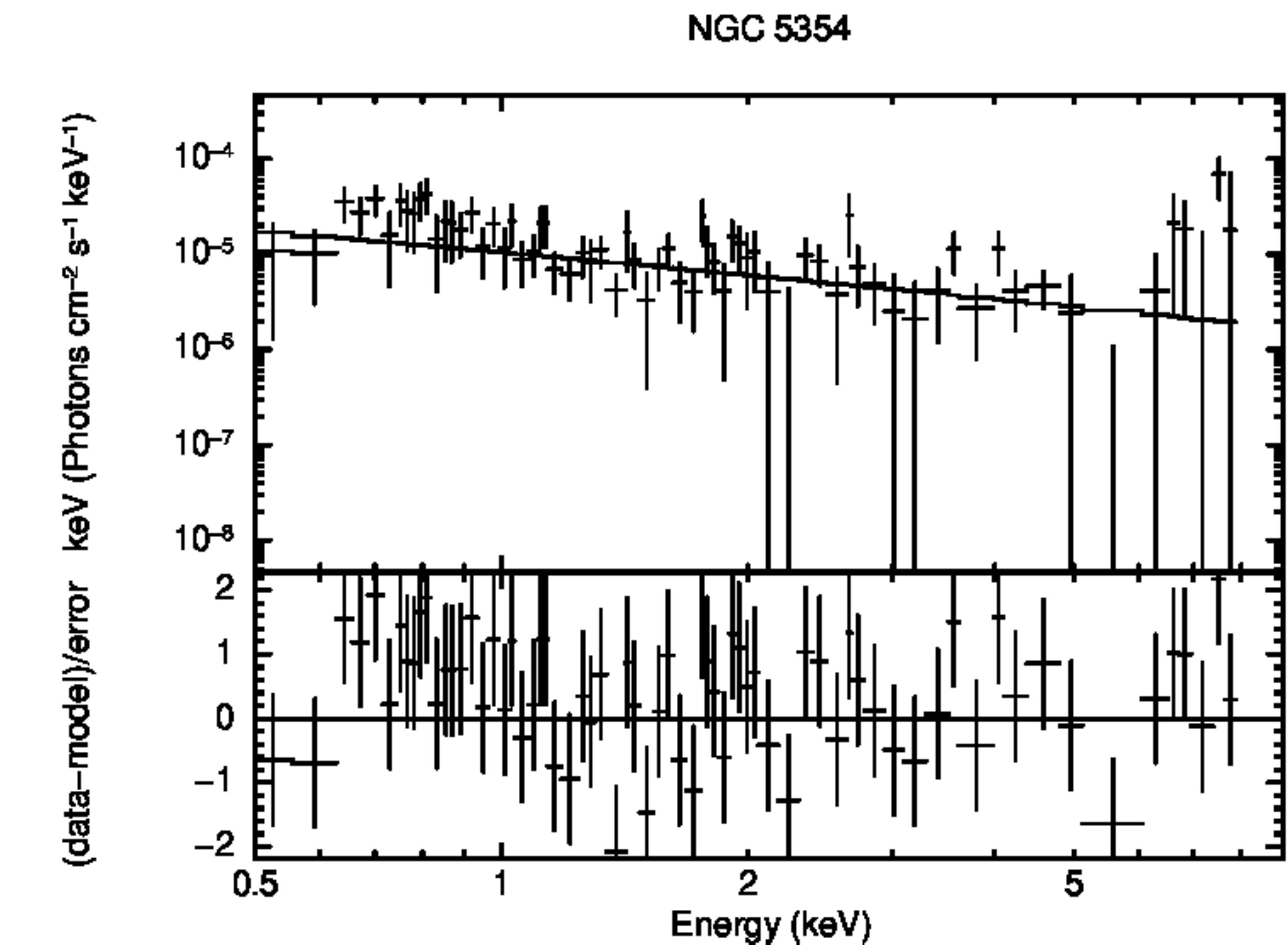}

\end{figure}
\end{center}

%
	 

\begin{center}
 \begin{figure}
	\includegraphics[width=0.89\columnwidth]{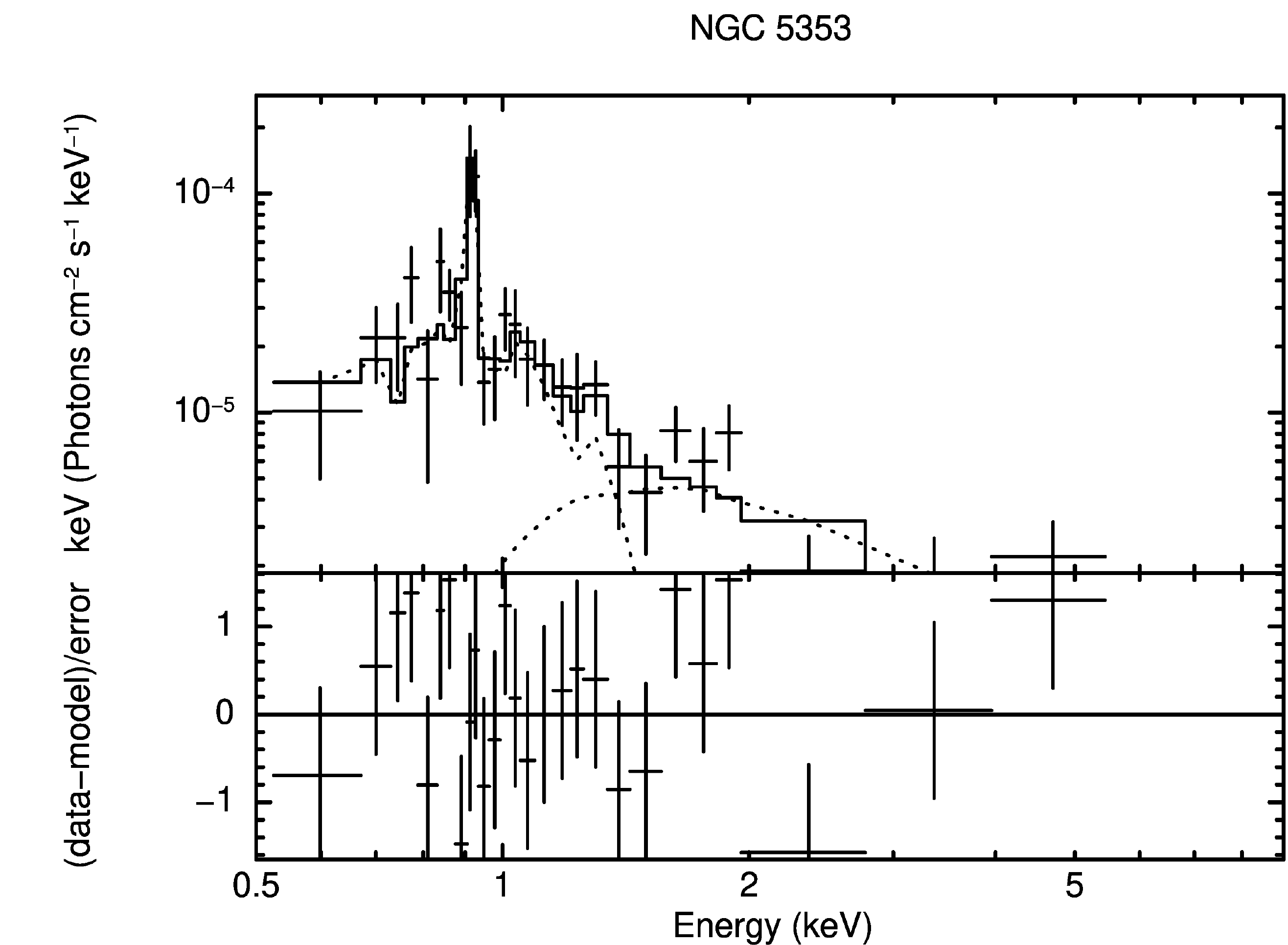}

\end{figure}
\end{center}

\begin{center}
 \begin{figure}
	\includegraphics[width=0.89\columnwidth]{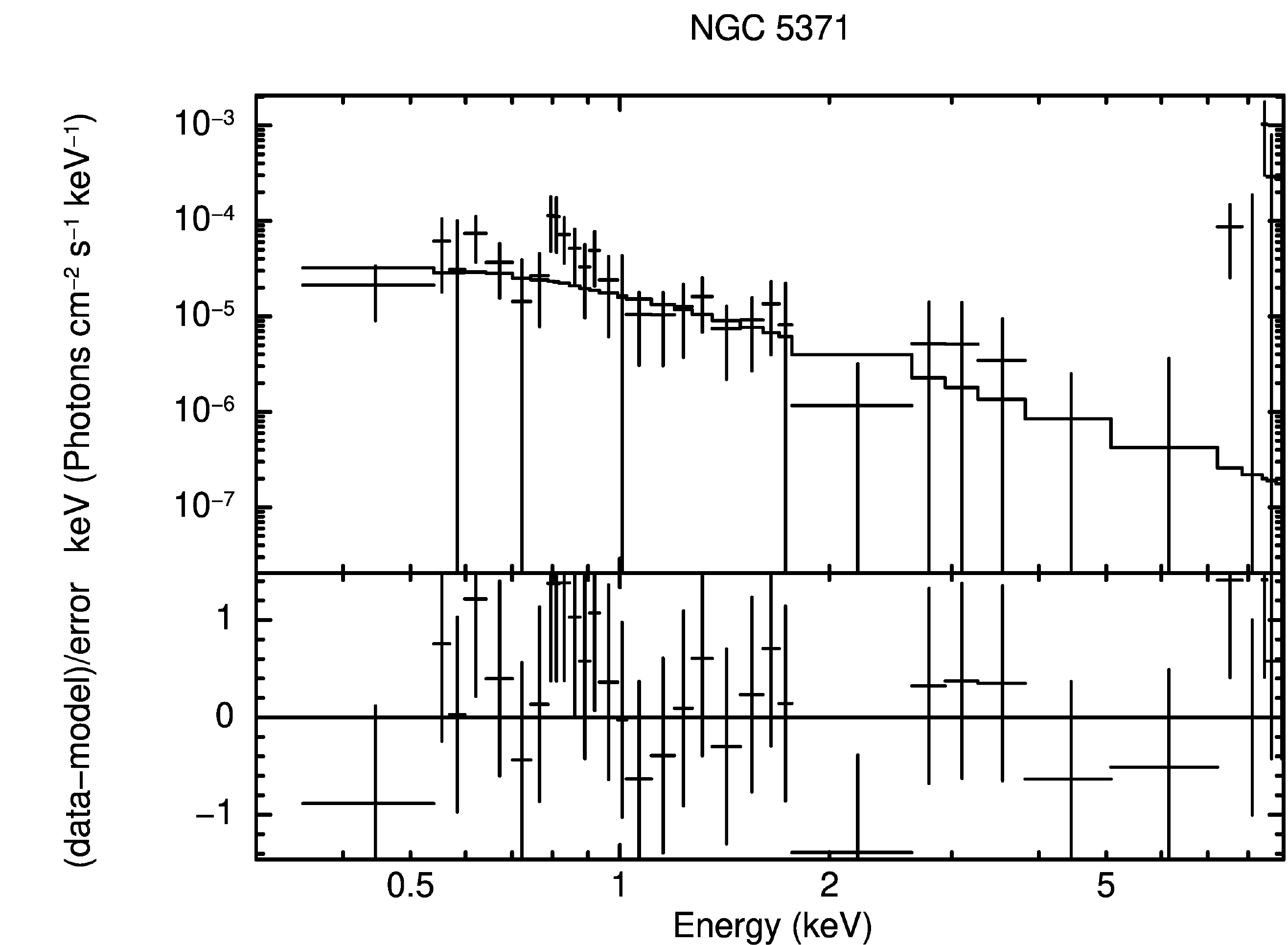}

\end{figure}
\end{center}

\begin{center}
 \begin{figure}
	\includegraphics[width=0.89\columnwidth]{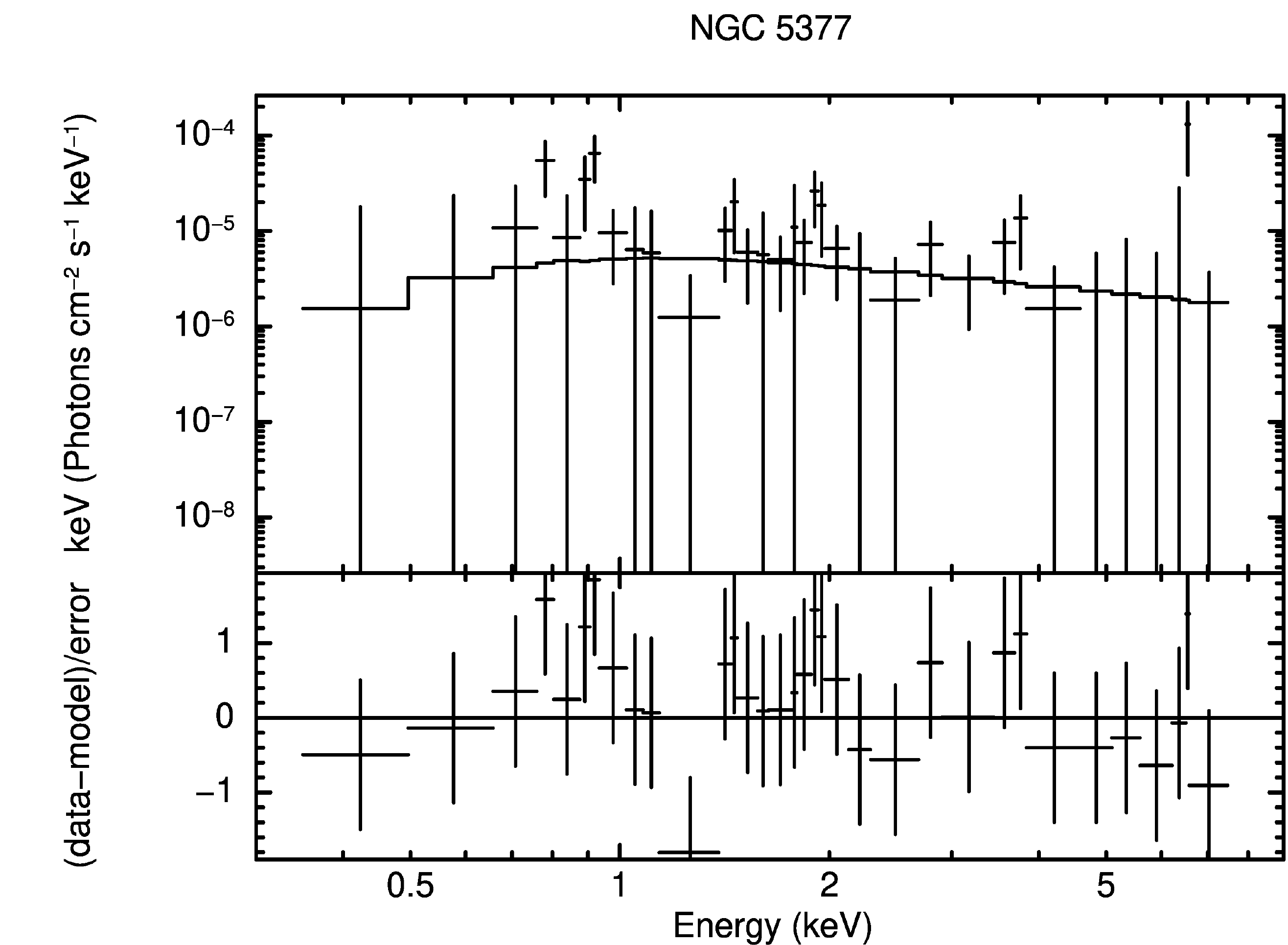}

\end{figure}
\end{center}

\begin{center}
 \begin{figure}
	\includegraphics[width=0.89\columnwidth]{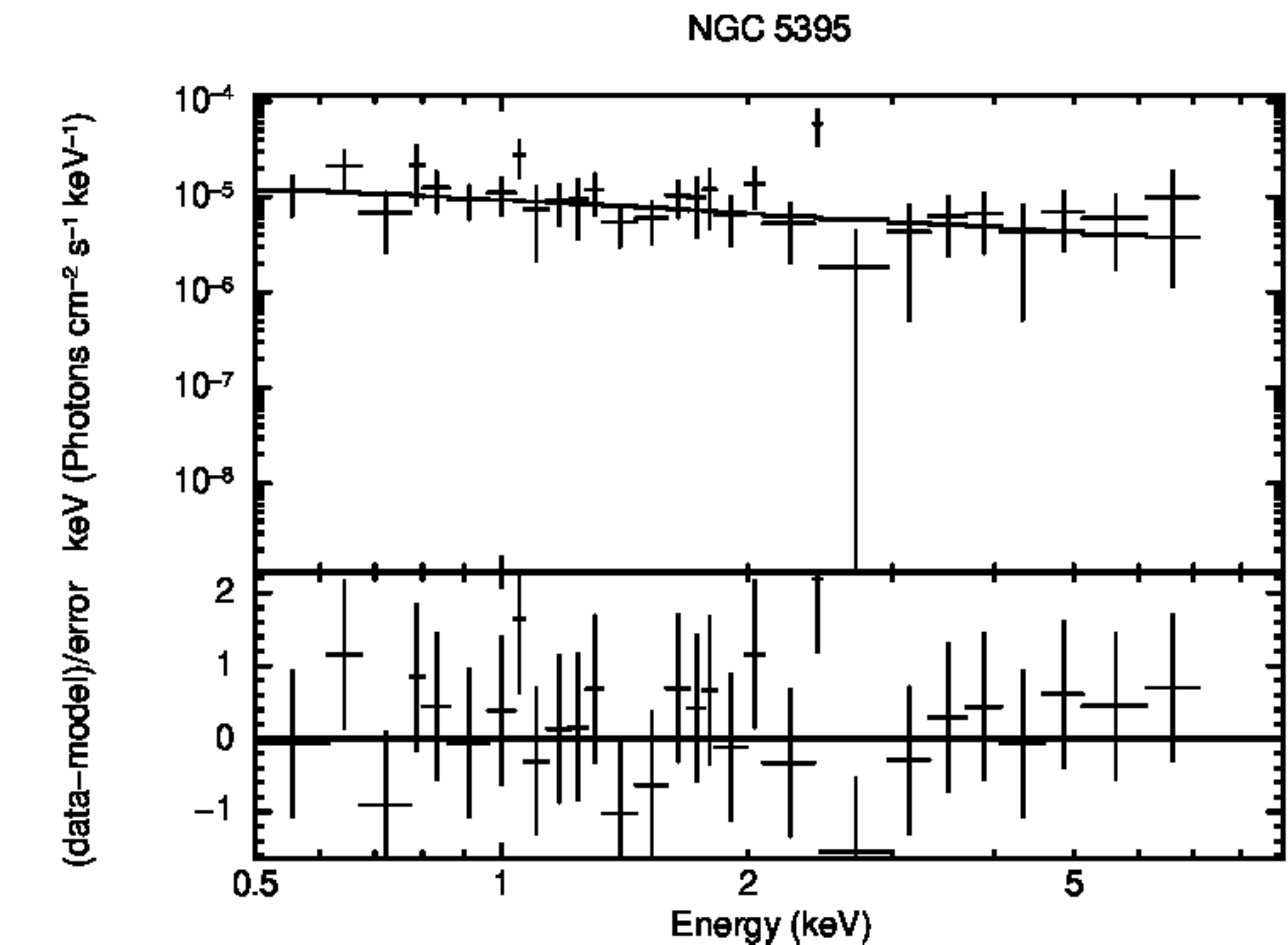}

\end{figure}
\end{center}

%
	 

\begin{center}
 \begin{figure}
	\includegraphics[width=0.89\columnwidth]{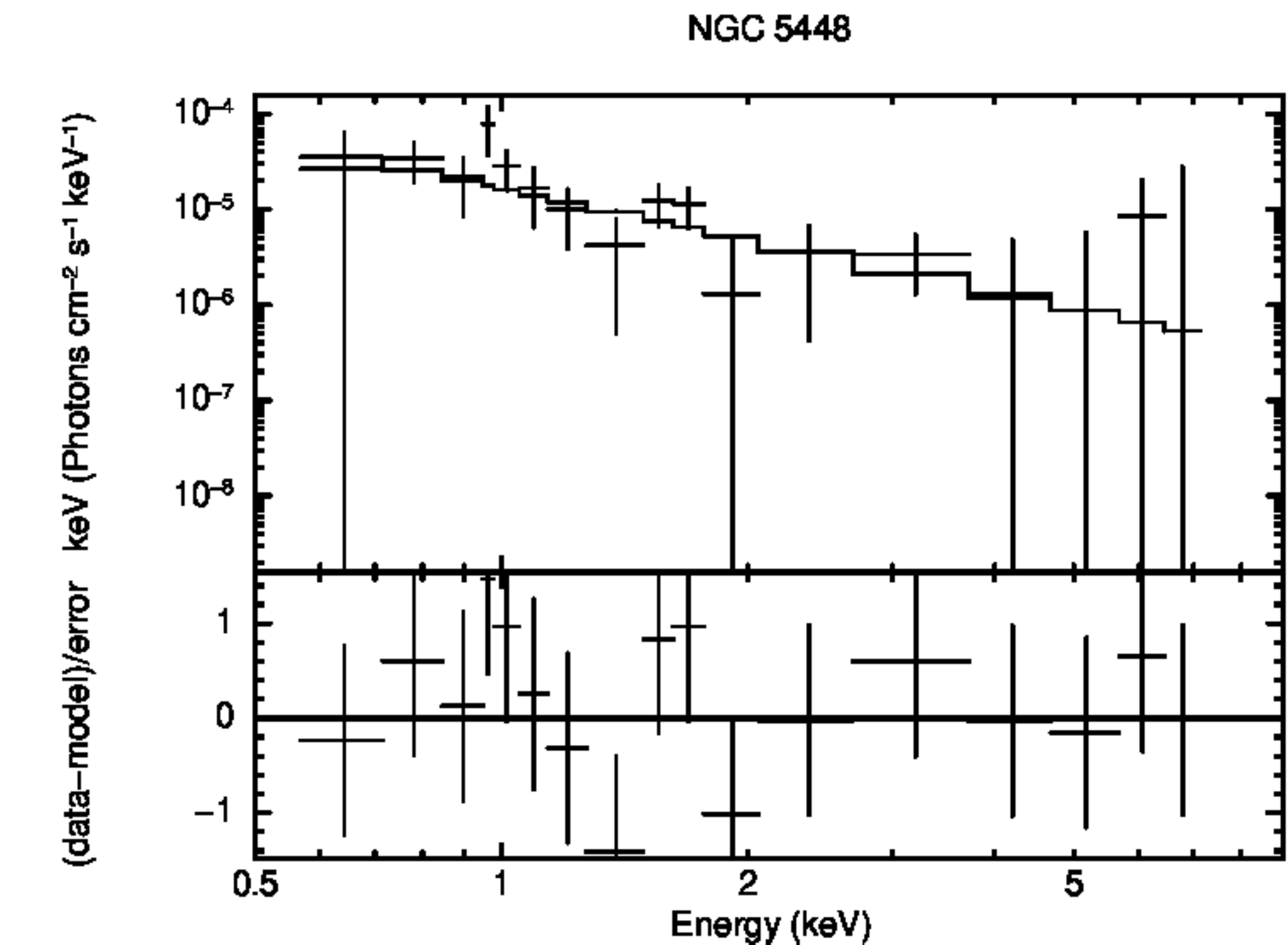}

\end{figure}
\end{center}

%
	 

\begin{center}
 \begin{figure}
	\includegraphics[width=0.89\columnwidth]{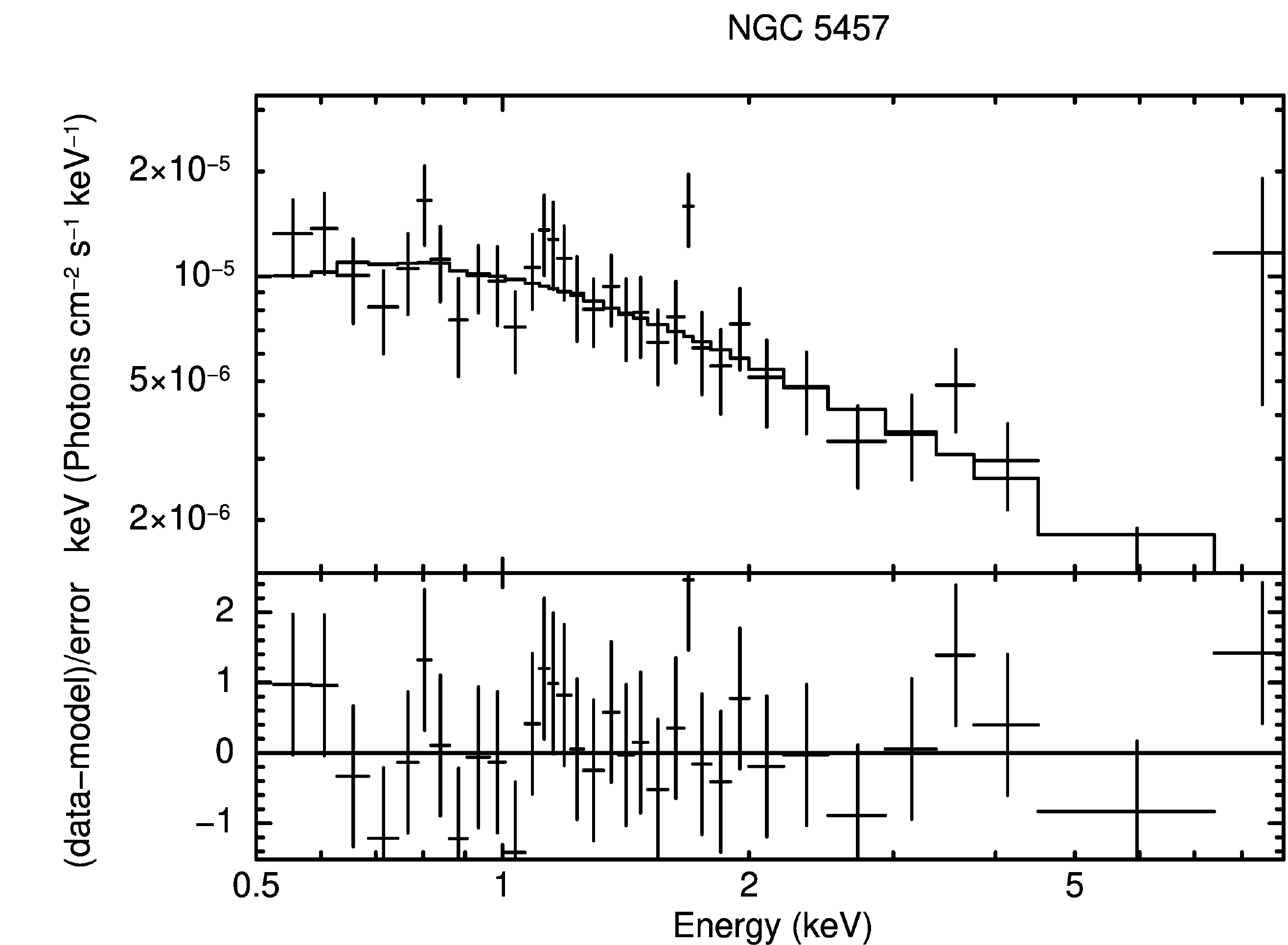}

\end{figure}
\end{center}

\begin{center}
 \begin{figure}
	\includegraphics[width=0.89\columnwidth]{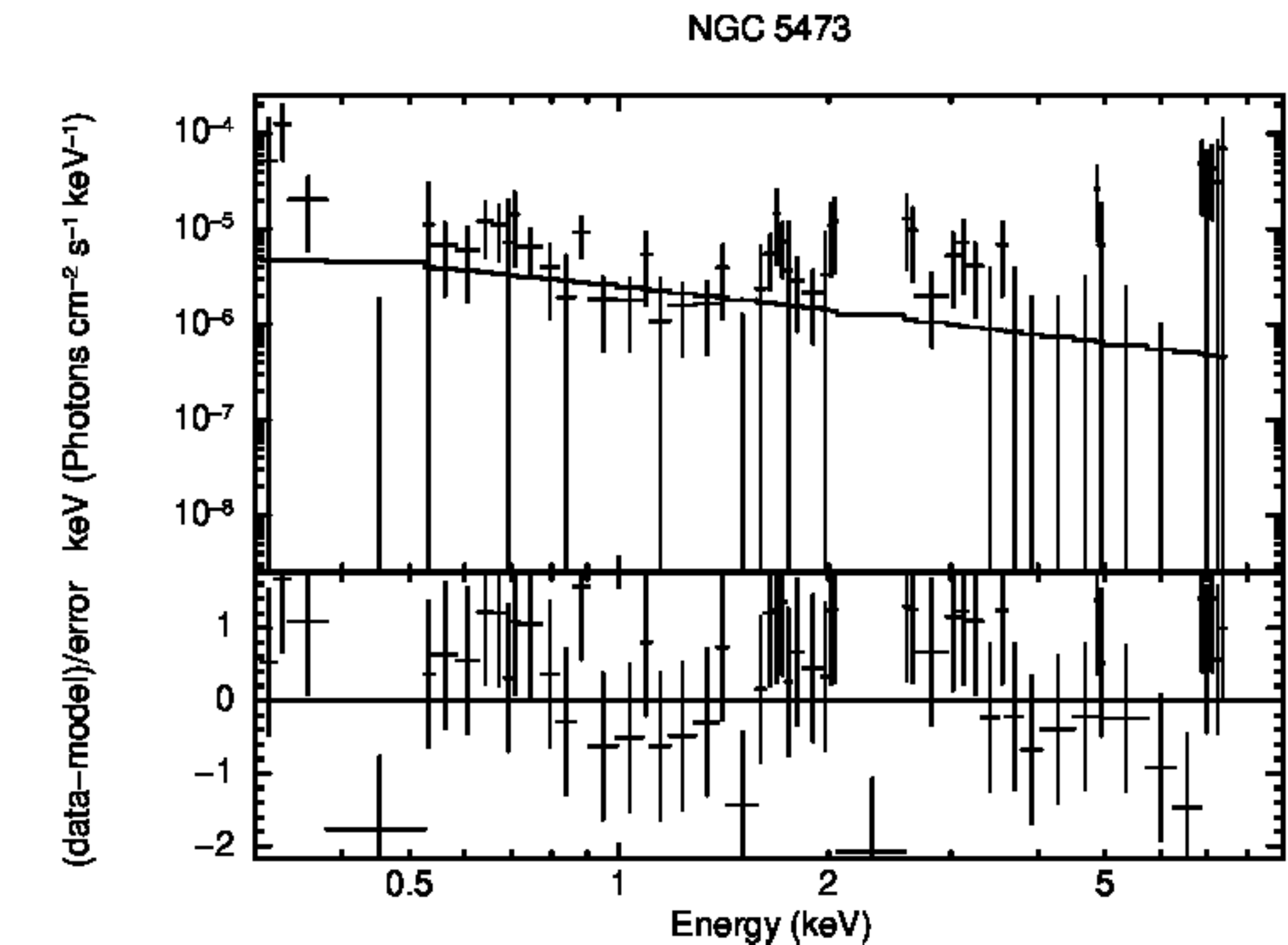}

\end{figure}
\end{center}

%
	 

\begin{center}
 \begin{figure}
	\includegraphics[width=0.89\columnwidth]{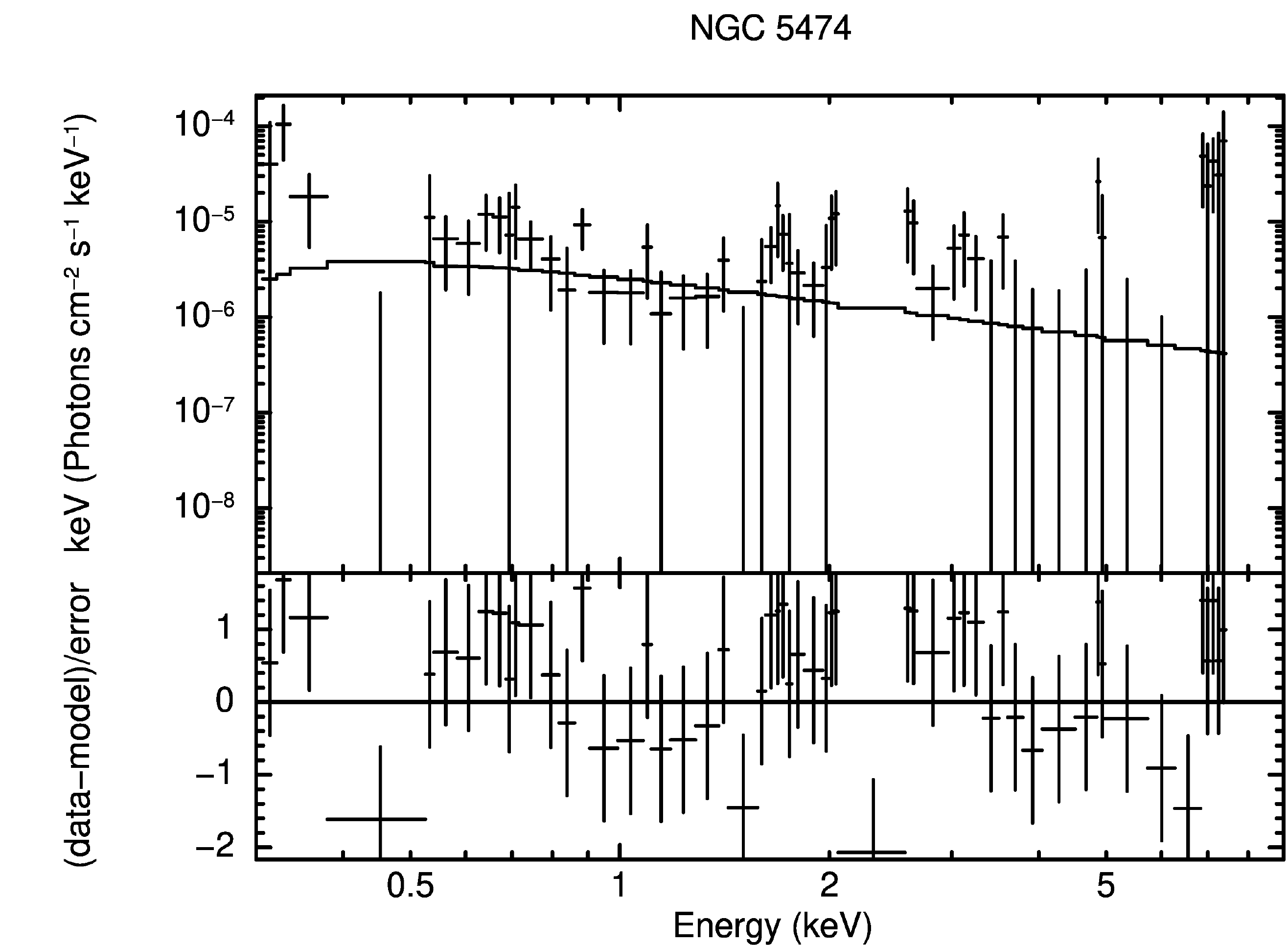}

\end{figure}
\end{center}

\begin{center}
 \begin{figure}
	\includegraphics[width=0.89\columnwidth]{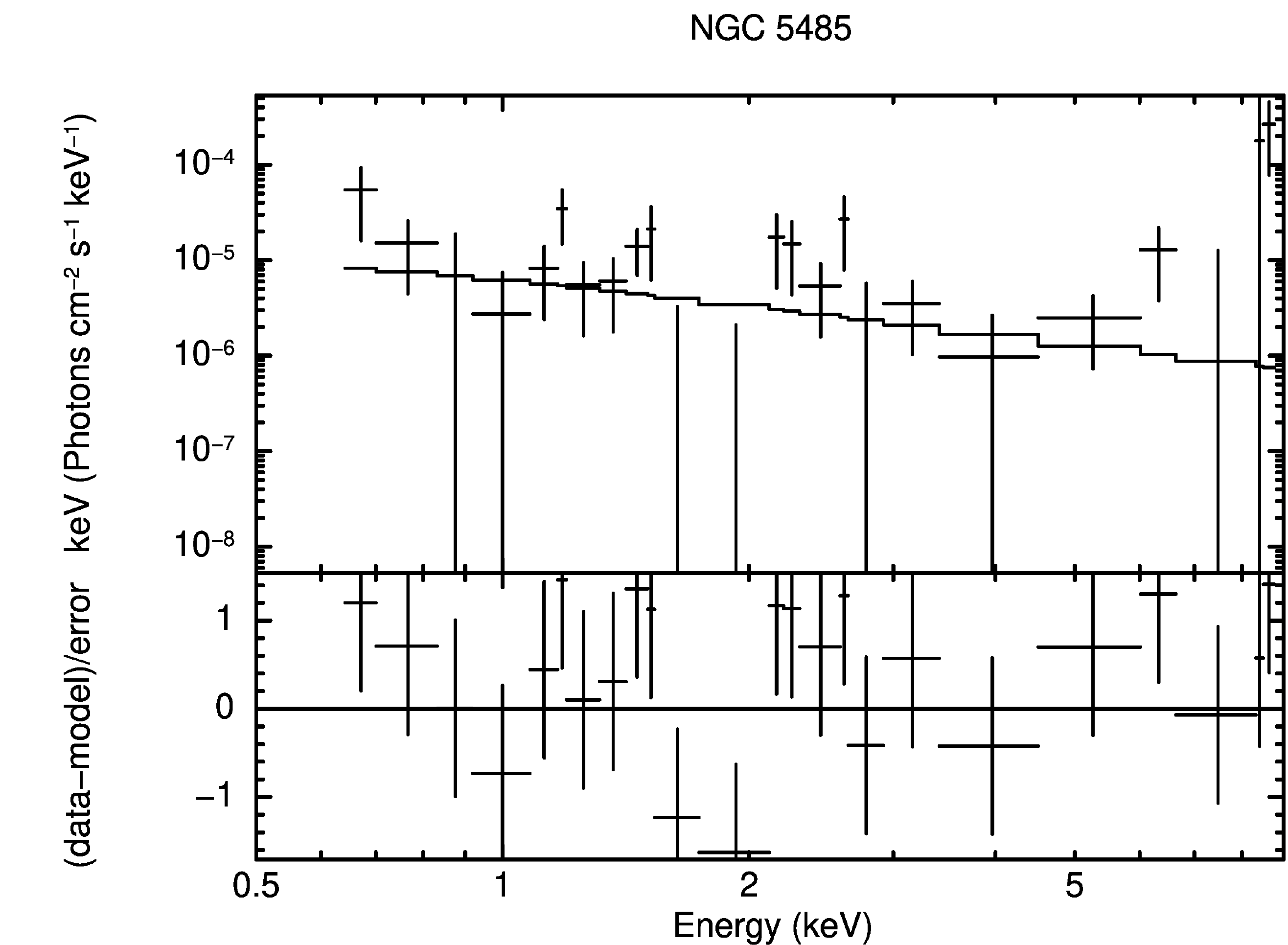}

\end{figure}
\end{center}

\begin{center}
 \begin{figure}
	\includegraphics[width=0.89\columnwidth]{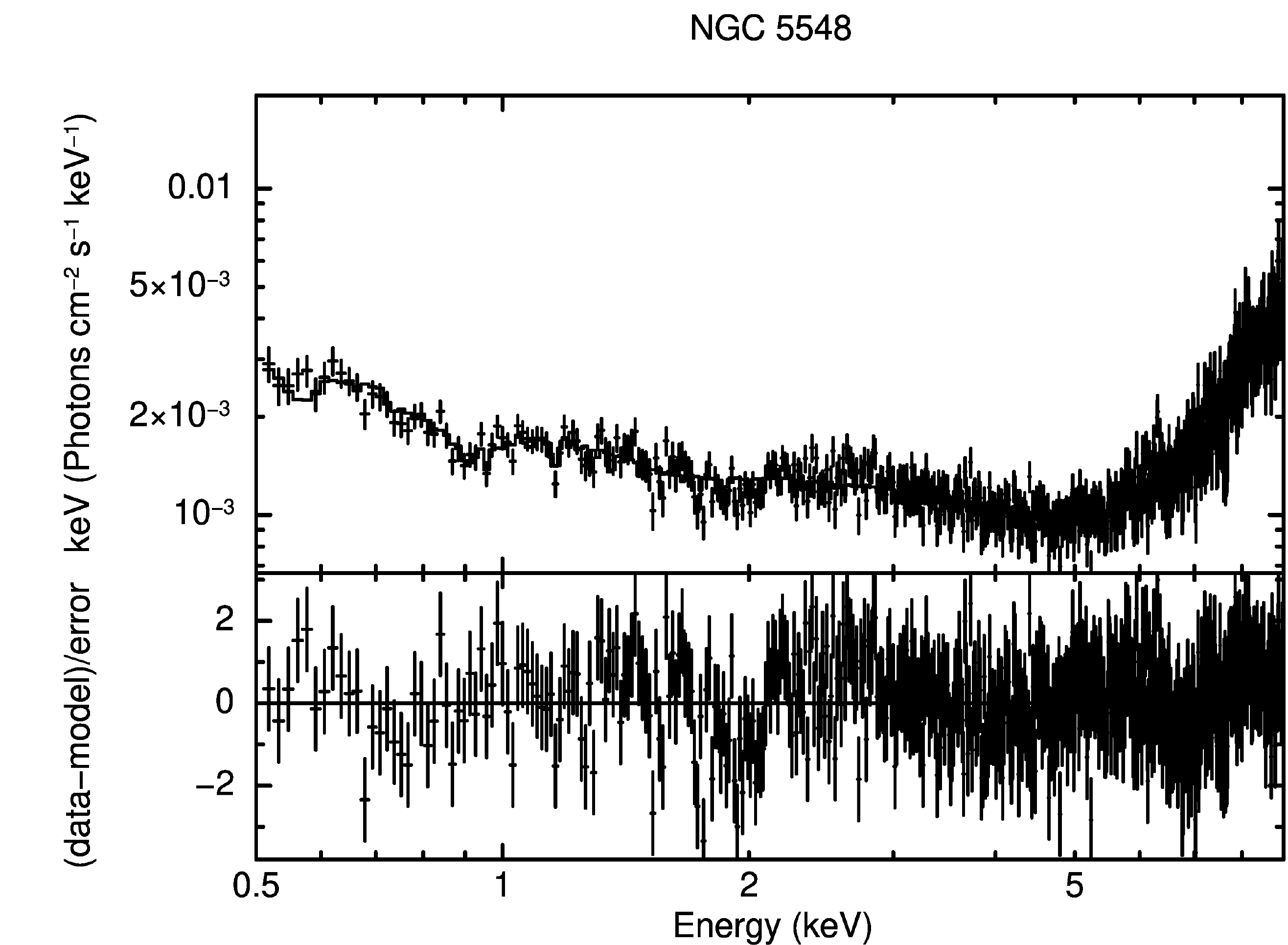}

\end{figure}
\end{center}

\begin{center}
 \begin{figure}
	\includegraphics[width=0.89\columnwidth]{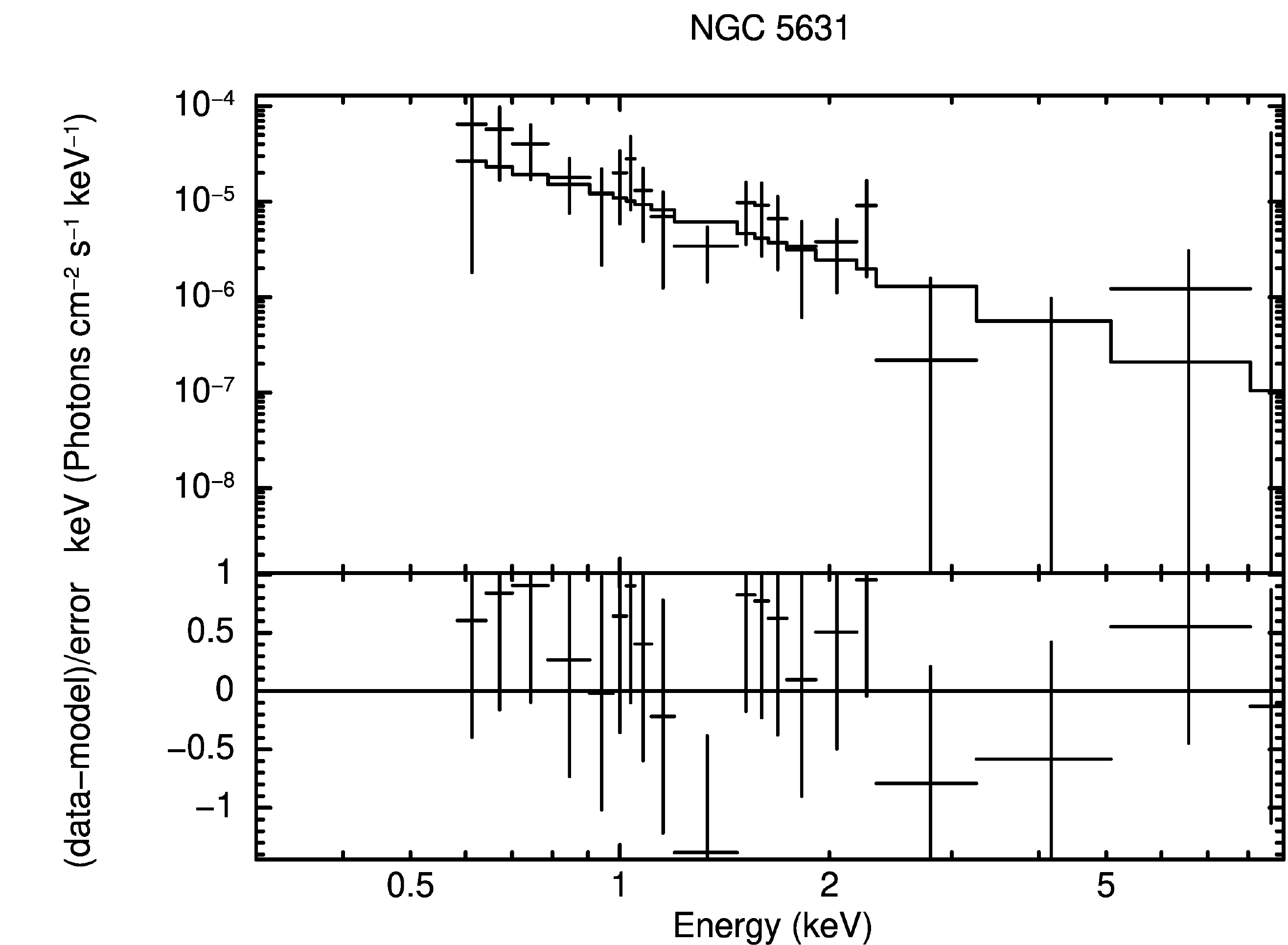}

\end{figure}
\end{center}

\begin{center}
 \begin{figure}
	\includegraphics[width=0.89\columnwidth]{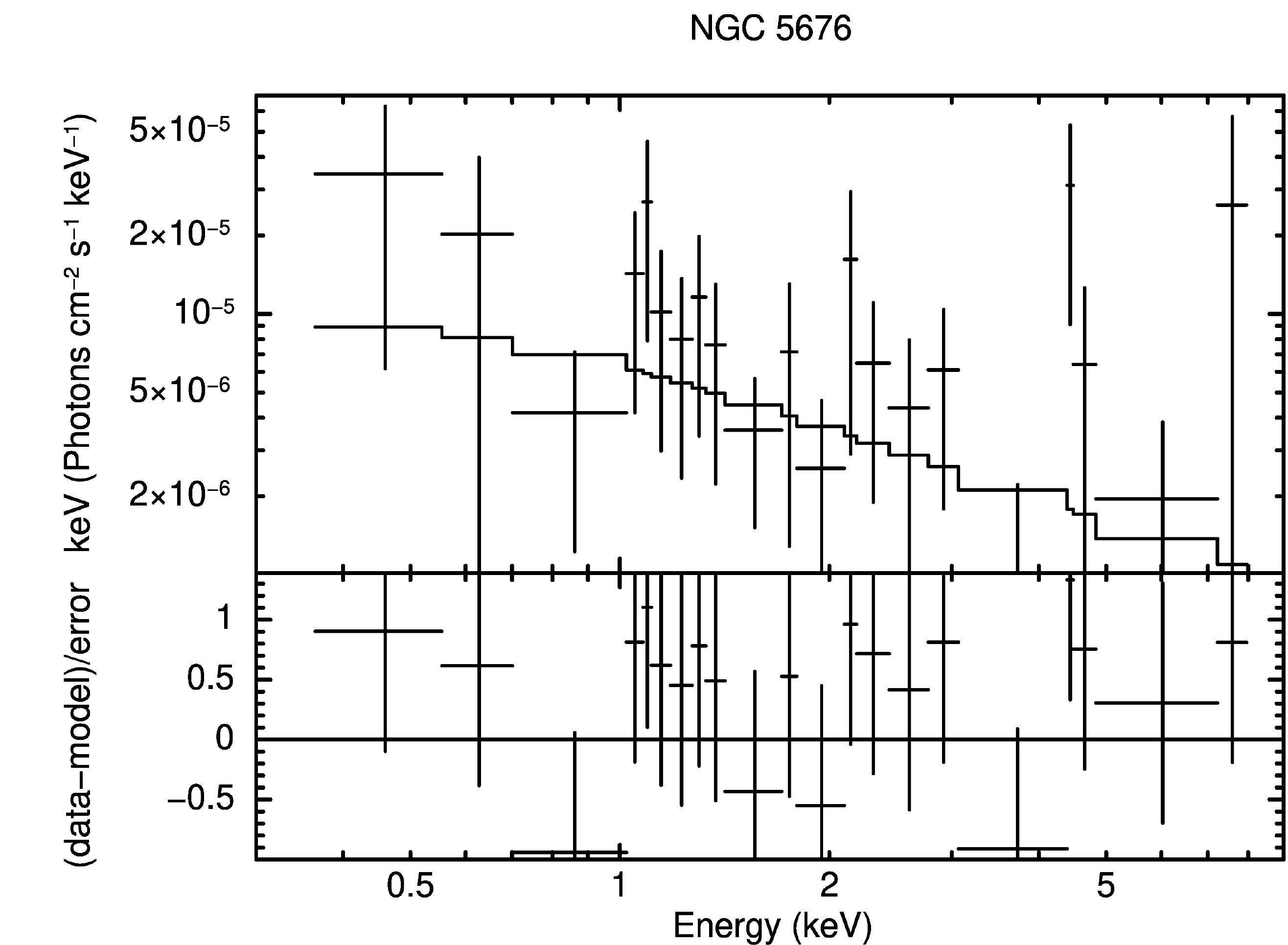}

\end{figure}
\end{center}

\begin{center}
 \begin{figure}
	\includegraphics[width=0.89\columnwidth]{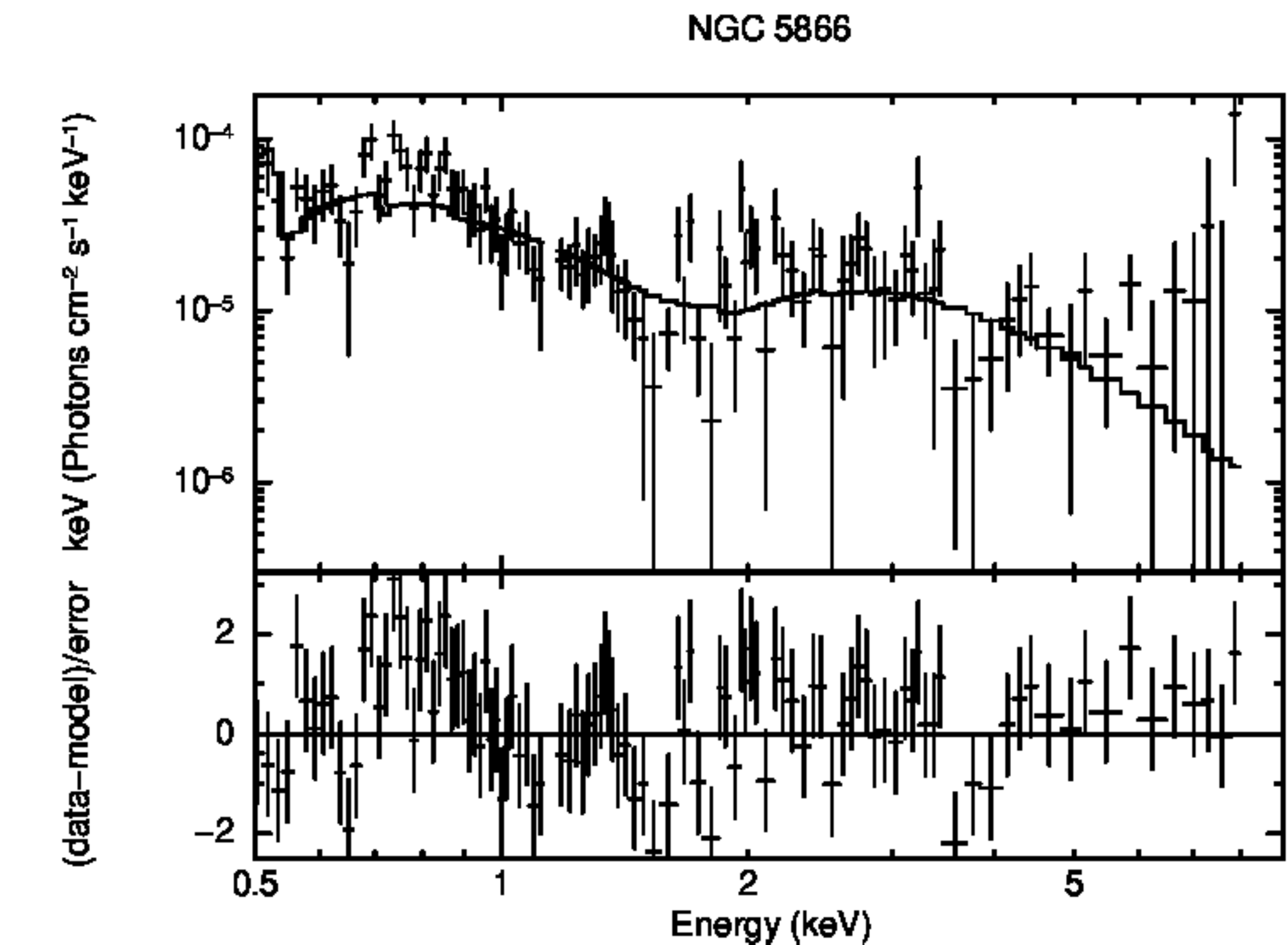}

\end{figure}
\end{center}

%
	 

\begin{center}
 \begin{figure}
	\includegraphics[width=0.89\columnwidth]{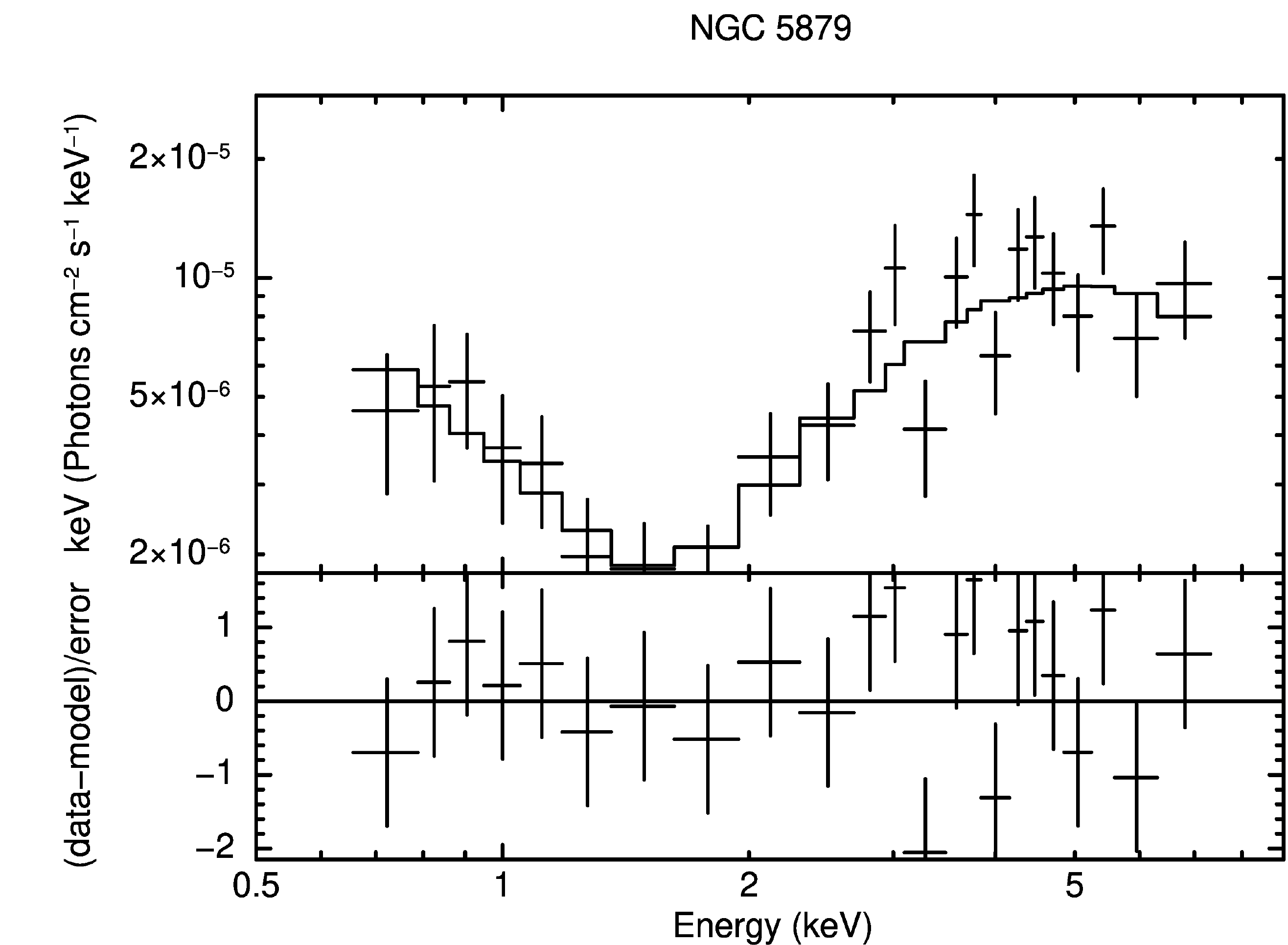}

\end{figure}
\end{center}\begin{center}
 \begin{figure}
	\includegraphics[width=0.89\columnwidth]{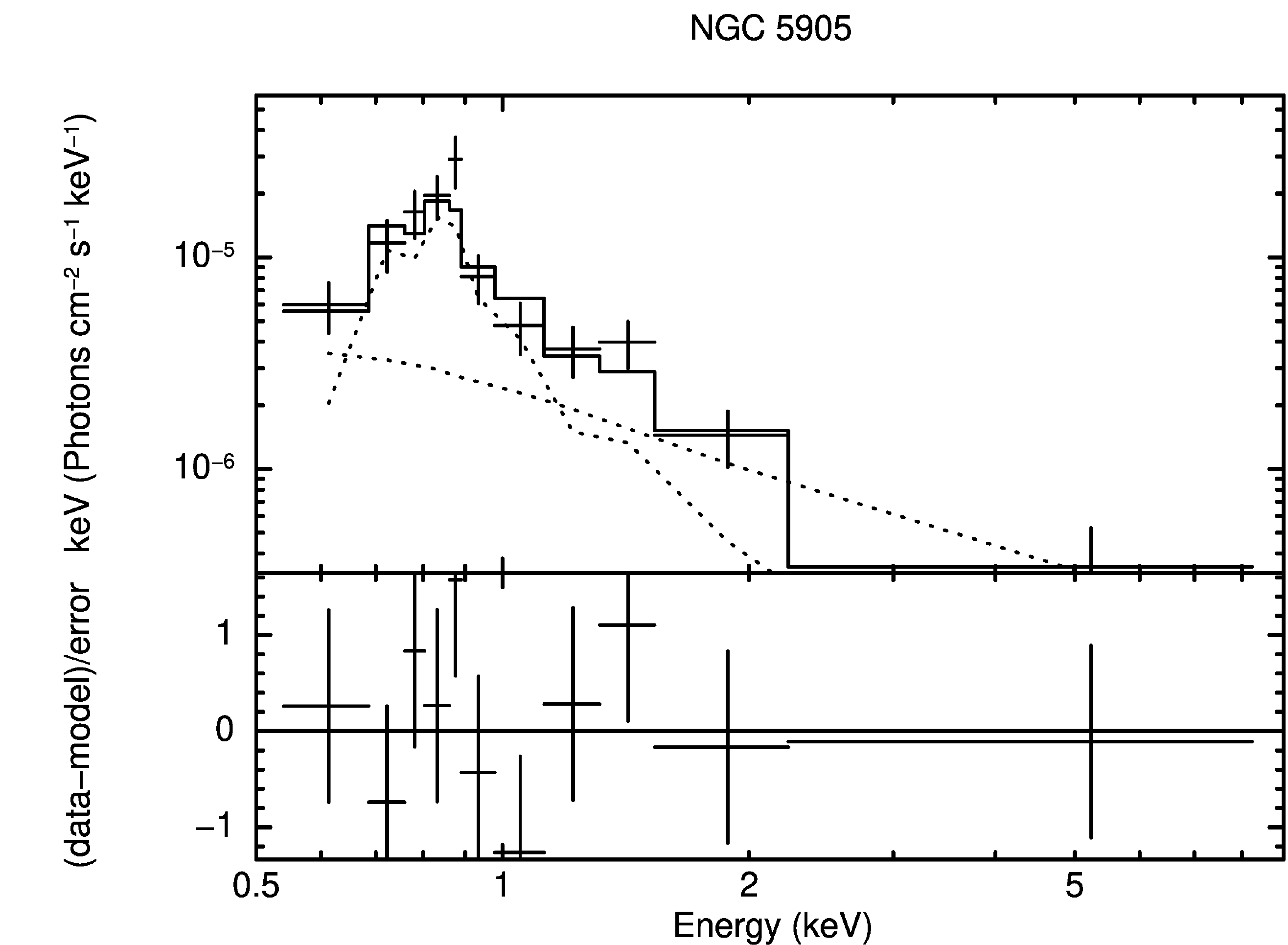}

\end{figure}
\end{center}

\begin{center}
 \begin{figure}
	\includegraphics[width=0.89\columnwidth]{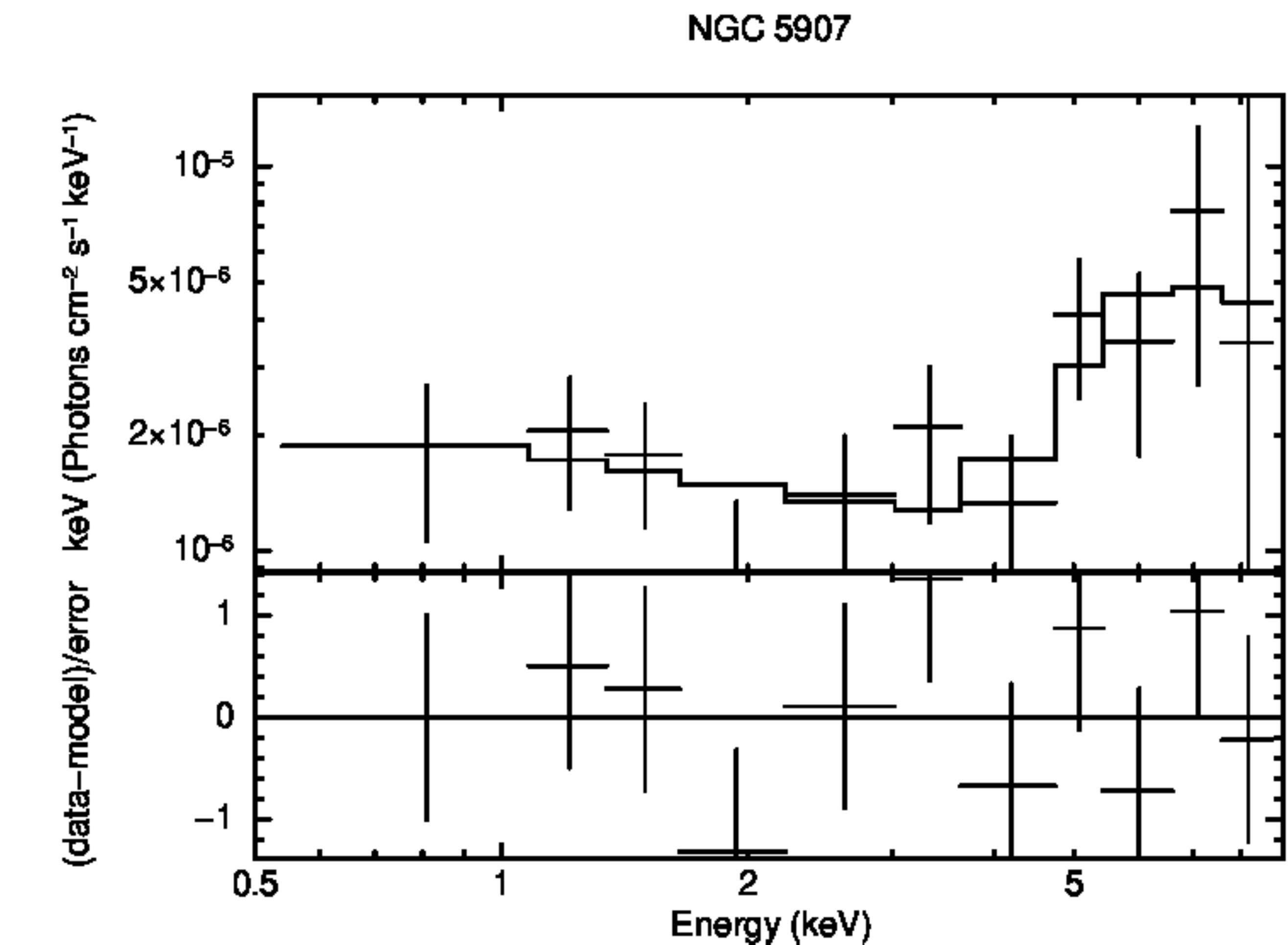}

\end{figure}
\end{center}

%
	 

\begin{center}
 \begin{figure}
	\includegraphics[width=0.89\columnwidth]{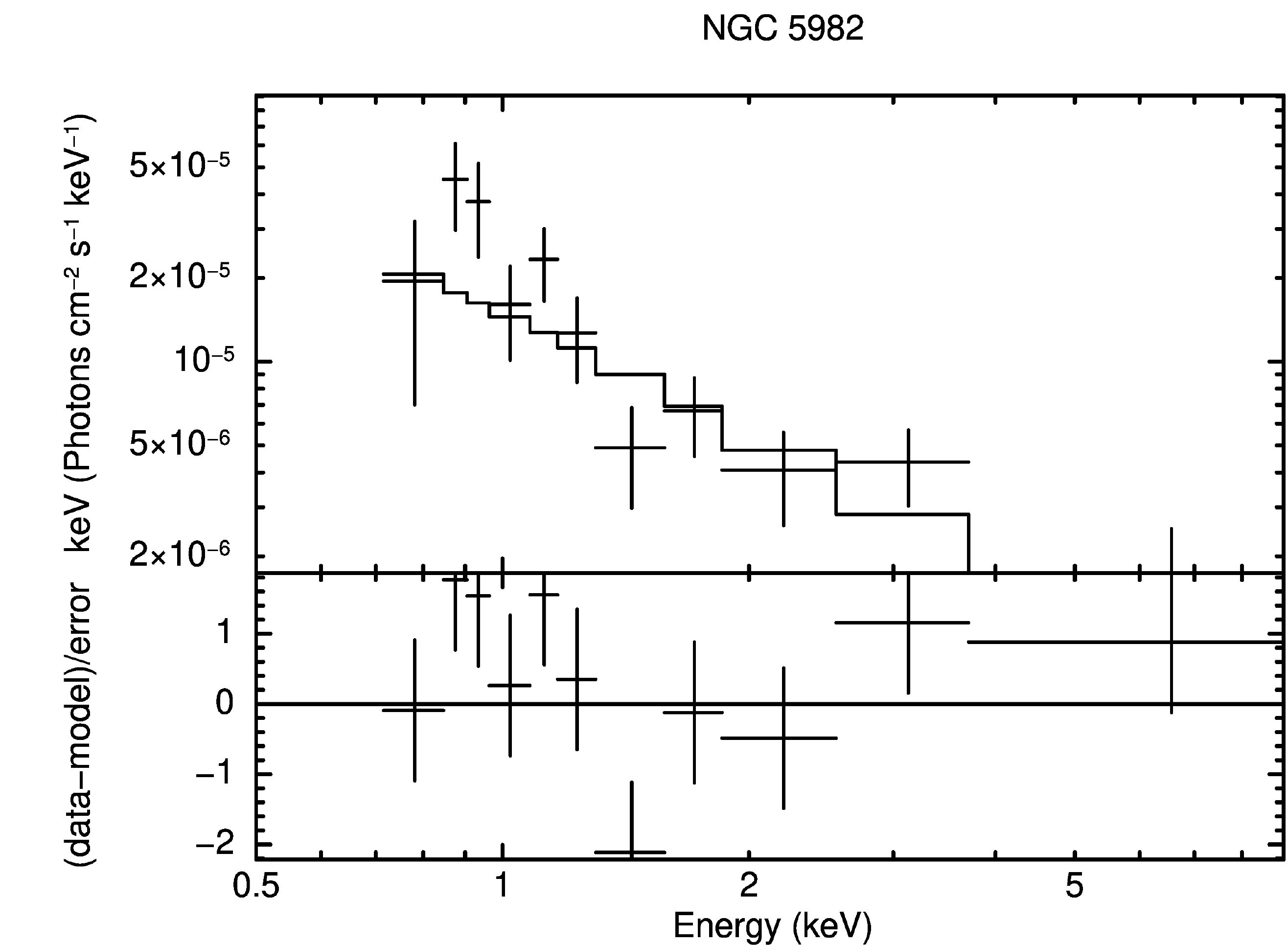}

\end{figure}
\end{center}

\begin{center}
 \begin{figure}
	\includegraphics[width=0.89\columnwidth]{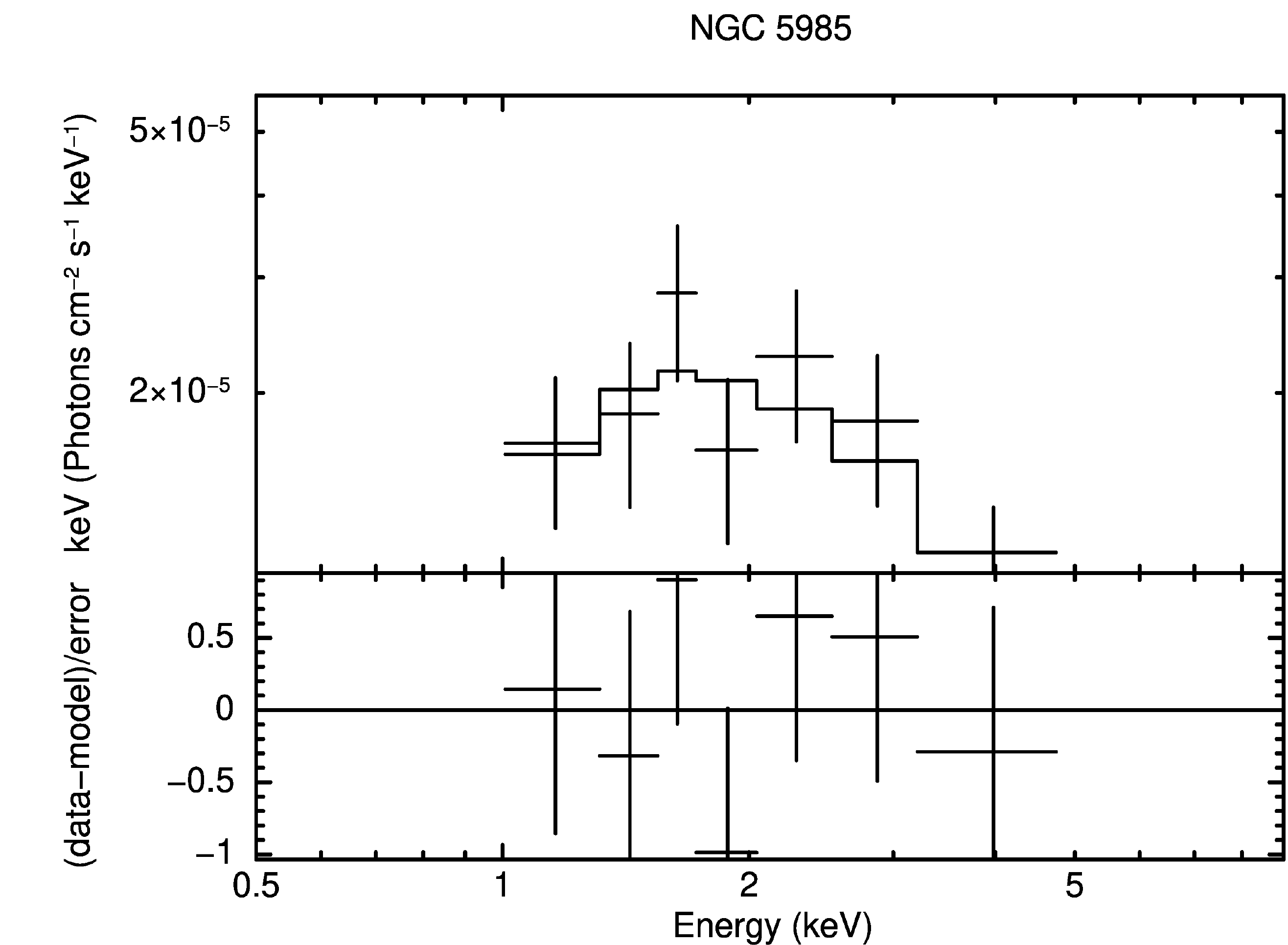}

\end{figure}
\end{center}

\begin{center}
 \begin{figure}
	\includegraphics[width=0.89\columnwidth]{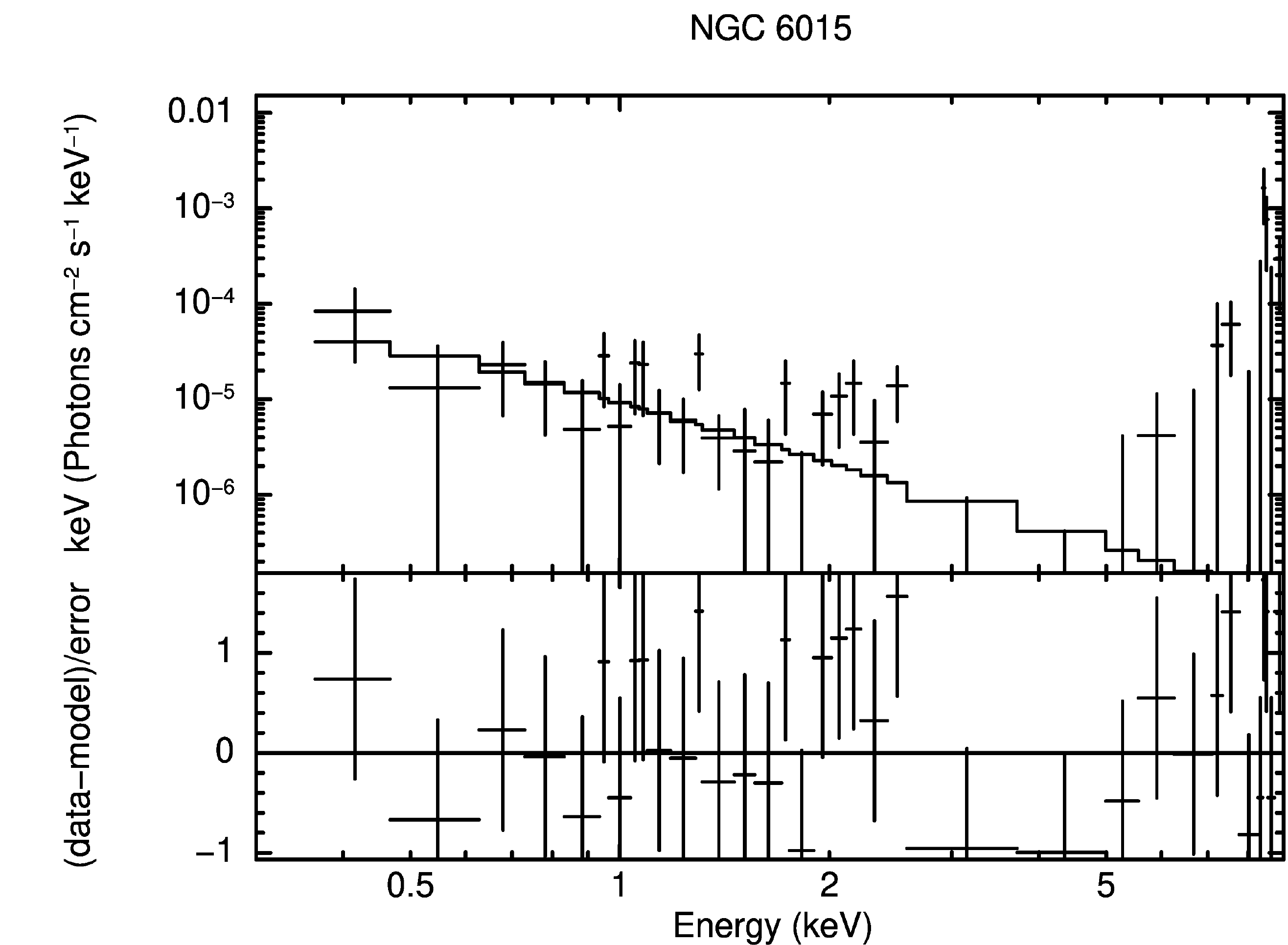}

\end{figure}
\end{center}

\begin{center}
 \begin{figure}
	\includegraphics[width=0.89\columnwidth]{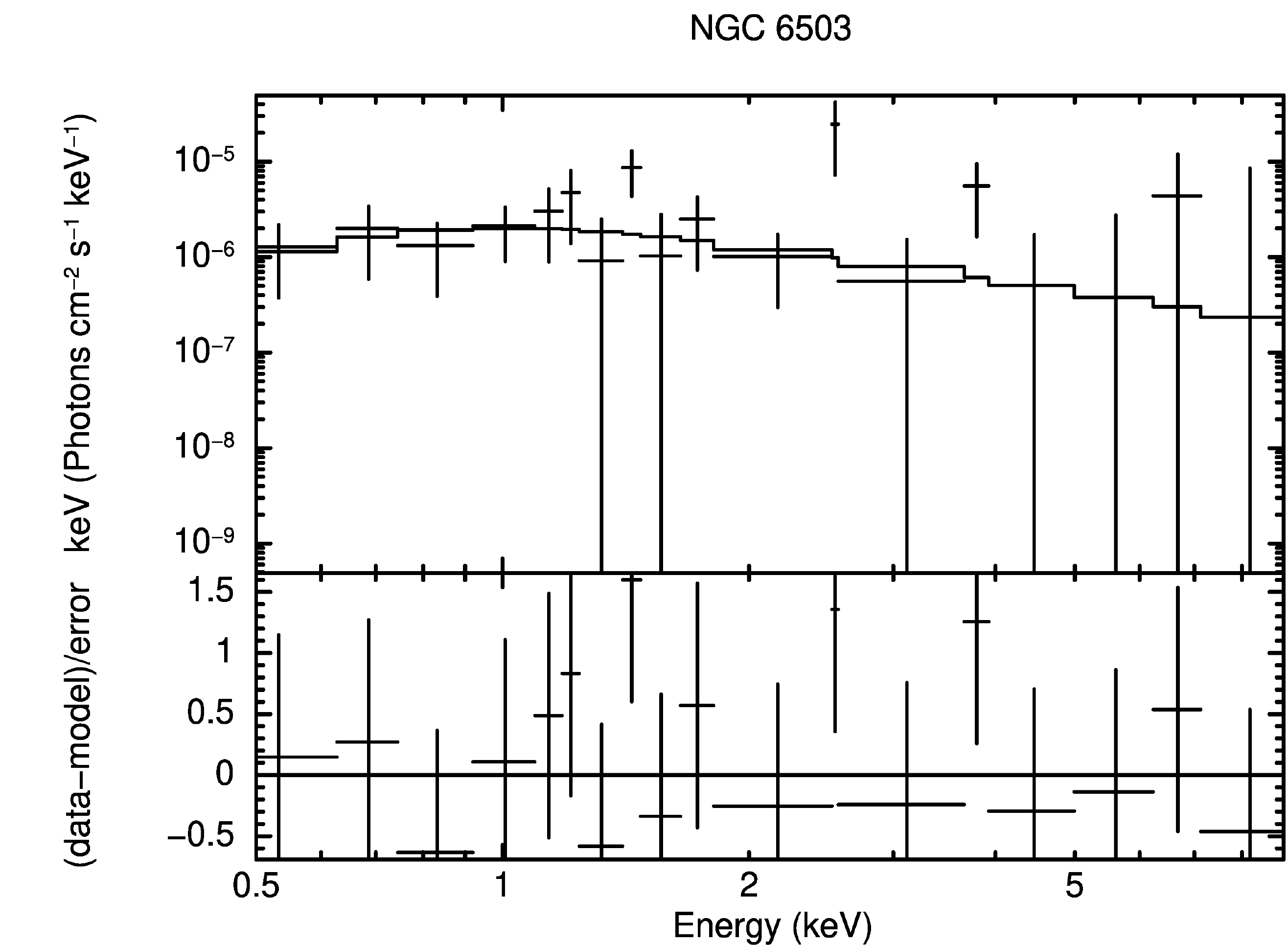}

\end{figure}
\end{center}

\begin{center}
 \begin{figure}
	\includegraphics[width=0.89\columnwidth]{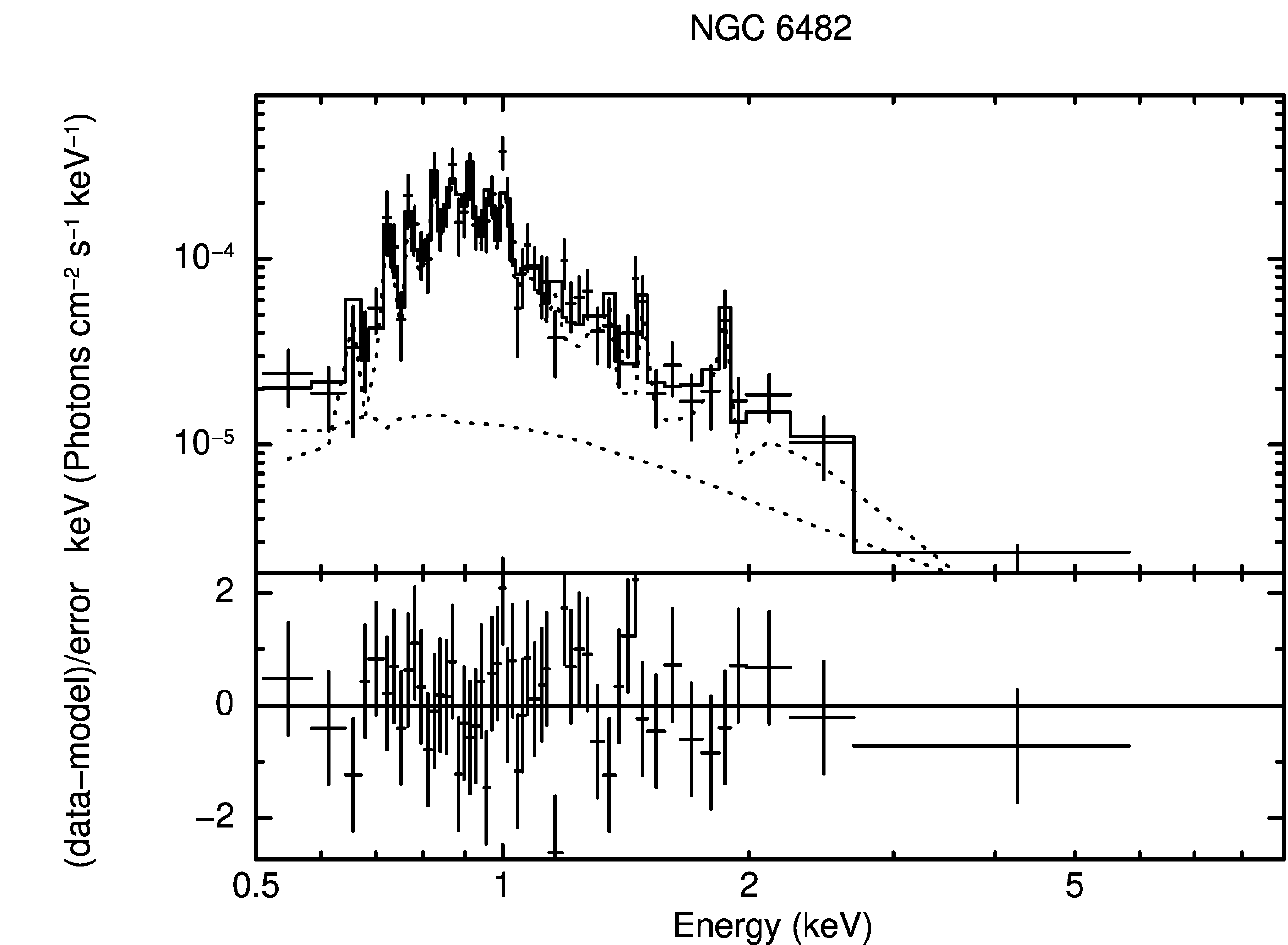}

\end{figure}
\end{center}

\begin{center}
 \begin{figure}
	\includegraphics[width=0.89\columnwidth]{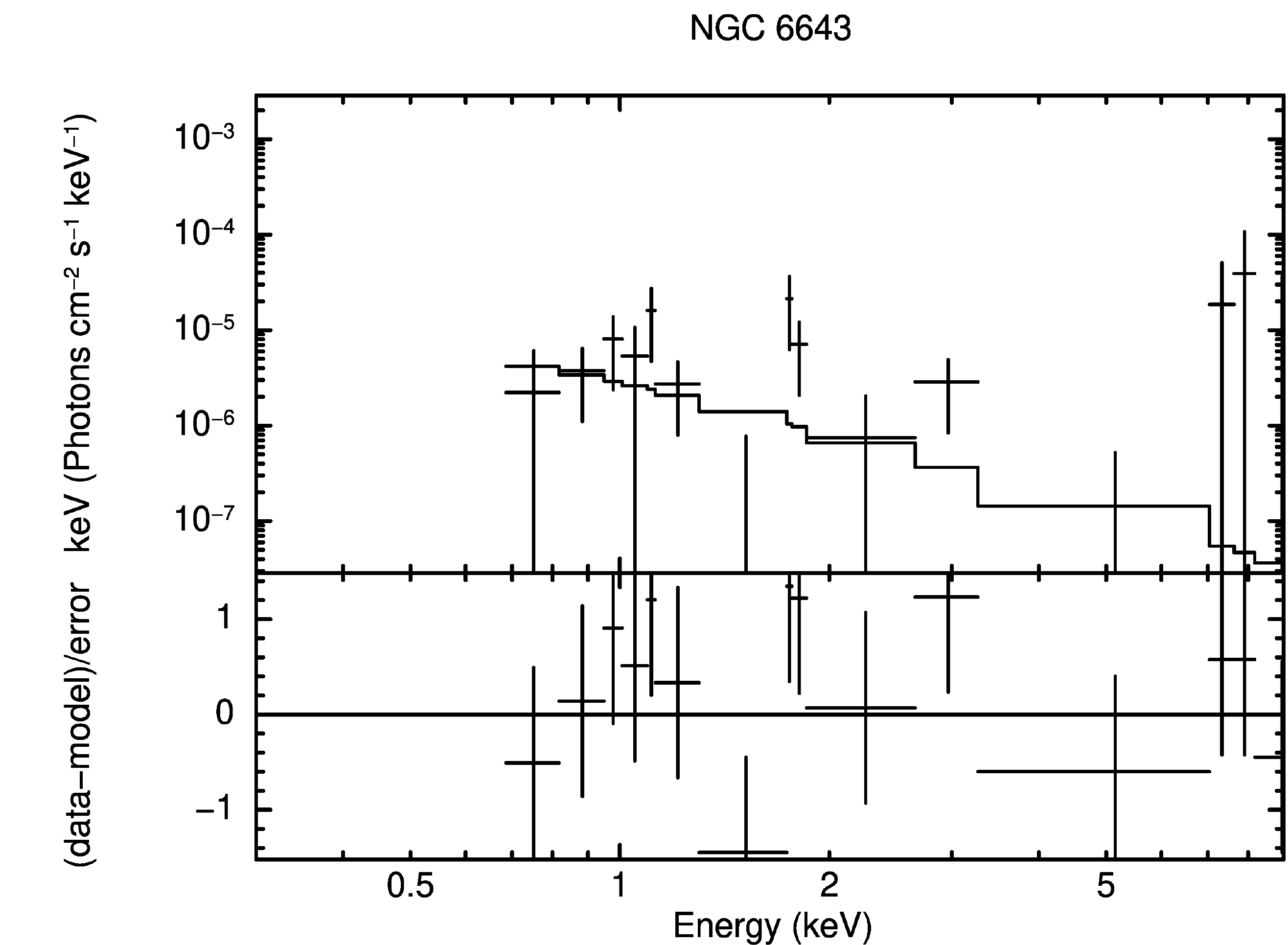}

\end{figure}
\end{center}

\begin{center}
 \begin{figure}
	\includegraphics[width=0.89\columnwidth]{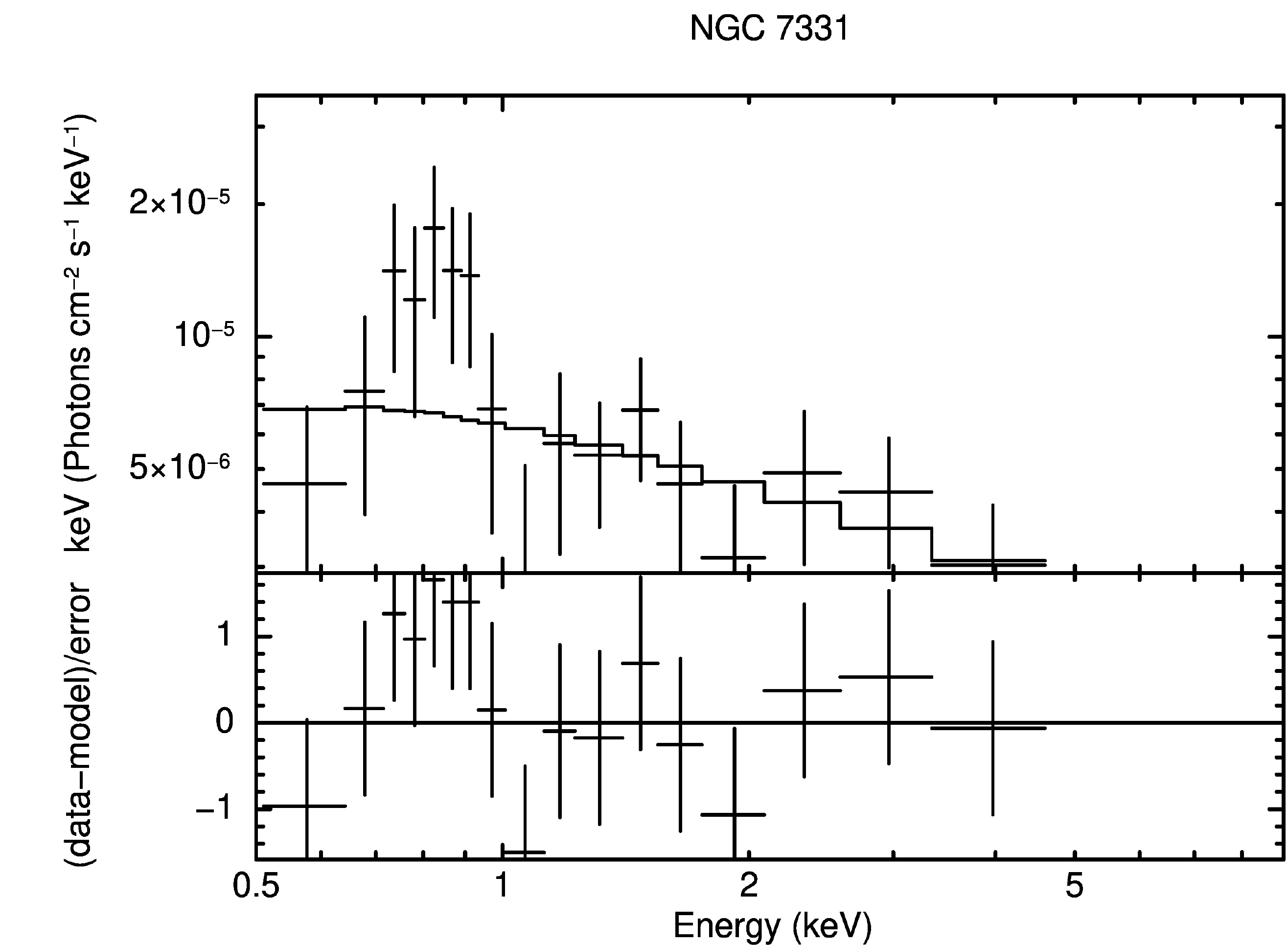}

\end{figure}
\end{center}

\begin{center}
 \begin{figure}
	\includegraphics[width=0.89\columnwidth]{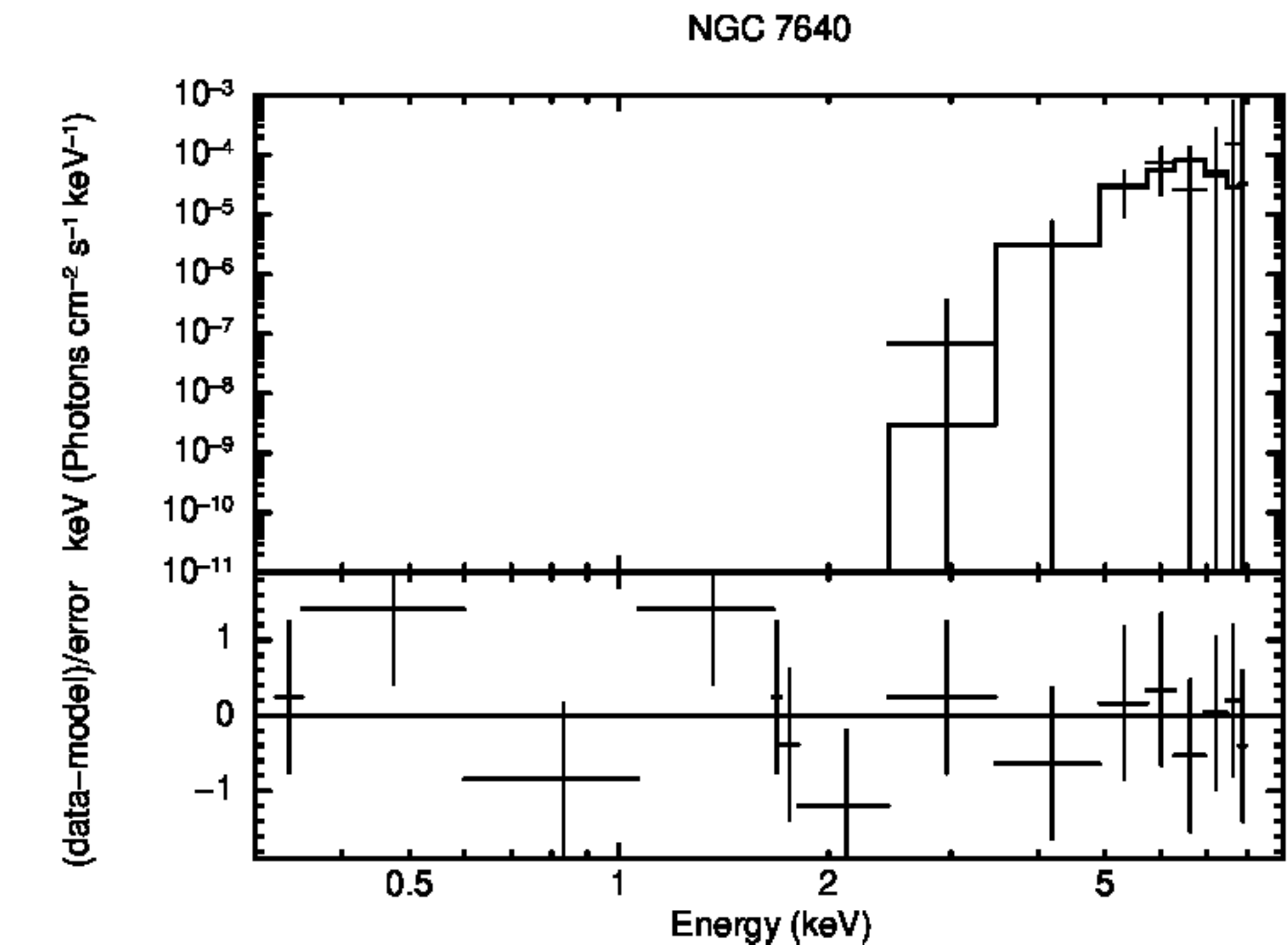}

\end{figure}
\end{center}
\begin{center}
 \begin{figure}
	\includegraphics[width=0.89\columnwidth]{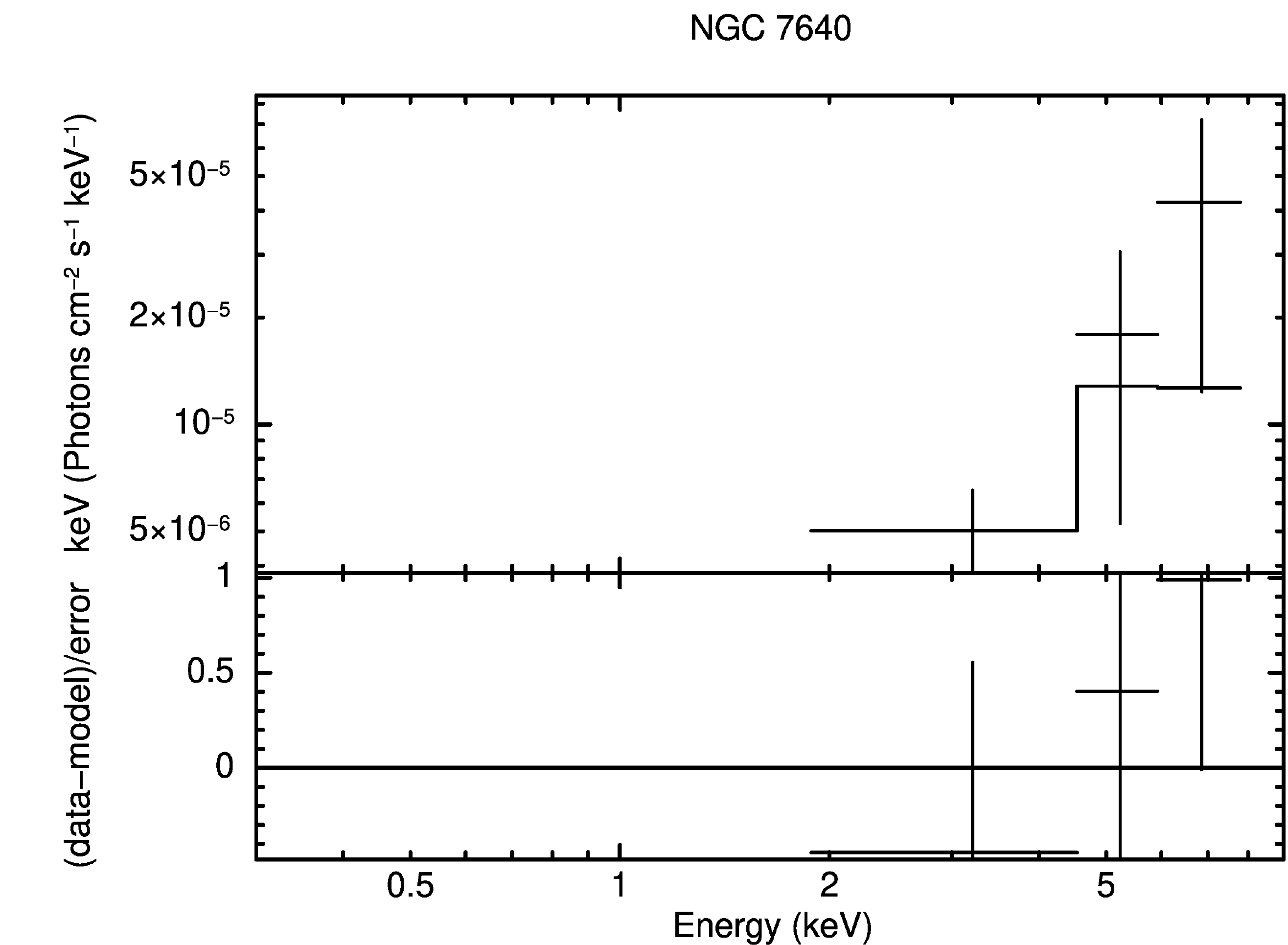}

\end{figure}
\end{center}


\bsp	
\label{lastpage}
\end{document}